\documentclass[11pt,a4paper,oneside]{config/uolthesis}
\usepackage{url}                    
\usepackage{subfigure}
\usepackage{mathrsfs}
\usepackage{graphicx}
\usepackage{epstopdf}

\usepackage{lmodern}
\usepackage[T1]{fontenc}

\input{config/myConfig.cls}

\begin{document}


\title{Binary Black Holes in Modified Gravity}

\author{\textbf{Tiago França}}
\department{
School of Mathematical Sciences} 
\college{Queen Mary, University of London}
\degree{Doctor of Philosophy} \degreemonth{December} \degreeyear{2022}

\copyrightnoticetext{\copyright ~Queen Mary University of London, 2022}


\setcounter{page}{1}
\pagenumbering{roman} 

\maketitle


\blankpage

\thispagestyle{plain}
\begin{declaration}
I, Tiago França, confirm that the research included within this thesis is my own work or that where it has been carried out in collaboration with, or supported by others, that this is duly acknowledged below and my contribution indicated. Previously published material is also acknowledged below.

I attest that I have exercised reasonable care to ensure that the work is original, and does not to the best of my knowledge break any UK law, infringe any third party's copyright or other Intellectual Property Right, or contain any confidential material.

I accept that the College has the right to use plagiarism detection software to check the electronic version of the thesis.

I confirm that this thesis has not been previously submitted for the award of a degree by this or any other university.

The copyright of this thesis rests with the author and no quotation from it or information derived from it may be published without the prior written consent of the author.
\newline
\newline
\newline
Signature: 

Date: 27$^{\text{th}}$ December 2022
\end{declaration}

\thispagestyle{plain}
\begin{listpubs}

Chapter \ref{chapter:paper1} of this dissertation is based on the following co-authored publication:
\begin{itemize}
    \item \cite{Figueras:2020dzx}, P. Figueras and T. França, \emph{Gravitational Collapse in Cubic Horndeski Theories}, \emph{Classical and Quantum Gravity 37 (2020) 225009} \url{https://arxiv.org/abs/2006.094140}.
\end{itemize}
Chapter \ref{chapter:paper2} of this dissertation is based on the following co-authored publication:
\begin{itemize}
    \item \cite{Figueras:2021abd}, P. Figueras and T. França, \emph{Black hole binaries in cubic Horndeski theories}, \emph{Physical Review D 105 (2022) 124004} \url{https://arxiv.org/abs/2112.15529}.
\end{itemize}

Chapter \ref{chapter:eft} is based on on-going work in collaboration with Pau Figueras, Luis Lehner and Ramiro Cayuso:
\begin{itemize}
    \item \cite{Cayuso:2023aht}, R. Cayuso, P. Figueras, T. França and L. Lehner, \emph{Modelling self-consistently beyond General Relativity}, \url{https://arxiv.org/abs/2112.15529}.
\end{itemize}

Furthermore, I am one of the core developers of \texttt{GRChombo}, an open-source numerical relativity code. This resulted in much of the results of chapter \ref{chapter:ahfinder} and appendix \ref{appendix:grchombo}. In particular, sections \ref{appendix:grchombo:ah_location}, \ref{appendix:grchombo:tagging} and \ref{appendix:grchombo:weyl} are derived from the co-authored publications:
\begin{itemize}
    \item \cite{Andrade:2021rbd}, GRChombo Collaboration, \emph{GRChombo: An adaptable numerical relativity code for fundamental physics}, \emph{Journal of Open Source Software 6 (2021) 3703} \url{https://arxiv.org/abs/2201.03458}.
    \item \cite{Radia:2021smk}, M. Radia, U. Sperhake, A. Drew, K. Clough, P. Figueras, E. A. Lim, J. L. Ripley, J. C. Aurrekoetxea, T. França, T. Helfer, \emph{Lessons for adaptive mesh refinement in numerical relativity}, \emph{Classical and Quantum Gravity 39 (2022) 135006} \url{https://arxiv.org/abs/2112.10567}.
\end{itemize}

Finally, I participated in several other projects as co-author, not included in this thesis, namely:
\begin{itemize}
    \item \cite{Figueras:2022zkg}, P. Figueras, T. França, C. Gu and T. Andrade, \emph{The endpoint of the Gregory-Laflamme instability of black strings revisited} \url{https://arxiv.org/abs/2210.13501}.
    \item \cite{boosted_head_ons}, T. Helfer, K. W. K. Wong, M. H. Cheung, E. Berti, J. C. Aurrekoetxea, V. Baibhav, P. Figueras, T. França, C. Gu, E. Lim, M. Radia, J. Ripley, U. Sperhake, \emph{Ultrarelativistic head-on black hole collisions}, \emph{2023, in preparation}.
    \item \cite{deJong:2023}, E. de Jong, J. C. Aurrekoetxea, E. A. Lim, T. França, \emph{Spinning primordial black holes in a matter dominated universe}, \emph{2023, in preparation}.
\end{itemize}

\end{listpubs}

%

\thispagestyle{plain}
\begin{abstract}

General relativity (GR) is the currently accepted classical theory of gravity, standing the test of time for more than 100 years. The recently Nobel-Prize-winning detection of gravitational waves (GWs) opens doors for direct search of new physics, by matching theoretical predictions to experimental data. Numerical relativity attempts to solve in computers strong gravity problems without analytical solutions. In this thesis, we use numerical relativity to investigate gravitational waves from binary black holes in extensions of GR.

We first study spherically symmetric gravitational collapse in cubic Horndeski theories of gravity. By varying the coupling constants and the initial amplitude of the scalar field, we determine the region in the space of couplings and amplitudes for which it is possible to construct global solutions to the Horndeski theories. Furthermore, we identify the regime of validity of effective field theory (EFT) as the sub-region for which a certain weak coupling condition remains small at all times.

We study black hole binary mergers in these cubic Horndeski theories of gravity, treating them fully non-linearly. In the regime of validity of EFT, the mismatch of the gravitational wave strain between Horndeski and GR (coupled to a scalar field) can be larger than $30\%$ in the Advanced LIGO mass range. Initial data and coupling constants are chosen so the theory always remains in the weakly coupled regime. We observe that the waveform in Horndeski theories is shifted by an amount much larger than the smallness parameter that controls initial data. This effect is generic and may be present in other theories of gravity involving higher derivatives.

We explore a higher-order curvature correction of GR. Guided by toy models, we develop systems capable of reproducing the low energy behaviour of many such theories with a fully nonlinear/non-perturbative approach. We evolve binary black holes, observing a shift in phase accumulated over time which is not statistically significant when compared to GR, for the methods and coupling used.

Finally, we present \texttt{AHFinder}, a flexible multi-purpose tool to find apparent horizons in the open-source numerical relativity code \texttt{GRChombo}.

\end{abstract}


%

\thispagestyle{plain}
\begin{acknowledgments}

I will cherish the Ph.D. experience of the past four years forever, dearly miss many moments and cherish many memories. This was possible due to all the people that celebrated with me on top of the mountain or supported me in the valley of despair. It is a good opportunity to recall the words of Antoine de Saint-Exupéry: \textit{``Those who pass by us, do not go alone, and do not leave us alone; they leave a bit of themselves, and take a little of us.''}. This dedication is for me to set in stone some gratitude words towards those people. 

I was given the opportunity to work with Pau Figueras. Besides the obvious - knowledge, patience, guidance, and support - Pau made me feel part of Science. He was a great mentor and I'm thankful for being introduced to his vast network, for being encouraged to work with different groups and projects, and for his kind warnings along the way, steering me in the right direction. I learnt from his example what it means to be successful in theoretical physics, with a great life balance and great academic initiative. Thank you.

I had the pleasure of working closely with amazing collaborators, Ramiro Cayuso and Luis Lehner, who have made invaluable contributions to the work in this thesis. A special thanks to the \texttt{GRChombo} team, who offered a very welcoming environment, great answers to my many questions and many invaluable questions that filled the gaps in my knowledge. I also acknowledge helpful conversations with Áron D. Kovács, Harvey S. Reall, Helvi Witek, Chenxia Gu, Llibert Aresté Saló, Tomas Andrade, Josu C. Aurrekoetxea, Katy Clough, Robin Croft, Eloy de Jong, Thomas Helfer, Cristian Joana, Kacper Kornet, Miren Radia, Justin Ripley, Ulrich Sperhake, Kaze Wong and Timothy Clifton. An enormous appreciation to Miren and Josu, who perhaps without realising were my academic older brothers during these years without whom I would have struggled significantly more and definitely miss some great memories.

Beyond this, the environment in Queen Mary University of London (QMUL) was surprisingly friendly, and I'm grateful to all staff support, way above my expectations. Thanks to the faculty members Shabnam Beheshti and Arick Shao for advice on my yearly progress, to Juan Valiente-Kroon for teaching me in my first year and to Martin Benning for giving me the opportunity to teach machine learning with you. Thanks to the Geometry and Analysis group for your effort in making our group meetings more productive and to all the Ph.D. colleagues who made the roller-coaster more enjoyable.

I could inappropriately fill a couple pages describing the names and deep gratitude I have to all my old friends that kept in contact with me, from school, from uni, from London. I cherish your relentless support, and I hope life will keep us close. A special thanks to Jin, Esma, Lorenzo, Enrico, Alejandro and Rajat, who transformed in one way or another my life in London. I am grateful to my teachers and tutors throughout my education. Thanks to my family, my parents and brother, for the unconditional love and advice. I don't know how you stand me. Whatever you did, you did well. And a warm thanks to the wonderful Ines.

My Ph.D. was supported by a PhD studentship from the Royal Society RS\textbackslash PhD\textbackslash 181177.  The simulations presented were made possible due to the use of the MareNostrum4 cluster at the Barcelona Supercomputing Centre\footnote{Grant No. FI-2020-2-0011, FI-2020-2-0016, FI-2020-3-0007, FI-2020-3-0010, FI-2021-3-0010.}, the GCS Supercomputer SuperMUC-NG at Leibniz Supercomputing Centre (www.lrz.de)\footnote{The resources provided by the Gauss Centre for Supercomputing e.V. (www.gauss-centre.eu) were made available under the PRACE Grant No. 2020235545.}, the Cambridge Service for Data Driven Discovery (CSD3)\footnote{Under projects DP128, DP214, TC011. CSD3 is in part operated by the University of Cambridge Research Computing on behalf of the STFC DiRAC HPC Facility (www.dirac.ac.uk). The DiRAC component of CSD3 was funded by BEIS capital funding via STFC capital grants ST/P002307/1 and ST/R002452/1 and STFC operations grant ST/R00689X/1. DiRAC is part of the National e-Infrastructure.}, the Athena at HPC Midlands+ cluster\footnote{Funded by EPSRC on the grant EP/P020232/1, as part of the HPC Midlands+ consortium.}. Some calculations also made use of the Cosma cluster\footnote{Under project DP092. This work used the DiRAC@Durham facility managed by the Institute for Computational Cosmology on behalf of the STFC DiRAC HPC Facility (www.dirac.ac.uk). The equipment was funded by BEIS capital funding via STFC capital grants ST/P002293/1, ST/R002371/1 and ST/S002502/1, Durham University and STFC operations grant ST/R000832/1.}, the Sulis Tier 2 HPC platform\footnote{Hosted by the Scientific Computing Research Technology Platform at the University of Warwick. Sulis is funded by EPSRC Grant EP/T022108/1 and the HPC Midlands+ consortium.}, Queen Mary's Apocrita HPC facility\footnote{Supported by QMUL ResearchIT \cite{king_thomas_2017_438045}.}, the Young Tier 2 HPC cluster at UCL\footnote{We are grateful to the UK Materials and Molecular Modelling Hub for computational resources, which is partially funded by EPSRC (EP/P020194/1 and EP/T022213/1).}, the GCS Supercomputer JUWELS at the Jülich Supercomputing Centre (JCS)\footnote{Used under PRACE resources with Grant Number 2020225359 through the John von Neumann Institute for Computing (NIC), funded by the Gauss Centre for Supercomputing e.V. (www.gauss-centre.eu).}, the DiRAC Data Intensive service at Leicester, DiAL3 cluster, operated by the University of Leicester IT Services\footnote{Under project DP092. DiAL3 is part of the STFC DiRAC HPC Facility (www.dirac.ac.uk). The equipment was funded by BEIS capital funding via STFC capital grants ST/K000373/1 and ST/R002363/1 and STFC DiRAC Operations grant ST/R001014/1.} and the ARCHER2 UK National Supercomputing Service\footnote{https://www.archer2.ac.uk}.

I would like to acknowledge the support of the ARCHER UK National Supercomputing Service and the Argonne Leadership Computing Facility, which offered me substantial training.

\end{acknowledgments}

\blankpage


{
    \setstretch{1.2} 
    \hypersetup{linkcolor=black}
    \tableofcontents   
}

\blankpage\blankpage

\begin{abbreviations}

\begin{itemize}
    \item GR  - General Relativity
    \item NR  - Numerical Relativity
    \item BH  - Black Hole
    \item BBH - Binary Black Hole
    \item AH  - Apparent Horizon
    \item GW  - Gravitational Wave
    \item WCC - Weak Coupling Conditions
    \item PDE - Partial Differential Equation
    \item ODE - Ordinary Differential Equation
    \item AMR - Adaptive Mesh Refinement
    \item AHFinder - Apparent Horizon Finder
    \item CFL factor - Courant-Friedrichs-Lewy factor
\end{itemize}

\end{abbreviations}

\blankpage
\listoffigures     
\listoftables      
\blankpage
\blankpage

\newpage

\setcounter{page}{1} %
\pagenumbering{arabic}
\pagestyle{fancy}



\part{\done{Background material}}
\blankpage

\chapter{\done{Introduction}}\label{chapter:introduction}

\section{\done{The universe of today}}

Gravity is one of the fundamental forces of nature. From the apple on the tree to the tides on the sea, gravity puts weight on our shoulders, but holds the world together. While holding us down, it has intrigued humans to reach the sky and understand planetary orbits. While making us fall, its mysteries revealed it also warps time and bends light. 

Our understanding of gravity has undergone a revolution in the past century, ever since Albert Einstein proposed the groundbreaking theory of general relativity (GR) in 1915, describing it not as a force but as pure geometry, or curvature of spacetime, influenced by the presence of matter and energy in a four dimensional universe that puts time and space on more equal footing. This idea challenged the traditional Newtonian view of gravity as an instantaneous force acting between masses, in a universe with absolute time and frame of reference. Beautifully, Newton's law of gravitation fits perfectly within the umbrella of GR under the assumptions of small curvatures and velocities, making GR a perfect example of our expanding knowledge of the universe. It showed us a myriad of effects we could have never have predicted: expansion of the universe, existence of black holes and gravitational waves. It has also left us with many puzzles to understand, as dark matter and dark energy, the evolution of the early universe, the cosmological structure formation, the possibility of existence of higher dimensions, and, above all, the non-renormalisability of GR leaves little clues to unify it to quantum field theory for the ultimate understanding of spacetime, quantum gravity.

Nevertheless, this theory has shown predictive power and remarkable accuracy for all measurements and observations made to this day. Such tests entered the new era of gravitational wave astronomy when the LIGO/Virgo/KAGRA collaboration made the first detections of a gravitational waves \cite{Abbott:2016blz,TheLIGOScientific:2017qsa} as predicted long before by GR, which resulted in the Nobel Prize award in 2017. This extreme event was emitted by two black holes 1.4 billion light years away, each with a mass about 30 times bigger than our Sun and yet each with size of only about 50 km, emitting the energy of three solar masses in less than a second, in the form of gravitational waves we detected on Earth. Since then, the collaboration has made detections on a weekly basis \cite{LIGOScientific:2016aoc,LIGOScientific:2016sjg,LIGOScientific:2017vwq}, not only of coalescing black holes, but also binaries involving two neutron starts or a black hole and a neutron star.

Other evidences for the existence of black holes include the galactic X-ray source Cygnus X-1, the monitoring of stellar orbitals around galactic centres \cite{Ghez_1998,Gillessen_2009} and the observation of a black hole shadow by the Event Horizon Telescope \cite{Akiyama_2019}, which resulted in the Nobel Prize award in 2020.

\section{\done{The universe of tomorrow}}

We will soon enter the era of precision gravitational wave astronomy. Due to improvements in sensitivity and number of ground based observatories, the development of third-generation gravitational-wave observatory such as the Einstein Telescope \cite{Maggiore:2019uih}, and the future construction of the Laser Interferometer Space Antenna (LISA) \cite{eLISA:2013xep,LISA:2022kgy}, many more (and more precise) detections will allow us to put stronger constraints on the limits of gravity. The Einstein Telescope is a European project planned to be an underground infrastructure, bigger, more precise and with reduced noise. LISA, expected to launch in 2037, will be a space-based gravitational wave detector, consisting of three spacecrafts separated by 5 million kilometres that act as a gigantic interferometer antenna.

The present data suggests that the corrections to GR are small \cite{Johnson-McDaniel:2021yge}. Therefore, one may hope that there is a better chance to detect some deviations from GR in the strong field regime, where some effects may be enhanced, namely in the final collision of compact objects such as black holes, known as merger phase. The unprecedented amount of data giving us access to the strong regime of gravity\footnote{By the `strong field regime', we mean the regime in which the non-linearities of the theory are important.} offers the opportunity (and carries the duty) to better test GR and observe deviations from it. However, detection of gravitational waves requires matching the data against templates of theoretical predictions of waveforms, without which some deviations from GR may be undetected.

One common approach is to aim to test the self-consistency of GR, by making more (and more accurate) predictions hoping to find hints of failure that can guide us in a new direction. Though promising, one big challenge in the development of this approach is the vast dimension of the parameter space required to understand the gravitational wave landscape: mass ratio and total mass, relative velocities and separation, spin orientations, position and orientation relative to detectors, and equation of state if involving neutron stars.

Alternatively, one can take the effective field theory approach, by looking at GR as an effective theory of some more fundamental theory, valid up to some finite energy scale. We can then attempt to describe the low energy limit of such fundamental theory to probe emerging phenomena in the regime of high curvature, namely around black holes and compact binaries. In this approach, it is equally or more difficult to come up with templates of waveforms. The issue is we have no single preferred theory extending GR and, in fact, many (many!) have been proposed (see section \ref{sec:gr:modified_gravity} for details). Which theories should be picked and how should we simulate them?

Given some interesting theory, one possibility is to focus solely on those phases of the binary that can be treated using perturbation theory, namely the inspiral \cite{Berti:2018cxi} and the ringdown phases \cite{Berti:2018vdi,Cano:2020cao} respectively. So far the merger phase has been modelled phenomenologically \cite{Yunes:2009ke,Agathos:2013upa}, or by treating the deviations from GR perturbatively \cite{Okounkova:2017yby,Witek:2018dmd,Okounkova:2018pql,Okounkova:2019dfo,Okounkova:2020rqw,Okounkova:2019zjf}. Treating the modifications to Einstein’s gravity perturbatively may seem justified given that the present data indicates that they are small. In this case, there are no issues with the well-posedness of the equations and this is the approach that has been adopted in a number of papers \cite{Okounkova:2017yby,Witek:2018dmd,Okounkova:2018pql,Okounkova:2019dfo,Okounkova:2020rqw,Okounkova:2019zjf,Cano:2020cao,deRham:2020ejn}. However, it has some serious limitations: it is well-known that small effects can accumulate over time due to secular effects\footnote{There are recent interesting attempts to re-sum the perturbative series and hence alleviate these secular effects \cite{GalvezGhersi:2021sxs}.} and eventually lead to a breakdown of perturbation theory in a regime where it should still be valid. Furthermore, this approach is completely insensitive to certain non-perturbative effects encoded in the full non-linear theory which, even if very small, may be detectable over sufficiently many orbits of a binary. For instance, the non-linear perturbation theory around anti-de Sitter space breaks down precisely before a black hole forms \cite{Bizon:2011gg}.

Alternatively, one can demand that the theory satisfies some mathematical consistency requirements, namely having a well-posed initial value problem. This is not an issue if one is considering weak corrections to GR. Up until recently, only the so called scalar-tensor and scalar-vector-tensor theories of gravity have been considered in their full non-linear glory in all phases of the binary \cite{Healy:2011ef,Barausse:2012da,Hirschmann:2017psw,Sagunski:2017nzb}. In recent years, significant progress has been made in uncovering which theories can be formulated in a well-posed manner \cite{Kovacs:2019jqj,Kovacs:2020pns,Kovacs:2020ywu}, as well as studies of the strong field dynamics of certain alternative theories of interest using this approach \cite{Bezares:2021dma,East:2020hgw,East:2021bqk,East:2022rqi,Corman:2022xqg,AresteSalo:2022hua}. It is of interest to treat alternative theories of gravity fully non-linearly and uncover some of their (perhaps) unique physical effects that may break certain degeneracies.

In this thesis, we aim to study gravitational waves in interesting 
 and viable alternative theories of gravity, and compare them to GR, in order to probe generic deviations of GR one may expect to encounter. By interesting, we mean theories that extend GR non-trivially offering new phenomenology that satisfies the weak field assumptions compatible with small deviations. By viable, we mean theories where a formulation exists where the theory can be simulated fully non-linearly.

\section{\done{Outline of the thesis}}

This manuscript is split in three main parts: background material, research work and extra material.

The background material comprises chapters \ref{chapter:introduction}-\ref{chapter:ngr_overview} and provides an overview of the relevant material for the research work presented. Chapter \ref{chapter:introduction} lays out the motivation for the rest of the thesis. Chapter \ref{chapter:gr_overview} introduces general relativity and conventions used, as well as background concepts and recent developments required for the discussion that follows, namely scalar fields, modified gravity (Horndeski theories in particular), well-posedness and procedures to deal with higher derivative theories numerically. Chapter \ref{chapter:ngr_overview} reviews historical numerical evolution schemes and presents \texttt{GRChombo}, the numerical code used to perform the main simulations in this thesis. More details of such schemes are shown in appendix \ref{appendix:ngr_formulations}.

The research work in chapters \ref{chapter:paper1}-\ref{chapter:eft} constitutes the core of the thesis. In chapter \ref{chapter:paper1}, accompanied by appendix \ref{appendix:paper1}, we study gravitational collapse in cubic Horndeski theories in order to identify its regime of validity from the point of view of effective field theory. Chapter \ref{chapter:paper2}, accompanied by appendix \ref{appendix:paper2}, builds on this by exploring binary black holes in the regime of validity of such theories, with the goal of comparing them to GR. In chapter \ref{chapter:eft}, accompanied by appendix \ref{appendix:eft}, we analyse an extension of GR involving higher-order curvature corrections. We describe methods that allow a stable evolution of such theories and present results on binary black hole waveforms. In chapter \ref{chapter:ahfinder}, we present \texttt{AHFinder}, a flexible tool to find apparent horizons in multiple scenarios, integrated in the \texttt{GRChombo} open-source code.

We summarise our conclusions and propose future directions in chapter \ref{chapter:conclusions}. Finally, appendix \ref{appendix:grchombo} presents code development work and technical discussions relating all the simulations performed.

\chapter{\done{Overview of general relativity and modified gravity}}\label{chapter:gr_overview}

The goal of this background material chapter is to introduce some core ideas fundamental to the development of this thesis and which set the basis for the work done. Their importance lies on communicating conventions used and reviewing relevant literature work in the fields of modified gravity and effective field theory. Basic knowledge of general relativity is assumed\footnote{See details in Reall \cite{Reall:GR:Online}, Caroll \cite{Carroll:2004st} or d'Inverno \cite{dInverno:1992gxs}. Also a useful intuition behind dimensions was made by Volovik \cite{Volovik:2020rjz} and Mana \cite{Mana:2020bzs}.}, though a summary of definitions used throughout the thesis is laid out in \ref{gr_summary}.

\section{\done{Notation and conventions}}

We adopt notation as follows unless stated otherwise. Spacetime has $D=d+1$ dimensions, fixed to $d=3$, $D=4$ where explicitly mentioned. Greek letters ($\a$, $\b$, $\m$, $\n$, $\dots$) denote full spacetime indices, ranging from $0$ to $D$, and Latin letters ($i$, $j$, $k$, $l$, $\dots$) denote purely spatial ones, ranging from $1$ to $D$. We adopt the mostly plus metric signature ($-,+,+,+$) and we use geometric units, with $(G=c=1)$, where $G$ is Newton's gravitational constant and $c$ is the speed of light. This implies that length, time, mass and energy all have the same dimension\footnote{For conversion to conventional SI units, see Wald \cite[p.~471]{Wald:1984rg}.}. All these are expressed in a geometrised unit mass $M$, which can be freely chosen. 

Partial derivatives with respect to any variable $x^\mu$, $\frac{\partial}{\partial x^\mu}(\cdot)$, are written as $\partial_\mu(\cdot)$. The symbols $(\dots)$ and $[\dots]$ around tensor indices (e.g. $W_{[ab]}$) denote, respectively, total symmetrisation and antisymmetrisation of the indices within them. We use Einstein's summation convention, where repeated indices are summed over all their possible values (i.e. $U^\a V_\a$ means $\sum_{\a=0}^{D}U^\a V_\a$).

\newpage
\begin{summary}{\done{Key concepts of general relativity}}\label{gr_summary}
The key formal ideas that motivated GR were: the \textit{Principle of General Covariance}, the \textit{Principle of Equivalence} and \textit{Mach's Principle}. \textit{Spacetime curvature} is described by a symmetric $\binom 02$ tensor, the \textit{metric} $g_{\mu\nu}$, from which \textit{proper distance} $ds^2$ (invariant interval between infinitesimally close points separated by $\{dx^\alpha\}$) and \textit{proper time} $d\tau$ (time experienced by massive objects in their reference frame) can be calculated:
\vspace{-7pt}
\begin{equation}
    ds^2 = d\tau^2 = -g_{\mu\nu}dx^\mu dx^\nu\,.\vspace{-7pt}
\end{equation}
Due to curvature, basic \textit{vectors} and \textit{co-vectors} change between points. The notion of derivative of a $\binom mn$ tensor $T$ is replaced by \textit{covariant derivative}, $\grad$, using the \textit{Christoffel symbols}\footnote{Not a tensor.}, $\Gamma^\rho_{~\mu\nu}$:
\begin{align}
    \label{summ:covder}\nonumber
    \grad_\alpha T^{\mu_1\dots\mu_m}_{~~~~~~~~\nu_1\dots\nu_n} \equiv
    &~ T^{\mu_1\dots\mu_m}_{~~~~~~~~\nu_1\dots\nu_n~,\alpha} +\Gamma^{\mu_1}_{~~~\alpha\beta}~T^{\beta\mu_2\dots\mu_m}_{~~~~~~~~~\nu_1\dots\nu_n} + \dots + \Gamma^{\mu_m}_{~~~\alpha\beta}~T^{\mu_1\dots\mu_{m-1}\beta}_{~~~~~~~~~~~~\nu_1\dots\nu_n}-\\
    &- \Gamma^{\beta}_{~~\alpha\nu_1}~T^{\mu_1\dots\mu_m}_{~~~~~~~~\beta\nu_2\dots\nu_n} - \dots - \Gamma^{\beta}_{~~\alpha\nu_n}~T^{\mu_1\dots\mu_m}_{~~~~~~~~\nu_1\dots\nu_{n-1}\beta}\,,\\
    \Gamma^\rho_{~\mu\nu} \equiv&~ \tfrac{1}{2}g^{\rho\sigma}\br{g_{\mu\sigma,\nu}+g_{\sigma\nu,\mu}-g_{\mu\nu,\sigma}}\,.
\end{align}
Free massive particles\footnote{Characterised by \textit{timelike separation}: $ds^2<0$. For massless objects, travelling at the speed of light, $ds^2=0=d\tau^2$, one can write $u^\m = \tfrac{dx^\m}{d\l}$, with $\lambda$ an affine parameterisation of \textit{null geodesics}.}, at coordinates $\{x^\mu\}$ with \textit{spacetime velocity} $u^\mu = \frac{dx^\mu}{d\tau}$, follow \textit{timelike geodesics}, the equivalent of ``straight lines'' in curved spacetime, according to:
\begin{equation}\label{eq:geodesic}
    \frac{d^2x^\mu}{d\tau^2} + \Gamma^\mu_{~\rho\sigma}~\frac{dx^\rho}{d\tau}~\frac{dx^\sigma}{d\tau}=0 ~\Longleftrightarrow~ u^\nu\grad_\nu u^\mu = 0\,.
\end{equation}
To analyse curvature, one defines the \textit{Riemman curvature tensor}, $\mathcal{R}^\alpha_{~\beta\mu \nu }$, the \textit{Ricci tensor}, $\mathcal{R}_{\mu \nu }$, and \textit{Ricci scalar}, $\mathcal{R}$, as follows. The symmetries of the Riemann tensor are as in \eqref{RiemmanSyms}.
\vspace{-3pt}
\begin{align}
    \mathcal{R}^\alpha_{~\beta\mu\nu} & \equiv \partial_\mu \Gamma^\alpha_{~\beta\nu} - \partial_\nu \Gamma^\alpha_{~\beta\mu} + \Gamma^\alpha_{~\rho\mu} \Gamma^\rho_{~\beta\nu} - \Gamma^\alpha_{~\rho\nu} \Gamma^\rho_{~\beta\mu}\,,\\
    \mathcal{R}_{\mu\nu} & \equiv \mathcal{R}^\lambda_{~\mu\lambda\nu}\,,\\\label{RicciScalar}
    \mathcal{R} & \equiv \mathcal{R}^\mu_{~\mu}\,,\\\label{RiemmanSyms}
    \mathcal{R}_{\alpha[\beta\mu\nu]} &= 0\,,\qquad \mathcal{R}_{\alpha\beta\mu\nu}=-\mathcal{R}_{\beta\alpha\mu\nu}=-\mathcal{R}_{\alpha\beta\nu\mu}=\mathcal{R}_{\mu\nu\alpha\beta}\,,\\
    \grad_\m\grad_\n T^\a &= \grad_\n\grad_\m T^\a + R^\a_{~\b\m\n}T^\b\,.
\end{align}
\vspace{-3pt}Geometry of spacetime, described by the \textit{Einstein tensor}, $G_{\mu \nu }$, and distribution of matter, described by the \textit{energy-momentum tensor}, $T_{\mu \nu }$, are related by \textit{Einstein's field equations}:
\vspace{-10pt}
\begin{gather}\label{eq:einstein}
    G_{\mu\nu} \equiv \mathcal{R}_{\mu\nu} - \tfrac{1}{2}g_{\mu\nu}\mathcal{R} = \tfrac{\k}{2} ~T_{\mu\nu} - \Lambda g_{\mu\nu}\,,
\end{gather}
\vspace{-5pt}
where $\Lambda$ is the \textit{cosmological constant} and $\k=16\p G$. The \textit{Bianchi identities/contracted Bianchi identities} are:
\vspace{-10pt}
\begin{equation}\label{eq:bianchi}
    \mathcal{R}_{\alpha\beta[\mu\nu;\lambda]} = 0 ~\Longrightarrow~\grad^\m G_{\mu\nu} =0=\grad^\m T_{\mu\nu}\,.
\end{equation}
\end{summary}

\section{\done{Black holes and gravitational waves}}\label{sec:gr:bh_and_gws}

In this section follows a one page overview of BHs and GWs. See Reall \cite{Reall:BH:Online} and Centrella and Baker \cite{Centrella:2010mx} for a review.

The trivial solution to Einstein's equations and where special relativity `lives' is not the familiar geometry of Euclidean space. The flat spacetime metric solution can be written as $\eta_{\m\n} = \text{diag}(-1,1,1,1)$, and such spacetime is called Minkowski spacetime.

In vacuum, without any matter, i.e. $T_{\m\n}=0=G_{\m\n}$, the most fundamental solutions of Einstein's equations are black holes, regions of ``no escape'' delimited by an horizon (see details about horizons in \ref{sec:ahfinder:horizons}). Surprisingly, black hole solutions typically characterise fully empty spacetimes, apart from an infinitely dense point or ring at the centre. The simplest black hole is the Schwarzschild black hole, describing a static and spherically symmetric spacetime. It can be written as the metric:
\begin{equation}
    ds^2 = -\br{1-\frac{2M}{r}} dt^2 + \br{1 - \frac{2M}{r}}^{-1} dr^2 + r^2\br{d\q^2 + \sin^2\q d\f^2}\,,
\end{equation}
where $M$ is the mass of the black hole. The Schwarzschild solution also describes the exterior of stars. In the limit of $M\to0$ or $r\to\infty$, Minkowski spacetime is recovered.

On the other hand, consider perturbations of flat space, taking some metric:
\begin{equation}
    g_{\m\n} = \eta_{\m\n} + h_{\m\n}\,,
\end{equation}
where $h_{\m\n}$ is a small perturbation, $|h_{\m\n}|\ll1$. Linearising Einstein equations for $h_{\m\n}$ yields a wave equation:
\begin{equation}\label{eq:GWs_wave}
    \square \bar{h}_{\m\n} = -\k T_{\m\n}\,,
\end{equation}
where $\bar{h}_{\m\n}=h_{\m\n} - \frac{1}{2}\br{\eta^{\a\b}h_{\a\b}}\eta_{\m\n}$ and where we use the Lorentz gauge, $\pd_\n\bar{h}^{\n\m}=0$. This has two fascinating consequences. The first one is that for small velocities, $\square \approx \grad^2$ and $T^{00}\approx \r$, where $\r$ is the energy density. This implies that:
\begin{equation}
    \grad^2 \bar{h}^{00} = -\k \r\,,
\end{equation}
which is precisely the same as Newton's law of gravitation $\grad^2\f = 4\p G \r$ if one takes $h^{00}=h^{ii}=-2\f$. This is a brief description revealing how GR allows the recovery of Newton's law of gravitation under the assumption of weak gravitational force and velocities.

Moreover, the wave equation \eqref{eq:GWs_wave} indicates perturbations of flat space propagate as waves, and is the basis for the prediction of gravitational waves, \textit{ripples in the fabric of space and time}.

\section{\done{Scalar fields}}

Scalar fields describe concepts like temperature, a concept with a number assigned to each point in space. Fundamental fields are not macroscopic properties such as temperature, but are seen as elemental parts of nature, such as the electron. The only currently empirically verified fundamental scalar field in nature is the Higgs boson \cite{PhysRevLett.13.508}. Other fundamental scalar fields are hypothesised to exist, such as the inflaton, a scalar field candidate to extend the Big Bang theory and explain problems such as the homogeneity and isotropy of the CMB, Cosmic Microwave Background. Beyond this, scalar fields can be useful as effective theories, describing the low energy behaviour of other fundamental theories or representations of higher dimensions, or as toy models to understand phenomena in complex systems, such as critical collapse in strong gravity settings. Classical scalar fields are an approximation to the fundamental description via QFT, Quantum Field Theory, in a regime where occupation numbers are high and wavelengths are larger than Compton wavelengths of the quanta of the field. This does not allow analysing quantum behaviours such as quantum tunnelling, but allows to study effects of gravity over large scales.

The evolution of a classical scalar field, $\f$ with some potential $V(\f)$ is modelled via the action:
\begin{equation}
    \mathcal{S} := \int d^4x \sqrt{-g} \br{X - V(\f)}\,,
\end{equation}
where $X = -\frac{1}{2}\grad_\m\f\grad^\m\f$ is the kinetic term of the scalar field. This leads to the equations of motion:
\begin{equation}
    \square \f = \pd_\f V\,.
\end{equation}
For a potential $V(\f)=\frac{1}{2}m^2\f^2$, this is the well-known Klein Gordon equation where $m$ is the mass of the field. Without extra interactions, the field has a tendency to follow the gradient of the potential towards its minima.

\subsection{\done{Critical collapse}}\label{subsec:gr:critical_collapse}

Gravitational collapse of a scalar field minimally couple to GR is an area with very active interest. It was first studied analytically by Christodoulou \cite{Christodoulou:1993uv,Christodoulou:1987vv,Christodoulou:1991yfa} and numerically by Goldwirth and Piran \cite{Goldwirth:1987nu}. Soon after, Choptuik \cite{Choptuik:1992jv} discovered a critical phenomena associated with gravitational collapse, briefly described below.

Start from a gravity environment where we place a bubble of scalar field of some shape. This scalar field, if coupled to gravity, will invariably follow gravitational collapse by its own attraction. If the profile of such bubble is faint and disperse, this collapse will propagate just as a typical wave in a wave equation, rebound and disperse to infinity. However, if sufficiently dense, it is understandable that under the action of gravity it can be dense enough to distort spacetime and form a black hole\footnote{It may be useful for the reader to see these effects with the visual animation of our simulations of such effects: \href{https://www.youtube.com/watch?v=cfXF1wIcIJc}{https://www.youtube.com/watch?v=cfXF1wIcIJc}.}. These two behaviours are generic for any configuration.

What Choptuik found was a critical point at which the transition between the two states occur, the critical collapse. Interestingly, for any family of configurations parameterised by one parameter $p$ (e.g. amplitude of the profile, width, etc.), the approach to the critical point $p^*$ reveals a universal scaling:
\begin{equation}
    M \propto |p - p^*|^{\g_\f}\,,
\end{equation}
where $M$ is the mass of the black hole formed close to the critical point. The constant $\g_\f\approx 0.37$ is universal for any one parameter families of minimally coupled real scalar fields (it is different for instance for vector fields). Among other interesting phenomena such as discrete self-similarity in the solutions, the universality of this behaviour suggest the generic appearance of black holes with mass approaching zero. This leads to the possibility of creation of naked singularities that violate the cosmic censorship conjecture, later proved false for some classes of initial data \cite{Christodoulou:1999ve}. For comprehensive reviews and recent developments, see \cite{Gundlach:2007gc,Jimenez-Vazquez:2022fix,Wang:2001xba,Veronese:2022fnv}. 

\section{\done{Modified gravity overview}}\label{sec:gr:modified_gravity}

As of the writing of this thesis, gravitational wave observations of BBHs and neutron stars by the LIGO/Virgo/KAGRA collaboration
have been consistent with the predictions of GR \cite{LIGOScientific:2016aoc,LIGOScientific:2016sjg,LIGOScientific:2017vwq}. Yet, there has been an extensive effort by many researchers to model the dynamics of binary black holes and neutron stars in modified (non-GR) theories of gravity.

\begin{figure}[h]
\centering
\hspace{-30mm}\includegraphics[width=1.2\textwidth]{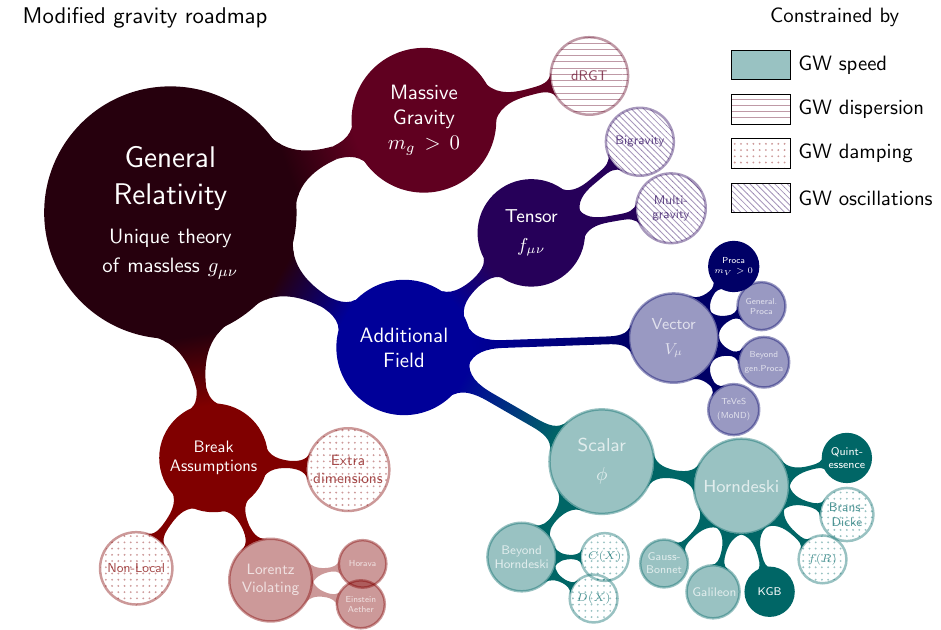}
\caption{Modified gravity roadmap summarising possible extensions of GR, highlighting constraints obtained from the speed, dispersion, damping and oscillations of gravitational waves. Figure taken from Ezquiaga and Zumalac\'arregui \cite{Ezquiaga:2018btd}.}
\label{fig:gr:modified_gr_roadmap}
\end{figure}

There are many topics in fundamental physics beyond the current standard models of gravity, cosmology and particle physics being researched to sharpen our chances of discovering new physics. Research topics range from probing for new physics (dark matter, dark energy, screening models, primordial black holes, echoes, speed of light, parity violations, etc.) to improving methodologies (better models, control of systematics, model-independent and dependent tests, etc.) \cite{LISA:2022kgy}. On the front of new and better models of gravity, extending general relativity has resulted in a diversity hard to keep up with. A schematic attempting to categorise this is shown in figure \ref{fig:gr:modified_gr_roadmap}. Following \cite{CANTATA:2021ktz}, the distinctions can be broken down in the following main categories:
\begin{itemize}
    \item Adding new fields: often considered the easiest approach, adding a new field often results in simple models. Some of these theories can be considered simply exotic matter and not modified gravity, as detailed further in section \ref{subsec:Horndeski}. This class of theories can be broken down further:
    \begin{itemize}
        \item Scalar fields: this includes both real and complex scalar fields, and some examples include Horndeski gravity, Beyond Horndeski gravity or Cherns-Simons gravity.;
        \item Vector fields: such as Einstein-Aether gravity or generalised Proca gravity;
        \item Tensor fields: such as massive gravity, bigravity and bimetric MOND (Modified Newtonian Dynamics) gravity.
    \end{itemize}
    \item Adding invariants: these include breaking certain assumptions like adding extra dimensions, such as Lovelock theories (which are the diffeomorphism invariant generalisations of the Hilbert-Einstein action for higher dimensions), or adding higher order terms to the action, such as $f(R)$ gravity, some of which break locality.
    \item Changing geometry: examples of non-Riemannian geometries are Finsler geometry, non-commutativity gravity and teleparallel theories. It is often debated whether problems such as dark matter will be solved by adding particles and fields or by modifying geometry \cite{Martens:2020lto}.
    \item Quantising gravity: quantum arguments can be used to extend gravity, mainly using string theory. Other examples are Loop quantum gravity, causal sets, Horava-Lifshitz gravity or spin foams.
\end{itemize}

Many of these extensions or modifications of gravity should be seen as effective field theories, truncations of the full UV-complete quantum gravity theory we aspire to find, with the hope of finding constraints to each different type of parameterisation and the trust that higher derivative operators are suppressed for low energy physics. There are many more extensions or modifications of gravity not mentioned above and at the moment there is no theoretical consensus or experimental evidence that favours a particular theory. The hope is that each modification of GR should be reflected in a unique way in the corresponding waveform to allow for exploration in analysis of gravitational waves. However, in many of these theories it is not known whether the initial value problem is well-posed. Without a well-posed initial value problem, one cannot possibly simulate the non-linear regime of the theory on a computer and obtain the desired waveforms. 

In the next sections, we will describe the idea of well-posedness in more detail and two theories of gravity that can be written in a well-posed formulation.

\section{\done{Well-posedness and hyperbolicity}}\label{subsec:gr:well_posedness}

Following Alcubierre \cite{alcubierre}, in classical physical theories, the behaviour of a system is formulated in terms of functions and fields governed by a system of differential equations. General solutions to these equations depend on free parameters or free functions, which are fixed by imposing initial conditions and boundary conditions. Setting up appropriate initial conditions allows for a deterministic evolution. This is called an initial value problem or Cauchy problem.

Mathematical stability and consistency ultimately demands from these systems a set of hyperbolic partial differential equations and well-posedness. These conditions arise from the expectation that solutions are unique, but also that small perturbations on the initial data lead to small perturbations in the solution at later times. Considering these expectations, a system is well-posed if (i) there are unique solutions and (ii) the solutions depend continuously on the initial data, or more strictly, solutions do not increase more rapidly than exponentially. For some defined norm $||\cdot||$, the solution vector $\textbf{u}$ must satisfy:
\begin{gather}
    ||\textbf{u}(t,x^i)|| \le k\,e^{\a\,t}||\textbf{u}(0,x^i)||\,,
\end{gather}
where $k$, $\a$ are constants independent of the initial data. Often for generic initial data one can only prove local well-posedness, which means existence, uniqueness and continuous dependence on initial data for only a finite non-zero amount of time.

\subsection{\done{Notions of hyperbolicity}}

When considering small perturbations, any $n$-dimensional partial differential equation system can be written in first order form as:
\begin{gather}
    \pd_t \textbf{u} + \textbf{M}^j \pd_j \textbf{u} = \textbf{S}(\textbf{u})\,,
\end{gather}
where \textbf{S} is a source vector dependent on \textbf{u} but not its derivatives, $\textbf{M}^i$ are matrices, one for each direction $i$. Second order PDEs can also be considered \cite{Kovacs:2021vdk}, and for fully non-linear PDEs the concepts of hyperbolicity often gets blurry. Note that most systems can be written like this if one considers re-writing higher derivatives as first derivatives of new auxiliary variables. The vector \textbf{u} should be seen as an ordered list of evolution variables without any geometric meaning.

Focusing on the principal part of the equation (terms involving the highest degree in derivatives of \textbf{u}, as these are the ones that involve higher frequencies that enter the analysis of hyperbolicity), one can ignore the source term as if $\textbf{S}=0$. Consider an arbitrary unit vector $n^i$ and construct the principal symbol or characteristic matrix $\textbf{P}(n_i) = M^i n_i$. From the decomposition of this matrix into its eigenvectors and eigenvalues, one can classify the system as:
\begin{itemize}
    \item Weakly hyperbolic, if \textbf{P} has real eigenvalues for any $n_i$, but not a complete set of eigenvectors,
    \item Strongly hyperbolic, if \textbf{P} has real eigenvalues and a complete set of eigenvectors for all $n_i$. If this is the case, then one can diagonalise \textbf{P} into a diagonal matrix \textbf{$\L$} (with the eigenvalues $\l$ in the diagonal) using the matrix of column eigenvectors:
    \begin{equation}
        \mathbf{\L} = \textbf{R}^{-1}\textbf{P}\textbf{R}\,.
    \end{equation}
    One can then define an positive definite matrix $\textbf{H} = (\textbf{R}^{-1})^T \textbf{R}^{-1}$, that satisfies:
    \begin{equation}
        \textbf{H}\textbf{P} - \textbf{P}^T \textbf{H}^T = (\textbf{R}^{-1})^T \mathbf{\L} \textbf{R}^{-1} - (\textbf{R}^{T})^{-1} \mathbf{\L} \textbf{R}^{-1} = 0\,.
    \end{equation}
    This can be used to define the \textit{energy norm} of \textbf{u} and its adjoint (complex-conjugate transpose):
    \begin{equation}
        ||\textbf{u}||^2 = \textbf{u}^\dagger \textbf{H} \textbf{u}\,.
    \end{equation}
    Using a Fourier mode of the form $\textbf{u}(t,x^i) = \ti{\textbf{u}}(t)e^{ik\,x^in_i}$, one can estimate the growth in the energy norm:
    \begin{equation}
        \pd_t ||\textbf{u}||^2 = ik \ti{\textbf{u}}^T\br{\textbf{P}^T \textbf{H} - \textbf{H}\textbf{P}}\ti{\textbf{u}} = 0\,,
    \end{equation}
    meaning that the energy norm is constant in time and, as required, the system is well-posed. We then see that strong hyperbolicity implies well-posedness. It is also possible to show that weakly hyperbolic systems are not, meaning that strong hyperbolicity and well-posedness are equivalent properties.
\end{itemize}

To study well-posedness of a system, the methodology consists of looking at the principal part of the system of equations and trying to build a linearly independent set of eigenvectors. If attained, the system is hyperbolic, and these eigenvectors, combinations of the original evolution variables, will evolve through simple advection equations with their own characteristic speed.

\subsection{\done{Well-posedness in gravity}}

The original formulation of general relativity as a system of PDEs used harmonic coordinates developed by Choquet-Bruhat \cite{10.1007/BF02392131,bruhat1962}, later evolving to generalised harmonic coordinates \cite{Lindblom:2005qh,Pretorius:2004jg}. A popular alternative is the $3+1$ approach, first leading to the \textit{ADM} system (see section \ref{sec:ngr:adm}). This turned out to result in weakly hyperbolic evolution equations, meaning lack of well-posedness and unstable numerical behaviour. This led to the development of more complex formulations that satisfy well-posedness, such as the BSSNOK \cite{Nakamura:1987zz,Shibata:1995we,Baumgarte:1998te} and CCZ4 \cite{Alic:2011gg, Alic:2013xsa, Bona:2003fj, Bona:2003qn} formulations (see sections \ref{sec:ngr:bssnok} and \ref{sec:ngr:ccz4}).

In the realm of modified theories of gravity, most theories are either not well-posed or not known to be well-posed (often they are non-linear theories for which the concept of hyperbolicity analysis is not well understood). There have been some recent efforts that have successfully managed to construct well-posed formulations of certain modified theories of gravity of physical interest, such as cubic Horndeski theory \cite{Kovacs:2019jqj}. Earlier works studied the well-posedness of Lovelock and certain Horndeski theories, such as Einstein-scalar-Gauss-Bonnet gravity, and found that the equations of motion are weakly hyperbolic (and hence not well-posed) in a certain class of generalised harmonic gauges \cite{Papallo:2017qvl,Papallo:2017ddx}. A new modified generalised harmonic gauge was then found to circumvent these problems for all Lovelock and Horndeski theories \cite{Kovacs:2020pns,Kovacs:2020ywu,AresteSalo:2022hua}. Alternatively, \cite{Cayuso:2017iqc,Allwright:2018rut} have proposed to find well-posed formulations of alternative theories of gravity extending the M\"uller-Israel-Stewart formalism of viscous relativistic hydrodynamics \cite{Muller:1967aa,Israel:1976tn,Israel:1976213,Israel:1979wp} to those theories of gravity. Very recently \cite{Cayuso:2020lca} succeeded in applying this formalism to theories of gravity with higher curvature corrections assuming spherical symmetry. This method has also been applied to scalar-tensor theories with second order equations of motion \cite{Bezares:2021yek,Lara:2021piy,Franchini:2022ukz,Gerhardinger:2022bcw}.

Even for some of the theories proven to be well-posed, it has been shown that hyperbolicity can fail if the spacetime curvature and/or the derivatives of the scalar field become too large in the future development of the solution \cite{Reall:2014pwa,Ripley:2019aqj,Ripley:2019hxt,Bernard:2019fjb,Ripley:2019irj,Ripley:2020vpk,Bezares:2020wkn,Figueras:2020dzx,R:2022hlf}.

\section{\done{Horndeski}}\label{subsec:Horndeski}

Having in mind the importance of well-posedness, we focus on particular modified theory of gravity which is known to have a well-posed initial value problem: Horndeski theory\footnote{This theory was first found by Horndeski \cite{Horndeski:1974wa} and rediscovered in other works \cite{Nicolis:2008in,Deffayet:2009wt-discovery_of_Horndeski_Galileon,Deffayet:2011gz}.} \cite{Kovacs:2019jqj,Kovacs:2020pns,Kovacs:2020ywu}.\footnote{Rendall \cite{Rendall:2005fv} had previously proven well-posedness of the initial value problem for the so called $k$-essence theories, which are a subclass of the Horndeski theories considered in these papers.} This is the most general theory satisfying 1) a metric tensor coupled to a scalar field, 2) with second order equations of motion, 3) arising from a diffeomorphism invariant action\footnote{Also known as general covariance, this property means invariance of the physical laws under arbitrary coordinate transformations. Simplistically, this equates to expressing systems with tensor fields.}, 4) in four spacetime dimensions. The general action for this 4D theory is\footnote{The Teleparallel gravity version of this theory has been recently worked out by Bahamonde et al. \cite{Bahamonde:2019shr,Bahamonde:2019ipm}. While this version may offer a phenomenologically attractive avenue to explore, the well-posedness of the initial value problem in these theories has not been established.}
\begin{equation}\label{eq:Horndeski_general}
    \mathcal{S} := \int d^4x \sqrt{-g} \br{\mathcal{L}_1+\mathcal{L}_2+\mathcal{L}_3+\mathcal{L}_4+\mathcal{L}_5}\,,
\end{equation}
with
\begin{equation}
    \begin{split}
        \mathcal{L}_1 =&~ \mathcal{R} + X - V(\f)\,,\\
        \mathcal{L}_2 =&~ G_2(\f,X)\,,\\
        \mathcal{L}_3 =&~ G_3(\f,X)\,\square\f\,,\\
        \mathcal{L}_4 =&~ G_4(\f,X)\,\mathcal{R} + \partial_X G_4(\f,X)\,\sbr{\br{\square\f}^2-\br{\grad_\m\grad_\n\f}\br{\grad^\m\grad^\n\f}}\,,\\
        \mathcal{L}_5 =&~ G_5(\f,X)G_{\m\n}\grad^\m\grad^\n\f - \tfrac{1}{6}\partial_X G_5(\f,X)\,\big[\br{\square\f}^3 -3\square\f\br{\grad_\m\grad_\n\f}\br{\grad^\m\grad^\n\f} \\
        & \hspace{65mm} + 2\br{\grad_\m\grad_\n\f} \br{\grad^\n\grad^\r\f}\br{\grad_\r\grad^\m\f}\big]\,,
    \end{split}
\end{equation}
where $\f$ is a dimensionless scalar field; $X := -\tfrac{1}{2}(\grad_\mu\phi)(\grad^\mu\phi)$ and $V(\f)$ are the usual scalar kinetic and potential terms in the standard action for a minimally coupled scalar field; $G_i$ ($i=2,3,4,5$) are freely specifiable functions, and $\mathcal{R}$ and $G_{\m\n}$ are the Ricci scalar and Einstein tensor of the spacetime metric $g_{\m\n}$, respectively. We have explicitly separated the canonical kinetic and potential terms from $G_2$ so that $G_i$ parameterise only the higher derivative terms and non-minimal couplings of the scalar field to gravity. Having only second-order equations is essential to avoid a physical instability known as Ostrogradsky instabilities, detailed in the section below \ref{subsec:gr:ostrogradsky}. This theory has found numerous other applications to cosmology; the literature on the subject is vast and we will not attempt to review it here. We do not provide a detailed review of the current observational and theoretical constraints on the Horndeski gravity theories, nor do we discuss methods to construct solutions to the Horndeski gravity theories, such as Post-Newtonian methods. We refer the reader to the recent reviews \cite{Quiros:2019ktw,Kobayashi:2019hrl,Clifton:2011jh,Ripley:2022cdh,Carson:2019fxr}.

Restricting the theory to the class $G_4=G_5=0$, the action \eqref{eq:Horndeski_general} reduces to \textit{cubic Horndeski theories}, which will be extensively discussed in chapters \ref{chapter:paper1} and \ref{chapter:paper2} and comprise several well-known particular cases that have been extensively studied in other contexts, mostly cosmology \cite{Quiros:2019gbt,Quiros:2019ktw,Kobayashi:2019hrl}.

Following Quiros et al. \cite{Quiros:2019gbt} and Ripley \cite{Ripley:2022cdh}, one can summarise a non-extensive list of main sub-classes of Horndeski studied in the literature:
\begin{itemize}
    \item \textit{quintessence}, which consists of a simple scalar field minimally coupled to GR; this model is obtained by setting $G_2=G_3=G_4=G_5=0$.
    \item \textit{k-essence}, setting $G_3=G_4=G_5=0$, with the common choice of $G_2(\f,X)=f(\f)g(X)$ for arbitrary functions $f$ and $g$ of their arguments \cite{terHaar:2020xxb}. This can be extended to a \textit{g-essence} theory by adding a fermionic field its corresponding non-abelian kinetic term \cite{Razina:2011wv}.
    \item \textit{Galileons}, also known as \textit{kinetic gravity braiding} \cite{Deffayet:2010qz,Nucamendi:2019uen} or cubic galileons \cite{Deffayet:2009wt, Deffayet:2009mn,Frusciante:2019puu,Appleby:2020dko,Appleby:2020njl,Korolev:2020yyy}, are obtained by choosing $G_3\neq0$ with $G_4=G_5=0$; this class of models is often simplified to the shift symmetric case \cite{Schmidt:2018zmb}, corresponding to $G_3(\f,X)=g(X)$, for an arbitrary function $g$. This class evades many constraints to Horndeski theories \cite{Ezquiaga:2017ekz,Creminelli:2017sry,Baker:2017hug,Tattersall:2018map,Tahura:2018zuq}.
    \item $4\pd ST$ (4 derivative scalar-tensor) is a special case with the action\footnote{This action does not seem to appear in the form \eqref{eq:Horndeski_general}, but through field re-definitions can be transformed into the $G_i$ form \cite{Kobayashi:2011nu} and, as expected, leads to second order equations of motion.}:
    \begin{equation}
        \mathcal{S}_{4\pd ST} := \int d^4x \sqrt{-g} \br{\mathcal{R} + X - V(\f) + \a(\f)X^2 + \b(\f)\mathcal{G}}\,,
    \end{equation}
    where $\mathcal{G} = \mathcal{R}^2 - 4\mathcal{R}_{\m\n}\mathcal{R}^{\m\n} + \mathcal{R}_{\a\b\m\n}\mathcal{R}^{\a\b\m\n}$ is the Gauss-Bonnet invariant. This theory has attracted attention because for some coupling functions $\b(\f)$ black holes can have scalar hair, and because it can be shown to be an expected truncated theory for lower energies from an effective field theory perspective \cite{Weinberg:2008hq}. The theory with $\a(\f)=0$ is known as EsGB (Einsten scalar Gauss-Bonnet) \cite{East:2022rqi}.
    \item \textit{Brans-Dicke theory}, choosing $G_3=G_5=0$, $G_2=\br{\frac{w}{\f}-1}X$ and $G_4=\f$, leading to the action:
    \begin{equation}
        \mathcal{S}_{BD} := \int d^4x \sqrt{-g} \br{f(\f) \mathcal{R} + \frac{w}{\f} X}\,.
    \end{equation}
    Due to the conformal coupling of the scalar field to the Ricci scalar, this action is said to be in the Jordan frame \cite{Faraoni:1999hp,Galaverni:2021xhd,Bamber:2022eoy,Flanagan:2004bz,Hawking:1972qk}. Through a suitable conformal transformation redefining the metric $g_{\m\n}\to \W(\f)\ti{g}_{\m\n}$, this action can be re-written as:
    \begin{equation}
        \mathcal{S}_{BD} := \int d^4x \sqrt{-\ti{g}} \br{\ti{\mathcal{R}} + \ti{X}}\,,
    \end{equation}
    where $\ti{\mathcal{R}},\ti{X}$ are computed using the conformal metric $\ti{g}$. This is called the Einstein frame.
\end{itemize}

As a final note, there is a nuance to be made on the distinction between modified gravity and exotic matter, as simple theories such as Galileons can be considered as part of both classifications. These theories have been used as one of many ways of modifying the behaviour of gravity and potentially address certain phenomena. Hence, for the purposes of this thesis, such extensions of GR are considered modified gravity, leaving the term exotic matter to refer to matter not in the standard model of particle physics that, for instance, violates energy conditions, such as having negative energy density or pressure.

\subsection{\done{Ostrogradsky instability}}\label{subsec:gr:ostrogradsky}

Having only second-order equations is essential to avoid a physical instability known as Ostrogradsky instability \cite{Woodard:2015zca,Motohashi:2014opa,Donoghue:2021eto}, which we summarise below. Theories with higher than second derivatives correspond to Hamiltonians unbounded from below, typically associated with a new degree of freedom of the theory, the Ostrogradsky ghost. A Hamiltonian unbounded from below implies an empty state can decay into an unstable collection of unbounded positive and negative energy excitations, often entropically favoured. Below we give an example of one such Hamiltonian.

The instability can be circumvented by breaking non-degeneracy of the Lagrangian canonical coordinates \cite{Aoki:2020gfv,Ganz:2020skf,Crisostomi:2017ugk}. For scalar-tensor theories, this results in the DHOST (Degenerate Higher-Order Scalar-Tensor) theories \cite{Langlois:2015cwa,Langlois:2018dxi,Heisenberg:2018vsk,Jana:2020vov}, which are the largest generalisation of scalar-tensor theories with higher-order derivatives (higher than second) that avoid the Ostrogradsky instability. These obviously include Horndeski as a subclass, as well as \textit{Beyond Horndeski} theories and disformal transformations of Horndeski theories. Note that the absence of this instability does not guarantee well-posedness, and the presence of ghosts, in general, does not exclude well-posedness either.

Let us see this effect taking place with a simple explicit example \cite{Woodard:2015zca}. Take the Lagrangian of higher derivative deviations from an harmonic oscillator, parameterised by $\e$:
\begin{equation}
    \mathcal{L} = -\frac{\e\,m}{2\w^2}\ddot{x}^2 + \frac{m}{2}\dot{x}^2 - \frac{m\,\w^2}{2}x^2\,,
\end{equation}
which leads to the Hamiltonian:
\begin{equation}
    \mathcal{H} = \frac{\e\,m}{\w^2}\dot{x}\dddot{x}-\frac{\e\,m}{2\w^2}\ddot{x}^2 + \frac{m}{2}\dot{x}^2 + \frac{m\,\w^2}{2}x^2\,.
\end{equation}
The Euler-Lagrange equation and its general solution are,
\begin{align}
    0 &= -m\br{\frac{\e}{\w^2}\ddddot{x} + \ddot{x} + \w^2x}\,,\\
    x(t) &= C_+ \cos{\br{k_+ t}} + S_+ \sin{\br{k_+ t}} + C_- \cos{\br{k_- t}} + S_- \sin{\br{k_- t}}\,,
\end{align}
where $k_\pm = \w\sqrt{\frac{1\mp\sqrt{1-4\e}}{2\e}}$ are constants and $C_\pm$ and $S_\pm$ are determined by initial conditions (a fourth order equation requires four initial conditions). Replacing this solution back into the Hamiltonian:
\begin{equation}
    \mathcal{H} = \frac{m}{2}\sqrt{1-4\e}\, k_+^2\br{C_+^2 + S_+^2} - \frac{m}{2}\sqrt{1-4\e}\,k_-^2\br{C_-^2 + S_-^2}\,.
\end{equation}
The last form makes it clear that the $C_+$ and $S_+$ carry positive energy modes, and $C_-$ and $S_-$ carry negative energy modes. Moreover, the negative energy modes show the Hamiltonian is unbounded from below, leading to problems such as vacuum decay when considering interactions \cite{Woodard:2015zca}. This is the Ostrogradsky instability. These unbounded modes are called Ostrogradsky ghosts.

\section{\done{Fixing higher derivative effective field theories}}\label{sec:gr:fixing_hd_theories}

\begin{figure}[h]
\centering
\includegraphics[width=0.8\textwidth]{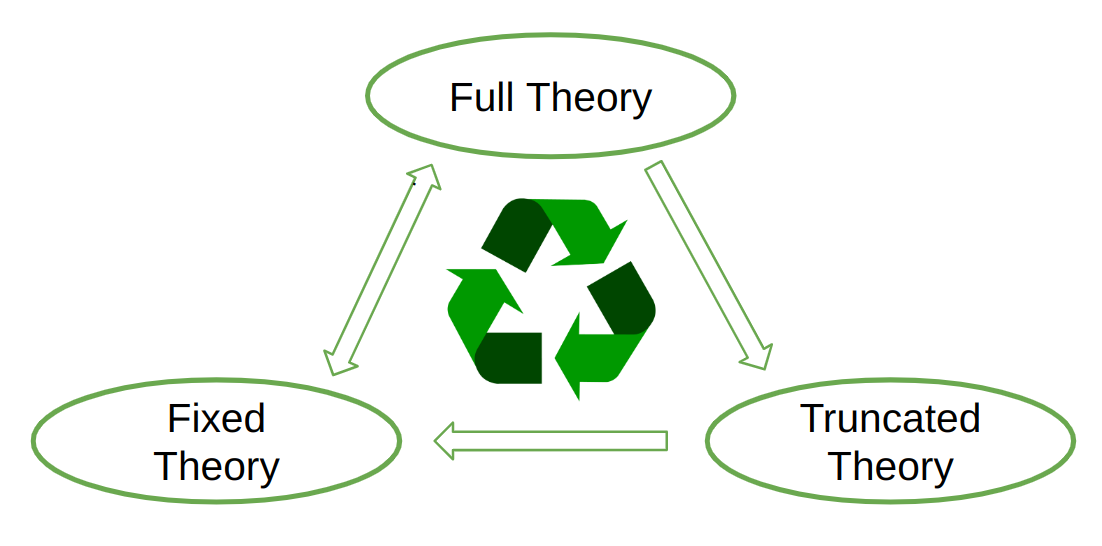}
\caption{Schematic of three possible theories: an unknown \textit{full theory} of nature free of pathologies, a \textit{truncated theory} probing low energy phenomenology, yet with pathologies, and a \textit{fixed theory} constructed from the truncated one, free of pathologies and now able to inform about the full theory.}
\label{fig:gr:fixed_theories}
\end{figure}

In section \ref{subsec:gr:well_posedness} we discuss the hyperbolicity required to assess the well-posedness in a traditional way. There are theories for which such analysis is hard to develop or for which partial differential equation theory is still to be developed, giving us no mathematical guidance to enquire about the validity of an initial value problem. This is the case for theories with higher than second derivatives or with non-linear terms such as second derivatives multiplied together. The issue often lies not in the \textit{full theory} these effective field theories are trying to reproduce, but in strong energy cascades and characteristic shocks resulting from a \textit{truncated theory} with a certain energy cutoff, as we shall see in an example below. In reality, sensitivity to high frequency behaviour affecting the physics of a system hints the solution is away from the regime of applicability of the theory in the first place. As intended in figure \ref{fig:gr:fixed_theories}, one can then hope to develop suitable techniques to develop a \textit{fixed theory} that can capture the low energy behaviour of an extension of GR while being insensitive to uncontrolled growth one naively would find \cite{Allwright:2018rut, Cayuso:2017iqc}.

In the next sections we will apply the ideas above to a sample toy model that exemplifies how a truncated theory can introduce problems the full theory does not share, and present the generic method that can be used to fix an extended class of theories with higher derivatives in the equations of motion.

\subsection{\done{Toy model, from full theory to truncated theory}}

Let us consider some complete theory and truncate it in some effective field theory sense \cite{Burgess:2014lwa, Allwright:2018rut}. Consider the 4D full theory action:
\begin{equation}\label{eq:subsubsec:toy_example:full_theory1}
    S = - \int d^4x\sbr{\br{\pd_\m\f^*}\br{\pd^\m\f} + V_\f\br{\f^*\f}}\,,
\end{equation}
where the potential is given by:
\begin{equation}
    V_\f\br{\f^*\f} = \tfrac{\l}{2}\br{\f^*\f - \tfrac{v^2}{2}}^2\,,
\end{equation}
with a minimum for $\f^*\f = \frac{v^2}{2}$. Decomposing the complex scalar field into a real phase and amplitude expanded around the minimum, $\q$ and $\r$, one can write:
\begin{equation}
    \f(x) = \tfrac{v}{\sqrt{2}}\sbr{1+\r(x)}e^{i\,\q(x)}\,.
\end{equation}
The action \eqref{eq:subsubsec:toy_example:full_theory1} becomes:
\begin{equation}
    \frac{S}{v^2} = - \int d^4x \sbr{\tfrac{1}{2}\br{\pd_\m\r}\br{\pd^\m\r} + \tfrac{1}{2}\br{1 + \r}^2\br{\pd_\m\q}\br{\pd^\m\q} + V_\r(\r)}\,,
\end{equation}
with the potential $V_\r$ re-written as:
\begin{equation}
    V_\r(\r) = \frac{M^2}{2}\br{\r^2 + \r^3 + \frac{\r^4}{4}}\,,
\end{equation}
where $M^2 = \frac{v^2\l}{2}$. $\r$ represents a massive field and $\q$ a Goldstone boson. The equations of motion are:
\begin{equation}
    \begin{split}
        \square \r &= (1+\r)\br{\pd_\m\q}\br{\pd^\m\q} + V'(\r)\,,\\
        \square \q &= -2\br{1+\r}^{-1} \br{\pd_\m\r}\br{\pd^\m\q}\,.
    \end{split}
\end{equation}
In the limit where $M$ is very large compared with the energies of interest we can integrate out the $\r$ field to determine its leading-order effects on the low-energy physics of $\q$-particles. As described by Burguess and Williams \cite{Burgess:2014lwa}, this yields:
\begin{equation}
    \frac{S}{v^2} \approx -\int d^4x \br{\tfrac{1}{2}\br{\pd_\m\q}\br{\pd^\m\q} - \tfrac{1}{2M^2}\sbr{\br{\pd_\m\q}\br{\pd^\m\q}}^2 + \tfrac{2}{M^4}\br{\pd_\m\pd_\n\q}\br{\pd^\m\pd^\s\q}\br{\pd^\n\q}\br{\pd_\s\q}}\,.
\end{equation}
The last term in this action leads to equations of motion that clearly have third and fourth order derivatives. In the original theory, the equations of motion for $\r$ and $\q$ only had up to second order derivatives. We then see how a perfectly healthy full theory can be truncated into a pathological one.

\subsection{\done{Fixing truncated theories}}

In the context of modified gravity, let us explore extensions of Einstein's equation by considering an energy-momentum tensor that depends on derivatives higher than second order in the metric, without extra fields. Take the Einstein tensor as a function of the metric up to second derivatives, $G_{\m\n} = G(g^a, \pd_\m g^a, \pd_\m^2 g^a)$, where $g^a$ are the components of the metric $g_{\m\n}$ (here and below $a,b,...$ represent generic indices counting the variables, not spacetime indices). Following \cite{Cayuso:2020lca}, write Einstein's extended equation generically as:
\begin{equation}
    G^b(g^a,\pd_\m g^a, \pd_\m^2 g^a) = \e\, S^b(g^a,\pd_\m g^a, \pd_\m^2 g^a, \pd_\m^3 g^a, \pd_\m^4 g^a) + \mathcal{O}(\e^2)\,,
\end{equation}
for some small parameter $\e$. The symbol $S^b$ encodes contributions of extensions to GR. We consider up to fourth order terms as generic terms that may appear in lowest order EFT. On the left hand side one can factor out second time derivatives to obtain:
\begin{equation}
    \pd_t^2 g^b = \ti{G}^b(g^a, \pd_\m g^a, \pd_i\pd_\m g^a) + \e\, S^b(g^a,\pd_\m g^a, \pd_\m^2 g^a, \pd_\m^3 g^a, \pd_\m^4 g^a) + \mathcal{O}(\e^2)\,,
\end{equation}
where $\pd_i^2 g^a$ and $\pd_i\pd_t g^a$ indicate purely spatial and mixed time and spatial derivatives, respectively, and $\ti{G}^b$ is related to the original $G^b$ functions. The higher order time derivatives clearly make this problem intractable at first sight. We will now see how to deal with higher order time derivatives and then with higher order spatial derivatives:

\subsubsection{\done{Time derivative order reduction}}\label{subsubsec:gr:time_reduction}

Take the previous equation now only up to $\mathcal{O}(\e)$:
\begin{equation}\label{eq:gr:dt2b}
    \pd_t^2 g^b = \ti{G}^b(g^a, \pd_\m g^a, \pd_i\pd_\m g^a) + \mathcal{O}(\e)\,.
\end{equation}
From this, notice how higher order derivatives can be written as:
\begin{equation}\label{eq:gr:dt3g}
\begin{split}
    \pd_t^3 g^b &= \br{\pd_t\ti{G}}^b(g^a, \pd_\m g^a, \pd_\m^2 g^a, \pd_i^2\pd_t g^a, \pd_i\pd_t^2 g^a) + \mathcal{O}(\e)\,,\\
    \pd_t^4 g^b &= \br{\pd_t^2\ti{G}}^b(g^a, \pd_\m g^a, \pd_\m^2 g^a, \pd_t\pd_\m^2 g^a, \pd_i^2\pd_t^2 g^a, \pd_i\pd_t^3 g^a) + \mathcal{O}(\e)\,.
\end{split}
\end{equation}
The right hand side of \eqref{eq:gr:dt3g} depends on second time derivatives of $g^a$, which can be re-expressed (recurrently as necessary, since we are discarding terms of $\mathcal{O}(\e)$) using equation \eqref{eq:gr:dt2b} to yield:
\begin{equation}\label{eq:gr:dt3g_v2}
\begin{split}
    \pd_t^3 g^b &= \br{\pd_t\ti{G}}^b(g^a, \pd_\m g^a, \pd_i\pd_\m g^a, \pd_i^2\pd_\m g^a) + \mathcal{O}(\e)\,,\\
    \pd_t^4 g^b &= \br{\pd_t^2\ti{G}}^b(g^a, \pd_\m g^a, \pd_i\pd_\m g^a, \pd_i^2\pd_\m g^a, \pd_i^3\pd_\m g^a) + \mathcal{O}(\e)\,.
\end{split}
\end{equation}
This procedure expresses all second or higher order time derivatives with spatial derivatives (of higher order), keeping only time derivatives up to first order, which are typically reduced to auxiliary variables when decomposing the system into a first order form. One can now use these re-written higher derivatives to re-write $S^b$ into $\ti{S}^b$, satisfying:
\begin{equation}
    \e\, S^b = \e\, \ti{S}^b + \mathcal{O}(\e^2)\,.
\end{equation}
This yields the ``time reduced'' evolution equations, equal to the original up to $\mathcal{O}(\e)^2$:
\begin{equation}
    \pd_t^2 g^b = \ti{G}^b(g^a, \pd_\m g^a, \pd_i\pd_\m g^a) + \e\, \ti{S}^b(g^a,\pd_\m g^a, \pd_i\pd_\m g^a, \pd_i^2\pd_\m g^a, \pd_i^3\pd_\m g^a) + \mathcal{O}(\e^2)\,.
\end{equation}

\subsubsection{\done{Dealing with higher spatial derivative, ``Fixing the equation''}}\label{subsubsec:gr:fixing}

The existence of higher order spatial derivatives, even without higher order time derivatives, are known to bring uncontrolled high frequency modes \cite{Cayuso:2017iqc}. Allwright and Lehner \cite{Allwright:2018rut} mention two solutions to this problem: ``Fixing the equations'' and ``Reduction of Order'' (not to be confused with the time reduction of order used in the previous section). Reduction of order is an iterative method where the source is assumed to be sub-leading, allowing to iterate until convergence the solution using the source evaluated on previous iterations: $G^b(g_{i}^a,\pd_\m g_i^a, \dots) = \e\,S^b(g_{i-1}^a,\pd_\m g_{i-1}^a,\dots)$, effectively decoupling it from the principal part of the equation as it is treated as an independent source. This method was found to behave less robustly, so we will not describe it more extensively.

The ``fixing'' method, inspired by the Israel-Stewart formulation of relativistic viscous hydrodynamics \cite{ISRAEL1979341, Baier:2007ix}, was found to behave extremely well. It consists of writing an \textit{ad hoc} system with an auxiliary variable that replaces the badly behaved terms in the source of the evolution equation. This auxiliary variable is dynamic and ``tracks'' the original source. However, its evolution equation is written in such a way that it captures the low frequency behaviour of the source, damping away any high frequency spurious modes. This allows us to achieve our original goal of reproducing the low energy behaviour of the full theory that our truncated equations are trying to probe, without introducing pathological behaviour.

Let us look at a simple version of such alternative system by introducing independent variables $\P^b$ that ``track'' the source $\ti{S}^b$:
\begin{equation}
\begin{split}
    \pd_t^2 g^b &= \ti{G}^b(g^a, \pd_\m g^a, \pd_i\pd_\m g^a) + \e\,\P^b\,,\\
    \pd_t\P^b &= -\tfrac{1}{\t}\br{\P^b - \ti{S}^b(g^a,\pd_\m g^a, \pd_i\pd_\m g^a, \pd_i^2\pd_\m g^a, \pd_i^3\pd_\m g^a)}\,,
\end{split}
\end{equation}
where we ignore $\mathcal{O}(\e)^2$ terms and $\t$ is a constant. The variables $\P^b$ follow a damped equation converging to the source $\ti{S}^b$ on a timescale $\t$. If the long wavelength physics of the original equations can be decoupled from short wavelength modes, then the particular form of this equation should not be unique as long as it retains the same long distance behaviour. Moreover, the physics obtained should be also insensitive to the choices of $\t$, within a wide range of timescales. For specific problems, these auxiliary variables can be engineered to completely remove all higher order spatial derivatives, as we will demonstrate in sections \ref{sec:eft:toymodels} and \ref{eft:subsec:c_system}.

Using time order reduction to reduce the system to first order time derivatives, and dealing with higher order spatial derivatives by ``fixing'' the equations with auxiliary variables, one can arrive at the fixed theory, with stable equations ready for numerical evolution. Cayuso and Lehner \cite{Cayuso:2020lca} make an explicit application of this, which was the basis for chapter \ref{chapter:eft}. Toy models analysed by Cayuso et al. \cite{Cayuso:2017iqc} apply this technique to various equations. Its toy model example labelled $n=4$ is analysed in more detail in section \ref{sec:eft:toymodels}, as well as extended to a more ``gravitatonal'' setting useful to understand the theory simulated with black hole binaries in chapter \ref{subsubsec:eft:alternative_systems_gravity}.

\chapter{\done{Overview of numerical general relativity}}\label{chapter:ngr_overview}

\section{\done{Introduction}}

GR requires the solution to Einstein's field equations: multidimensional, nonlinear, coupled partial differential equations in four dimensions. Due to its non-linearity, only a few idealised cases with high degree of symmetry (e.g. static solutions, isotropic solutions, etc.) can be solved exactly. This led to the birth of the field of Numerical Relativity (NR), developing computer algorithms using numerical techniques to finish the exciting task of finding solutions to Einstein's equations.

One of the main difficulties associated with NR is related to singularities, regions where the gravitational tidal forces, matter density and curvature all become infinite (where the theory breaks down), as present in BHs, potentially causing overflows in a simulation tied to finite numbers. On top of that, one encounters both theoretical challenges, in terms of initial data, gauge conditions and boundary conditions, as well as numerical implementation challenges, as the requirement of accurate resolution at small and large scales in a single simulation demands the use of parallel processing technology and computer clusters.

\begin{figure}[h]
\centering
\includegraphics[width=.8\textwidth]{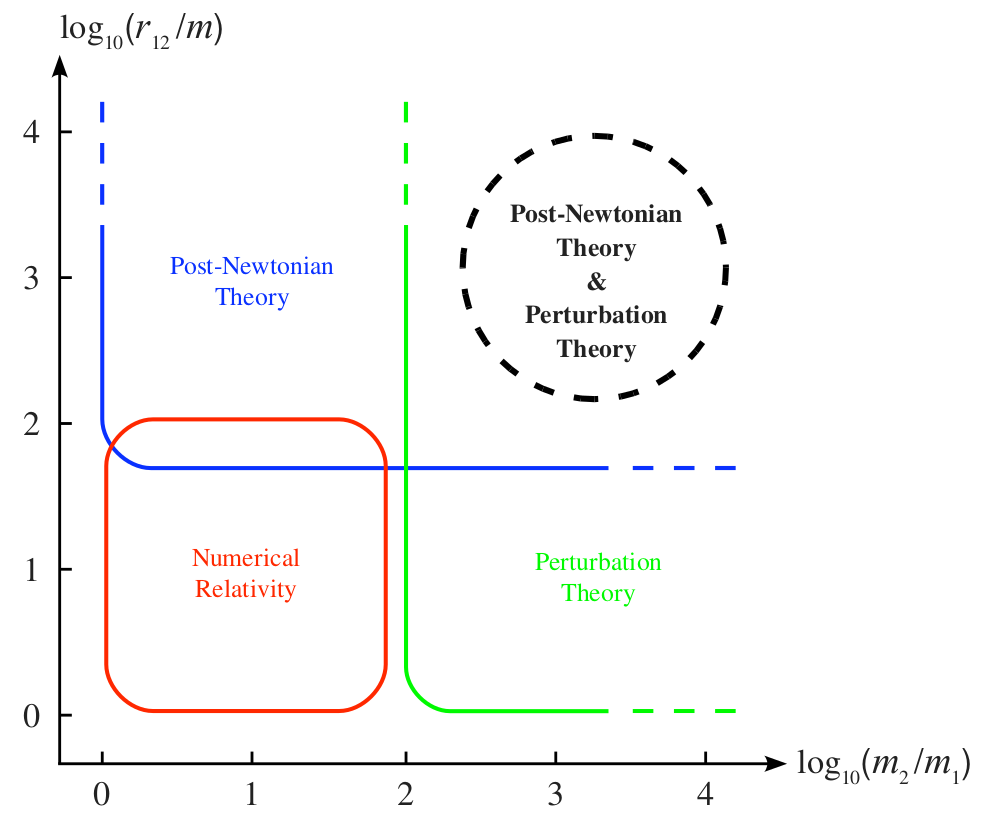}
\caption{Different schemes to study black hole binaries, depending on the mass ratio and distance between the bodies. Figure taken from Blanchet \cite{Blanchet:2013haa}. The representation of numerical relativity up to distances of $r_{12}\sim 10^2m$ overestimates current standards.}
\label{fig:ngr:pn_vs_ngr}
\end{figure}

One early result of NR was Choptuik's discovery of
critical phenomena in the gravitational collapse of a massless scalar field \cite{Choptuik:1992jv}, already discussed in section \ref{subsec:gr:critical_collapse}. For problems like the binary black hole spacetime, it took more than fifty years \cite{Sperhake:2014wpa} until the breakthrough in 2005, Pretorius \cite{Pretorius:2005gq} achieved the first ever numerical evolution of a binary black hole spacetime with more than one orbit before merger. By now, about 90 compact binary mergers have been observed by the gravitational wave detector network \cite{LIGOScientific:2018mvr, LIGOScientific:2020ibl, LIGOScientific:2021djp}. Many numerical schemes can be developed to approximate the inspiral and the post-merger solution, such as post-Newtonian theory \cite{Blanchet:2013haa}, self-force \cite{Barack:2011yga} and effective one-body approximations \cite{Buonanno:1998gg}. Figure \ref{fig:ngr:pn_vs_ngr} illustrates the applicability of these in different regimes of approximation, such as extreme mass ratios or very far apart orbits. To evolve the full system without symmetry assumptions and with time dependence, numerical general relativity is the only path forward to this day.

There are several approaches to this problem and in this work the $d+1$ formalism will be the focus, due to its widespread use and importance for the applications developed. The gravitational field is thought to have a future evolution in ``time'', splitting time from the 3-dimensional space. Let us recall that the Einstein's field equations provide no clear distinction between space and time, which, even though it was intentional and natural from the perspective of differential geometry, is not always the ideal picture for the desired outcome: give certain initial data and obtain the subsequent evolution of the gravitational field. Therefore, it turns to a problem of separating the equations into space and time to suit this goal. It is important to note that unlike other Cauchy problems, there is no preferred time direction when doing this split and no global concept of time. Each point on a spatial slice evolves according to a local time coordinate, which leaves up to us the freedom to choose the path of each of such point ``observer'', commonly called \textit{gauge freedom}. Einstein's equations \eqref{eq:einstein} only fix the Ricci curvature, not the full metric components, which are dependent on the coordinate system.

In the next sections we discuss 3 possible standard formulations that perform the split of space and time and generate physical solutions, their intuition, advantages and disadvantages, each with important consequences for numerical stability: the ADM formalism, the first step in the life of any numerical relativist which results in evolution equations that are weakly hyperbolic and cannot be numerically evolved; the BSSN formalism \cite{Nakamura:1987zz,Shibata:1995we,Baumgarte:1998te}, a conformal reformulation with strong hyperbolicity properties, and the CCZ4 formulation \cite{Alic:2011gg, Alic:2013xsa, Bona:2003fj, Bona:2003qn}, which extends BSSN with added stability properties. We will also discuss gauge conditions (freely specifiable conditions), initial data and boundary conditions, essential ingredients in any Cauchy problem. Finally, we will review \grchombo, the numerical code we use for all numerical simulations. For reviews of numerical relativity, see Alcubierre \cite{alcubierre}, Baumgarte and Shapiro \cite{shapiro} and Gourgoulhon \cite{Gourgoulhon:2007ue}.

\vspace{10mm}\begin{summary}{\done{ADM formulation}}\label{adm_summary}
    Standard $d+1$ split into gauge functions (the \textit{lapse function}, $\alpha$, and the  \textit{shift vector}, $\beta^i$) and the \textit{spatial metric}, $\gamma_{ij}$:
    \begin{equation}
        ds^2 = \br{-\alpha^2+\beta_i\beta^i}dt^2 + 2\beta_i~ dtdx^i + \gamma_{ij}~dx^idx^j\,.
    \end{equation}
    This results in the following unit vector normal to spatial hypersurfaces: $n^\mu = \br{\frac{1}{\alpha},-\frac{\beta^i}{\alpha}}$.
    
    The \textit{extrinsic curvature} is introduced as the projections of the gradients of the normal vector into the spatial slice:
    \begin{equation}
        K_{\mu\nu} \equiv -\gamma_\mu^{~\alpha} \gamma_\nu^{~\beta}\grad_\alpha n_\beta\,.
    \end{equation}
    \textit{Matter source terms} are introduced as:
    \begin{equation}\label{adm_source}
        \rho \equiv n^\m n^\n T_{\m\n}~,~~~~~~~S^i \equiv - \g^{i\m} n^\n T_{\m\n}~,~~~~~~~S^{ij}\equiv \g^{i\m}\g^{j\n}T_{\m\n}~,~~~~~~~S=S_i^{~i}\,.
    \end{equation}
    The \textit{ADM variables} are: $\{\gamma_{ij},K_{ij}\}$. Einstein's equations are decomposed into the \textit{Hamiltonian and Momentum constraint},
    \begin{align}\label{eq:H_ADMf}
        \mathcal{H} &= \tfrac{1}{2} \sbr{R + K^2 - K_{ij}K^{ij}} - \tfrac{\k}{2}\rho - \L = 0\,,\\\label{eq:M_ADMf}
        \mathcal{M}_i &= D^j\sbr{K_{ij}-\gamma_{ij}K}-\tfrac{\k}{2} S_i=0\,,
    \end{align}
    where $R$ and $D^j$ are the Riemann tensor and covariant derivative associated with $\gamma_{ij}$, and $K\equiv K_i^{~i}$; and \textit{first order evolution equations for the spatial metric and extrinsic curvature},
    \begin{align}\label{eq:gammaf}
        \br{\pd_t - \b^k\pd_k}& \gamma_{ij} = -2\alpha K_{ij} + 2\gamma_{k(i}\partial_{j)}\beta^k\,,\\\label{eq:Kf}
        \br{\pd_t - \b^k\pd_k}& K_{ij}  = 2K_{k(i}\partial_{j)} \beta^k - D_iD_j\alpha + \alpha\sbr{R_{ij} + K K_{ij} - 2K_{ik}K_j^{~k}}\\\nonumber
        &~~~~~~~  + \tfrac{\a\,\k}{2}\sbr{\gamma_{ij}\tfrac{S-\rho}{d-1}-S_{ij}} - \tfrac{2\,\a}{d-1}\g_{ij}\L\,,
    \end{align}
    where $R_{ij}$ is the Ricci tensor associated with $\g_{ij}$. See details in appendix \ref{sec:ngr:adm}.
\end{summary}

\newpage
\begin{summary}{\done{BSSNOK formulation}}\label{BSSNOK_summary}
    The spatial metric of the $d+1$ equations, $\gamma_{ij}$, is decomposed into a \textit{conformal metric}, $\ti{\gamma}_{ij}$ with unit determinant, assuming Cartesian coordinates, and a \textit{conformal factor}, $\chi$. The extrinsic curvature is also decomposed into its \textit{trace} and \textit{traceless part}, with the latter conformally transforming as the metric. New variables are added as independent, the \textit{conformal connection functions}, $\ti{\Gamma}^i$, adding a new constraint to the system.
    \vspace{-6mm}
    \begin{multicols}{2}
        \begin{equation}
            \ti{\gamma}_{ij} \equiv \chi\gamma_{ij}\,,
        \end{equation}
        
        \begin{equation}\label{eq:bssn:kij_definition}
            K_{ij} \equiv \tfrac{1}{\chi}\br{\ti{A}_{ij} +\tfrac{1}{d}\ti{\gamma}_{ij}K}\,.
        \end{equation}
    \end{multicols}
    The \textit{BSSNOK variables} are: $\{\chi, \ti{\gamma}_{ij}, K, \ti{A}_{ij}, \ti{\Gamma}^i\}$. Adding to the constraints the \textit{conformal connection functions equation}, the constraints of the BSSNOK system are:
    \vspace{-5pt}
    \begin{equation}\label{eq:BSSNOK_constraints}
        \text{det}(\ti{\gamma}_{ij})=1\,,\qquad
        \text{tr}(\ti{A}_{ij})=0\,,\qquad
        \ti{\Gamma}^i = \ti{\gamma}^{jk}\ti{\Gamma}^i_{~jk}=-\partial_j\ti{\gamma}^{ij}\,,
    \end{equation}
    \vspace{-30pt}
    \begin{align}\label{eq:H_BSSNf}
        \mathcal{H} &= \tfrac{1}{2}\sbr{\chi\ti{R}+\br{d-1}\ti{D}^i\ti{D}_i\chi-\tfrac{(d+2)(d-1)}{4\chi}\ti{D}^i\chi\ti{D}_i\chi+\tfrac{d-1}{d}K^2 - \ti{A}_{ij}\ti{A}^{ij}} - \tfrac{\k}{2}\rho - \L =0\,,\\\label{eq:M_BSSNf}
        \mathcal{M}_i &= \ti{D}^j\ti{A}_{ij} - \tfrac{d}{2\chi} \ti{A}_{ij} \ti{D}^j \chi -\tfrac{d-1}{d}\partial_i K- \tfrac{\k}{2} S_i\,,
    \end{align}
    where $\ti{\Gamma}^i_{~jk}$, $\ti{R}$, $\ti{D}_i$ are, respectively, the Christoffel symbols, Ricci scalar and covariant derivatives associated (and raised/lowered) with $\ti{\gamma}_{ij}$. The \textit{first order evolution equations} become\footnote{The matter source terms are the same as for the ADM equations as in \ref{adm_source}. See details in appendix \ref{sec:ngr:bssnok}.}:\vspace{-1mm}
    \begin{align}
        \br{\partial_t-\b^k\pd_k}&\chi = \tfrac{2}{d}\chi\br{\alpha K-\partial_i\beta^i}\,,\\
        \br{\partial_t-\b^k\pd_k} & \ti{\gamma}_{ij} =  -2\alpha \ti{A}_{ij}+2\ti{\gamma_{k(i}}\partial_{j)}\beta^k-\tfrac{2}{d}\ti{\gamma_{ij}}\partial_k\beta^k\,,\\\label
        {eq:K_B}
        \br{\partial_t-\b^k\pd_k} & K = -\gamma^{ij}D_jD_i\alpha +\alpha\br{\ti{A}_{ij}\ti{A}^{ij}+\tfrac{1}{d}K^2}\\\nonumber
        &~~~~~~ +\tfrac{\a\,\k}{2(d-1)}\br{S+(d-2)\rho}-\tfrac{2}{d-1}\a\L\,,\\
        \br{\partial_t-\b^k\pd_k} & \ti{A}_{ij} = \chi\br{-D_iD_j\alpha +\alpha R_{ij}-\tfrac{\a\,\k}{2} S_{ij}}^{TF}+\alpha \br{K\ti{A}_{ij}-2\ti{A}_{ik}\ti{A}^k_{~j}}\\\nonumber
        &~~~~~~~ +2\ti{A}_{k(i}\partial_{j)} \beta^k -\tfrac{2}{d}\ti{A}_{ij}\partial_k\beta^k\,,\\\nonumber
        \br{\partial_t-\b^k\pd_k} & \ti{\Gamma}^i = -2\ti{A}^{ij}\partial_j\alpha +2\alpha\br{\ti{\Gamma}^i_{~jk}\ti{A}^{kj}-\tfrac{d}{2}\ti{A}^{ij}\tfrac{\partial_j\chi}{\chi}-\tfrac{d-1}{d}\ti{\gamma}^{ij}\partial_jK-\tfrac{\k}{2}\ti{\gamma}^{ik}S_k}\\
        &~~~~~~ -\ti{\Gamma}^j\partial_j\beta ^i +\tfrac{2}{d}\ti{\Gamma }^i\partial _j\beta ^j+\tfrac{d-2}{d}\ti{\gamma }^{ki}\partial _k\partial_j\beta^j + \ti{\gamma}^{kj}\partial _j\partial_k\beta^i\,,
    \end{align}
    \vspace{-1mm}To calculate the covariant derivative, $D_i$, and the Ricci tensor, $R_{ij}$, the following is useful:
    \begin{align}\label{BSSNOK_help}
        \Gamma^i_{~jk} &= \ti{\Gamma}^i_{~jk} - \tfrac{1}{2\chi}\br{\delta_j^i\partial_k\chi + \delta_k^i\partial_j\chi-\gamma_{jk}\gamma^{il}\partial_l\chi}\,,\\
        R_{ij} &= \ti{R}_{ij} + R^\chi_{ij}\,,\\
        \ti{R}_{ij} & = -\tfrac{1}{2}\ti{\gamma}^{lm}\partial_l\partial_m\ti{\gamma}_{ij}+\ti{\gamma}_{k(i}\partial_{j)}\ti{\Gamma}^k+\ti{\Gamma}^k\ti{\Gamma}_{(ij)k}+\ti{\gamma}^{lm}\sbr{2\ti{\Gamma}^k_{~l(i}\ti{\Gamma}_{j)km}+\ti{\Gamma}^k_{~im}\ti{\Gamma}_{klj}}\,,\\\label{eq:R_3}
        R^\chi_{ij}&=\tfrac{d-2}{2\chi}\br{\ti{D}_i\ti{D}_j\chi - \tfrac{1}{2\chi}\ti{D}_i\chi\ti{D}_j\chi} + \tfrac{1}{2\chi}\ti{\gamma}_{ij}\br{\ti{D}^k\ti{D}_k\chi   -\tfrac{d}{2\chi}\ti{D}^k\chi\ti{D}_k\chi}\,.
    \end{align}
\end{summary}

\newpage
\begin{summary}{\done{CCZ4 formulation}}\label{ccz4_summary}
    For improved control of constraint violations, Einstein's field equations are extended covariantly, introducing a vector $Z_\mu$ and two damping coefficients, $\kappa_1$ and $\kappa_2$:
    \begin{equation}\label{eq:gr:ccz4_eom}
        \mathcal{R}_{\mu\nu} + 2\grad_{(\mu}Z_{\nu)} - \kappa_1 \sbr{2n_{(\mu}Z_{\nu)}-\tfrac{2}{D-2}(1+\kappa_2)g_{\mu\nu}n_\sigma Z^\sigma} - \tfrac{2}{D-2}\L g_{\m\n} = \tfrac{\k}{2}\br{T_{\mu\nu}-\tfrac{1}{D-2}g_{\mu\nu}T}\,.
    \end{equation}
    $Z_\mu$ assesses deviations from Einstein's equations, measuring constraint violations. It is decomposed into spatial and normal components, $Z_i$ and $\Theta$. A conformal transformation is made as for BSSNOK equations (\ref{BSSNOK_summary}). The conformal connection functions (\ref{eq:BSSNOK_constraints}) are re-defined as $\hat{\Gamma}^i$.
    \vspace{-8mm}
    \begin{multicols}{2}
    \begin{equation}
        \Theta \equiv -n_\mu Z^\mu = \alpha Z^0\,,
    \end{equation}
    
    \begin{equation}
        \hat{\Gamma}^i \equiv \ti{\Gamma}^i+2\ti{\gamma}^{ij}Z_j\,.
    \end{equation}
    \end{multicols}
    \vspace{-2mm}
    Effective damping occurs for $\kappa_1>0$ and $\kappa_2>-1$. And extra damping parameter $\kappa_3$ is added to the evolution equation of $\hat{\Gamma}^i$ to exponentially suppress constraint violating modes.

    The \textit{CCZ4 variables} are: $\{\chi, \ti{\gamma}_{ij}, K, \ti{A}_{ij}, \Theta, \ti{\Gamma}^i\}$. The algebraic constraints of the CCZ4 system are\footnote{The energy-momentum constraints can still be computed using (\ref{eq:H_ADMf}, \ref{eq:M_ADMf}) or (\ref{eq:H_BSSNf}, \ref{eq:M_BSSNf}).}:
    \vspace{-10mm}
    \begin{multicols}{2} 
        \begin{equation}\label{det1}
            \text{det}(\ti{\gamma}_{ij})=1\,,
        \end{equation}
        
        \begin{equation}\label{tr0}
            \text{tr}(\ti{A}_{ij})=0\,,
        \end{equation}
    \end{multicols}\vspace{-20pt}
    \begin{equation}\label{eq:z0}
            Z_\mu = 0 ~~~\Longrightarrow~~~ \hat{\Gamma}^i = \ti{\gamma}^{jk}\ti{\Gamma}^i_{~jk}=-\partial_j \ti{\gamma}^{ij}\,.\vspace{-3mm}
    \end{equation}
    The \textit{first order evolution equations} become\footnote{To calculate the covariant derivative, $D_i$, and the Ricci tensor, $R_{ij}$, equations \ref{BSSNOK_help}-\ref{eq:R_3} apply. The matter source terms are the same as for the ADM equations as in \ref{adm_source}. See details in appendix \ref{sec:ngr:ccz4}.}:
    \begin{align}
        \br{\pd_t - \b^k \pd_k} & \chi = \tfrac{2}{d}\chi\br{\alpha K -\partial_i\beta^i}\,,\\
        \br{\pd_t - \b^k \pd_k} & \ti{\gamma}_{ij} =  -2\alpha \ti{A}_{ij}+2\ti{\gamma_{k(i}}\partial_{j)}\beta^k-\tfrac{2}{d}\ti{\gamma_{ij}}\partial_k\beta^k\,,\\
        \br{\pd_t - \b^k \pd_k} & K = -\gamma^{ij}D_jD_i\alpha +\alpha\br{R +2D_i Z^i +K(K-2\Theta)}\\\nonumber
        &~~~~~~ -\tfrac{2d}{d-1}\alpha \kappa_1\br{1+\kappa _2}\Theta+\tfrac{\a\,\k}{2(d-1)}\br{S-d\rho}-\tfrac{2d}{d-1}\a\L\,,\\
        \br{\pd_t - \b^k \pd_k} & \ti{A}_{ij} = \chi \sbr{-D_iD_j\alpha +\alpha\br{R_{ij}+2D_{(i}Z_{j)}-\tfrac{\k}{2} S_{ij}}}^{TF}\\\nonumber
        &~~~~~~~ +\alpha \sbr{\br{K-2\Theta}\ti{A}_{ij}-2\ti{A}_{ik}\ti{A}^k_{~j}} +2\ti{A}_{k(i}\partial_{j)} \beta^k-\tfrac{2}{d}\ti{A}_{ij}\partial_k\beta^k\,,\\
        \br{\pd_t - \b^k \pd_k} & \Theta = \tfrac{\a}{2}\br{R +2D_i Z^i -2K\Theta +\tfrac{d-1}{d} K^2 -\ti{A}_{ij}\ti{A}^{ij}-\k\rho-2\L}\\\nonumber
        &~~~~~~ -Z^i\partial_i\alpha-\alpha \kappa_1\br{2+\kappa_2}\Theta\,,\\
        \br{\pd_t - \b^k \pd_k} & \hat{\Gamma}^i = -2\ti{A}^{ij}\partial_j\alpha +2\alpha\br{\ti{\Gamma}^i_{~jk}\ti{A}^{kj}-\tfrac{d}{2}\ti{A}^{ij}\tfrac{\partial_j\chi}{\chi}-\tfrac{d-1}{d}\ti{\gamma}^{ij}\partial_jK-\tfrac{\k}{2}\ti{\gamma}^{ik}S_k}\\\nonumber
        &~~~~~~ -\ti{\Gamma}^j\partial_j\beta ^i +\tfrac{2}{d}\ti{\Gamma }^i\partial_j\beta ^j+\tfrac{d-2}{d}\ti{\gamma }^{ki}\partial _k\partial_j\beta^j + \ti{\gamma}^{kj}\partial _j\partial_k\beta^i-2\alpha \kappa _1 \ti{\gamma}^{ij}Z_j\\\nonumber
        &~~~~~~ +2\ti{\gamma }^{ij}\br{\alpha \partial _j\Theta -\Theta \partial _j\alpha -\tfrac{2}{d}\alpha K Z_j}+2\kappa _3 \br{\tfrac{2}{d}\ti{\gamma }^{ij}Z_j \partial_k \beta^k-\ti{\gamma }^{jk}Z_j \partial _k \beta ^i}\,,
    \end{align}
\end{summary}

\section{\done{Gauge conditions}}\label{sec:ngr:gauge}

As mentioned previously, there are $D$ parameters freely specifiable throughout the evolution, the lapse function and shift vector, which dictate how grid points (constant coordinates on the numerical grid) move in the spatial hypersurfaces. The choices for these degrees of freedom are called slicing condition and shift condition, as the lapse function dictates the advancement of proper time between spatial slices in relation to the coordinate time (possibly differently at each point in space) and the shift vector shapes how spatial points at rest with respect to normal observers are relabelled when moving to neighbouring hypersurfaces. These conditions are tied to the coordinate system and do not affect physical results, but their choice is fundamental to achieve long term robust numerical evolutions. One such example is the presence of black holes: we want to slow down time around black holes to prevent grid observers from ever reaching the physical singularity, as this would obviously bring a divergence. Following Alcubierre \cite{alcubierre}, we examine a few slicing and shift conditions in the sections below.

\subsection{\done{Slicing conditions}}

The most natural choice for the lapse corresponds to matching coordinate time with proper time, $\a=1$. This corresponds to asking for the coordinate time $t$ to coincide with the proper time of observers following the direction normal to the hypersurface. To see this analytically, one can compute the acceleration vector by looking at how much the normal vector changes in the normal direction:
\begin{gather}
    a_\m = n^\n\grad_\n n_\m = D_\m \ln{\a}\,.
\end{gather}
One can now clearly see that a constant lapse means no acceleration, i.e., normal observers are free falling following geodesics. This is known as \textit{geodesic slicing}. This an inadequate condition as hinted by the black hole problem mentioned above. Geodesics tend to focus in areas of high density and decrease coordinate volume potentially to zero, leading to singular systems. Analytically, the problem lies in having $n^\m\grad_\m K > 0$, leading to ever-growing $K$. From equation \eqref{eq:ngr:K4Ddef} one can see that $K = -\grad_\m n^\m$, the negative divergence of the normal vector, implying that large growing $K$ means converging normal observers.

The natural solution to this issue is trying to impose the condition $K=0=\pd_t K$ at all times, called \textit{maximal slicing}, which ensures volume preservation along the normal direction. Equation \eqref{eq:K_B} implies:
\begin{gather}\label{eq:ngr:maximal_slicing}
    \gamma^{ij}D_jD_i\alpha = \alpha\br{\ti{A}_{ij}\ti{A}^{ij}+\tfrac{1}{d}K^2+\tfrac{\k}{2(d-1)}\br{S+(d-2)\rho} - \tfrac{2}{d-1}\L}\,.
\end{gather}
The main advantage of this condition is called \textit{singularity avoidance}, which means preventing observers from getting too close to either coordinate or physical singularities. This happens by an effect known as \textit{collapse of the lapse}, as this condition forces the lapse to approach zero close to singularities. The interpretation of this is a slow down of the clock of observers falling into singularities, effectively freezing time inside horizons, such that the singularity is never reached in finite coordinate time while numerically the exterior solution is still evolved. The main disadvantage of maximal slicing is that equation \eqref{eq:ngr:maximal_slicing} is an elliptic equation that needs to be solved for each spatial hypersurface at each timestep to determine the lapse at each point, which is extremely expensive numerically. Another side effect of maximal slicing is called \textit{slice stretching}: the slices get distorted as time passes. As more and more grid points evolve to approach the black hole, where the simulation is freezing, the horizon expands and eventually all grid points are inside the black hole, grid points become increasingly far apart in terms of physical distance and the components of the metric develop large gradients.
\begin{figure}[h]
\centering
\includegraphics[width=.6\textwidth]{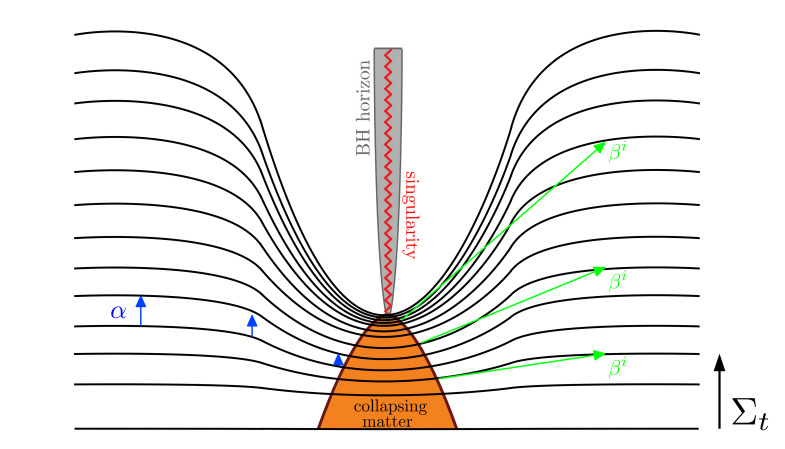}
\caption{Trumpet slicing. The lapse collapses to zero close to a singularity, allowing the exterior spacetime to evolve. The shift also increase close to the centre to fix slice distortion due to slice stretching, discussed in section \ref{subsec:ngr:shift_conditions}. Figure taken from Aurrekoetxea \cite[p.~46]{JosuThesis}.}
\label{fig:ngr:slice_stretching}
\end{figure}
As an alternative to solving an elliptic equation, while trying to preserve the \textit{collapse of the lapse} property, one can use an hyperbolic slicing condition that collapses the lapse to zero if $K$ becomes large (representing collapsing volumes). These are referred to as Bona-Masso family of slicing conditions \cite{Bona:1994dr}, or hyperbolic K-driver, and the most adopted version is called $1+\log$ slicing\footnote{Recent developments in slicing conditions beyond the Bona-Masso family have been proposed by Baumgarte and Oliveira \cite{Baumgarte:2022auj} and Baumgarde and Hilditch \cite{Baumgarte:2022ecu}:
\begin{gather}
    \pd_t \a = -2\a (K-2\Q) + \b^i \pd_i \a\,,
\end{gather}
where the shift term allows the region of small lapse to move with the singularity, and the $2\Q$ term maintains consistency between the CCZ4 and BSSN formulation in the continuum limit (recall that the BSSN equation of motion for $K$ implied subtracting a multiple of the Hamiltonian constraint, represented by $\Q$). The factor of $2$ originates from trying to stabilise simulations. It applies to $4D$ and may require different values in higher dimensions to achieve the same stability.} like a critical collapse of a scalar field, which can either collapse to a black hole or disperse to infinite (details in Alcubierre \cite[p.~136]{alcubierre}). An hyperbolic condition requires only local quantities and derivatives at each point, being well suited to numerical evolution.

With this new condition, we are almost ready to evolve black hole spacetimes. We must only use a clever condition on the shift to overcome the slice stretching problem alluded to previously. There are well known conditions such as harmonic slicing, but they do not bring useful details to the current discussion.

\subsection{\done{Shift conditions}}\label{subsec:ngr:shift_conditions}

Even though often a zero shift condition, $\b^i=0$, is sufficient, for black hole spacetimes this turns out to be a poor choice due to the slice stretching described in the previous section. To tackle it, one needs a shift vector pointing outwards that actively prevents coordinate's timelines to fall into the black hole, as depicted in figure \ref{fig:ngr:slice_stretching}. With this, we can ensure that the size of the black hole stays the same throughout the evolution, i.e. the horizon stays at a finite distance from the centre\footnote{One might thing that taking derivatives across the singular point at the centre is problematic, but as long as the singularity does not coincide with an actual grid point, this in practice does not cause any issues.}. Note that grid points do not have to follow physically allowed paths and can travel faster than the speed of light. Similarly to the maximal slicing, an appropriate shift condition would entail $\ti{\G}^i = 0 = \pd_t\ti{\G}^i$, but this also results in elliptic equations unfeasible to be solved on each slice. We can instead drive $\ti{\G}^i$ to zero dynamically. The most popular way of achieving this is the \textit{Gamma-driver} shift condition \cite{Campanelli:2005dd,Baker:2005vv}:
\begin{equation}
\begin{split}\label{eq:ngr:gamma_driver}
    \pd_t \b^i &= F B^i\,,\\
    \pd_t B^i &= \pd_t \hat{\G}^i - \eta B^i\,,
\end{split}
\end{equation}
where $F$ and $\eta$ are constants. $\hat{\G}^i$ is replaced for $\ti{\G}^i$ when using BSSN instead of CCZ4. This is an hyperbolic condition that arises from noticing that the principal part of the equation \eqref{eq:ngr:bssn_gamma_eom} involves second derivatives of the shift. The choice of $F$ is typically tied to enforcing that the longitudinal gauge modes propagate at the speed of light, resulting in fixing $F = \frac{d}{2(d-1)}$ \cite{Kunesch:2018jeq}. The parameter $\eta$ is typically chosen to be of order $\eta \approx M^{-1}$, where $M$ is the spacetime mass, in order to damp unwanted oscillations of the shift over long timescales. The Gamma-driver shift condition above can be integrated to yield a form that does not require the auxiliary variable $B^i$, but instead requires storing the initial value of the $\hat{\G}^i$ variables \cite{vanMeter:2006vi}. As a final note, the superluminal gauge speeds that this condition generates can lead to numerical instabilities. These can be overcome by computing advection term derivatives, $\b^k\pd_k$, using one-sided derivatives.

The Gamma-driver shift condition together with $1+\log$ slicing condition, is called the moving puncture gauge. This has been one of the key developments in the field of numerical relativity, as it allowed effective evolution of black hole spacetimes in the BSSN and CCZ4 formulations \cite{Campanelli:2005dd,Baker:2005vv}. Other conditions, such as the \textit{minimal distortion} \cite{Smarr:1977uf,Smarr:1978dia}, the \textit{Gamma freezing} or the \textit{generalised harmonic} shift conditions will not be detailed.

\section{\done{Initial data}}\label{sec:ngr:initialData}

Given a coordinate grid on some spatial slice, one must specify initial data at each point for the spatial metric, $\g_{ij}$, and the extrinsic curvature, $K_{ij}$, together with the matter components $\r,S^i,S^{ij}$. This data is not completely free, as it must satisfy the Hamiltonian and Momentum constraints \eqref{eq:H_ADMf}-\eqref{eq:M_ADMf}. These are elliptic equations that constraint the fields in space in each instant of time. The variables $(\g_{ij}, K_{ij})$, as $d$ dimensional symmetric tensors, have in total $d(d+1)$ components that must be specified for the initial time slice, while satisfying the $d+1$ constraint equations, leaving $d^2-1$ free degrees of freedom to specify. The main challenge is understanding how to choose these variables according to the physical knowledge about the system (e.g. two black holes, a scalar field cloud, etc.).

A common choice for initial data involve setting the conformal metric to be flat, $\g_{ij} = \d_{ij}$, or the extrinsic curvature or its trace to be zero, $K_{ij}=0$ or $K=0$. This may not be a natural initial state (i.e. though physically possible, represents an artificial distortion into something that does not form dynamically), potentially leading to spurious burst of gravitational wave emissions (known as \textit{junk radiation}) on the initial time steps of a simulation while the simulation settles down\footnote{This can be improved with methods proposed by Gleiser et al. \cite{Gleiser:1999hw,Gleiser:1997ng}.}. For example, superposing initial data of black holes or scalar field clouds that have $K_{ij}=0$, i.e. are stationary, is not natural as clearly two bodies do not pop our of nowhere close to each other without an existent gravitational pull driving an inward velocity. One can either solve more complex equations to find more accurate solutions or find a configuration with acceptable junk radiation that does not affect significantly physical observables like initial mass or total energy radiated.

About the gauge variable lapse and shift, even though they commonly do not affect the constraints (only if the matter terms depend on them due to dependence on the coordinate system), a common choice is simply $\a=1$ and $\b^i=0$. With correct gauge evolution equations, such as the moving puncture gauge, these quickly settle into dynamically appropriate values\footnote{Typically in the timescale of $\sim 20-30M$, where $M$ is the mass of the spacetime. See appendix \ref{appendix:grchombo:ah_location} for an example.}, such as having the lapse collapsed and the shift pointing radially outwards. This often leads to a lot of non-physical yet intense dynamics during the initial stage of a simulation until the gauge reaches equilibrium. One can minimise this with cleverer choices, such as a pre-collapsed lapse, $\a=\sqrt{\chi}$.

Several methods have been developed to attempt to separate the constrained parts of the fields into some form that simplifies finding a solution for the initial data (see Cook \cite{Cook:2000vr} for a review). Among others, the most commonly used ones are the \textit{conformal transverse-traceless decomposition} and the \textit{conformal thin-sandwich decomposition} (see Baumgarte and Shapiro \cite[chapter 3]{shapiro} and Alcubierre \cite[chapter 3]{alcubierre}). We shall describe an adaptation of the former in the next section, as it proves relevant for the discussions of chapter \ref{chapter:paper1}-\ref{chapter:eft}. Additionally, for data only violating the Hamiltonian constraint (when the Momentum constraint can be trivially satisfied), one can use a simple yet inefficient method of relaxing the conformal factor using a decay equation as $\pd_t \chi = c_R \mathcal{H}$, where $c_R$ is the relaxation speed constant. Alternatively, for complex systems where exact initial data is hard to find but where almost constraint-satisfying solutions exist (e.g. using a black hole solution in a modified theory of gravity slightly perturbing GR), one can use the CCZ4 formulation (see section \ref{sec:ngr:ccz4}), which quickly damps any violations away and generically provides very robust results. Finally, a recent breakthrough in full initial condition solvers has been achieved with the CCTK method \cite{Aurrekoetxea:2022mpw}.

\subsection{\done{Conformal transverse-traceless decomposition}}\label{subsec:ngr:conformalTT}

Intentionally breaking apart degrees of freedom, we start by decomposing $(\g_{ij}, K_{ij})$ into the conformal metric and trace of extrinsic curvature, $(\chi,\ti{\g}_{ij},K,\ti{A}_{ij})$, as described in section \ref{sec:ngr:bssnok}. This is where we differ from the presentation by Baumgarte and Shapiro \cite[section 3.2]{shapiro}, as the rescaling of $\ti{A}_{ij}$ uses a distinct power of the conformal factor. The traceless tensor $\ti{A}_{ij}$ can be split into two components, using a scalar-vector-tensor decomposition, by splitting it into a transverse-traceless divergenceless part and a symmetric traceless gradient of a vector:
\begin{equation}\label{eq:ngr:AijConformalTT}
    \ti{A}^{ij} = \ti{A}_{TT}^{ij} + \ti{A}_V^{ij}\,,
\end{equation}
where the component $(\ti{A}_{TT}^{ij}, \ti{A}_V^{ij})$ satisfy:
\begin{gather}
    \ti{D}_j\ti{A}_{TT}^{ij} = 0\,,\\\label{eq:ngr:AijToW}
    \ti{A}_V^{ij} = 2\,\ti{D}^{(i}W^{j)} - \tfrac{2}{d}\ti{\g}^{ij}\ti{D}_k W^k\,,
\end{gather}
where $W^i$ is a vector potential. Note the original $\frac{d(d+1)}{2}-1$ degrees of freedom of $\ti{A}^{ij}$ are split into $\frac{d(d-1)}{2}-1$ degrees of freedom in $\ti{A}_{TT}^{ij}$ and $d$ for $W^i$.
Relating back to the momentum constraints \eqref{eq:M_BSSNf}, one can write:
\begin{equation}
    \ti{D}_j \ti{A}^{ij} = \ti{D}_k\ti{D}^k W^i + \tfrac{d-2}{d}\ti{D}^i\br{\ti{D}_jW^j} + \ti{R}^i_{~j}W^j := \br{\ti{\D}_V W}^i\,,
\end{equation}
where $\ti{R}^i_{~j}$ is the Ricci tensor associated with the conformal metric $\ti{\g}_{ij}$ and $\ti{\D}_V$ called the \textit{vector Laplacian}. Finally, one can re-write the system of constraint equations to solve as:
\begin{align}
    \mathcal{H} &= \tfrac{1}{2}\sbr{\chi\ti{R} + (d-1)\ti{D}^i\ti{D}_i \chi - \tfrac{(d+2)(d-1)}{4\chi}\ti{D}^k\chi\ti{D}_k\chi + \tfrac{d-1}{d}K^2 - \ti{A}^{ij}\ti{A}_{ij} - \k\r - 2\L} = 0\,,\\\label{eq:ngr:initialData:momentum}
    \mathcal{M}_i &= \br{\ti{\D}_V W}_i - \tfrac{d}{2}\ti{A}_{ij}\ti{D}^j\ln{\chi} - \tfrac{d-1}{d}\pd_i K - \tfrac{\k}{2} S_i = 0\,.
\end{align}
These equations can be solved for the $d+1$ variables $(\chi,W^i)$, while $\ti{\g}_{ij}$, $K$ and $\ti{A}_{TT}^{ij}$ have $d^2-1$ freely specifiable parameters. From $W^i$, one can reconstruct both $\ti{A}_{ij}$ or $K_{ij}$ and obtain the physical initial data needed. We do not go into details of specific solutions, but some applications, such as spinning and boosted black holes, and simplification\footnote{Note that the different rescaling of $\ti{A}_{ij}$ done in Baumgarte and Shapiro \cite[p.~64]{shapiro} is helpful to eliminate the second term in equation \eqref{eq:ngr:initialData:momentum}. This allows to solve the equation explicitly for a couple of black hole spacetimes. For our purposes, the Horndeski equations, with non-vacuum and non-trivial $\r$ and $S_i$, result in Hamiltonian and Momentum coupled equations that we solve numerically, and such simplifications are not relevant.} can be found in chapter 3.2 and appendix B of Baumgarte and Shapiro \cite{shapiro}. For instance, the solution to a boosted black hole found therein is used in chapter \ref{chapter:eft}. Alternatively, one can use numerical methods, by providing an initial guess $(\chi_0, W^i_0)$, and solving perturbatively for $\chi = \chi_0 + \d\chi$ and $W^i = W^i_0 + \d W^i$ \cite{JosuThesis, Aurrekoetxea:2022mpw}.

\section{\done{Boundary conditions}}

Boundary conditions is an essential ingredient, but has proven to have easier solutions to reach good numerical stability and accuracy. Many codes use unphysical exterior boundary conditions that are easy to implement numerically, such as periodic boundary conditions (which in $3+1$ dimensions is equivalent to evolving a spacetime with the topology of a 3-torus), or symmetric\footnote{Often called \textit{reflective} as well.} boundary conditions, but one can typically get away with this by using a big enough numerical grid, such that effects from the boundary either never affect the centre of the grid due to the finite speed of light or do physically reach it but with minor to no effects. Symmetric boundary conditions can be truly useful to evolve only part of the grid; for instance evolving one eighth of the $3d$ grid for an equal mass head on black hole collision that has axisymmetry over the collision axis.

A common condition in numerical relativity is the Sommerfeld or radiative boundary condition \cite{Alcubierre:2002kk}, which enforces that outgoing waves are not reflected back into the computational domain. For this effect, the evolution equations at the boundary cells of the computational domain are changed to:
\begin{equation}
    \pd_t \xi = -\frac{x^i}{r}\pd_i\xi - \frac{\xi - \xi_\infty}{r}\,,
\end{equation}
where $\xi$ represents any evolution variable and $\xi_\infty$ represents the desired asymptotic value for it, typically the value it would take in Minkowski space. Other used boundary conditions include static and extrapolating boundary conditions.

\section{\done{\grchombo numerical code}}\label{sec:grchombo_scheme}


\grchombo \cite{Clough:2015sqa, Andrade:2021rbd, Radia:2021smk} is an open-source code for performing numerical relativity evolutions, built on top of the publicly available \textit{Chombo} software \cite{Adams:2015kgr} for the solution of PDEs. This is the software used to perform the simulations in this thesis, and in this chapter we describe its core characteristics. It is one of the premier codes for numerical relativity, focused on physics problems that require high flexibility and adaptability of grid structure and ease of code modification. It is this feature that greatly distinguishes it from other numerical relativity codes such as PAMR/AMRD \cite{Pretorius:2004jg, Pretorius:2005ua}, \textsc{LEAN} \cite{Sperhake:2006cy}, \textsc{BAM} \cite{Thierfelder:2011yi}, \textsc{NRPy+} \cite{Ruchlin:2017com} or pseudospectral codes for gravitational wave templates such as \textsc{SpEC} \cite{Pfeiffer:2002wt}, among others\footnote{For an overview of numerical relativity codes, see Sperhake \cite{Sperhake:2014wpa}.}. \grchombo allows simulating non-trivial structures \cite{Andrade:2021rbd}, such as ring-like configurations \cite{PhysRevD.99.104028}, cosmological spacetimes \cite{Aurrekoetxea_2020}, higher dimensional black rings and black strings \cite{Andrade:2020dgc,Bantilan:2019bvf,Figueras:2017zwa,Figueras:2015hkb}, or modified gravity systems \cite{Figueras:2020dzx, Figueras:2021abd}. Videos of simulations using \grchombo can be viewed via the website \href{https://www.youtube.com/@grchombo1458}{https://www.youtube.com/@grchombo1458} and more information can be found in \href{https://www.grchombo.org/}{https://www.grchombo.org/} or \href{https://github.com/GRChombo/GRChombo/wiki}{https://github.com/GRChombo/GRChombo/wiki}.

\subsection{\done{Numerical features}}

\grchombo is primarily written in C\texttt{++}, heavily using class structure object oriented programming and templating, in order to make the code modular for the various processes and computer cores.

\subsubsection{\done{Standardised output and visualisation}}

\chombo uses the HDF5 output format \cite{hdf5} for loading and writing files in parallel, which is supported by many popular visualisation tools such as VisIt \cite{Childs_VisIt_An_End-User_2012} or ParaView \cite{ahrens2005paraview}, as well as data analysis tools, such as python YT \cite{YT_ref}. In addition, the output files can be used as input files if one chooses to continue a previously stopped run.

\subsubsection{\done{Space and time discretisation}}\label{subsubsec:ngr:space_time_discretization}

\grchombo evolves PDEs using the method of lines with fourth-order Runge-Kutta method for time discretisation. For spatial discretisation, it uses fourth or sixth order centred stencils, except for advection terms, which use lopsided stencils. \grchombo uses typically a CFL (\textit{Courant-Friedrichs-Levy}) condition \cite{Courant_1928} of $\a_C = \frac{\D t}{\D x} = \frac{1}{4}$. In appendix \ref{appendix:grchombo:scaling} we show convergence of these methods.


\subsubsection{\done{Berger-Rigoutsos adaptive mesh refinement}}

AMR (Adaptive Mesh Refinement) is a technique used to tackle the problem of simulations with a large range of dynamic spatial and temporal scales combined with the limit of computational resources available. Numerical relativity requires a mesh that dynamically adjusts itself in response to the underlying physical system, following a certain refinement criteria. The numerical scheme to increase resolution in regions of interest is called \textit{mesh refinement}. If the mesh and its size is specified in advance, this is typically called \textit{moving box refinement}, whereas if it is dependent on some dynamic criterion, it is called \textit{adaptive mesh refinement}. If the refinement regions can have arbitrary shape and topology, the implementation is called \textit{fully adaptive mesh refinement}. The later is only needed for complex dynamics problems, such as the ones \grchombo was designed for.

In block-structured AMR, first described and implemented by Berger et al. \cite{BERGER1984484}, the computational domain is built from a hierarchy of increasingly fine levels, with each one containing a set of (not necessarily contiguous) boxes of meshes, with the only condition being that a finer mesh must lie on top of one or possibly more meshes from the next coarsest level. \grchombo uses Berger-Oliger style AMR \cite{BERGER198964} with Berger-Rigoutsos block-structured grid generation \cite{BERGER120081}. For a simple explanation, see Clough \cite{Clough:2017ixw} and Kunesch \cite{Kunesch:2018jeq}. Although AMR brings great flexibility and efficiency, it creates a new set of problems, such as unwanted interpolations for regridding\footnote{Intermediate values required by Runge-Kutta timestepping actually involve third order polynomial interpolation in time, potentially generating errors that can effect the whole convergence order of simulations from fourth to third order convergence.}, prolongation errors and additional ghost cells required for interpolation between coarse and fine grids, and artificial boundaries that create spurious unphysical reflections. As hinted before, AMR also requires a fine-tuned user-defined criteria to determine the tagging of grid cells for refinement, which in itself can also be hard to control if one wants to keep all regions of interest accurately resolved.

\begin{figure}[h]
\centering
\begin{multicols}{2}
\includegraphics[width=.45\textwidth]{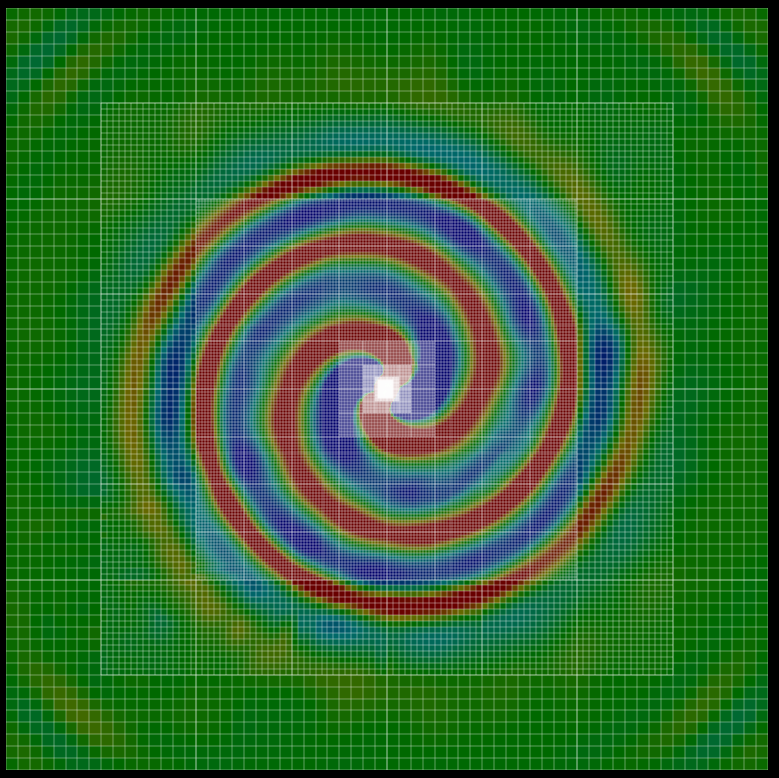}
\includegraphics[width=.45\textwidth]{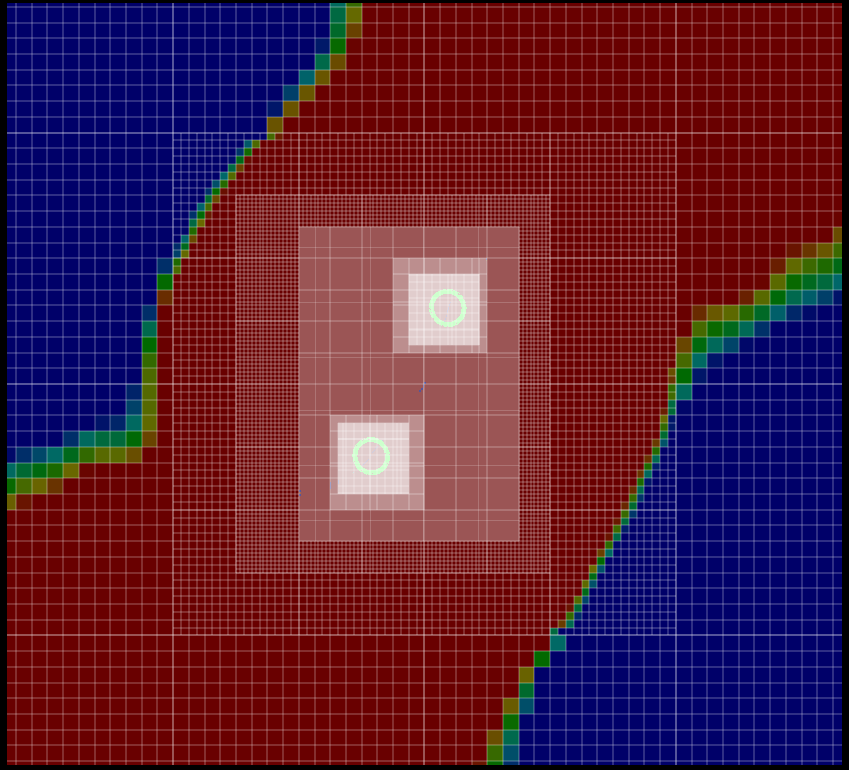}
\end{multicols}
\caption{Display of multiple levels in mesh refinement in a BBH simulation. The colour scheme represents the real part of the Weyl scalar $\Psi_4$. On the right, we see a zoom in of the left figure with yet more refinement levels, and two circumferences representing the horizons of the two black holes. The tagging used is $\chi$ and puncture tagging mentioned in \ref{appendix:grchombo:tagging}.}
\label{fig:ngr:mesh_amr}
\end{figure}

In \grchombo, the hierarchy of levels consists of $l_{\text{max}}+1$ refinement levels labelled $l=0,...,l_{\text{max}}$, with grid spacing doubling from level to level $\D x_l = \D x_0 / 2^l$. Each level has its mesh split into boxes distributed between CPUs as described in section \ref{subsubsection:ngr:parallelization}. This hierarchy of levels can be seen in figure \ref{fig:ngr:mesh_amr}. Note that each level, having a smaller grid spacing $\D x$, must also have a smaller timestep $\D t$ to preserve the CFL condition, implying that deeper levels run $2^l$ timesteps for every full timestep of the coarsest level 0.

Tagging criteria are a Boolean flag defined by the user on each cell (e.g. tag if the derivative of the conformal factor is bigger than some threshold $\t_R$; see more information in section \ref{appendix:grchombo:tagging}). Block-structured AMR partitions the mesh based on tagged cells, taking into account inflection points and clusters of cells before deciding whether to regrid and how to split the new grid into several boxes. The minimum and maximum size of boxes, how frequently to regrid each level, how many cells need to be tagged in a region to proceed to regridding and buffer regions of forced tagging around a given identified tagged cell\footnote{This is useful to set mesh boundaries further apart from each other and reduce errors from high frequency resonances bouncing off neighbouring boundaries.} are some of the parameters a user can tune.

\subsubsection{\done{Parallelisation}}\label{subsubsection:ngr:parallelization}

\begin{figure}[h]
\centering
\includegraphics[width=.85\textwidth]{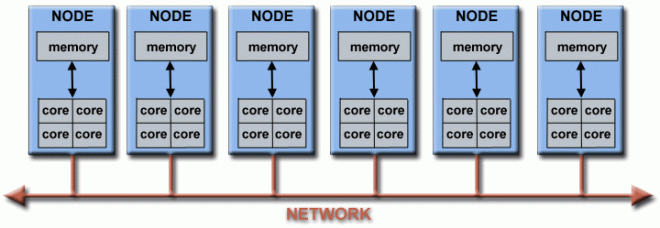}
\caption{Diagram displaying the layout of a computing networks, where multiple stand-alone nodes, each with multiple cores and shared memory, communicate to make a larger parallel computer clusters. Figure taken from \href{https://hpc.llnl.gov/documentation/tutorials/introduction-parallel-computing-tutorial}{https://hpc.llnl.gov/documentation/tutorials/introduction-parallel-computing-tutorial}.}
\label{fig:ngr:mpi_openmp}
\end{figure}

To take advantage of performance scaling of modern computing architectures, \grchombo makes use of parallelism at different scales: each AMR level is split into boxes, distributed between cores of computing nodes using Message Passing Interface (MPI); evaluations of the right hand side of evolution equations for all cells within a box are threaded over the $y$ and $z$ directions using OpenMP, taking advantage of several CPUs available to a given MPI process; and finally over the $x$ direction, \grchombo relies on \textit{vectorisation} (also known as SIMD or vector intrinsics), templated away from the user using C\texttt{++}14 templates, to utilise the full vector-width of the target architecture\footnote{Typically compilers will automatically do this, but the complexity of the CCZ4 equations typically results in a failure of auto-vectorisation.} \cite{Kunesch:2018jeq}. Represented in figure \ref{fig:ngr:mpi_openmp}, the main difference between MPI and OpenMP is the use of shared memory between cores, for OpenMP, versus distributed memory for MPI. The later requires more independent work for each node since communication incurs less efficiency. This hybrid OpenMP/MPI parallelism with explicit vectorisation via intrinsics makes \grchombo an efficient strong scaling code up to thousands of CPU cores (see appendix \ref{appendix:grchombo:scaling}).

Load balancing seeks to avoid the situation where most nodes are waiting for a subset of other nodes to finish their computational work before proceeding to the next timestep. This is performed following the Morton ordering algorithm, which splits work among nodes assuming work is proportional to the number of cells in a box, and which also performs an exchange phase to maximises neighbouring box communication. Note that each cell can evaluate derivatives numerically using neighbouring cell in the same box, or ghost cells in their box whose values are exchanged with other boxes by interpolation at the end of each step.

For efficient load balancing, the decision of more or less MPI process and OpenMP threads does not follow easy general rules. Using less but bigger boxes is more efficient\footnote{Given the use of fourth order stencils, each box requires three ghost cells. This implies that a box of side $N^3$ has an effective side of $(N+6)^3$, making the volume with physical cells only $38\%$ for $N=16$ or $60\%$ for $N=32$.}, but limits how many nodes the code can scale to and may not be suitable for systems with less memory available per node. Using more and smaller boxes can make use of more nodes and MPI processes (and hence make the code faster), but also increases the percentage of ghost cells per box which, given the added communication and computation required, may worsen performance. \texttt{GRChombo} performs best for boxes of side $16$ or $32$, and typically more than $2$ or $4$ OpenMP threads does not improve performance, meaning that more MPI processes is typically preferred over adding OpenMP threads \cite{Kunesch:2018jeq}.

\subsubsection{\done{Dissipation}}\label{subsub:ngr:dissipation}

Finite difference methods can often introduce spurious high-frequency modes, particularly when using adaptive mesh refinement and regridding. To control such high-frequency noise, from both truncation and regridding interpolation errors, \grchombo uses $N=3$ Kreiss-Oliger dissipation \cite[p.~343]{alcubierre} \cite{Kreiss:1973,Calabrese:2003vx}, which adds a diffusion operator of amplitude $\s$ to the right hand side of each evolution variable. This is linearly stable if $\s < \frac{2}{\a_C}$, where $\a_C$ is the CFL condition as in section \ref{subsubsec:ngr:space_time_discretization}. It is conventional wisdom to use $N$ such that $2N-1 \ge m$, where $m$ is the order of the finite difference scheme \cite[p.~345]{alcubierre}. For \grchombo, even with sixth order spatial stencils, time stepping is still fourth order, and hence $N=3$ dissipation operator does not affect stability properties. Some projects have found that having a level dependent $\s$ is relevant for stability \cite{Bozzola:2021elc}, even though this is not part of the standard \texttt{GRChombo} code.


\subsubsection{\done{Convergence}}

Appendix \ref{appendix:grchombo:convergence} discusses basic convergence concepts to evaluate the order a finite differences scheme has in practice. In AMR it proves difficult to test and obtain convergence, because of the lack of predictability of where meshes are and behave exactly in a given region of space. Due to the discrete side length that level boxes have, low resolutions simulations may not have the exact same mesh as high resolution simulations which have smaller boxes (in physical length, though potentially the same in number of cells). Besides this, differences in truncation error may by themselves lead to different cells being tagged. One must tune the tagging criteria parameters to attempt to make refinement regions similar over different resolutions. Nevertheless, if one ensures that physically significant regions have approximately the same refinement, convergence can typically be demonstrated, as shown in appendixes \ref{appendix:paper1:convergence} and \ref{appendix:paper2:convergence}. Some alternatives are using truncation error itself as a tagging criteria, or enforcing specific regions to regrid (e.g. a sphere of fixed radius). For details of these and other approaches, see Radia et al. \cite{Radia:2021smk}. Taking into account sections \ref{subsubsec:ngr:space_time_discretization} and \ref{subsub:ngr:dissipation}, in \grchombo one should expect between third to fourth order convergence.


\subsection{\done{Physical methods}}

\subsubsection{\done{Evolution equations}}

\grchombo implements the BSSN/CCZ4 formalism with moving puncture gauge. It implements periodic, Sommerfeld/radiative, extrapolating and symmetric/reflective boundary conditions. It provides analytical and semi-analytical initial data for black hole binaries, Kerr black holes and scalar matter. The code also modified the open-source code TwoPunctures \cite{PhysRevD.70.064011} for accurate spinning black hole binary data.

The equations of motion are evolved with \grchombo \cite{Clough:2015sqa, Andrade:2021rbd, Radia:2021smk, Andrade:2021rbd}, using MPI (Multi Processor Interface), OpenMP and templated vector intrinsics/SIMD (Single Instruction, Multiple Data) to obtain a good performance in the most common architectures. \grchombo uses the \texttt{Chombo} adaptive mesh refinement libraries \cite{Adams:2015kgr}. 

\grchombo implements the usual puncture gauge (i.e., $1+\log$ slicing plus Gamma driver \cite{Bona:1994dr,Campanelli:2005dd,Baker:2005vv}) for the evolution of the gauge variables and $N=3$ \textit{Kreiss-Oliger} (KO) dissipation.

\subsubsection{\done{Observables}}\label{subsubsection:ngr:observables}

\grchombo includes key diagnostics necessary to analyse numerical relativity simulations. These include:
\begin{itemize}
    \item \textit{Puncture tracking}: to track the movement of punctures when moving black holes exist in the grid, \grchombo uses the shift to store the positions $x^i_{\text{punc}}$ of each puncture \cite{Bruegmann:2006ulg}, by integrating:
    \begin{equation}
        \pd_t x^i_{\text{punc}} = - \b^i(x^i_{\text{punc}})\,.
    \end{equation}
    \item \textit{Apparent Horizon Finder}: this feature was developed during this thesis and made public in the code. It allows to easily track the position of multiple black holes over time and determine their properties, such as mass and spin. See details in appendix \ref{chapter:ahfinder}.
    \item \textit{Interpolation}: using an AMR grid interpolator, users can interpolate grid variables and its derivatives at any arbitrary position in the computational domain. This can be done up to arbitrary order in interpolation accuracy, but typically fourth order Lagrange polynomial interpolation is used \cite{Tunyasuvunakool:2017wdi}. Given this, one can perform extraction of values in $d-1$ dimension surfaces such as spheres, planes or cylinders, and then integrate them using a custom integration method (trapezium rule, Simpson's rule, Boole's rule, etc.).
    \item \textit{Wave extraction}: using spherical extractions with the interpolator, one can use the Newman-Penrose formalism \cite{Newman:1961qr} to compute the Weyl scalar $\Psi_4$ (include Z4 terms as described in appendix \ref{appendix:grchombo:weyl}) on multiple spheres of fixed radius and compute the modes $\psi_{\ell m}$ with respect to spin-weight -2 spherical harmonics $_{-2}Y^{\ell m}$ \cite[appendix D]{alcubierre} using:
        \begin{equation}
            r_{\text{ex}}\psi_{\ell m} = \oint_{S^2} r_{\text{ex}} \Psi_4\big|_{r=r_{\text{ex}}}\cdot \br{_{-2}Y^{\ell m}} d\W\,,
        \end{equation}
    where $d\W=\sin{\q}\,d\q\, d\f$ is the area element over the unit sphere $S^2$ and $r_{\text{ex}}$ is the extraction radius. \grchombo uses trapezium rule for the integration over $\f$ (since the periodicity means that any quadrature converges exponentially \cite{Poisson:1827}) and Simpson’s rule for the integration over $\q$.
    \item \textit{Extracting mass and momenta}: even though in general relativity there is no globally well defined law of conservation of energy and momenta from both matter and gravitational sectors, one can find approximations to define the mass $M$, momentum $P^i$ and angular momentum $J^i$ of a spacetime.
    
    One can measure the energy from the matter sector using the matter decomposition of the stress-energy tensor, $\r$ and $S^i$ as:
    \begin{equation}
        M = \int \r \sqrt{\g}\, dV\,\quad P^i = \int S^i \sqrt{\g}\, dV\,\quad J^i = \int \e^{ijk}x_jS_k \sqrt{\g} dV\,,
    \end{equation}
    where $\e^{ijk}$ is the Levi-Civita symbol.
    
    An alternative method consists of computing the ADM measures, defined as $d-1$ dimensional surface integrals at infinity. See appendix \ref{appendix:paper1:adm_mass} for details on the ADM mass and \cite[p.~83]{shapiro} for more details on linear and angular momentum.
    
    Even though these integrals should be evaluated up to infinity, \grchombo measures them as far out in the computational domain as possible, and post-processing analysis allows to extrapolate their values to infinity (see appendix \ref{appendix:sec:richardson_extrapolation} for details on extrapolation).
    \item \textit{ADM Constraints}: the Hamiltonian and Momentum constraints are also diagnostics that can be computed over the grid, and also its $L_1$ or $L_2$ norm over a part or the full computational domain.
\end{itemize}

\subsubsection{\done{Parameter choices}}\label{subsec:ngr:grchombo_params}

To carry out the numerical simulations presented in chapter \ref{chapter:paper1}-\ref{chapter:eft}, we used \grchombo, a multipurpose numerical relativity code \cite{Clough:2015sqa}\footnote{See also \href{www.grchombo.org}{\texttt{www.grchombo.org}}.} that implements the BSSNOK \cite{Nakamura:1987zz,Shibata:1995we,Baumgarte:1998te} or CCZ4 \cite{Alic:2011gg,Alic:2013xsa,Bona:2003fj,Bernuzzi:2009ex,Bona:2003qn} formulations of the Einstein equations. 

\grchombo implements Bona-Masso slicing conditions \cite{Bona:1994dr} and \textit{gamma-drive} shift conditions as its gauge equations, generalised with the following parameterisation:
\begin{align}
    \partial_t\alpha & = -\mu_{\alpha_1}\alpha^{\mu_{\alpha_2}} \br{K-2\Theta} + \mu_{\alpha_3}\beta^k\partial_k\alpha\,,\\
    \partial_t \beta^i & = FB^i + \mu_B\beta^k\partial_k\beta^i\,,\\
    \partial_t B^i & = \partial_t \hat{\Gamma}^i - \eta B^i + \mu_B (\beta^k\partial_k B^i - \beta^k\partial_k\hat{\Gamma}^i)\,.
\end{align}
Common choice of parameters are $\mu_{\alpha_1}=2$, $\mu_{\alpha_2}=1$, $\mu_{\alpha_3}=1$ for the slicing condition ($1+\log$ slicing) and the usual $F=\tfrac{3}{4}$, $\eta=1$, $\mu_B=0$ for the shift condition (usually with $\eta\approx\tfrac{1}{M_{ADM}}$).

Other typically parameter choices in \grchombo are $G=1$, Courant factor (also known as CFL condition, the ratio between the discretisation in space and time $\frac{\D t}{\D x}$) of $0.25$ and, as CCZ4 parameters, $\kappa_1 = \tfrac{0.1}{\alpha}$, $\kappa_2 = 0$, $\kappa_3 = 1$\footnote{This is the default in the code used, namely \grchombo, and it greatly improves the stability of black hole spacetimes, see Andrade et al. \cite{Andrade:2021rbd} for detailed studies.}. These can be assumed to be the choices in most simulations, except if made explicit otherwise.

\blankpage 
\part{\done{Research work}}
\blankpage

\chapter{\done{Gravitational collapse in cubic Horndeski theories}}\label{chapter:paper1}

\section{\done{Introduction}}\label{sec:paper1:introduction}

In this chapter we study the non-linear regime of a subclass of Horndeski theories for which \cite{Kovacs:2019jqj} found a well-posed CCZ4 formulation of the Einstein equations. In this chapter, unlike \cite{Okounkova:2017yby,Witek:2018dmd,Okounkova:2019dfo,Okounkova:2020rqw}, we consider the theory in its full non-linear splendour, which allows us to explore its distinctive non-perturbative physics; our goal is to identify the weakly coupled regime of the theory so that it can be consistently treated as a valid EFT from which one can obtain meaningful predictions.  Rather than studying a specific phenomenologically viable theory, our ultimate goal is to identify general features in the waveforms that do not depend on the details and that can be attributed to the higher derivatives and non-linearities in the action. Therefore, we treat it as a toy model that can give us a glimpse of the  type of effects that one can expect in more complicated theories which involve higher derivatives of the spacetime metric tensor.

In detail, we study gravitational collapse and black hole formation in cubic Horndeski theory. Our goal is to identify the region in the space of couplings for which the Horndeski theories under consideration are weakly coupled throughout the evolution.

\subsection{\done{Summary of the results}}

We consider gravitational collapse in Horndeski theories using as initial data a spherically symmetric lump of scalar field \eqref{eq:paper1:phiData}. Even though the initial data is spherically symmetric, we evolve it using a 3+1 evolution code based on \texttt{GRChombo} \cite{Clough:2015sqa}, without symmetry assumptions. We have also considered gravitational collapse of some non-spherical scalar field configurations but we did not observe significant differences from the spherically symmetric case. However, a thorough study of gravitational collapse beyond spherical symmetry in Horndeski theories is beyond the scope of this chapter.  

Before we describe our results, we comment on previous works that are directly related to ours. Gravitational collapse and black hole dynamics in spherical symmetry in Einstein-dilaton-Gauss-Bonnet (EdGB) theory has been studied before \cite{Ripley:2019hxt,Ripley:2019irj,Ripley:2019aqj,Ripley:2020vpk,R:2022hlf}. This theory can be considered to be a member of the Horndeski class, but the mapping between the two is highly non-trivial \cite{Kobayashi:2011nu}. In these papers mentioned, its authors study, among other things, the hyperbolicity of the equations of motion in various regions of the spacetime, including the interior of black holes, as a function of the coupling. They show that for large enough couplings the equations of motion can change character from hyperbolic to elliptic, even outside black holes, in which case one cannot solve them as an evolution problem. In a related work, Bernard et al. \cite{Bernard:2019fjb} consider the conditions under which one may be able to construct global solutions of Horndeski theories. In this paper, the authors study in detail the hyperbolicity of the equations of motion and the pathologies that may arise during the evolution in some specific examples. They also perform  numerical simulations of spherically symmetric scalar field collapse to illustrate the breakdown of the hyperbolicity at strong coupling in different situations. Our work can be considered as an extension of these papers in different directions, as we now explain.

In this article we consider the so called cubic Horndeski theories \eqref{eq:cubic_horndeski}, for which Kovács \cite{Kovacs:2019jqj} showed that they have a well-posed initial value problem in the CCZ4 formulation of the Einstein equations and in puncture gauge. Because we are not particularly interested in a specific theory but rather in 
identifying general features of the non-linear dynamics of Horndeski theories, we consider two particularly simple and illustrative cases, see equation \eqref{eq:relevantGs}. In fact, from the point of view of EFT, the $G_2$ theory considered here, equation \eqref{eq:relevantGs}, is the most general scalar matter term up to four derivatives that one can include to the action \cite{Weinberg:2008hq}. In order for these theories to make sense as EFTs, the Horndeski terms have to be suitably small compared to the GR terms. Indeed, the well-posedness result of Kovács \cite{Kovacs:2019jqj} only holds if a certain weak field condition is satisfied. For the class of theories that we consider, the relevant weak field conditions are given by \eqref{eq:paper1:WFC_2}. The main goal of this chapter is to identify the region in the space of initial conditions and couplings for which the weak field conditions \eqref{eq:paper1:WFC_2} are small at all times.  

In our simulations of scalar field collapse we keep the radius $r_0$ and width $\omega$ of the initial Gaussian lump fixed, and vary both the amplitude $A$ and Horndeski coupling ($g_2$ or $g_3$ depending on the theory under consideration). For every pair $(A,g_2)$ or $(A,g_3)$, we monitor both the character of the equations of motion of the scalar field\footnote{The evolution equations for the metric are given by the CCZ4 equations which are strongly hyperbolic.} and the weak field conditions \eqref{eq:paper1:WFC_2} everywhere in spacetime, except in a certain region of the interior of black holes when they form.  It seems reasonable to accept the breakdown of EFT in a region sufficiently close to a singularity as long as this region is covered by a horizon. In this case, there is no loss of predictivity since this region is causally disconnected from the Universe outside the black hole, where EFT remains valid. The same criterion was adopted by Ripley and Pretorius \cite{Ripley:2019aqj}.  

\begin{figure}[h]
\centering
\hspace*{-10mm}\includegraphics[width=1.15\textwidth]{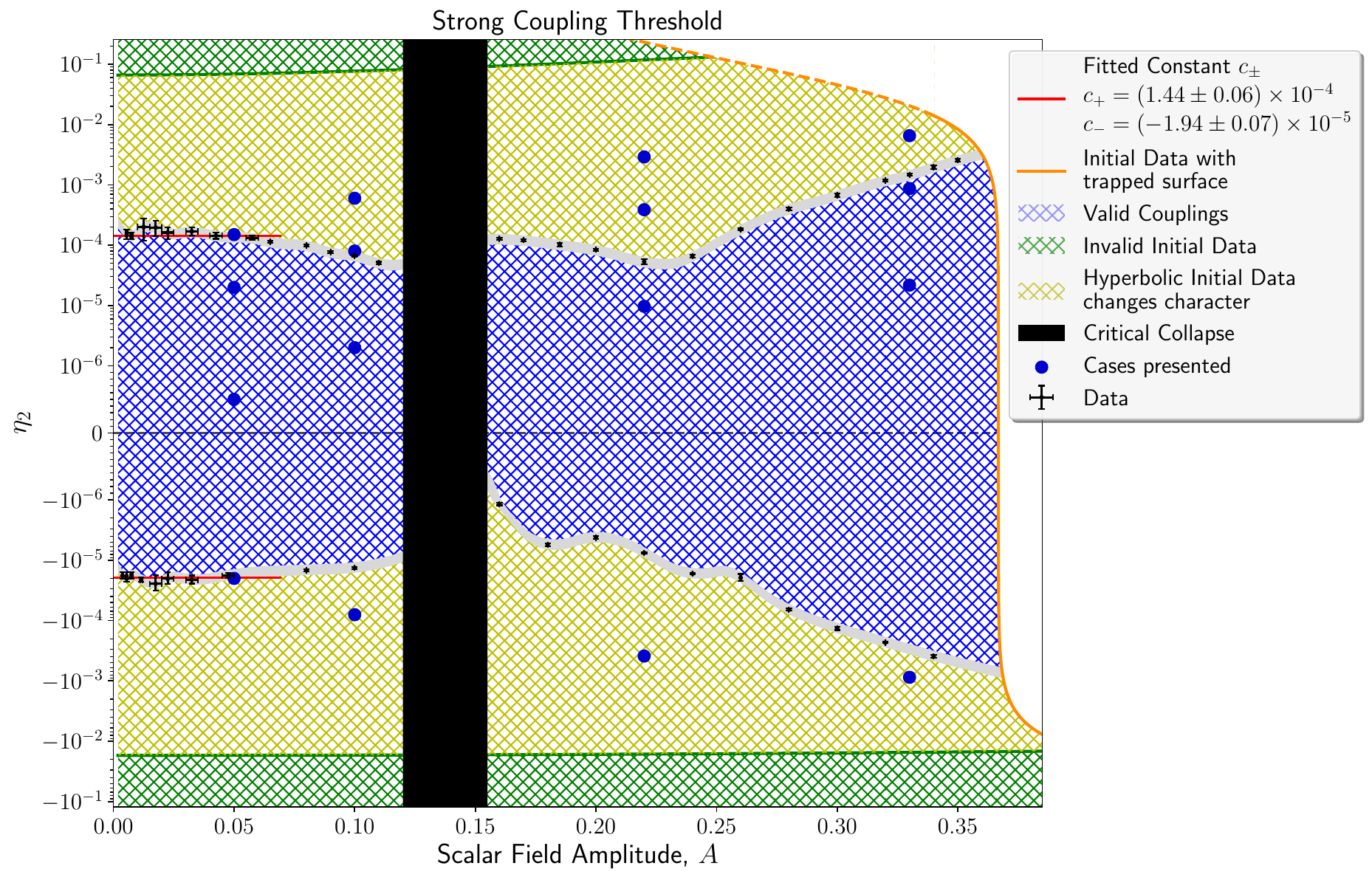}
\caption{Dynamical regimes of the $G_2 = g_2\,X^2$ theory as a function of the initial amplitude $A$ and the dimensionless coupling constant $\eta_2$, see equation \eqref{eq:eta2}. The black band denotes the region near critical collapse; black holes form to the right of this band. The orange curve on the right marks the region where the initial data contains a trapped surface. The scalar equation is hyperbolic at all times in the blue region; EFT is valid in the interior of this region. In the yellow region, the scalar equation is initially hyperbolic but it changes character during the evolution. In the green region the initial value problem is not well-posed.}
\label{fig:results_eta2}
\end{figure}

Our main results for the $G_2\neq 0$, $G_3=0$ theory are summarised in figure \ref{fig:results_eta2}.  The analogous figure for the $G_3\neq 0$, $G_2=0$ theory is qualitatively similar and can be found in section \ref{sec:G3_neq_0}, figure \ref{fig:results_eta3}. For the sake of definiteness, in the following we shall focus our discussion on the $G_2\neq 0$, $G_3=0$ theory, but essentially the same conclusions apply to the $G_3\neq 0$, $G_2=0$ theory. 

The dimensionless coupling constants $\eta_2$ and $\eta_3$, see equations \eqref{eq:eta2}-\eqref{eq:eta3}, control the future development for our initial data; in other cubic Horndeski theories one should be able to define analogous dimensionless couplings, and therefore the conclusions of this chapter should apply to those theories as well. In figure \ref{fig:results_eta2} we show the various dynamical regimes of the $G_2$ theory  as a function of  $\eta_2$ and the initial scalar amplitude $A$. As one would expect, the weakly coupled regime of the theory corresponds to  suitably small values of $\eta_2$, but the boundary of this region depends non-trivially on the scalar amplitude.

The blue region in figure \ref{fig:results_eta2} denotes the values of $(A,\eta_2)$ for which the scalar equation is hyperbolic at all times. The yellow region corresponds to the values of $(A,\eta_2)$ for which the scalar equation is initially hyperbolic, and hence the initial value problem is well-posed, but it changes character during the evolution, signalling a breakdown of the formulation of the theory\footnote{The boundary between the blue and yellow region was obtained numerically by finding the boundary $\eta_2$ for a fixed $A$ for which the simulations crashed or not.}. The green region corresponds to the values of $(A,\eta_2)$ for which the scalar equation is not hyperbolic on the initial data slice and hence the initial value problem is not well-posed\footnote{Following section \ref{sec:paper1:hyperbolicity}, this could be approximated by the value of $\eta_2$ for a fixed $A$ for which $g_2 X \sim \mathcal{O}(1)$ (or $\t g_3 \sim \mathcal{O}(1)$ for the case of $G_2 = 0$, $G_3\neq0$). It was obtained numerically by finding, for a fixed amplitude, the value of $\eta_{2,3}$ for which the character of the equation at $t=0$ was already not hyperbolic, by checking whether $\det\br{h^\m_{~\n}}$ or $h^{00}$ were zero.}. The black band in figure \ref{fig:results_eta2} corresponds to the range of amplitudes, numerically determined, for which the future development of the initial data gets close to Choptuik's critical solution \cite{Choptuik:1992jv}, which is a naked singularity. This band splits the figure into two regions corresponding to the small and large data regimes: for initial data in the blue region to the left of the black band, the scalar field disperses to infinity. On the other hand, initial data in the blue region to the right of the black band collapses into a black hole. The orange curve on the right marks the region where the initial data contains a trapped surface\footnote{If the initial data contains a trapped surface, empirically this corresponds to the impossibility of solving the Hamiltonian and Momentum constraints using the method described in appendix \ref{appendix:paper1:initialData}. However, this region was determined numerically making use of Kip Thorne's Hoop Conjecture \cite{hoop_conjecture, Flanagan:1993zz}, which, for spherical symmetry, roughly says that a black hole forms when a mass $M$ is compacted in a region with radius $R \le 2M$. For a fixed $\eta_{2,3}$, we determined the amplitude for which the initial data has a radius $r=2M_r$, where $M_r$ is the mass up to radius $r$. $M_r$ can be estimated using $M_r = \int_{V_r} d^3x \r \sqrt{\g}$, where $\r$ is the scalar energy density, $V_r$ a ball of radius $r$ and $\g$ the spatial metric.}.

For initial data in any of blue regions in figure \ref{fig:results_eta2} it is possible to construct global solutions to the $G_2=g_2\,X^2$ Horndeski theory. Away from the boundary of this region, the deviations from GR are ``small'' everywhere on and outside black holes (if there are any) for all times. By ``small'' here we mean that the weak field condition \eqref{eq:paper1:WFC_2} is satisfied. Therefore, we identify the interior of the blue region as the regime of validity of EFT for the corresponding Horndeski theory. Of course, when black holes form during the evolution, EFT will break down near the singularity, just as GR does. In this case, we excise a portion of the interior of the black hole since it is causally disconnected from external observers.  For values of $(A,\eta_2)$ close to the boundary of the blue region, the weak field condition \eqref{eq:paper1:WFC_2}  can become $\mathcal{O}(1)$ during the evolution while the scalar equation remains hyperbolic. In this case one may argue that even though the theory has a well-posed initial value problem, higher derivative corrections not included in the action \eqref{eq:cubic_horndeski} should become important and hence one should not trust the theory as it stands. 

As mentioned above, the yellow region in figure \ref{fig:results_eta2} denotes the values of $(A,\eta_2)$ for which the evolution breaks downs due to the change of character of the scalar equation and this breakdown cannot be hidden behind a horizon. The change of character of the scalar equation is typically associated  to the weak field conditions becoming $\mathcal{O}(1)$ or larger, but this is not always the case. Indeed, for certain values of $(A,\eta_2)$, and in particular for $\eta_2<0$, the weak field condition can be $\mathcal{O}(10^{-2})$ during the evolution and yet the equations change character. Beyond this point it is no longer possible to solve the theory as an initial value problem. However, we note that whenever the equations change character, the weak field condition is much larger than the dimensionless coupling $\eta_2$ determined from the initial data. Therefore, in a certain sense, the theory becomes strongly coupled right before it breaks down.  In section \ref{sec:paper1:results}  we study in detail how and where in spacetime the loss of hyperbolicity of the scalar equation happens depending on the Horndeski couplings and we correlate it to the weak field conditions \eqref{eq:paper1:WFC_2}.  For $0<A\lesssim 0.05$, the boundary between the blue and yellow regions  is given by constant values of $\eta_2\sim (1.44\pm0.06)\times 10^{-4}$ and  $\eta_2\sim(-1.94\pm 0.07)\times 10^{-5}$. This is non-trivial since the location of this boundary is obtained from the non-linear evolution of the initial data.  As we will see in section \ref{sec:G2_neq_0}, for $\eta_2>0$ the breakdown of the evolution happens through a Tricomi-type-of transition while for $\eta_2<0$ the transition is of the Keldysh type.

A zoom in of figure \ref{fig:results_eta2} near the black band would show that $\eta_2\to 0$ as one approaches the critical regime from both sides. This is expected since for $A$ near the critical amplitude $A_\ast=0.13\pm 0.01$, the gradients of both the metric tensor and the scalar field become very large as the solution approaches the critical solution, which leads to a change of character of the scalar equation unless $g_2\to 0$ as $A\to A_\ast$.  Since the regime of validity of EFT is essentially the empty set at the critical solution, in the rest of the chapter we will purposely avoid the region near criticality.\footnote{Gannouji and Baez \cite{Gannouji:2020kas} study critical collapse in $k$-essence models. We thank Eugeny Babichev for bringing our attention to this article.} For values of $A>A_\ast$, a black hole forms during the evolution of the initial data. Interestingly, the larger the value of $A$, the larger the black hole that forms and the sooner it forms. Since larger black holes result in lower curvatures on the horizon scale, larger values of the couplings are allowed and yet the theory remains weakly coupled on and outside the black hole. This is the reason why $\eta_2$ increases for larger $A$. This is a very important effect, otherwise one could think that for any value of the couplings, one could always find some high enough amplitude that leads to a breakdown of hyperbolicity: this is not true, because eventually black holes start forming and absorbing the region where hyperbolicity breaks. Finally, for sufficiently large $A$, the initial data already contains a trapped surface. Since we are interested in studying gravitational collapse, we do not consider those values of $A$. 

It is clear from the previous discussion that our weak field conditions \eqref{eq:paper1:WFC_2} bear some relation with the hyperbolicity condition of the scalar equation of motion \eqref{eq:scalar} but such a relation is not a direct one. It is possible that one can come up with  refined and sharp weak field conditions that also capture the change of character of the equations when they are violated but finding them is beyond the scope of this chapter. It follows from our analysis that the regime of validity of EFT corresponds to the weak field conditions \eqref{eq:paper1:WFC_2} being satisfied (to justify that higher derivative terms in \eqref{eq:cubic_horndeski} can be neglected) \textit{and} that the initial value problem is well-posed, i.e., the scalar equations of motion are hyperbolic everywhere in spacetime, perhaps except in a small region inside black holes. These two conditions are satisfied in the interior of the blue region in figures \ref{fig:results_eta2} and \ref{fig:results_eta3}. For initial data in this region, the Horndeski theories that we have considered are valid EFTs and global solutions can be constructed. We note that whilst the conditions for hyperbolicity and the weak field conditions \eqref{eq:paper1:WFC_2} overlap near the GR limit, the latter are not necessarily contained in the former far away from GR.\footnote{We thank Harvey Reall for discussions on this issue.}

\subsection{\done{Outline}}

The rest of the chapter is organised as follows. In section \ref{sec:paper1:methodology} we present the theories that we consider and in \ref{sec:paper1:hyperbolicity} we analyse the corresponding hyperbolicity conditions. In section \ref{sec:paper1:results} we present and analyse the results of our numerical simulations. Section \ref{sec:G2_neq_0} discusses in detail the dynamics of the $G_2\neq 0$ theories, while the $G_3\neq 0$ theories are dealt with in subsection \ref{sec:G3_neq_0}. We conclude with some final remarks in section \ref{sec:paper1:conclusions}. We have relegated some technical details to the appendices. In appendix \ref{appendix:paper1:EOM} we write down the equations of motion for scalar field and the effective scalar metric in a 3+1 form. We collect some technical results in appendix \ref{appendix:paper1:effective_metric}, details of initial data in appendix \ref{appendix:paper1:initialData}, the convergence tests are presented in appendix \ref{appendix:paper1:convergence} and some notes on numerical implementation are in appendix \ref{appendix:paper1:implementation}. Appendix \ref{appendix:paper1:other} contains the results of certain numerical simulations that are also relevant for the main text. Finally, some useful computations are left in appendix \ref{appendix:paper1:adm_mass} and \ref{appendix:paper1:variation_g3}, regarding the ADM mass in spherical symmetric and the variation of the cubic Horndeski action, respectively.

\section{\done{Methodology}}\label{sec:paper1:methodology}

In this section, we describe the specific Horndeski theories that we have studied, its equations of motion, our initial data, the analysis of the hyperbolicity of the scalar equations and the weak field regime conditions that are necessary as to allow the evolution on and outside of the black hole horizon.

\subsection{\done{Cases explored}}\label{subsec:paper1:cases_explored}

We consider the special subset of Horndeski theories for which \cite{Kovacs:2019jqj} proved well-posedness of the initial value problem in both the BSSN and CCZ4 formulations of the Einstein equations in the usual gauges used in numerical relativity. This subclass of theories is given by setting $G_4=G_5=0$ in the general Horndeski action \eqref{eq:Horndeski_general}, which has been discussed in section \ref{subsec:Horndeski} to be very rich and with multiple applications to gravitational physics and cosmology. This is the so called cubic Horndeski theories described by the 4D action
\begin{equation}\label{eq:cubic_horndeski}
    \mathcal{S} := \tfrac{1}{\k}\int d^4x \sqrt{-g} \big[\mathcal{R} + X - V(\f) + G_2(\f , X) + G_3(\f , X)\square \f\big].
\end{equation}
In our work we are not interested in a particular model but rather in exploring general features of the non-linear physics encoded in cubic Horndeski theories. From the point of view of EFT, one would expect \eqref{eq:cubic_horndeski} to be valid when the $G_2$ and $G_3$ terms are suitably small, which corresponds to $X$ being small.    Therefore, one can consider Taylor-expanding some general (smooth) functions $G_2$ and $G_3$ for small $X$ and keep only the leading order terms. With this in mind,  we therefore focus on the simplest non-trivial functions $G_2$ and $G_3$:
\begin{equation}
    \begin{aligned}\label{eq:relevantGs}
        G_2(\f,X) &= g_2\, X^2,\\
        G_3(\f,X) &= g_3\, X,
    \end{aligned}
\end{equation}
where $g_2$ and $g_3$ are arbitrary coupling constants with dimensions of $\text{Length}^2$ that we can tune.  These or similar choices have been considered in the literature before, namely in models of dark energy \cite{Kase:2018iwp, Kase:2015zva,Myrzakulov:2014iza,Shahalam:2016kkg, Deffayet:2010qz,Peirone:2019au}, and in studies of the fate of the Universe in cosmological bounces or inflationary models \cite{Ijjas:2016tpn,Ijjas:2016vtq,Shahalam:2016kkg, Kobayashi:2010cm}, among others \cite{Ijjas:2018cdm, Emond:2019myx}. As we noted in the introduction \ref{sec:paper1:introduction}, from the point of view of EFT our choice for $G_2$ in \eqref{eq:relevantGs} corresponds to the most general scalar term that can be added to the action up to four derivatives \cite{Weinberg:2008hq}. The choice of $G_3$ is a matter of simplicity and the convenience of being able to use the standard BSSN/CCZ4 formulation.

\subsection{\done{Equations of motion}}\label{subsec:paper1:EOM}

Applying the variational principle to the action \eqref{eq:cubic_horndeski}, the resulting Einstein equations are\footnote{Appendix \ref{appendix:paper1:variation_g3} has the details of this calculation for the $G_3$ term.}:
\begin{align}\label{eq:einstein_cubic_horndeski}
        G_{\m\n} =&~ g_{\mu\nu} \bigl(G_2 + X - V + 2\, X\, \partial_\phi G_3\bigr) +(\nabla_{\mu}\phi)(\nabla_{\nu}\phi) \bigl(1 + \partial_X G_2 + 2\, \partial_\phi G_3\bigr)\\
        &+\partial_X G_3\bigl[(\square\phi)(\nabla_{\mu}\phi) (\nabla_{\nu}\phi) - 2\,(\nabla^{\rho}\phi)(\nabla_{(\mu}\phi) \nabla_{\nu)}\nabla_{\rho}\phi  + g_{\mu\nu} (\nabla^{\rho}\phi)(\nabla^{\sigma}\phi) \nabla_{\rho}\nabla_{\sigma}\phi \bigr],\nonumber
\end{align}
where $G_{\m\nu}$ is the Einstein tensor. The equation of motion for the scalar field is:\footnote{The direct variation of the action with respect to the scalar field yields a term $\partial_X G_3\, \mathcal{R}_{\m\n}(\grad^\m\f)(\grad^\n\f)$; one can use the metric equation of motion to replace $\mathcal{R}_{\m\n}$ in this term and obtain \eqref{eq:scalar} (see Kovács \cite{Kovacs:2019jqj} for details).}
\begin{equation}
    \begin{aligned}\label{eq:scalar}
    &-\square\phi \Bigl(1 + \partial_X G_2 + 2 \partial_\phi G_3 - 2 X \partial^2_{\phi X}G_3\Bigr)- \partial_\phi G_2 + \partial_\f V\\
    &+2\,X(\partial^2_{\phi X}G_2 + \partial^2_{\phi\phi} G_3)  + (\partial^2_{XX} G_2 + \partial^2_{\phi X}G_3)(\nabla^{\mu}\phi)( \nabla^{\nu}\phi) \nabla_{\mu}\nabla_{\nu}\phi \\
    &+X\,\partial_X G_3\bigl(G_2 - V + X\,\br{2 + \partial_X G_2 + 4\, \partial_\phi G_3}\bigr)\\
    &+ X\,\bigl(\partial_X G_3\bigr)^2\big[X\,\square\phi + 2\,(\nabla^{\mu}\phi)(\nabla^{\nu}\phi)\nabla_{\mu}\nabla_{\nu}\phi\big] \\
    &+\partial^2_{XX}G_3(\nabla^{\mu}\phi)(\nabla^{\nu}\phi)\bigl[(\square\phi) \nabla_{\mu}\nabla_{\nu}\phi - (\nabla_{\mu}\nabla^{\rho}\phi) \nabla_{\rho}\nabla_{\nu}\phi \bigr] \\
    &- \partial_X G_3\Big[\bigl( \square\phi\bigr)^2 - (\nabla^{\mu}\nabla^{\nu}\phi) \nabla_{\mu}\nabla_{\nu}\phi\Big]  = 0\,.
    \end{aligned}
\end{equation}
We write down equations \eqref{eq:einstein_cubic_horndeski} and \eqref{eq:scalar} in the usual 3+1 conformal decomposition and implement the CCZ4 form of the Einstein equations that is suitable for the numerical simulations. The equations that we have implemented in our code as well as the details of the numerical simulations are given in appendix \ref{appendix:paper1:EOM}.

\subsection{\done{Initial data}}\label{subsec:paper1:initial_data}

For the present analysis, motivated by the objective of studying gravitational collapse, we choose a family of initial data for the  scalar field $(\f,\P)$ modelling a spherically symmetric bubble centred at $\ve{c}$:
\begin{equation}\label{eq:paper1:phiData}
    \phi(t,\ve{x})\bigg\rvert_{t=0} = A \br{\frac{r^2}{r_0^2+2\,\omega^2}}e^{-\frac{1}{2}\br{\tfrac{r-r_0}{\omega}}^2},
\end{equation}
where $\ve{r}=\ve{x}-\ve{c}$ and $r=||\ve{r}||_2$ with the Euclidean 2-norm. 
Notice that the class of theories in \eqref{eq:relevantGs} have a reflection symmetry  $\f\to-\f\,,\,g_3\to -g_3$ ($g_2$ unchanged) and hence, we can choose $A>0$ without loss of generality. In the $\w \ll r_0$ limit, this profile has its peak at $r_{max} = r_0\br{1+2\br{\frac{\omega}{r_0}}^2 + \mathcal{O}\br{\frac{\w}{r_0}}^4}\approx r_0$, with a value of $\phi(r_{max})\approx A\br{1-2\br{\frac{\omega}{r_0}}^4+ \mathcal{O}\br{\frac{\w}{r_0}}^6}\approx A$. The weird normalisation denominator $r_0^2 + 2\omega^2$ is to make $A$ the maximum of $\phi$ up to $\mathcal{O}\br{\frac{\omega}{r_0}}^4$, useful for numerical purposes. The width at half height of the Gaussian profile is $\D r\approx 2\,\w\,\br{1+2\br{\frac{\omega}{r_0}}^2+ \mathcal{O}\br{\frac{\w}{r_0}}^4}$.

Regarding the scalar momentum, having zero momentum generates an in-going and an out-going pulse of equal amplitude. This is unphysical and undesired. Instead, assuming an approximately Minkowski initial background, we choose an in-going wave pulse\footnote{This is such that we have a wave-like equation for $\phi$ in spherical coordinates Minkowski space: $\partial_t^2 \phi_0 = \partial_t\Pi_0 = \frac{1}{r}\partial_r\br{r\Pi_0} = \frac{1}{r}\partial_r^2\br{r\phi_0}=\grad^2\phi$, i.e. $\Box \phi_0 = (-\partial_t^2 + \grad^2) \phi_0 = 0$.}:
\begin{align}\label{eq:paper1:ini_mom}
    \Pi(t,\ve{x})\bigg\rvert_{t=0} = \tfrac{1}{r}\partial_r\br{r\phi}\bigg\rvert_{t=0}\,.
\end{align}
To explore the relevant phenomenology of these theories, we have studied many different scenarios. Using a full 3D code, we were able to verify that all the features hereafter described are not a peculiarity of spherical symmetry, and also occur when the symmetry is broken, without any seemingly interesting new features emerging. However, we have not attempted to carry out a thorough analysis of non-spherically symmetric scalar field collapse.  Hence, in the following we only present the results for the spherically symmetric case. 

With the choices \eqref{eq:paper1:phiData} and \eqref{eq:paper1:ini_mom} for the initial scalar profile and momentum, we obtain the constraint satisfying initial data for the metric by solving the Einstein constraints using the conformal transverse-traceless decomposition (details in section \ref{subsec:ngr:conformalTT}). We choose a conformally flat initial metric and vanishing trace and transverse-traceless part of the extrinsic curvature. Hence, we solve for the conformal factor of the spatial metric and three leftover degrees of freedom of the traceless part of the extrinsic curvature (which reduce to one in spherical symmetry). More details are presented in appendix \ref{appendix:paper1:initialData}.

To get some intuition about how the modifications of GR affect our initial data, we can expand the initial spacetime matter mass for small amplitudes and couplings around a Minkowski background. This is done by performing $\int d^3x\, \r$, where $\r$ is the energy density for Horndeski (see equation \eqref{eq:appendix:energy_density}) assuming a flat metric. We find,
\begin{align}
    M_{\text{matter}} \approx&~ \frac{\bar{\rho}}{16\,\pi}\,\Bigg\{1+\tfrac{13}{2}\br{\tfrac{\omega}{r_0}}^2 + \mathcal{O}\bigg(\br{\tfrac{\omega}{r_0}}^4\bigg) \label{eq:rho0} \\
    &\hspace{0.5cm}+\bigg[\br{m^2\omega^2}+\br{\tfrac{5}{4\sqrt{2}}}\br{\tfrac{g_2A^2}{r_0^2}}+\bigg(\tfrac{28}{9}\sqrt{\tfrac{2}{3}}\bigg)\br{\tfrac{g_3A\omega^2}{r_0^4}}\bigg]\bigg[1+\mathcal{O}\bigg(\br{\tfrac{\omega}{r_0}}^2\bigg)\bigg]\Bigg\}\,, \nonumber
\end{align}
where $\bar{\rho}=\frac{\sqrt{\p}}{8}\frac{A^2r_0^2}{\omega}$ and we have included the contribution of the in-going momentum \eqref{eq:paper1:ini_mom} and of a mass term in the scalar potential $V(\f)=\frac{1}{2}m^2\f^2$. It is interesting to note that this energy is twice as big due to the in-going momentum contribution, when compared to the zero momentum rest case. From \eqref{eq:rho0} we see that for our initial data, the strength of the modifications of GR due to the cubic Horndeski terms $G_2$ and $G_3$ is measured by the dimensionless couplings:
\begin{equation}
\eta_2 = \frac{g_2A^2}{r_0^2}\,, \label{eq:eta2}
\end{equation}
and,
\begin{equation}
\eta_3 = \frac{g_3A\omega^2}{r_0^4}\,, \label{eq:eta3}
\end{equation}
respectively. These dimensionless couplings play an important role in predicting the future development of the initial data and determine the weakly coupled regime of these theories.
\begin{figure}[h]
\centering
\includegraphics[width=.8\textwidth]{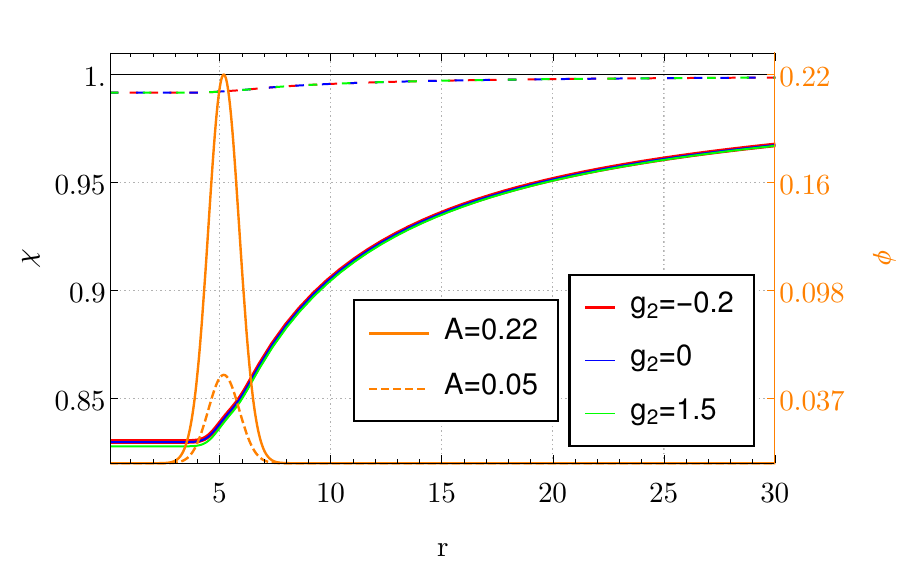}
\vspace{-3mm}
\caption{Initial conformal factor $\chi$ (left $y$-axis) for a given scalar field profile (in orange, right $y$-axis) with ingoing momentum, for different choices of $g_2$. The scalar profile in this figure corresponds to Cases 1 (dashed line) and 3 (solid line) of section \ref{sec:paper1:results}; the shown values of coupling $g_2$ correspond to the GR case ($g_2=0$) and the cases presented in figure \ref{fig:strong_g2_pos_det_h00} ($g_2=1.5$) and figure \ref{fig:strong_g2_neg_det_h00} ($g_2=-0.2$).}
\label{fig:paper1:initial_data}
\end{figure}

In figure \ref{fig:paper1:initial_data} we show the initial conformal factor $\chi$ and scalar profile $\phi$ for some representative cases.  From this figure we see that even for relatively large amplitudes within the range that we have considered, the conformal factor has a very small dependence on the Horndeski couplings. For the specific case of $A=0.22$, the difference between $g_2=1.5$ and GR at $r=0$ is $0.2\%$, which is in accordance with the fact that for this case the dimensionless coupling is small $(\eta_2\sim 3\times 10^{-3})$. One can also notice that for sufficiently small amplitude, the conformal factor is almost 1 for any reasonable value of $g_2$.

\subsection{\done{Numerical scheme}}

On top of what was already described in section \ref{subsec:ngr:grchombo_params}, for the simulations presented in this chapter, we use Kreiss-Oliger numerical dissipation with $\sigma=0.3$. Typical simulations used a Courant factor of $0.2$ (small value chosen to better resolve Keldysh-type-of transitions), a coarse grid resolution of $\D x = 1$ and up to 7 additional refinement levels, and a box size of $L = 96$ with Sommerfeld boundary conditions. We use the gradients of $\f$ and $\chi$ as well as contours of $\chi$ to tag cells for regridding (see appendix \ref{appendix:grchombo:tagging} for more details). Last but not least, we use the symmetry of the system to only simulate one octant of the full domain, which reduces the computational cost of the problem by a factor of 8.

\section{\done{Breakdown of hyperbolicity}}\label{sec:paper1:hyperbolicity}

From sections \ref{subsec:paper1:effective_metric}-\ref{subsec:paper1:WCCs}, we learn how the effective metric of the scalar equation has strong implications in the stability and existence of the solutions, how the breakdown of the evolution can be explained by analysing scalar speeds of propagation, and what are the necessary conditions we must satisfy to guarantee the physical relevance of the theory. In sections \ref{subsec:paper1:case_g2}-\ref{subsection:paper1:case_g3} we apply these lessons to the case of $G_2$ and $G_3$ theories. Finally, in section \ref{subsection:paper1:excision}, we discuss numerical implications of the hyperbolicity problems discussed below.

\subsection{\done{Effective metric}}\label{subsec:paper1:effective_metric}

To identify the regime of validity of EFT, we need to first determine the character of the equations of motion for the scalar field \eqref{eq:scalar} and the conditions under which they are hyperbolic. To do so, we consider the principal part of the scalar equation \eqref{eq:scalar}, which is a wave equation governed by an effective metric \cite{Deffayet:2010qz}:
\begin{equation}
\begin{aligned}
    h^{\mu\nu} =&~~~g^{\mu\nu}\,\Big[1 + \partial_X G_2 + 2\, \partial_\phi G_3 + 2\,\partial_X G_3 \,\Box\phi - X^2 \bigl(\partial_X G_3\bigr)^2 \\
    &\hspace{1.2cm}- \partial^2_{XX} G_3 (\grad^\rho\phi)(\grad^\sigma\phi)\grad_\rho\grad_\sigma\phi - 2\,X\,\partial^2_{\phi X} G_3\Big]\\
    & -(\grad^\mu\phi)(\grad^\nu\phi)\sbr{2\,X\bigl(\partial_X G_3\bigr)^2 + \partial^2_{XX} G_2 + \partial^2_{XX} G_3\,\Box\phi + 2\,\partial^2_{\phi X} G_3}\\
    & +~2\,\partial^2_{XX}G_3(\grad^\rho\phi)(\grad^{(\mu}\phi)\grad_\rho\grad^{\nu)}\phi - 2\,\partial_X G_3 \grad^\mu\grad^\nu\phi\,.
    \label{eq:eff_metric_up}
\end{aligned}
\end{equation}
The eigenvalues of $h^{\mu\nu}$ determine the character of the equation: if the product of the eigenvalues is negative then the equation is hyperbolic; if the product is positive then the equation is elliptic, and if it is zero the equation is parabolic. 

Having a well-posed initial value problem, is the minimum requirement that we should demand on any classical theory; therefore, the breakdown of hyperbolicity of the scalar equation in this case can be associated to the breakdown of the formulation of the theory itself. As \cite{Babichev:2016hys,Bernard:2019fjb} noted, the fact that the effective metric \eqref{eq:eff_metric_up} depends on the scalar field itself and its gradients implies that shocks can generically form from smooth initial data, at which point the well-posedness is lost.  Therefore, the local character of the scalar equation is a useful proxy to establish the regime of validity of the theory and to measure the size of the non-linearities and deviations from GR \cite{Bernard:2019fjb,Ripley:2019hxt,Ripley:2019irj,Ripley:2019aqj}. We will come back to this point below.  

When considering spacetimes containing black holes, the evolution of the spatial slices in puncture gauge is such that the determinant of the inverse spacetime metric goes to zero near the puncture, i.e., $\det(g^{\m\n})=-\frac{\chi^3}{\a^2}\to0$ (see section \ref{sec:ngr:adm}). Consequently the same happens for the effective metric \eqref{eq:eff_metric_up}. To distinguish this effect from an actual breakdown of the hyperbolicity of the scalar equation, we note that $h^{\mu\nu} = g^{\mu\rho}h^{\nu}_{~\rho}$ and therefore:
\begin{equation}
    \text{det}\br{h^{\mu\nu}} = \text{det}\br{h^\nu_{~\rho}}\,\text{det}\br{g^{\mu\rho}} = -\tfrac{\chi^3}{\alpha^2}\,\text{det}\br{h^\mu_{~\nu}},
    \label{eq:dethuu}
\end{equation}
with $\text{det}\br{h^\mu_{~\nu}}=1$ in GR. Clearly, deviations of this quantity from 1 encode the dynamics of the Horndeski theories and hence we will focus our attention on $\text{det}\br{h^\mu_{~\nu}}$.

\subsection{\done{Characteristic speeds}}\label{subsec:paper1:characteristic_speeds}

The characteristic speeds, also called front velocities, are important since they correspond to the local speed of propagation of the scalar modes and hence they tell us about the effective causal cone that the scalar field ``sees''.\footnote{Recall that the characteristic speeds do not coincide in general with the phase or group velocity, which do not have a direct relation with the causal structure.} The characteristics are given by the zeros of the characteristic polynomial which, for the scalar field equation, is
\begin{equation}
    Q(x,\xi) = h^{\mu\nu}\xi_\mu\xi_\nu=0\,,
\end{equation}
for some covector $\xi_\m$ that defines the characteristic surface. Physically this corresponds to considering the high frequency and small amplitude limit of a wave with wave vector $\xi_\mu$. To calculate the propagation speed $v$ without symmetry assumptions, we specify $\xi=\br{v,n^i}$, with $n^i$ a direction of propagation suitably normalised $n_i n_j \d^{ij}=1$, where $\delta_{ij}$ is the Euclidean 3D metric (we can take space as locally flat). Then, the speed of propagation in the $n^i$ direction is:
\begin{gather}
    h^{00}v^2 + 2h^{0i}n_i ~v + h^{ij}n_i n_j = 0 \quad \Rightarrow\quad
    v_{\pm} = \frac{-h^{0i}n_i \mp \sqrt{\br{h^{0i}n_i}^2 - h^{00}h^{ij}n_i n_j}}{h^{00}}.
\end{gather}
In spherical symmetry one can naturally use a radial vector for the direction of propagation, $n^i=\{\tfrac{x}{r},\tfrac{y}{r},\tfrac{z}{r}\}$, which gives \cite{Babichev:2018uiw,Bernard:2019fjb},
\begin{equation}
v_{\pm}=\frac{-h^{0r}\mp\sqrt{\br{h^{0r}}^2-h^{00}h^{rr}}}{h^{00}}\,.
\end{equation}
In our conventions, $v_+$ and $v_-$ correspond to the ingoing and outgoing modes of the scalar field respectively and by construction they tend to $+1$ and $-1$ at infinity. When $v_-\geq 0$ and $v_+\geq 0$ in a certain region, scalar modes cannot reach asymptotic observers; the boundary $v_-=0$ of this region is the sound horizon \cite{Babichev:2006vx,Akhoury:2011hr}.  

As discussed by Ripley and Pretorius \cite{Ripley:2019irj} and Bernard et al. \cite{Bernard:2019fjb}, the equations can change character from hyperbolic to parabolic and elliptic in a manner which is qualitatively similar to what happens in the two standard equations of mixed type, namely the Tricomi equation,
\begin{equation}
    \partial_y^2u(x,y)+y\,\partial_x^2 u(x,y)=0\,,
\end{equation}
and the Keldysh equation,
\begin{equation}
    \partial_y^2u(x,y)+\tfrac{1}{y}\,\partial_x^2 u(x,y)=0\,.
\end{equation}
Both equations are hyperbolic for $y<0$ and they change character at the transition line $y=0$. Related to this change of character are the appearance of ghosts, gradient instabilities and formation of caustics \cite{Babichev:2020tct,DeFelice:2010gb,Babichev:2016hys}. 

In the case of the Tricomi equation, the characteristic speeds go to zero at $y=0$ where the equation becomes parabolic. If the characteristic speeds of both the ingoing and outgoing modes vanish then the evolution freezes. This can happen because of the choice of gauge; for instance, in coordinates that are not horizon penetrating, the lapse asymptotically goes to zero at the horizon, effectively resulting in zero characteristic speeds.  However, in this case the freezing of the evolution is a consequence of the gauge choice and it does not correspond to a breakdown of EFT. Therefore, in the case of a Tricomi-type-of transition, we also need to check that the deviations from GR are suitably large to conclude that the loss of hyperbolicity corresponds to a breakdown of the theory. The determinant $\text{det}\br{h^\mu_{~\nu}}$ mentioned in the previous section can aid in this distinction.

On the other hand, a Keldysh-type-of transition involves diverging characteristic speeds at $y=0$,\footnote{At least in some direction in full 3D space, which is non-trivial to determine without spherical symmetry.} which will typically signal a breakdown of EFT. This case is more difficult to handle numerically since  one is forced to take prohibitively small time steps.\footnote{In fact, the degree of regularity of the solutions of these equations typically differs, with solutions of the Tricomi equation enjoying higher regularity \cite{Ripley:2019irj}.} Note from \eqref{eq:eff_metric_up} that $h^{00}$ has a factor of $-\tfrac{1}{\alpha^2}$ coming from $g^{00}$, and hence the deviations from GR are measured by $-\alpha^2 h^{00}$, for which this factor is $1$. Therefore, the Keldysh-type-of transition without symmetry assumptions is signalled by $-\a^2 h^{00}\to0$, which implies that the $t=\textrm{const}.$ hypersurface being evolved is no longer spacelike with respect to the scalar effective metric \cite{Babichev:2007dw}. We associate this breakdown of the evolution to a Keldysh-type-of transition since the characteristic speeds diverge.  However, strictly speaking, at this point the equation may not have changed character yet but the two effects go hand in hand.\footnote{We would like to thank Luis Lehner for discussions on these issues.}  We discuss in detail the different types of transitions in the $G_2\neq0$ and $G_3\neq0$ cases in subsections \ref{subsec:paper1:case_g2} and \ref{subsection:paper1:case_g3}. 

\subsection{\done{Weak coupling conditions}}\label{subsec:paper1:WCCs}

The previous discussion only relates to the existence of a well-posed initial value problem but it does not fully address the issue of whether the theory under consideration makes sense as a truncated EFT \cite{Solomon:2017nlh}. We now turn to this point. As mentioned by Kovács \cite{Kovacs:2019jqj}, local well-posedness is only guaranteed in the weak field regime, meaning that the Horndeski terms are small compared to GR ones. One possible weak field condition that compares the size of the Horndeski terms versus GR is:
\begin{equation}
\begin{aligned}\label{eq:paper1:WFC_1}
    &\left|\partial^k_X\partial^l_\phi G_2\right| \ll L^{2k-2} ~~~~~~~~~~~~~~~ k=0,1,2;~l=0,1;\\
    &\left|\partial^k_X\partial^l_\phi G_3\right| \ll L^{2k} ~~~~~~~~~~~~~~~~~~~~~~~~~~~ k,l=0,1,2.
\end{aligned}
\end{equation}
where $L$ is a length scale estimate for the system: $L^{-1}=\text{max}\{\left|\mathcal{R}_{\alpha\beta\mu\nu}\right|^{\tfrac{1}{2}},\left|\grad_\mu\phi\right|,\left|\grad_\mu\grad_\nu\phi\right|^{\tfrac{1}{2}}\}$ in all orthonormal bases. For the cases \eqref{eq:relevantGs}, this is explicitly\footnote{The conditions \ref{eq:paper1:WFC_1} result in more conditions than \ref{eq:paper1:WFC_2}, like $|g_i X| \ll 1$, but ultimately the gradients of $X$ are encapsulated in the length scale $L$ via $\grad_\m\f$ and all these conditions are, in order of magnitude, equivalent to simply \ref{eq:paper1:WFC_2}.}:
\begin{gather}\label{eq:paper1:WFC_2}
     |g_2 \, L^{-2}| \ll 1\,,  \quad\quad  |g_3\, L^{-2}| \ll 1\,.
\end{gather}
In order for the Horndeski theories under consideration \eqref{eq:relevantGs} to be in the regime of validity of EFT, in this analysis we require that the evolution equation of the scalar field is hyperbolic \textit{and} that \eqref{eq:paper1:WFC_2} is satisfied. These two conditions ought to be imposed on and outside black hole horizons, should there be any in the spacetime. The nuance of the interior of black holes is addressed in section \ref{subsection:paper1:excision}.

\subsection{\done{Case of $G_2 \neq 0, G_3 = 0$}}\label{subsec:paper1:case_g2}

To monitor the character of the scalar equation, we compute the determinant of the scalar effective metric. Even though it is possible to find an analytic expression for the full determinant (using Cayley–Hamilton’s theorem and Newton’s identities), for simplicity we consider the $G_2\neq 0$, $G_3=0$ and the $G_3\neq 0$, $G_2=0$ cases separately. 

As explained in the discussion surrounding equation \eqref{eq:dethuu}, we only need to consider the determinant of the effective metric with one index up and one index down. For the $G_2\neq 0$ case, we have
\begin{gather}
    h^{\mu}_{~\nu} = \delta^{\mu}_{~\nu}\br{1 + \partial_X G_2} - (\grad^\mu\phi)(\grad_\nu\phi)~\partial^2_{XX} G_2\,.
\end{gather}
Realising that, up to scalars, this metric is the identity plus the tensor product of two vectors, one can use the Weinstein–Aronszajn identity to calculate the determinant of the full 4D metric without assuming any symmetries. We find:
\begin{equation}
    \begin{aligned}\label{eq:det_h00G2}
    \text{det}\br{h^\mu_{~\nu}}=&~\br{1+\partial_X G_2}^3\br{1 + \partial_X G_2 + 2\,X\partial^2_{XX} G_2}\\
    =&~\br{1+2\,g_2X}^3\br{1+6\,g_2\,X}\,,
    \end{aligned}
\end{equation}
where in the last line we have used that $G_2=g_2\,X^2$. Knowing the determinant is the product of the eigenvalues, we can double check the consistency of this result by double checking these. For a vector $v^\nu$, note that:
\begin{equation}
    h^\mu_{~\nu}v^\nu = v^\mu\br{1 + \partial_X G_2} - \br{\partial^2_{XX}G_2v^\nu\grad_\nu\phi }\grad^\mu\phi\,,
\end{equation}
so we conclude that $\grad^\mu\phi$ is an eigenvector with eigenvalue $\br{1 + \partial_X G_2 + 2\,X\partial^2_{XX} G_2}$. The other three eigenvectors are any tetrad orthogonal to the 4-vector $\nabla^\m\f$ and have degenerate eigenvalues equal to $\br{1+\partial_X G_2}$, in accordance to \eqref{eq:det_h00G2}.

Besides the determinant of the effective metric, as discussed in section \ref{subsec:paper1:characteristic_speeds}, to monitor a Keldysh-type-of transition, we have to compute $-\a^2h^{00}$. For the $G_2\neq 0$, $G_3=0$ case, this is given by,
\begin{equation}
\begin{aligned}
    -\a^2 h^{00}&=1+\partial_X G_2 + \Pi^2~\partial^2_{XX}G_2 \\
        &=1+6\,g_2X+2\,g_2\,\Pi^i\Pi_i\,,\label{eq:h00G2}
\end{aligned}
\end{equation}
where $\P=n^\m\grad_\m\f$ is the scalar momentum, and in the last line we have used that $G_2=g_2\,X^2$ and $\Pi^2 = 2\,X + \Pi^i\Pi_i$, with $\P_i=D_i\f$ (see appendix \ref{appendix:paper1:EOM}). All in all, for $G_2$ as in \eqref{eq:relevantGs}, the two quantities that inform us about the breakdown of the initial value problem for the scalar equation are \eqref{eq:det_h00G2} and \eqref{eq:h00G2}.
The scalar equation is hyperbolic as long as these two quantities are non-negative.\footnote{Note that we have pulled out a minus sign in \eqref{eq:dethuu} so $\det\br{h^\mu_{~\nu}}>0$ corresponds to $h^{\m\n}$ having one negative eigenvalue and three positive ones, as it should for a hyperbolic equation.} These are the same conditions found by Babichev et al. \cite{Babichev:2007dw}, and  it is evident that if the weak field conditions \eqref{eq:paper1:WFC_2} are satisfied then the scalar equation is hyperbolic.  

\begin{figure}[h]
\centering
\includegraphics[width=.6\textwidth]{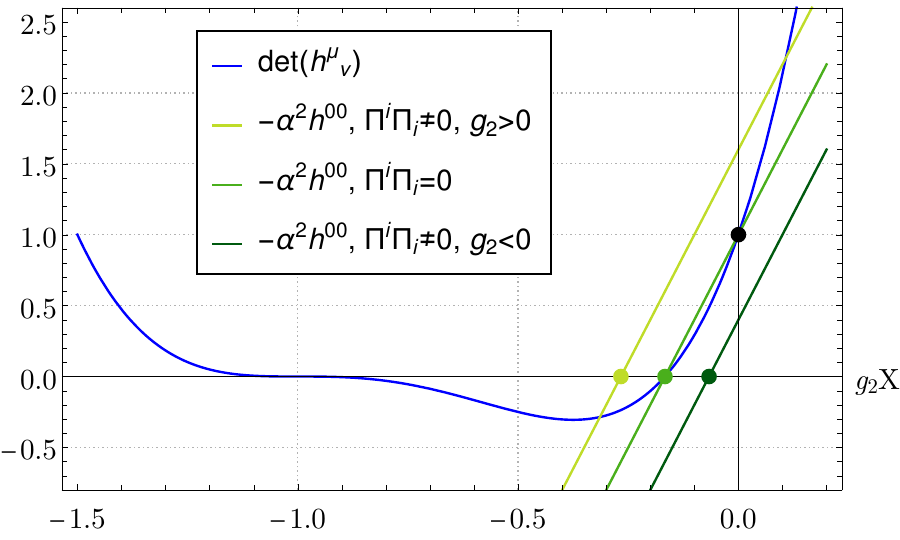}
\vspace{-3mm}
\caption{Sketch of the possible changes of character of the scalar equation in the $G_2\neq 0$ theory depending on the sign of the coupling constant $g_2$. For $g_2>0$, $\det(h^\m_{~\n})$ (blue curve) vanishes before $-\a^2h^{00}$ does (light green line), leading to a Tricomi-type-of transition. On the other hand, for $g_2<0$, $-\a^2h^{00}$ will vanish first (dark green line), leading to a Keldysh-type-of transition. In this case, the time coordinate $t$ is no longer a global time function and the scalar equation cannot be evolved further in this gauge. }
\label{fig:plot_g2X_det_and_h00}
\end{figure}
For a non-constant scalar profile $\f$, $\P^i$ satisfies $\P_i\P^i\ge0$, but on the other hand $X$ can be either positive or negative, depending on the balance between scalar gradients and  momentum. In a dynamical evolution, both can become large. As this happens, $g_2\,X$ can decrease to make either  \eqref{eq:det_h00G2} or \eqref{eq:h00G2} zero, see figure \ref{fig:plot_g2X_det_and_h00}. But interestingly, we can still conclude which transition will happen when. If $g_2>0$, the fact that $\P_i\P^i>0$ implies that $\text{det}\br{h^\mu_{~\nu}}$ will reach zero before $-\a^2h^{00}$, and the equation will become parabolic on a co-dimension one surface, where at least one of the  characteristic speeds goes to zero while the others remain bounded. This will correspond to a Tricomi-type-of transition.  On the other hand, if $g_2<0$ the opposite is true and $-\a^2h^{00}$ may become zero before $\text{det}\br{h^\mu_{~\nu}}$ does, leading to infinite speeds of propagation and a very abrupt termination of the evolution associated to a Keldysh-type-of transition. 
The transition is also expected to happen for small absolute values of $g_2$. Both behaviours were identified by Bernard et al. \cite{Bernard:2019fjb}.\footnote{Bernard et al. \cite{Bernard:2019fjb} use a coupling $g$ with the opposite sign as our $g_2$.} 

The changes of character described in the previous paragraph can only occur if $|g_2\,X|$ is suitably large and hence outside the weak field regime. While generically one can expect that weak data eventually enters the strong field regime,  one question that we need to address  is whether or not the region where EFT breaks down can be hidden inside a black hole. If the answer is positive, then one can hope that classical observers at infinity will be protected from any potential pathologies that arise in the scalar equations and EFT will retain its predictive power. The technical details on how we have dealt with the loss of hyperbolicity and the violations of the weak field condition \eqref{eq:paper1:WFC_2} inside black holes are given in section \ref{subsection:paper1:excision}.

\subsection{\done{Case of $G_3 \neq 0, G_2 = 0$}}\label{subsection:paper1:case_g3}

In equation \eqref{eq:det_full_g3} of appendix \ref{appendix:paper1:effective_metric} we present the full analytic form of the determinant of the scalar effective metric in the $G_3\neq 0$, $G_2=0$ case. For clarity, in this subsection we analyse \eqref{eq:det_full_g3} for small $g_3$, which is the relevant limit in the weak field regime. 

To obtain the expansion of \eqref{eq:det_full_g3} for small $g_3$, we use the scalar equation of motion (several times if necessary) to replace $\Box\f$ in \eqref{eq:det_full_g3} by $V'(\f)$ and terms which are higher order in $g_3$, in the spirit of order reducing schemes. We then obtain, up to second order:
\begin{align}\nonumber
    \text{det}\br{h^\m_{~\n}} =&~ 1 + 6\,g_3\,V'(\f) + g_3^2\sbr{-6\,V(\f)\,X + 8\,V'(\f)^2 + 12\,X^2 + 4\br{\grad_\m\grad_\n\f}\br{\grad^\m\grad^\n\f}} \\
    &+\mathcal{O}\br{g_3^3}\,.
\end{align}
Similarly, we find:
\begin{equation}
    -\a^2 h^{00} = 1+2\,\t g_3 - g_3^2 \br{X^2 - 2\,\P^2 X}+\mathcal{O}(g_3^3)\,,
\end{equation}
where $\t=K\,\P+D^i\P_i$ is independent of $g_3$, see \eqref{eq:auxvars}. 
Let us focus on the case of zero scalar potential, $V\br{\f}=0$, which is the relevant one for the analysis carried out in this chapter. In this case, the correction to GR in $\det(h^\m_{~\n})$ comes at $\mathcal O(g_3^2)$, while in $ -\a^2 h^{00}$ it comes at leading order. Because $\tau$ does not have a definite sign, then regardless of the sign of $g_3$, there will be regions in spacetime where $ -\a^2 h^{00}$ will vanish before $\text{det}\br{h^\m_{~\n}}$ does, resulting in a Keldysh-type-of transition. This should be the generic behaviour in the $V\br{\f}=0$ case for the $G_3\neq 0$, $G_2=0$ theory, and it is  indeed what we observe in our numerical simulations, see section \ref{sec:G3_neq_0}. The picture changes for $V(\f)\neq 0$ (for instance $V(\f)=\tfrac{1}{2}m^2\f^2$ implies $V'(\f)=m^2\f$); then  $\text{det}\br{h^\m_{~\n}}$ receives a contribution to leading order in $g_3$ and the type of transition will depend on the details of the scalar potential and the initial data. 

\subsection{\done{Excision}}\label{subsection:paper1:excision}

As described before, inevitably the evolution will exit the regime of validity of Horndeski theories inside black holes. To deal with strong field regime inside black holes, in practice we excise a portion of the interior of the black hole. In this section we provide the details of our implementation. 

Rather than performing proper excision, i.e., cutting out a region of the domain, we found that it was easier to modify the evolution equations inside the black hole. The result should be the same, since information cannot escape from this region and affect the physics in the domain of outer communications. Note that in certain Horndeski theories, depending on the sign of the couplings, the scalar field can propagate faster than light and consequently the associated scalar apparent horizon will be inside the black hole horizon \cite{Akhoury:2011hr}. As long as the weak coupling conditions hold, the sound horizons should be close to the usual metric horizon. Nevertheless, to avoid unphysical effects leaking out of the black hole, any modification of the equations of motion should be done in a region contained well within the metric apparent horizons.

Since puncture gauge can handle singularities very well in GR, in practice we turn off all Horndeski terms in a certain region inside the black hole and evolve the standard GR equations there, effectively interpolating from the full equations of motion to GR. To do so, we first define a smooth transition function, valued between 0 and 1, such as the sigmoid-like function:
\begin{equation}
    \sigma(x; \bar{x}, w) = \frac{1}{1 + e^{-\frac{2}{w}\br{\frac{x}{\bar{x}}-1}}}\,,
\end{equation}
where $\bar{x}$ represents the transition point, and $w$ represents the transition width, relative to $\bar{x}$, such that $w\bar{x}$ is the actual width of the transition.\footnote{Roughly, $\sigma${ \footnotesize$\lesssim$ }$0.1$ for $x<\bar{x}(1-w)$ and $\sigma${ \footnotesize$\gtrsim$ }$0.9$ for $x>\bar{x}(1+w)$, and $\sigma$ decays very fast to 0 or 1 outside of this interval.} The metric apparent horizon can be accurately tracked during simulations, but for a sense of what is ``well within the black hole'', contours of the conformal factor $\chi$ are, in puncture gauge, an excellent measure. For example, for a Schwarzschild black hole, after puncture gauge settles, the apparent horizon corresponds to a contour of $\chi$ around $0.25$, reducing to lower values as spin increases along the Kerr family of solutions (see details in appendix \ref{appendix:grchombo:ah_location}). For reasonable choices of Horndeski couplings in the regime of validity of the theory, the scalar apparent horizon is close to the metric horizon. Therefore,  the region inside a certain sufficiently small contour of $\chi$ will contain all apparent horizons.  Denoting by $W$ the maximum of all the weak field conditions \eqref{eq:paper1:WFC_1}, we define the excision function $e(\chi,W)$ as:
\begin{equation}
    e(\chi,W) = \sigma(\chi; \bar{\chi}, -w_\chi)\,\sigma(W; \bar{W}, w_W)\,,
    \label{eq:excision}
\end{equation}
where $w_\chi$ and $w_W$ are two adjustable parameters\footnote{Note that $w_\chi$ is accompanied by a negative sign such that the sigmoid $\s$ goes to $1$ as $\chi\to0$, while it also tends to $1$ when $W\to\infty$.}. In our simulations we typically used $\bar{\chi}=0.08$, $w_\chi = 0.2$, $\bar{W}=1$, $w_W=0.1$. This choice is robust, in the sense that changing these barely affects the evolution across resolutions as long as $\bar{\chi}$ is well within the black hole, which is the case for this choice. It follows from the definition \eqref{eq:excision} that $e\to 1$ when $\chi<\bar{\chi}$ and $W>\bar{W}$, and $e\to0$ otherwise.  We then modify the right hand side of the evolution equations, collectively denoted by RHS, as:
\begin{equation}
    \text{RHS} = (1-e)\text{RHS}_\text{Horndeski} + e~\text{RHS}_\text{GR}\,,
\end{equation}
with $e$ given by \eqref{eq:excision}. This is equivalent to changing the Horndeski couplings of the functions \eqref{eq:relevantGs} to $g_i\to(1-e)g_i$. As a reminder, in practice, we are only modifying the equations of motion in a region where the weak field condition is large and where the theory should not be trusted in any case.

\section{\done{Results}}\label{sec:paper1:results}

In this section we present the results of our numerical simulations of the gravitational collapse of a single massless (no scalar potential) scalar bubble with initial data as in section \ref{subsec:paper1:initial_data}. In all our simulations we keep the radius $r_0$ and the width $\omega$ of the initial Gaussian profile \eqref{eq:paper1:phiData} fixed, and we vary both the amplitude $A$ and Horndeski coupling $g_2$ or $g_3$. The reason is that varying $r_0$ and $\omega$ leads to similar results and varying $A$ alone makes the analysis simpler. We choose $r_0=5$ and $\w=\sqrt{0.5}$, which set the length scale in our simulations. 

\begin{figure}[t!]
\centering
\includegraphics[width=\textwidth]{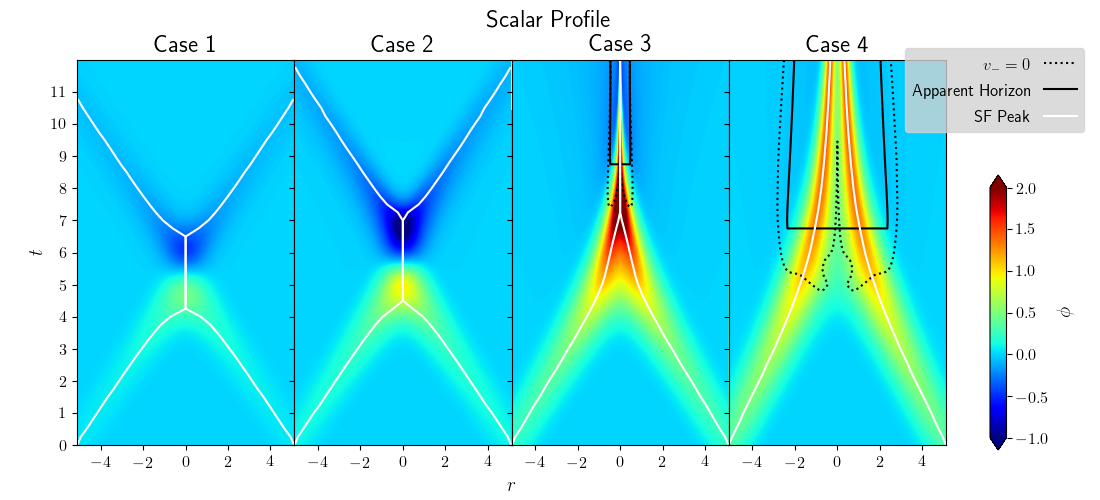}
\vspace{-5mm}
\caption{Evolution of the profile of a massless scalar field (SF) in GR for the Cases 1--4.}
\label{fig:GR_scalar_profile}
\end{figure}
Since we consider spherically symmetric scalar field collapse (even though we do not assume spherical symmetry in our simulations), there are essentially two relevant regimes depending on whether the initial data disperses to infinity (small data) or it collapses into a black hole (large data)\footnote{It may be useful for the reader to see these effects with the visual animation of our simulations of such effects: \href{https://www.youtube.com/watch?v=cfXF1wIcIJc}{https://www.youtube.com/watch?v=cfXF1wIcIJc}.}. We consider four representative values of the initial amplitude $A$, so that we can probe the regimes far and close to critical collapse for both small and large data:\footnote{The endpoints in Cases 1--4 below are obtained by evolving our initial data turning off \textit{all} Horndeski terms, see figure \ref{fig:GR_scalar_profile}. For large enough couplings, the equations may break down before the scalar field has either dispersed or collapsed into a black hole.}
\begin{itemize}
    \item \textbf{Case 1:} $A=0.05$ -- dispersion far from the critical regime, with initial ADM mass of $\approx0.022$.
    \item \textbf{Case 2:} $A=0.10$ -- dispersion closer to critical regime, with initial ADM mass of $\approx0.1$.
    \item \textbf{Case 3:} $A=0.22$ -- collapse into small black hole with initial ADM mass of $\approx0.5$.
    \item \textbf{Case 4:} $A=0.33$ -- collapse into a larger black hole with initial ADM mass of $\approx1.6$.
\end{itemize}
For each of these cases, we vary the Horndeski couplings ($g_2$ or $g_3$) while ensuring that the initial value problem is well-posed. We then  evolve the initial data  by solving the coupled equations of motion \eqref{eq:einstein}--\eqref{eq:scalar} numerically, and we monitor both the hyperbolicity of the scalar equation and the weak field conditions \eqref{eq:paper1:WFC_2}. In this way we can identify the regime of validity of the EFT for both small and large data.  We shall refer to the different cases as ``weakly'' or ``strongly'' coupled depending on the whether the hyperbolicity of the scalar equation breaks down at some point during the evolution; this breakdown is associated to the weak field conditions \eqref{eq:paper1:WFC_2} becoming large compared to the dimensionless couplings \eqref{eq:eta2} and \eqref{eq:eta3}.  The  evolution of the scalar field in GR (i.e., $g_2=g_3=0$) for the Cases 1--4 is shown in figure \ref{fig:GR_scalar_profile}.

It is  worth emphasising that Cases 2 and 3 above do not exhibit Choptuik's critical behaviour, as the amplitude $A$ is purposely chosen to be sufficiently `far' from the critical amplitude $A_*\approx 0.13\pm0.01$. The reason is that Choptuik's critical solution is a naked singularity and EFT will necessarily break down close to it. Indeed, zooming in near the black band in  figure \ref{fig:results_eta2} and  figure \ref{fig:results_eta3} would show that the coupling constants have to be tuned down to maintain the hyperbolicity of the scalar equation as we approach the critical regime from both sides. In addition, the weak field conditions \eqref{eq:paper1:WFC_2} become large the closer we get to the critical solution, as expected.

Since the $G_2\neq 0$ and $G_3\neq 0$ theories do not exhibit significant qualitative differences in terms of the dynamics of collapse of the scalar field, in the next subsection we will focus the discussion on the $G_2\neq 0$ theory considering different values and signs of the coupling constant $g_2$. In subsection \ref{sec:G3_neq_0} will only highlight the main differences in the $G_3\neq 0$ case.

\subsection{\done{$G_2$ theories}}\label{sec:G2_neq_0}

In the following subsections we will discuss gravitational collapse in Horndeski theories with $G_2 = g_2\,X^2$ for different values of the coupling constant $g_2$.  For our scalar field initial data, during collapse a positive and negative peak in $X$ form; these peaks grow as the evolution progresses and the scalar shell approaches the origin. After reaching the origin, they bounce back and smaller peaks of opposite signs form,  eventually resulting in the formation of a black hole or dispersion to infinity. See figure \ref{fig:GR_scalar_profile} for the evolution of the scalar field profile in GR; in the Horndeski theories it is qualitatively similar. With an initial ingoing momentum, as in our initial data, momentum dominates over spatial gradients and the positive peak will be much larger in amplitude than the negative or the subsequent peaks that form after the bounce. Considering the expressions for $\det(h^\m_{~\n})$ and $-\a^2\,h^{00}$ in \eqref{eq:det_h00G2} and \eqref{eq:h00G2} for the $G_2\neq 0$ theory, this implies that generically a negative $g_2$ will lead to a breakdown of the hyperbolicity of the equations for a significantly smaller $|g_2|$ and it will happen sooner than for a positive $g_2$. Furthermore, as described in section \ref{subsec:paper1:case_g2}, for $g_2<0$ the change of character will be of the Keldysh type while for $g_2>0$ it will be of the Tricomi type. 

\subsubsection{\done{Weak coupling}}
\label{sec:g2_pos_weak_coupling}

We first consider the case of a small and positive coupling constant $g_2$; we choose $g_2=0.005$ as a representative example.  This is a case of a theory that remains in the regime of validity of EFT throughout the whole evolution, both for small and large initial data. For this choice of parameters, the maximum of the weak field condition \eqref{eq:paper1:WFC_2} is small everywhere on and outside horizons (if they form) at all times. 

\begin{figure}[t!]
\centering
\includegraphics[width=\textwidth]{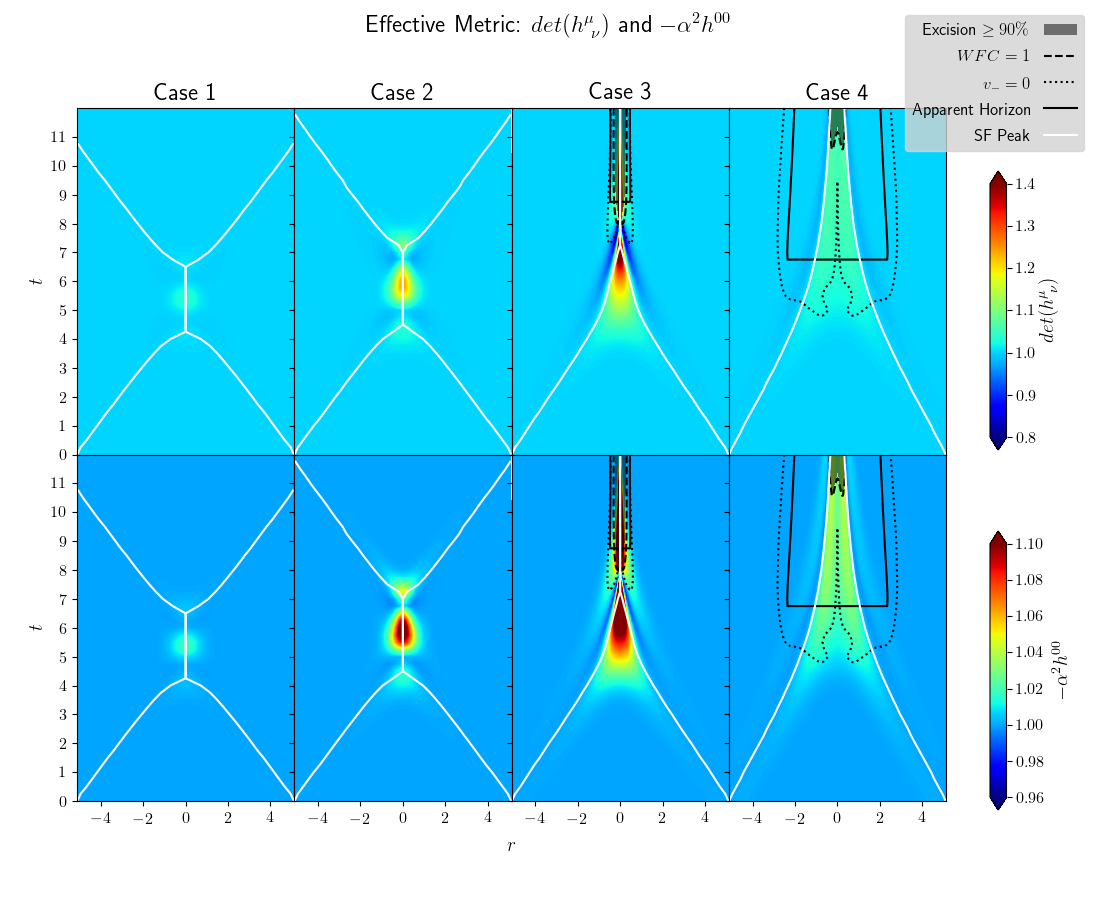}
\vspace{-5mm}
\caption{$\det (h^\m_{~\n})$ (\textit{top}) and $-\a^2\, h^{00}$ (\textit{bottom}) for $g_2=0.005$. The corresponding values of $\eta_2$ are, from left to right, $5\times 10^{-7},\,2\times 10^{-6},\,9.7\times 10^{-6},\,2.2\times 10^{-5}$ respectively.}
\label{fig:weak_g2_pos_det_h00}
\end{figure}
%

%
\begin{figure}[t!]
\centering
\includegraphics[width=\textwidth]{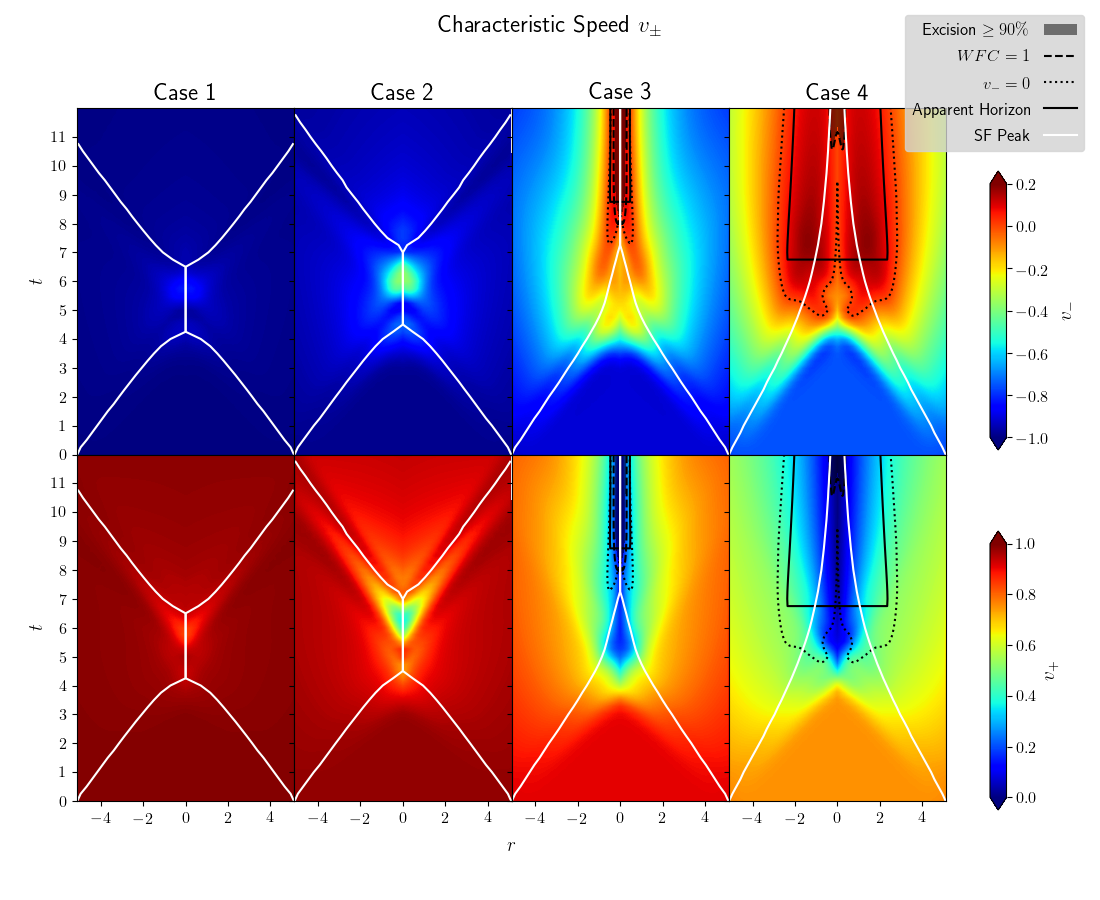}
\vspace{-5mm}
\caption{Outgoing (\textit{top}) and ingoing (\textit{bottom}) scalar characteristic speeds for $g_2=0.005$. The evolution freezes inside black holes as a consequence of the $1+\log$ slicing condition that we use. Sounds horizons form for large initial data.}
\label{fig:weak_g2_pos_speeds}
\end{figure}

In figure \ref{fig:weak_g2_pos_det_h00} we display $\det(h^\m_{~\n})$ (top) and $-\a^2\,h^\m_{~\n}$ (bottom) for Cases 1--4. The white lines in these plots indicate the trajectories of the initial scalar field peak and serve to guide the eye.  In Cases 1 and 2, the scalar field bounces at the origin and eventually disperses to infinity; as the amplitude increases from Case 1 to Case 2, the scalar field spends more time near the origin where gravitational focusing is stronger. For sufficiently large amplitudes (Cases 3 and 4) it collapses into a black hole. In all cases, both $\det (h^\m_{~\n})>0$ and $-\a^2\, h^{00}>0$ throughout the evolution so the scalar equations are hyperbolic at all  times.  The long dashed line in figure \ref{fig:weak_g2_pos_det_h00} indicates the contour where the maximum of weak field condition \eqref{eq:paper1:WFC_2} is equal to one; as we can see, for Cases 1 and 2 the weak field condition is always less than one everywhere in spacetime, while in Cases 3 and 4 and the weak field condition is greater than one only inside the apparent horizon (solid black line). Only in Case 3 there is a small region near the origin where the weak field condition is greater than one and for a short period of time is not covered by an apparent horizon. Note however that this region is already cloaked by the sound horizon (dotted black line), so the scalar modes emanating from this region cannot reach asymptotic observers, and likely by an event horizon as well. For Cases 2--4, $\det(h^\m_{~\n})$ can significantly deviate from 1 (its GR value) when the scalar field is most contracted at the origin. Likewise, the bottom plots in figure \ref{fig:weak_g2_pos_det_h00} show that $-\a^2\, h^{00}$ also exhibits some deviation from its GR value near the origin but it never gets anywhere close to $0$.  Therefore, despite the weak field condition being small at all times, the Horndeski terms can have a significant impact on the dynamics of the system, especially near the origin where the gravitational focusing is strongest. 

In figure \ref{fig:weak_g2_pos_speeds} we display the characteristic speeds for both the outgoing (top) and the ingoing (bottom) modes. Notice that in Case 2, both speeds approach zero at the origin when the scalar field collapses but their sign does not change.  This is indicative of strong gravitational dynamics, as one would expect since Case 2 is ``close'' to the critical regime. Also, note that there are no scalar horizons in this case and all the scalar field eventually disperses to infinity. The dynamics changes in Cases 3 and 4, where a black hole forms. First, notice that $v_-$ changes sign inside the black hole, from negative to positive; this implies that inside the black hole, outgoing modes travel inwards, as expected. Eventually both speeds become close to zero in the region near the singularity. This is just a consequence of using $1+\log$ slicing in our simulations, which effectively freezes the evolution inside black holes. Second, we do observe the formation of scalar horizons, where $v_-=0$ and $v_+>0$.  
In both Cases 3 and 4, the characteristic speed of the outgoing modes is small in the vicinity of the sound horizon; consequently, even though the scalar field can eventually reach infinity, it will  remain near the black hole for a long time, thereby interacting with itself and with the black hole.  

It is apparent from the results shown here that even though the weak field condition \eqref{eq:paper1:WFC_2} is small everywhere, the scalar field still exhibits strong dynamics, such as the dynamical formation of scalar horizons.
Evidently, if the couplings are small then the scalar horizon will be close to the metric horizon. In the case of a black hole binary in a Horndeski theory of gravity, even if the effects of the strong scalar dynamics are locally small, over a sufficiently long time they can lead to significant deviations from GR that may be observable \cite{Figueras:2021abd}. 

\subsubsection{\done{Strong coupling}}

\begin{figure}[th!]
\centering
\includegraphics[width=\textwidth]{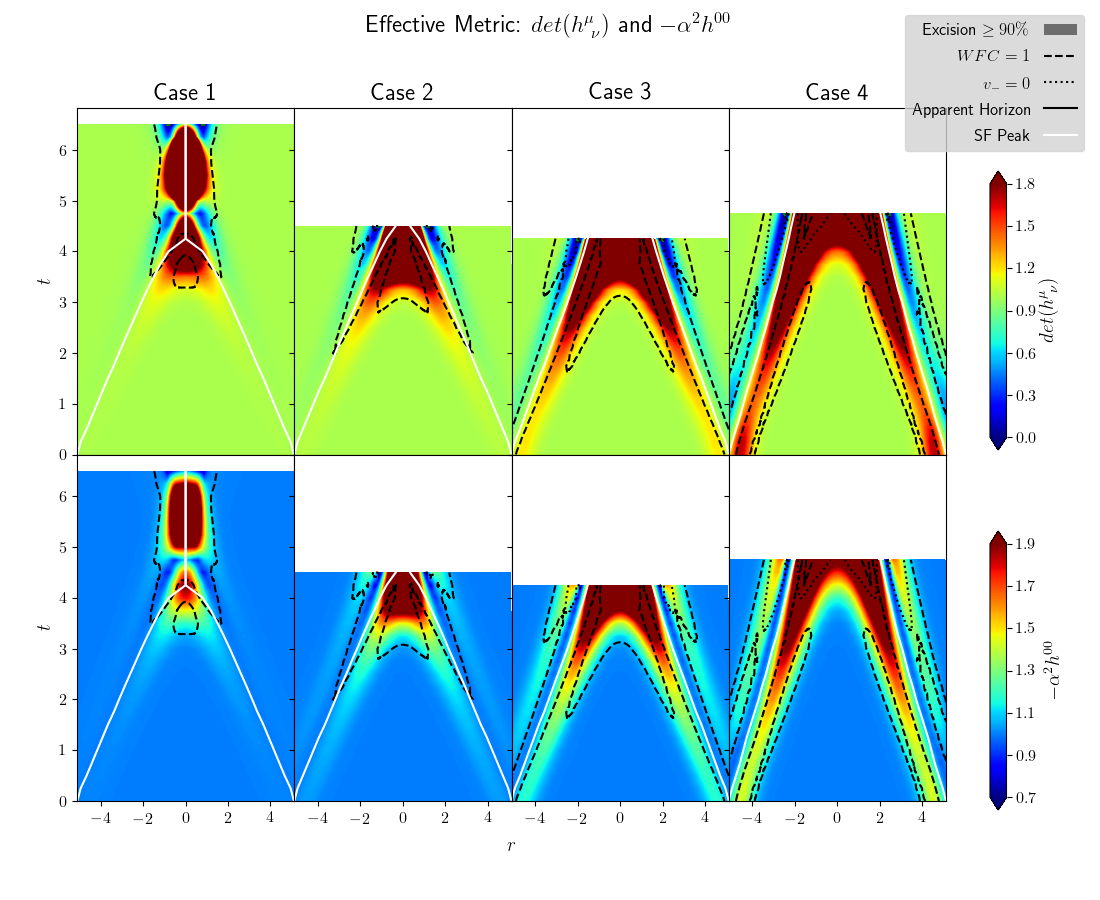}
\vspace{-7mm}
\caption{$\det (h^\m_{~\n})$ (\textit{top}) and $-\a^2\, h^{00}$ (\textit{bottom}) for strong and positive coupling $g_2=1.5$. The corresponding values of $\eta_2$, from left to right, are: $1.5\times 10^{-4},\,6\times 10^{-4},\,2.9\times 10^{-3},\,6.5\times 10^{-3}$.}
\label{fig:strong_g2_pos_det_h00}
\end{figure}

In this subsection we analyse the case for which the $g_2$ coupling is large and positive. We choose $g_2=1.5$ as a representative example. For this value of the coupling constant, the weak field condition \eqref{eq:paper1:WFC_2} can be $\mathcal{O}(1)$ for large initial data, see figure \ref{fig:strong_g2_pos_det_h00} Cases 3 and  4. Therefore, strictly speaking, in these cases the theory is already outside the regime of validity of EFT even though the initial value problem is well-posed. Nevertheless, we choose this value of the coupling constant as an illustrative example of the dynamics of Horndeski theories for large and positive $g_2$. 

%

In figure \ref{fig:strong_g2_pos_det_h00} we display  $\det(h^\m_{~\n})$ and $-\a^2 h^{00}$ during the evolution for our four cases. Unsurprisingly, this figure shows that in all cases the evolution breaks down at some point. For this choice of $g_2$ (and all other strong coupling values of $g_2>0$), the reason why the simulations crash is because $\det(h^\m_{~\n})\to 0$ in a certain region at some instant of time and hence the scalar equation changes character, becoming parabolic. Beyond this point it is not possible to solve the equations as an initial value problem. For this value of the Horndeski coupling, for all Cases 1--4 the weak field condition \eqref{eq:paper1:WFC_2} has become large before the equations change character. Also, note that for large initial data (Cases 3 and 4), the evolution breaks down before an apparent horizon has had time to form and hence the pathology in the scalar equations of motion cannot be hidden behind the horizon. Figure \ref{fig:strong_g2_pos_det_h00} (bottom) shows that in all cases $-\a^2 h^{00}$ deviates significantly from its GR value, but remains well above zero up until the breakdown of the evolution. Likewise, we observe that in these simulations the characteristic speeds of both the ingoing and outgoing modes remain bounded at all times. Therefore, the loss of hyperbolicity for the  $g_2>0$ theories is due to a Tricomi-type-of transition, in accordance with the discussion in section \ref{subsec:paper1:case_g2}. 

By lowering the coupling constant a bit, it is possible to hide the strong scalar field dynamics that causes the breakdown of the hyperbolicity of the equations inside a large enough black hole. This is illustrated in appendix \ref{appendix:paper1:other}, figure \ref{fig:appendix:intermediate_g2_pos}. For such ``intermediate'' couplings, the evolution still breaks down in Cases 2 and 3, while in Case 4 the pathologies that develop in the scalar equation can be hidden behind the horizon. In this case, one can continue the evolution without encountering any issues. Moreover, the weak field condition in Case 4 remains small on and outside the black hole horizon despite the fact that $g_2$ is large. This contains an important lesson. One could fear that any of these theories were doomed to failure given that for strong enough initial data or coupling (which, in the end, is partially a matter of the length scale chosen) they break down. But, our hope is partially restored when realising that is not the case: even for strong coupling, if our initial data is strong enough (big amplitude in this case) then a trapped surface forms fast enough such that any pathology stays well hidden behind an horizon and the theory does not breakdown.

\subsubsection{\done{Negative coupling}}

\begin{figure}[t!]
\centering
\includegraphics[width=\textwidth]{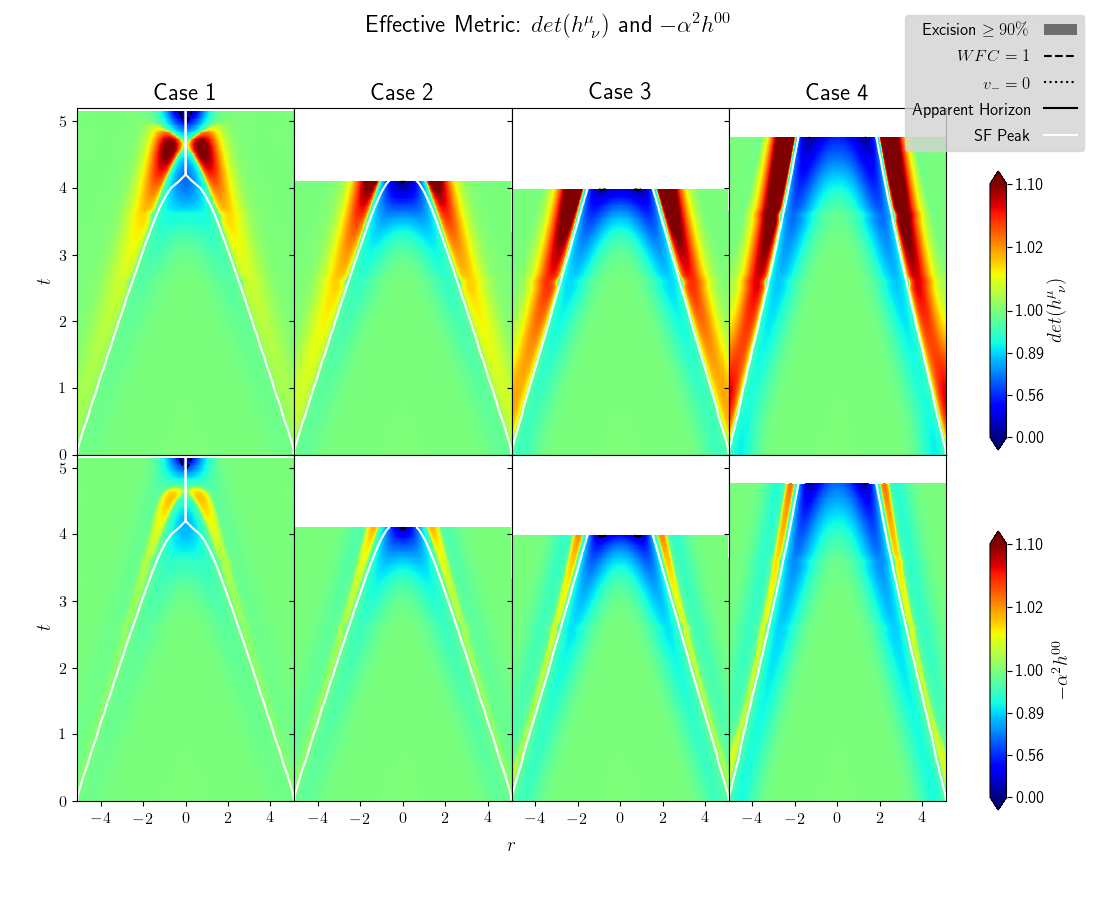}
\vspace{-7mm}
\caption{$\det (h^\m_{~\n})$ (\textit{top}) and $-\a^2\, h^{00}$ (\textit{bottom}) for a strong and negative coupling $g_2=-0.2$. The corresponding values of $\eta_2$, from left to right, are: $-2\times 10^{-5},-8\times 10^{-5},-3.9\times 10^{-4},-8.7\times 10^{-4}$.}
\label{fig:strong_g2_neg_det_h00}
\end{figure}

In this subsection we discuss the case of a strong and negative coupling constant $g_2$. As an illustrative example, we consider $g_2=-0.2$. 

\begin{figure}[t!]
\centering
\includegraphics[width=\textwidth]{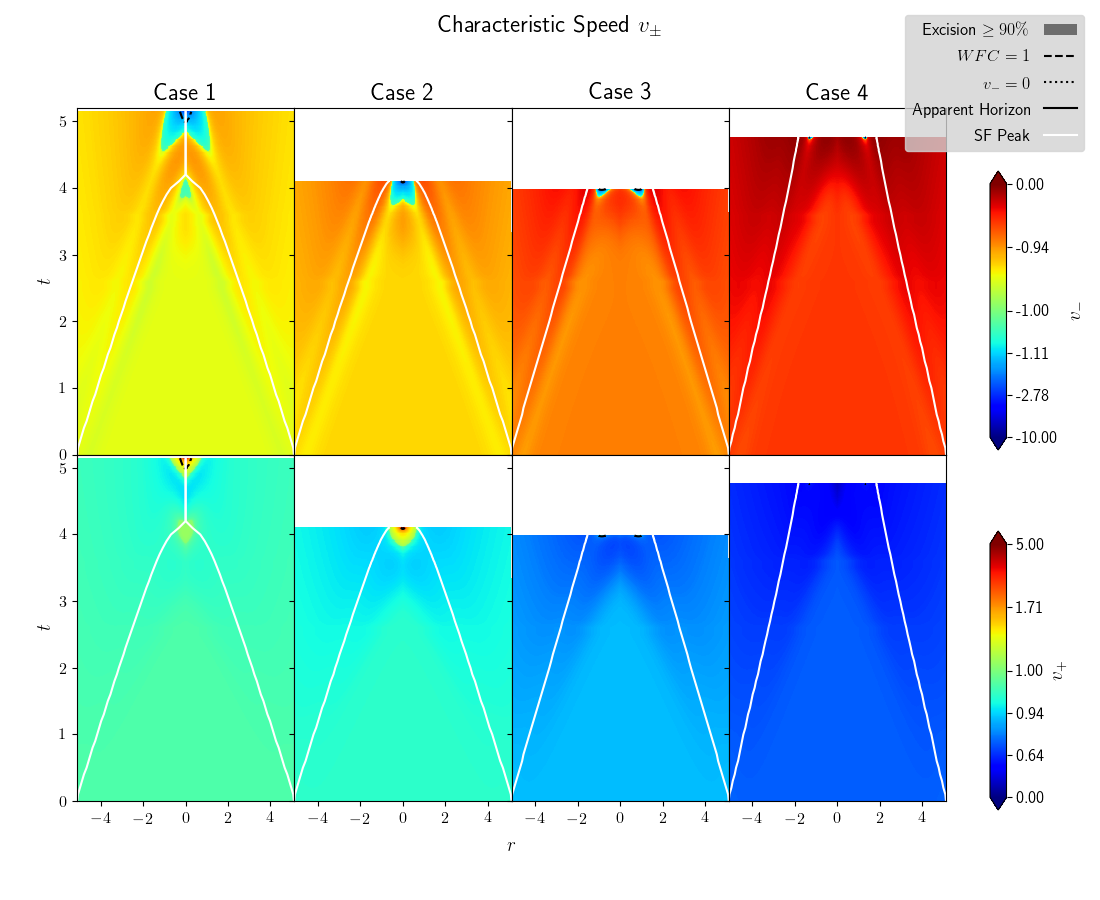}
\vspace{-7mm}
\caption{Characteristic speeds of the outgoing (\textit{top}) and ingoing (\textit{bottom}) scalar modes for $g_2=-0.2$. Both characteristic speeds simultaneously diverge when the evolution breaks down, but $v_-$ does so faster than $v_+$. This behaviour is in accordance with a Keldysh-type-of transition.}
\label{fig:fig:strong_g2_neg_speeds}
\end{figure}

As anticipated in section \ref{subsec:paper1:case_g2}, the dynamics of the scalar field changes quite significantly for negative couplings. First, a smaller absolute value of $g_2$ is enough to cause a breakdown of the hyperbolicity of the scalar equations for both small and large data. The results are shown in figures \ref{fig:strong_g2_neg_det_h00} and \ref{fig:fig:strong_g2_neg_speeds}. In all cases we find that $-\a^2 h^{00}\to 0$ before $\det (h^\m_{~\n})\to 0$, even though this is not easily seen from figure \ref{fig:strong_g2_neg_det_h00}. This implies that, in our gauge, the $t=\textrm{const.}$ surfaces are no longer spacelike with respect to the scalar effective metric before the scalar equation changes character. The fact that for $g_2<0$, $-\a^2 h^{00}\to 0$ first results in infinite characteristic speeds of propagation for both the ingoing and outgoing modes, see figure \ref{fig:fig:strong_g2_neg_speeds}. Therefore, we associate the breakdown of the hyperbolicity of the scalar equation to a Keldysh-type-of transition, in accordance to the discussion in section \ref{subsec:paper1:case_g2} (see also \cite{Bernard:2019fjb}). The diverging characteristic speeds near the transition point imply that the dynamics of the scalar field becomes increasingly fast right before it breaks down; to adequately resolve it, in our simulations we had to significantly reduce the Courant factor. However, at some point  it is no longer feasible in practice to keep reducing it, and numerical errors eventually build up until the simulation inevitably crashes. A possible way out would be to change our slicing conditions to ensure that the $t=\textrm{const}.$ hypersurfaces remain spacelike with respect to both $h^{\mu\nu}$ and $g^{\mu\nu}$, but we have not attempted to do so here.

Notice that for this value of the coupling constant, the weak field condition \eqref{eq:paper1:WFC_2} is always less than one everywhere in spacetime, including the region near the origin where gravitational focusing is strongest, except immediately before the breakdown. This is simply a consequence of the fact that  $|\eta_2|$ is small in all Cases 1--4. Related to this last observation, the breakdown occurs before either sound horizons or apparent horizons have had time to form, so the pathologies cannot be hidden from asymptotic observers. However, when the breakdown occurs, even though the weak field condition \eqref{eq:paper1:WFC_2} may be as small as $\mathcal{O}(10^{-2})$, this is still much larger than $\eta_2$, thereby suggesting that the system is strongly coupled. We expect that a refined weak field condition should be able to capture that this case indeed becomes strongly coupled in a precise sense before the breakdown of the evolution. 

Needless to say, for sufficiently small absolute values of $|g_2|$ the  scalar equations remain hyperbolic at all times for Cases 1--4. In this situations the evolution is qualitatively similar to the small and positive $g_2$ case that we have already discussed in subsection \ref{sec:g2_pos_weak_coupling}. Likewise, for a given $g_2<0$ and a sufficiently large $A$, the pathologies in the scalar equation can be hidden inside the black hole horizon, as seen for $g_2>0$ in the previous section.

\subsection{\done{$G_3$ theories}}\label{sec:G3_neq_0}

\begin{figure}[t!]
    \centering
    \hspace*{-10mm}\includegraphics[width=1.15\textwidth]{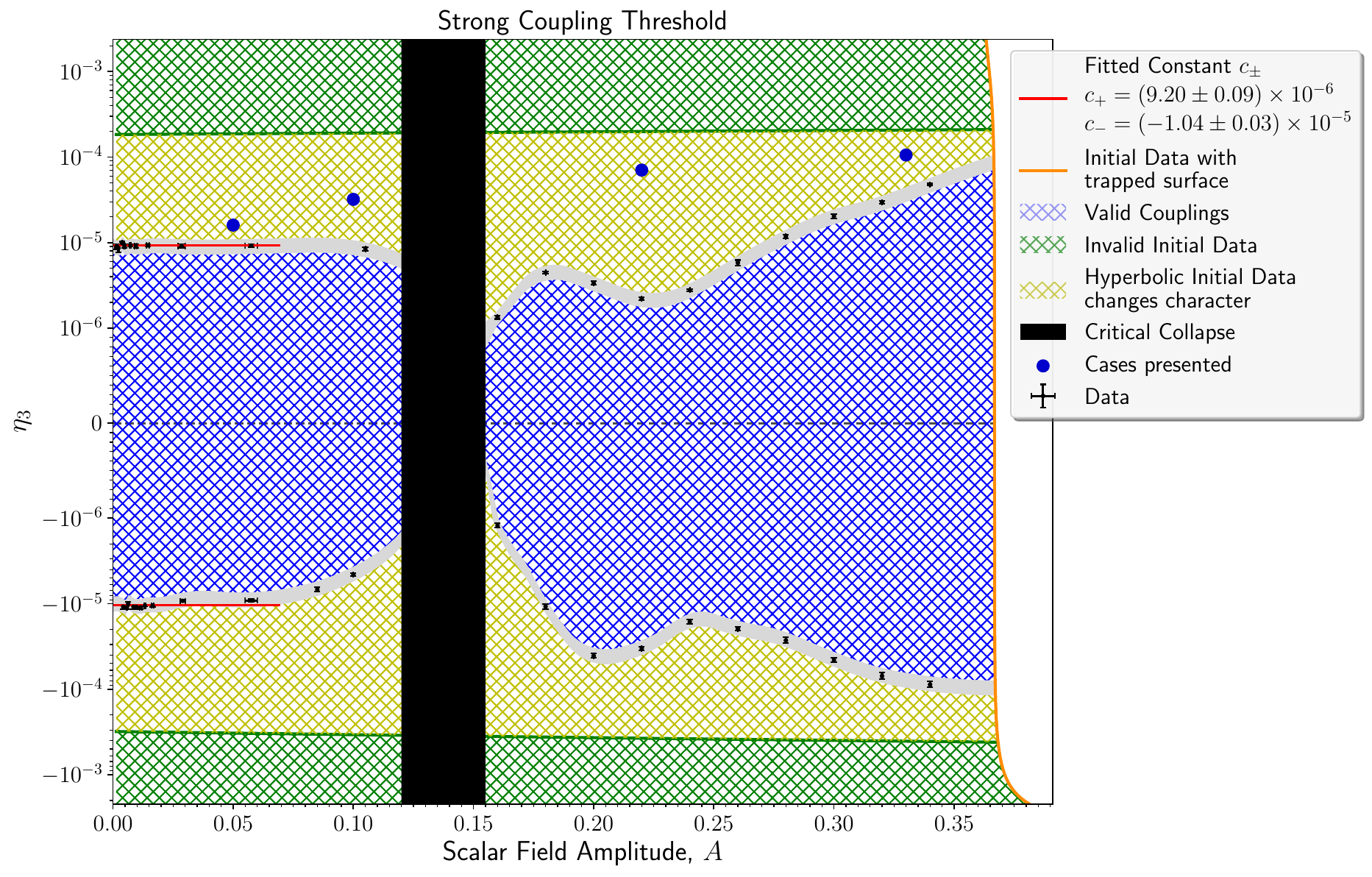}
    \caption{Dynamical regimes for the $G_3 = g_3\,X$ theory as a function of the initial amplitude $A$ and the dimensionless coupling constant $\eta_3$, see equation \eqref{eq:eta3}. The black band denotes the region near critical collapse; black holes form to the right of this band. The orange curve on the right marks the region where the initial data contains a trapped surface. The scalar equation is hyperbolic at all times in the blue region; EFT is valid in the interior of this region. In the yellow region, the scalar equation is initially hyperbolic but it changes character during the evolution. In the green region the initial value problem is not well-posed.}
    \label{fig:results_eta3}
\end{figure}

In this subsection we will briefly comment the dynamics in Horndeski theories with $G_3 = g_3\,X$ and $G_2=0$. In all cases that we have explored, either for $g_3>0$ or $g_3<0$, the dynamics is qualitatively similar to the $G_2=g_2\, X^2$ theories with $g_2<0$, so we will not go into much detail. 

As discussed in section \ref{subsection:paper1:case_g3}, we expect that for sufficiently small absolute values of $g_3$, the breakdown of the scalar evolution equations is expected to be due to a Keldysh-type-of transition. Our numerical simulations confirm that this is indeed the case for either signs of $g_3$. In figures \ref{fig:appendix:strong_g3_pos_det_h00} and \ref{fig:appendix:strong_g3_pos_speeds} of appendix \ref{appendix:paper1:other} we show the results for a representative case with $g_3=0.4$. In figure \ref{fig:appendix:strong_g3_pos_det_h00} we see that $-\a^2 h^{00}\to 0$ before $\det(h^\m_{~\nu})$ does, resulting in infinite characteristic speeds, as expected in a Keldsyh-type-of transition. In this case we observe that $v_-$ diverges as $-\a^2 h^{00}\to 0$ while  $v_+$ remains finite (see figure \ref{fig:appendix:strong_g3_pos_speeds}). Note that in this particular example, for large data (Cases 3 and 4) the evolution breaks down before the first apparent horizon appears.

Figure \ref{fig:results_eta3} summarises our results for the $G_3=g_3\,X$ theories. The colour code is the same as in figure \ref{fig:results_eta2} and the qualitative features are also the same. The black band corresponds to the range of $A$ for which the future development of initial data becomes close to Choptuik's critical solution. Black holes form for $A$ to the right of the black band while for $A$'s to left, the scalar field disperses.  As before, global solutions to this particular Horndeski theory can be constructed for values of $(A,\eta_3)$ in the blue region. The regime of validity of EFT corresponds to the interior of the blue region, away from its boundaries. For $0<A\lesssim 0.05$, the boundary between the blue and yellow regions is at a constant value of $\eta_3$  given by $\eta_3\sim(9.20\pm 0.09)\times 10^{-6}$ and $\eta_3\sim(-1.04\pm 0.03)\times 10^{-5}$. The two main lessons are similar to the $G_2\neq 0$ case: for low enough amplitude, $\eta_3$ is a perfect predictor of the stability of the evolution; for big amplitude, we realise that even if we increase $g_3$, or $\eta_3$, there is a big enough amplitude so that the theory remains in the regime of validity of EFT on and outside black holes, by engulfing the instability fast enough inside an horizon.

\section{\done{Conclusions}}\label{sec:paper1:conclusions}

In this chapter we have studied the regime of validity of certain cubic Horndeski theories of gravity that have a well-posed initial value problem. We have chosen two particularly simple cases, namely \eqref{eq:relevantGs}, but we expect that our results should extend as well to other generic models with an effective metric, at least in the weakly coupled regime which is where these theories should be valid EFTs. For instance, for a single massive scalar field the results are qualitatively unchanged during gravitational collapse. Nevertheless, one expects that a massive scalar field will stay trapped around the black hole for a much longer time, forming scalar clouds \cite{Clough:2018exo}. This can spark interesting results as, over long periods of time such as in a black hole binary inspiral, locally small deviations from GR introduced by Horndeski theories may accumulate, giving rise to significant deviations. 

For the particular class of models that we have studied, the reason why the evolution breaks down is because the scalar equation changes character. For the $G_2=g_2\,X^2$ theory the transition can be of the Tricomi type for $g_2>0$, while for $g_2<0$ the transition is of the Keldysh type. On the other hand, for the $G_3=g_3\,X$ theory, we have only observed a breakdown \textit{à la} Keldysh. However, this is not generic for the $G_3$ theories; other choices such as $G_3=g_3\,X^2$ can exhibit both behaviours. Furthermore, we have provided some level of analytic justification for the types of pathologies that may arise in each of the models that we have considered.  

In order for the initial value problem to be well-posed and the theory be a consistent (truncated) EFT, we need to impose that a certain weak field condition \eqref{eq:paper1:WFC_2} is suitably small. For certain choices of initial conditions and couplings (no fine-tuning required) the conditions in \eqref{eq:paper1:WFC_2} can be $\mathcal{O}(1)$ and yet the scalar equation of motion is perfectly hyperbolic. Conversely, the conditions in \eqref{eq:paper1:WFC_2} can be $\mathcal{O}(10^{-2})$ and yet the scalar equation changes character. In either case, the weak field conditions at the time of breakdown are much larger than the dimensionless couplings, \eqref{eq:eta2}-\eqref{eq:eta3}, of the initial data. Therefore, in a certain sense, the theory becomes strongly coupled by the time the hyperbolicity is lost. It would be very interesting to obtain a sharp condition that identifies the truly weakly coupled regime of the theory and provides  some analytic understanding of it, at least for certain classes of initial data. Using the tools of Reall \cite{Reall:2021voz} to analyse characteristic polynomials, it would also be interesting to analyse the case of cubic Horndeski in some appropriate generalised harmonic gauge, just as done for $G_4$ and $G_5$ theories like Gauss-Bonnet theory, to potentially understand whether the failure of hyperbolicity in the full theory occurs before the failure of hyperbolicity of the reduced theory.

Having identified the regime of validity of the Horndeski theories that we have considered,  we can proceed to study black hole binaries for initial data in this regime. These studies were presented in the companion paper \cite{Figueras:2021abd} (see chapter \ref{chapter:paper2}).

\chapter{\done{Black hole binaries in cubic Horndeski theories}}\label{chapter:paper2}

\section{\done{Introduction}}

Using the results of chapter \ref{chapter:paper1}, we study black hole binary mergers, treating the theory fully non-linearly while remaining the regime of validity of EFT in all phases of the binary.

When we discuss alternative theories of gravity, we refer to the `strongly coupled regime' of the theory as the regime in which the new terms in the equations of motion that modify GR are comparable (or even larger) to the original (two-derivative) terms. Conversely, by the `weakly coupled regime' we will mean the regime of the theory in which the modifications to the GR equations of motion are small. This is compatible with still being in the strong field regime of gravity. It is in the weakly coupled regime that alternative theories of gravity make sense as low energy effective field theories (EFTs).

Up until recently, only the so-called scalar-tensor and the scalar-vector-tensor theories of gravity had been studied fully non-linearly \cite{Healy:2011ef,Barausse:2012da,Hirschmann:2017psw,Sagunski:2017nzb}. The reason is that for this class of theories, it is straightforward to find a well-posed formulation. For other, more general, classes of theories involving higher derivatives and yet second order equations of motion, such as Horndeski or Lovelock theories, finding a suitable well-posed formulation turns out to be far more difficult. In fact, it has been shown that weak hyperbolicity can fail in Lovelock \cite{Reall:2014pwa} or Horndeski \cite{Ripley:2019aqj,Ripley:2019hxt,Bernard:2019fjb,Ripley:2019irj,Ripley:2020vpk,Bezares:2020wkn,Figueras:2020dzx,R:2022hlf} theories if the spacetime curvature and/or the derivatives of the scalar field become too large, i.e., in the strongly coupled regime. In a recent breakthrough, \cite{Kovacs:2020pns,Kovacs:2020ywu} showed that these theories can be strongly hyperbolic in certain modified generalised harmonic coordinates, also in the weakly coupled regime, i.e., when the deviations from GR are small. These theoretical developments have led to the first studies of the fully non-linear dynamics of black holes in a particular subset of these theories, namely scalar Einstein-Gauss-Bonnet theory \cite{East:2020hgw,East:2021bqk,East:2022rqi,Corman:2022xqg} or the more general four-derivative scalar tensor theory \cite{AresteSalo:2022hua}.

We should remark that even if a well-posed formulation can be found for a certain alternative theory of gravity, it is possible that during the dynamical evolution of certain classes of initial data, the hyperbolicity of the equations of motion is lost due to strong coupling effects.  In the context of Einstein-dilaton-Gauss-Bonnet and Horndeski theories, it has been shown that this loss of hyperbolicity is due to Tricomi or Keldysh transitions, in which characteristic speeds go to zero or diverge in finite time respectively, effectively changing the character of the scalar equation of motion from hyperbolic to parabolic or elliptic, after which time evolution cannot proceed \cite{Ripley:2019hxt,Ripley:2019irj,Ripley:2019aqj,Bernard:2019fjb,Ripley:2020vpk,Figueras:2020dzx}.

Therefore, given these recent theoretical developments, it is the right time to start probing the non-linear regime of alternative theories of gravity and infer predictions for black hole binary mergers. Building on our previous work \cite{Figueras:2020dzx} (see chapter \ref{chapter:paper1}), in this chapter we study black hole binary mergers in cubic Horndeski theories. The reason why we consider these theories in the present chapter and in chapter \ref{chapter:paper1} is because these theories are known to be well-posed in the standard gauges used in numerical GR \cite{Kovacs:2019jqj}. As we previously mentioned, the general case has also been shown to be well-posed by Kovàcs and Reall \cite{Kovacs:2020pns,Kovacs:2020ywu}, but in a modified version of the generalised harmonic coordinates. While here we only consider the cubic case for simplicity and convenience, one may expect that some of our conclusions hold for more general Horndeski theories. 

In chapter \ref{chapter:paper1} we studied gravitational collapse of a massless scalar in spherical symmetry in certain cubic Horndeski theories\footnote{Whilst in this chapter the initial data was chosen to be spherically symmetric, our code did not assume spherical symmetry.} given by the choices
\begin{align}
        G_2(\f,X) &= g_2\, X^2, \label{eq:choiceG2}\\
        G_3(\f,X) &= g_3\, X, \label{eq:choiceG3}
\end{align}
where $g_2$ and $g_3$ are arbitrary dimensionful coupling constants that we can tune. This particular choice of $G_2$ is well-motivated by EFT, since it is the leading order correction to GR minimally coupled to a scalar field \cite{Weinberg:2008hq}. On the other hand, this choice of $G_3$ is a matter of simplicity and the convenience of being able to use the standard BSSN/CCZ4 formulation. Both choices have been extensively considered in the literature (see e.g., \cite{Brahma:2020eqd, Ijjas:2018cdm, Shahalam:2016kkg, Deffayet:2010qz} and references therein).

One of the main results of chapter \ref{chapter:paper1} was to identify the region in the space of initial conditions and couplings such that the solution in the domain of dependence of the initial data surface remained in the weakly coupled regime of the theory on and outside black hole horizons if any are present, see section \ref{sec:paper2:methodology} for more details. This is relevant in the context of EFT to justify that one can consistently keep only the leading order terms beyond GR, i.e., Horndeski, and neglect the otherwise (presumably) infinite number of higher derivative corrections. At the same time, it is consistent to treat the theory fully non-linearly, as we do here. 

In the present chapter we consider the same theories as in chapter \ref{chapter:paper1}, namely \eqref{eq:choiceG2}--\eqref{eq:choiceG3}. For the initial data, we choose two boosted lumps of scalar field, with amplitudes chosen so that they quickly collapse into black holes, thus forming a black hole binary. Whilst most of the scalar field is absorbed by the black holes during the initial collapse stage, a scalar cloud remains in their vicinity throughout the lifetime of the binary. This scalar cloud can interact with itself and with the black holes and, over sufficiently long times, give rise to interesting effects. Furthermore, since spacetime curvature can source the scalar field via the Einstein equations, it is conceivable that when the spacetime curvatures are large, i.e., in the merger phase of a binary, one can observe sizeable deviations from GR. We mostly consider massive scalar fields and we restrict ourselves to a choice of potential $V(\phi)=\frac{1}{2}m^2\phi$, where $m$ is the mass of the scalar field. The reason is that the corresponding scalar cloud can remain in the vicinity of the black holes for longer \cite{Hui:2019aqm} and hence there is a greater chance of producing larger deviations from GR. The initial separations and velocities of the scalar lumps are tuned so that the black holes that form describe an eccentric binary that merges in 5 orbits. As we shall see in section \ref{sec:paper2:results}, eccentric binaries seem to be particularly well-suited to detect small deviations from GR since the system enters the strong field regime in every close encounter of the binary and not only in the merger phase. Circular binaries are more computationally costly but they exhibit a build up of the deviations from GR also during the inspiral phase and not only in the merger phase. As such, one can expect even larger deviations from GR.

Finally, we choose the coupling constants $g_2$ and $g_3$ such that on the initial data surface, the solution lies well inside the weakly coupled regime and we monitor that the solution remains in this regime throughout the evolution; this is necessary to ensure the consistency of the truncated EFT.

Note that for the cubic Horndeski theories, the natural frame to consider is the Einstein frame. This would not be the case had we considered more general Horndeski theories such as $\mathcal{L}_4$ \cite{Faraoni:1999hp}. On top of this, since massive scalar fields cannot propagate to the wave-zone\footnote{In a skematic way, for the flat space wave equation, solutions to $\square \f = 0$ decay as $\phi \propto \frac{1}{r}$, whereas solutions to the massive case $(\square-m^2)\phi=0$ decay exponentially faster as $\f \approx \frac{e^{-mr}}{r}$.}, the waveforms presented in section \ref{sec:paper2:results} would look the same in the Einstein and Jordan frames respectively.

\subsection{\done{Outline}}

The rest of the chapter is organised as follows. In section \ref{sec:paper2:methodology} we describe our methods, numerical techniques and construction of suitable initial data. Section \ref{sec:paper2:results} contains the main results of the chapter. In section \ref{subsec:paper2:waves}, we present the waveforms of elliptic binaries computed in various cubic Horndeski theories and we compare them to the waveforms obtained in GR coupled to a scalar field. In section \ref{subsec:paper2:scalar_cloud} we discuss the properties of the scalar cloud surrounding the black holes and in section \ref{subsec:paper2:resultsWCCs} we show that in our simulations, the weak coupling conditions are satisfied throughout the evolution of the binaries. In section \ref{subsec:paper2:mismatch} we analyse the mismatch between GR and Horndeski. In section \ref{subsec:paper2:circular_binaries}, we describe waveforms for circular binaries. In section \ref{sec:paper2:conclusions} we summarise the main results of the chapter and we discuss future directions for research. The convergence tests are presented in appendix \ref{appendix:paper2:convergence}. Appendix \ref{appendix:paper2:strain_time_vs_frequency} describes details in the computation of the gravitational strain.

\section{\done{Methodology}}\label{sec:paper2:methodology}

\subsection{\done{Equations of motion}}

The equations of motion and numerical implementation are the same as the ones of chapter \ref{chapter:paper1} (see section \ref{subsec:paper1:EOM}).

\subsection{\done{Initial data}}\label{subsec:paper2:initial_data}

For initial data, we consider the superposition of two equal boosted scalar field bubbles, similarly to what was done by Pretorius \cite{Pretorius:2005gq}. Each individual scalar bubble follows the setup of section \ref{subsec:paper1:initial_data}: spherically symmetric and with some in-going momentum, which prompts a quick collapse into a black hole without any outgoing scalar wave while leaving some leftover dynamical scalar hair surrounding the black hole. Whilst the individual scalar bubbles satisfy the Hamiltonian and momentum constraints, the superposition of the two does not; however, we place them sufficiently far apart so that the initial constraint violations due to superposing the two scalar bubbles are sufficiently small. Further details and explicit form of the constraint satisfying scalar bubble profiles used can be found section \ref{subsec:paper1:initial_data}.

For binary systems, we boost the individual profiles with a Galilean boost with velocity $\ve{v}$. This can be implemented in the scalar momentum by adding to it (with the appropriate sign) the Galilean boost\footnote{This follows simply from a transformation from unboosted to boosted coordinates $(t',\ve{x}')\to(t=t',\ve{x}=\ve{x}'+\ve{v}\,t')$. With the usual decomposition of the unit normal into the lapse function and shift vector, $n^\m=(\tfrac{1}{\a},-\tfrac{\b^i}{\a})$, then $\Pi = n^\m\grad_\m\f = \tfrac{1}{\a}\br{\frac{\pd}{\pd t} - \b^i\frac{\pd}{\pd x^i}}\f$. The shift transforms as a vector under Galilean boosts, leaving $\b^i\frac{\pd}{\pd x^i} = \b'^{i'}\frac{\pd}{\pd x'^{i'}}$ unchanged, whereas $\frac{\pd}{\pd t} = \frac{\pd}{\pd t'} - v^i\frac{\pd}{\pd x'^{i'}}$, resulting in equation \eqref{eq:pi_boosted}.}:
\begin{equation}\label{eq:pi_boosted}
    \P(t,\ve{x})\big|_{t=0} = \P_{\text{original}}(t,\ve{x})\big|_{t=0} -\tfrac{1}{\alpha}\ve{v}\cdot\ve{\grad}\phi(t,\ve{x})\big|_{t=0}\,,
\end{equation}
where $\a$ is the lapse function and $\Pi=\mathcal{L}_n\phi=n^\m\grad_\m\phi$ is the scalar momentum, where $n^\m$ is the unit normal 4-vector to spatial hypersurfaces. Such a boost is valid for small velocities and avoids the obstacle of having to evaluate the initial data at different times for distinct points, as one would need to do to implement a proper Lorentz boost. Doing the latter would be unfeasible for non-static initial data.

To superpose the initial data of two scalar bubbles, $A$ and $B$, boosted in opposite directions, we use \cite{Radia:2021smk}:
\begin{equation}
    \begin{aligned}
        \psi &= \psi_A + \psi_B - 1\,,\\
        K_{ij} &= \g_{m(i}\br{K_{j)n}^A\g_A^{nm} + K_{j)n}^B\g_B^{nm}}\,,\\
        \phi &= \phi_A + \phi_B\,,\\
        \Pi &= \Pi_A + \Pi_B\,,
    \end{aligned}
    \label{eq:init_dat}
\end{equation}
where $\psi$ is the conformal factor associated with the induced metric on the initial data surface, $\g_{ij}$, so that $\g_{ij} = \psi^4\ti{\g}_{ij}$ and $\tilde \gamma_{ij}$ has unit determinant. The initial data corresponding to each of the individual scalar bubbles is conformally flat, so $\tilde \gamma_{ij}=\tilde\gamma^A_{ij}=\tilde\gamma^B_{ij}=\delta_{ij}$, and satisfies $K^A=K^B=0$. For $K=0$ and a conformally flat spatial metric, the condition for $K_{ij}$ above reduces to $\ti{A}_{ij} = \ti{A}_{ij}^A + \ti{A}_{ij}^B$, where $\ti{A}_{ij}$ is the conformal traceless part of the extrinsic curvature given by $\tilde A_{ij}=\psi^{-4}K_{ij}$. In our simulations, we initialise the lapse and shift vector as $\a = 1$ and $\b^i = 0$, as the initial perturbation of the scalar field on the metric is mild.

Motivated by the similarity with previous studies of binary black holes mergers in scalar field environments, we also tried an alternative way of constructing initial data by superposing scalar bubbles as suggested in Helfer et al. \cite{Helfer:2021brt}. The construction of Helfer et al. \cite{Helfer:2021brt} produces initial data which is physically different from \eqref{eq:init_dat} but, for the large separations between the initial scalar bubbles as in this analysis, the amounts of initial constraint violations are comparable. We identified that the evolution of this initial data leads to qualitatively similar results after one orbit. Therefore, henceforth we will only discuss the evolution of the binaries constructed using \eqref{eq:init_dat}.

Using the notation of equation \eqref{eq:paper1:phiData}, we use $(A,r_0,\w) = \br{0.21, 5, \sqrt{\tfrac{1}{2}}}$ and a scalar mass parameter $m=0.5$ for the scalar potential. For an isolated scalar bubble, this configuration has an ADM mass of approximately $M\approx 0.52$. 
The mass term in the scalar potential accounts for about $10\%$ of the total ADM mass, while the contribution of the Horndeski terms for a coupling of $g_2=0.02$ or $g_3=0.05$ is of order $\mathcal{O}(10^{-5})$ (as expected from the associated values of $\eta_2$ and $\eta_3$).

The binaries with the eccentric orbits presented in this chapter are obtained by choosing a ``large'' initial separation\footnote{This ``large'' initial separation makes a circular binary unfeasible with our computational resources, but it helps to minimise the errors from the initial data superposition. A more circular binary is presented in section \ref{subsec:paper2:circular_binaries}.} between the centres of the scalar bubbles of $D=40$ and an individual scalar boost velocity of $|\ve{v}| = 0.17$; after the initial gravitational collapse, the resulting black holes have an initial velocity of $0.042$. For GR, we calculate numerically that the superposed data has a total ADM mass of $M=1.0346\pm0.0001$. The values quoted above are in code units, and the mass parameter $m$ and couplings $g_2$, $g_3$ will be referred without units henceforth (e.g., $g_2=0.02$ as opposed to $g_2\sim0.0214\,M^{-2}$ after taking into account that the total mass is $M=1.0346$).

\subsection{\done{Weak coupling conditions}}\label{subsec:paper2:wccs}

In order for the Horndeski theories to be valid EFTs, the basic requirement is that corrections to the two-derivative GR terms in the equations of motion on and outside black hole horizons (if there are any in the spacetime) should be small at all times. Inside black holes both GR and the Horndeski theories break down but classically this region is inaccessible to external observers. Therefore, in practice, we only monitor the weak coupling conditions (WCCs) on and outside black hole horizons.\footnote{While the weak cosmic censorship conjecture \cite{Penrose:1969pc,Christodoulou:1999ve} remains unproven in the astrophysical settings considered in this chapter, we will assume that it holds. We do not find any evidence against it in our setting.} 

For the cases considered in this chapter, the WCCs translate into the requirement of equations \eqref{eq:paper1:WFC_2}, with the addition of the mass parameter in the scalar field potential, $m$, to the length scale estimate of the system\footnote{It turns out that for our choice of couplings and scalar mass parameter $m$, near the black holes the contribution of $m$ to $L$ is always smaller than the metric and scalar curvature terms.}:
\begin{equation}\label{eq:paper2:lengthWCC}
    L^{-1}=\text{max}\{\left|\mathcal{R}_{\a\b\m\n}\right|^{\tfrac{1}{2}},\left|\grad_\m\f\right|,\left|\grad_\m\grad_\n\f\right|^{\tfrac{1}{2}},m\}\,.
\end{equation} 
We consider values for the couplings $g_2$ and $g_3$ based on the valid values of $\eta_2$ and $\eta_3$ displayed in figures \ref{fig:results_eta2} and \ref{fig:results_eta3}. In practice, for massive scalar fields, the scalar clouds that form near black holes tend to be more extended, have lower densities and smaller gradients than in the massless case, thus allowing for larger couplings without violating the WCCs that may lead to a loss of the hyperbolicity of the scalar equations.

\subsection{\done{Excision}}

We implement excision as explained in section \ref{subsection:paper1:excision}. We only implement this during the initial stages of the evolution, while $t \lesssim 40M$, namely during gravitational collapse and gauge re-adjustment phases.\footnote{The initial data is not in puncture gauge, so in the initial stages of the simulation there is a certain amount of artificial dynamics in grid variables due to the adjustment of the coordinates.} Once the black hole has stabilised, the matter density at its centre is very small and no loss of hyperbolicity of the scalar equations occurs.

\subsection{\done{Gravitational wave extraction}}\label{subsec:paper2:gravitational_wave_extraction}

We extract gravitational waves at finite radii by projecting the Weyl scalar $\Psi_4$ onto the spin-weighted spherical harmonics on multiple spheres of fixed coordinate radius in the standard way, obtaining the multipoles $\psi_{\ell m}$ (see section \ref{subsubsection:ngr:observables}). Data on each integration sphere is obtained from the finest available level in the numerical grid at a given extraction radius, using fourth-order Lagrange interpolation. We calculate the Weyl scalar $\Psi_4$ using the Newman-Penrose formalism \cite{Baker:2001sf} and the electric and magnetic parts of the Weyl tensor, $E_{ij}$ and $B_{ij}$ \cite[p.~289]{alcubierre}. The latter can be computed from our evolution variables using the following expressions adapted to the 3+1 CCZ4 formulation of the Einstein equations (see details in appendix \ref{appendix:grchombo:weyl}):
\begin{align}
    E_{ij} &= \big[R_{ij} + D_{(i}\Theta_{j)} + (K-\Theta)K_{ij} - K_{im}K^{m}_{\phantom{m}j} - \tfrac{\kappa}{4}S_{ij}\big]^{\text{TF}}\,, \label{eq:Eij}\\
    B_{ij} &= \epsilon_{mn(i}D^m K_{j)}^{\phantom{j)}n}\,,\label{eq:Bij}
\end{align}
where $R_{ij}$ is the 3 dimensional Ricci tensor, $K_{ij}$ the extrinsic curvature, $\Theta:=-n_\m Z^\m$ is the projection of the Z4 vector, $Z^\m$,\footnote{The components of the Z4 vector essentially correspond to the Hamiltonian and Momentum constraints, assumed to be zero in the BSSN formulation.} onto the timelike unit normal vector $n^\m$, $S_{ij}:=\g_i^{~\m}\g_j^{~\n}T_{\m\n}$ is the spatial projection of the stress-energy tensor, $\g_{\m\n} := g_{\m\n}+n_\m n_\n$ is the induced metric on the spatial hypersurfaces, $\e_{\m\n\r} = n^\s \e_{\s\m\n\r}$ is the volume form on such hypersurfaces and $[\cdot]^{TF}$ denotes the trace-free part of the expression in square brackets. Note that equations \eqref{eq:Eij}-\eqref{eq:Bij} guarantee that $E_{ij}$ and $B_{ij}$ are automatically trace-free and symmetric, unlike usual 3+1 ADM expressions \cite[p.~289]{alcubierre}, which require that the constraints are satisfied (see details in appendix \ref{appendix:grchombo:weyl}).

\subsection{\done{Gravitational strain}}\label{subsec:paper2:strain}

The natural observable measured in detectors and used when constructing waveform templates is the gravitational strain. In the conventions of Alcubierre \cite[p.~308]{alcubierre}, the strain of a gravitational wave, $h$, can be obtained from the $\Psi_4$ Weyl scalar using the transformation \cite[p.~308]{alcubierre} \cite{Radia:2021smk}:
\begin{equation}\label{eq:paper2:strain:def}
    \Psi_4 = -\ddot{h} = -\ddot{h}^+ + i \ddot{h}^\times\,,
\end{equation}
where the dot $\dot{\,}$ denotes a time derivative, and $h^+$ and $h^\times$ are the usual plus and cross polarisations of the wave. This gives the strain multipoles $\ddot{h}^+_{\ell m} = -\text{Re}(\psi_{\ell m})$ and $\ddot{h}^\times_{\ell m} = \text{Im}(\psi_{\ell m})$, where $\psi_{\ell m}$ are the amplitudes of each mode in the multipolar decomposition of the Weyl scalar $\Psi_4$. To avoid artefacts from finite length of the wave, discrete sampling and noisy data, we perform the double time integration in the frequency domain using a fixed low frequency filter \cite{Reisswig:2010di}, with a cutoff of $0.01M^{-1}$ for low frequencies\footnote{This choice affects the noise in the strain, but low frequencies have negligible effects in the final computation of the mismatch.} (see appendix \ref{appendix:paper2:strain_time_vs_frequency} for more details). Adding a cutoff for high frequencies resulted in no improvement. We taper the signal in the time domain with a Tukey window \cite{McKechan:2010kp} of width $\sim 40M$ on each side\footnote{This choice reduces noise, but it does not affect the results in any meaningful way, as the signal is essentially zero in this region.} and zero-pad to the nearest power of two. We further increase the length of the waveform data points by a factor of eight, zero-padding it, before applying the fast Fourier transform \cite{Varma:2018mmi}. This increases the frequency resolution of the discrete Fourier transform and reduces the noise for low frequencies introduced by the discretisation. Removing the initial junk radiation of the inspiral from the time domain did not result in any significant improvement. Finally, the signal at null infinity can be obtained by extrapolating the results at finite radii \cite{Scheel:2008rj}. Using the tortoise radius $r^*=r + 2M \log\left|\frac{r}{2M}-1\right|$, one first aligns separate extraction radii in retarded time, $u = t - r^*$, and then extrapolates the aligned waves assuming a Taylor series expansion in $\frac{1}{r^*}$ \cite{Sperhake:2010tu}. This is preferred over Richardson extrapolation which incurs high errors with noisy data.

\subsection{\done{Waveform mismatch}}\label{subsec:paper2:mismatch_theory}

In order to estimate the difference between two waveforms, we compute the mismatch between the strain resulting from each wave. First, given the strain of two waves, $h_1(t)$ and $h_2(t)$, one can compute the overlap, $\mathrm{O}$, using the frequency domain inner product \cite{Blackman:2017dfb, Owen:1995tm, Lindblom:2008cm,Cutler:1994ys,Giesler:2019uxc}:
\begin{equation}
\begin{aligned}\label{eq:paper2:inner_product_strain}
    \mathrm{O}(h_1, h_2) &= \frac{\text{Re}\br{\left<h_1,h_2\right>}}{\sqrt{\left<h_1,h_1\right>\left<h_2,h_2\right>}}\,,\\
    \left<h_1,h_2\right> &= 4\int_{f_{\text{min}}}^{f_{\text{max}}} \frac{\ti{h}_1^*(f)\ti{h}_2(f)}{S_n(f)}df\,,
\end{aligned}
\end{equation}
where $\ti{h}(f)$ denotes the Fourier transform of the function $h(t)$, $^*$ denotes complex conjugation, $S_n(f)$ is the power spectral density (PSD) of a detector's strain noise as a function of frequency $f$ (e.g., updated Advanced LIGO sensitivity design curve \cite{aLIGOupdatedantsens}), $f_{\text{min}}$ and $f_{\text{max}}$ is the lowest and highest frequency cutoffs of the PSD of the detector or the frequency minimum/maximum imposed by the timestep and duration of the simulation\footnote{Notice that the strain is a dimensionless quantity. Moreover the inner product \eqref{eq:paper2:inner_product_strain} can be computed in geometric units for flat PSD, but LIGO's PSD is in physical units, $\text{Hz}=s^{-1}$. To convert Hertz into geometric units, one can use the conversion factor $f_{G=c=1} = n_\odot \cdot g_\odot \cdot f_{\text{Hz}}$, where $g_\odot = \frac{GM_\odot}{c^3} = 4.927\,\m$s and $n_\odot$ is the number of solar masses considered for the system's total mass.}. Notice that for $h_1=h_2$, $\left<h_1,h_2\right>$ is real.

Then, we compute the mismatch by maximising the overlap, $\mathrm{O}$, over time and phase shifts of the second wave, $h_2^{\d t,\f}(t) = h_2(t+\d t)\,e^{i\f}$:
\begin{align}
    \text{mismatch} &= 1 - \underset{\d t,\f}{\text{max}}~\mathrm{O}(h_1, h_2^{\d t,\f})\,.
\end{align}
Noticing that $\ti{h}_2^{\d t,\f}(f) = \ti{h}_2(f) e^{i\f}e^{2\p i\d t}$ and
\begin{align}
    \mathrm{O}(h_1, h_2^{\d t, \f}) = \frac{\text{Re}\br{e^{i\f}e^{2\p i\d t}\left<h_1,h_2\right>}}{\sqrt{\left<h_1,h_1\right>\left<h_2,h_2\right>}}\,,
\end{align}
then maximising over phase shifts corresponds to simply taking the absolute value of $\left<h_1,h_2^{\d t}\right>$ as opposed to the real part (i.e. $\underset{\f}{\text{max}} ~\text{Re}(e^{i\f}z) = |z|$, for any complex number $z$). Maximising over time shifts is more subtle because in the discrete domain with a finite time window, the discrete Fourier transform changes by more than a mere phase $e^{2\p i\d t}$. Hence, we perform the time shift maximisation numerically. To allow for continuous time shifts, we interpolate the data $h_{1,2}(t)$ and re-sample appropriately after the time shift. The number of points when re-sampling the time series was checked not to affect the final result. When comparing two gravitational waves, the length of the time interval used for the Fourier transform and the size of the frequency domains used for integration are enforced to be the same for all waves (taking into account the time shifts). All in all:
\begin{align}
    \text{mismatch} &= 1 - \frac{\underset{\d t}{\text{max}} ~\left|\left<h_1,h_2^{\d t}\right>\right|}{\sqrt{\left<h_1,h_1\right>\left<h_2,h_2\right>}}\,.
\end{align}

\subsection{\done{Numerical scheme}}

On top of what was already described in section \ref{subsec:ngr:grchombo_params}, for the simulations presented in this chapter, we use a tagging criterion that triggers the regridding based on second derivatives of both the scalar field and the conformal factor, forcing certain levels around the location of apparent horizons and around gravitational wave extraction regions. We use Kreiss-Oliger numerical dissipation with fixed $\s=1$ in all our simulations. As for boundary conditions, we use Sommerfeld boundary conditions and take advantage of the reflective/bitant symmetry of the binary problem to evolve only half of the grid. Sixth order spatial stencils are used in order to improve phase accuracy of the binaries \cite{Husa:2007hp}. Time updates are still made with a fourth-order Runge-Kutta scheme, which implies that the global convergence order cannot be higher than four.\footnote{Notice that this allows us to still use the usual KO dissipation stencils that are commonly implemented with fourth order finite differences.} For the results presented in this chapter, we have a \textit{Courant-Friedrichs-Lewy} factor of $1/4$, a coarsest level resolution of $\D x = \frac{16}{7}$, with 8 additional refinement levels, and a computational domain of size $L = 1024$.

\section{\done{Results}}\label{sec:paper2:results}

In this section we present the results of our numerical simulations for the various Horndeski theories that we have considered and we compare them to GR (with a minimally coupled scalar field). To carry out the comparison, we consider standard GR coupled to a massive scalar field (with same mass parameter $m=0.5$ as in Horndeski). We comment on the massless scalar field case in section \ref{subsec:paper2:scalar_cloud}.
We have also considered the evolution of a black hole binary in vacuum GR with the same total ADM mass and initial velocities for the black holes. In this case, the binary describes many more orbits before merger, as expected, since no energy is transferred to the scalar field. We will not comment any further on this case since it is not relevant for the kind of comparisons that we carry out.

We have constructed superposed initial data for GR coupled to a massive scalar field and for Horndeski theories. One could question whether different results arise from small differences in the initial data. As discussed in chapter \ref{chapter:paper1}, the effect of the Horndeski terms in the initial data is proportional to $\frac{g_2A^2}{r_0^2}$ and $\frac{g_3Aw^2}{r_0^4}$ ($\eta_2$ and $\eta_3$) depending on the theory. For the values of the couplings $g_2$ and $g_3$ that we consider, this results in a difference of order $\mathcal{O}(10^{-5})$ between the Horndeski and the GR counterpart. To confirm that the small Horndeski corrections in the initial data do not affect the subsequent evolution, we evolved the equations of motion of the Horndeski theories using initial data constructed for GR. Clearly this procedure introduces extra initial constraint violations proportional to the Horndeski couplings. However, our results from the Horndeski theories initialised with GR initial data and those results obtained using proper Horndeski initial data do not exhibit any significant or quantitative difference. Therefore, we conclude that the differences observed between GR and Horndeski theories are caused by the evolution with distinct evolution equations and not by the extremely small differences in the initial data. Henceforth, for the Horndeski theories we will only present results obtained with Horndeski initial data.
\begin{figure}[h]
\centering
\includegraphics[width=0.45\textwidth]{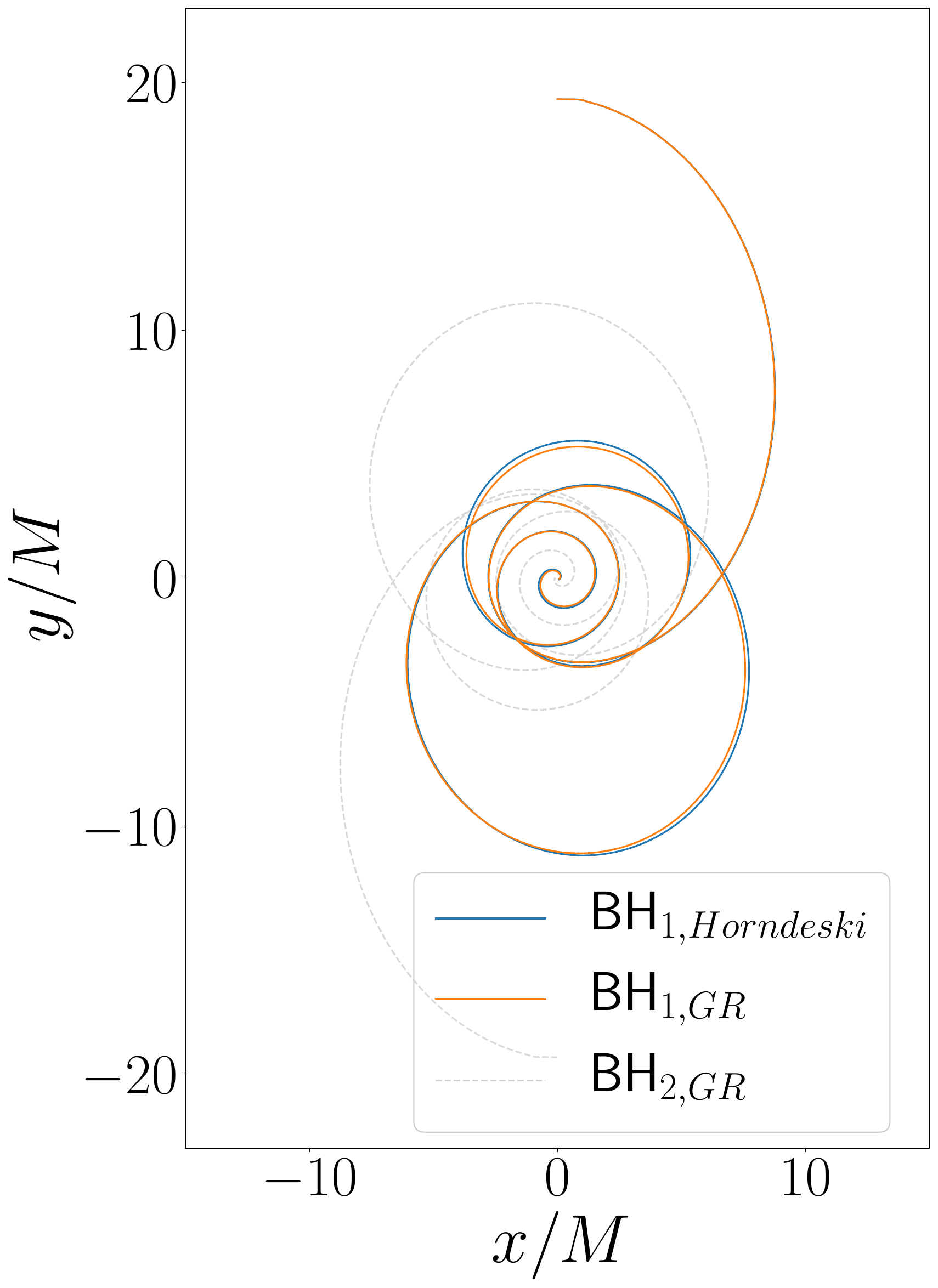}
\caption{Orbits of the two black holes in Horndeski for $g_2=0.02$ and GR. For clarity, in the Horndeski case we only show one of the black holes.}
\label{fig:AH_xy}
\end{figure}

In figure \ref{fig:AH_xy} we display the trajectories of the punctures on the orbital plane for GR and for a Horndeski theory with $g_2=0.02$. This figure shows that after the first close encounter of the binary, the trajectories that the black holes follow in GR and in Horndeski are visibly different. Interestingly, the black holes seem to recombine to the same trajectory in the final stages of binary. In the following subsections we will quantify the differences in other observables such as the gravitational strain.

\subsection{\done{Waveform strain}}\label{subsec:paper2:waves}

\begin{figure}[t]
\centering
\includegraphics[width=0.98\textwidth]{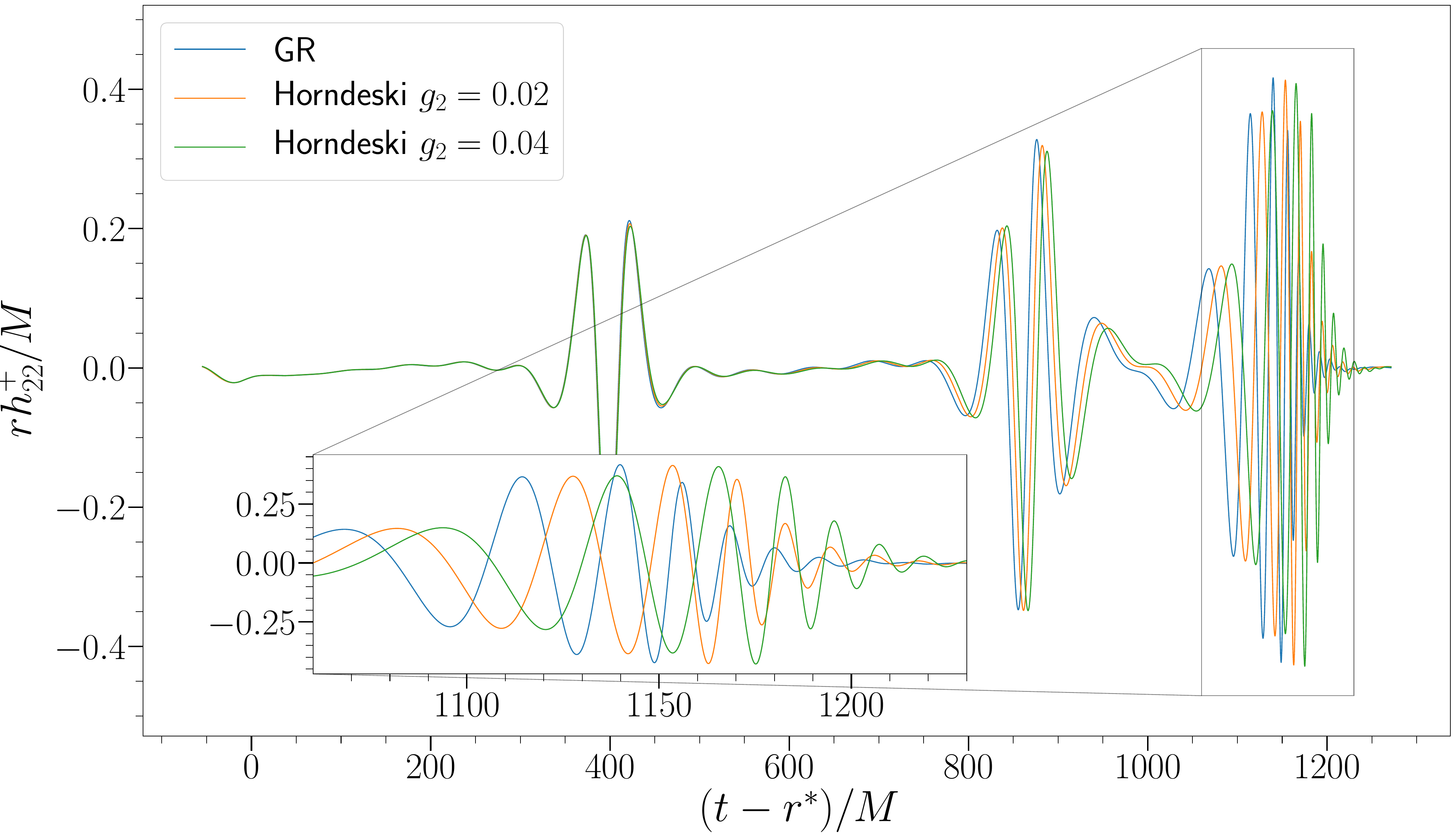}
\caption{\label{fig:g2_0.02_gw}Comparison of gravitational wave between Horndeski theory with $g_2=0.02$, $g_2=0.04$ and GR in retarded time, $u=t-r^*$, where $r^*$ is the tortoise radius. Displaying the $(\ell,m)=(2,2)$ mode of the plus polarisation of the strain, $h^+_{22}$, extrapolated to null infinity. There is a visible misalignment between GR and Horndeski that builds up over time, becoming larger during the merger phase.}
\end{figure}

\begin{figure}[t]
\centering
\includegraphics[width=0.98\textwidth]{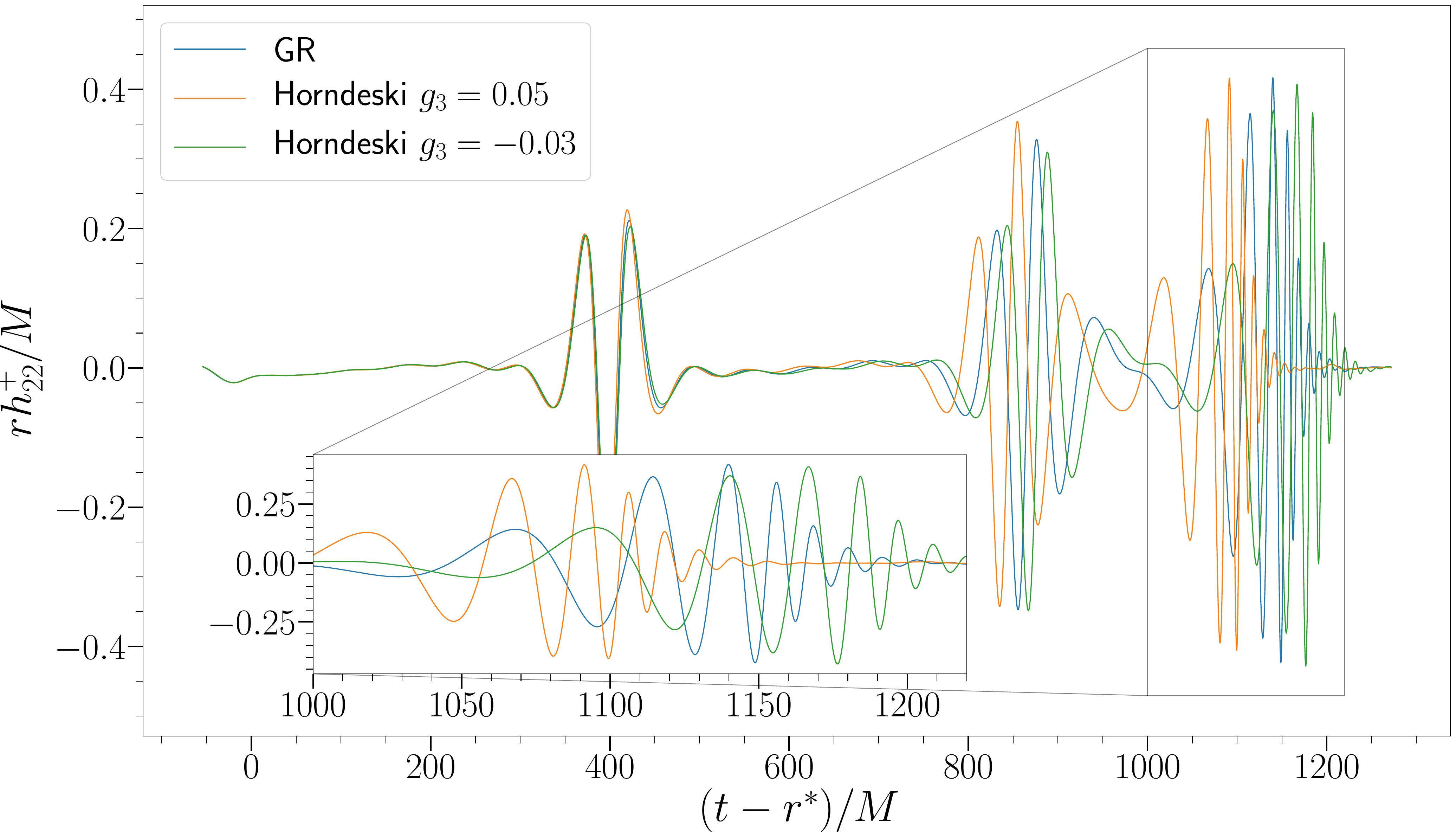}
\caption{\label{fig:g3_gw}Comparison of gravitational wave between GR and Horndeski theories with $g_3=0.05$ and $g_3=-0.03$, in retarded time, $u=t-r^*$, with $r^*$ the tortoise radius. Displaying the $(\ell,m)=(2,2)$ mode of the plus polarisation of the strain, $h^+_{22}$, extrapolated to null infinity. As in the $G_2$ theory, figure \ref{fig:g2_0.02_gw}, we observe a misalignment that builds up over time.}
\end{figure}

In this subsection we compare the waveform strain for eccentric binaries in GR and in different Horndeski theories. For the latter, we consider both the $G_2$ and the $G_3$ theories for different values and signs of the coupling constants. In figures \ref{fig:g2_0.02_gw} and \ref{fig:g3_gw} we present the $(\ell,m)=(2,2)$ mode of the plus polarisation of the strain, $h^+$, extrapolated to null infinity using 6 radii between $50-150M$, for the $G_2$ and $G_3$ theories respectively, using the method described in section \ref{sec:paper2:methodology}. The strains for higher $(\ell,m)$ modes have lower amplitude and more noise, but exhibit qualitatively similar features.

Referring to figures \ref{fig:g2_0.02_gw} and \ref{fig:g3_gw}, the two peaks that can be seen at $t\sim 400M$ and $t\sim 850M$ correspond to the bursts of radiation emitted during the first two close encounters of the eccentric binary\footnote{It may be useful for the reader to match the gravitational wave signal in these figures with the visual animation of one of our simulations: \href{https://www.youtube.com/watch?v=uOed4AG1ulg}{https://www.youtube.com/watch?v=uOed4AG1ulg}.} before the final merger phase. The latter starts at around $t\sim 1100M$ and ends by $t\sim 1200M$, depending on the theory and the value and sign of the coupling constants. As for the final state, to the best of our knowledge, it is not known if the class of theories that we consider admit stationary hairy black holes.\footnote{For the theories considered in this chapter and in the shift-symmetric case, Refs. \cite{Hui:2012qt,Sotiriou:2013qea,Sotiriou:2014pfa,Maselli:2015yva} have proven that no slowly rotating hairy black holes exist, but in our case the mass term explicitly breaks the scalar shift symmetry in the equations of motion.} However, we find evidence that the end state of the evolution is an approximately Kerr black hole surrounded by a scalar cloud that decays in time on a time scale much longer than that set by the initial ADM mass. This evidence is reinforced by the similarity between the final scalar cloud profiles between GR and Horndeski, as described in section \ref{subsec:paper2:scalar_cloud}. For the runs shown in figures \ref{fig:g2_0.02_gw} and \ref{fig:g3_gw}, the estimated parameters of the final black holes are summarised in Table \ref{tab:runs}. Note that any junk radiation caused by the initial constraint violations or choice of initial data is very small on the scale of these figures, but still visible in the first $\sim 50M$.

\begin{table}[h]
\centering
\begin{tabular}{cccccc}
\hline\hline
\textrm{Coupling}&
\textrm{Final Mass $M_F/M$}&
\textrm{Spin Parameter $a/M$}\\
\hline
$\text{GR}$ & $0.973\pm0.001$ & $0.676\pm0.001$ \\
$g_2=~~0.005$ & $0.973\pm0.001$ & $0.676\pm0.001$ \\
$g_2=~~0.02~$  & $0.973\pm0.001$ & $0.675\pm0.001$ \\
$g_2=~~0.04~$  & $0.973\pm0.001$ & $0.673\pm0.001$ \\
$g_3=~~0.05~$  & $0.975\pm0.001$ & $0.680\pm0.001$ \\
$g_3=-0.03~$  & $0.972\pm0.001$ & $0.672\pm0.001$ \\
\hline\hline
\end{tabular}
\caption{\label{tab:runs}%
Parameters of the final state Kerr black hole for each coupling $g_2$ and $g_3$. The mass and spin are estimated from the final apparent horizon since the ADM quantities are typically significantly nosier. The errors are estimated from the differences between the medium and high resolution runs.}
\end{table}

As figures \ref{fig:g2_0.02_gw} and \ref{fig:g3_gw} show, the waveforms obtained in GR and in the various Horndeski theories that we considered, coincide during the initial stages of the binary, but a clear misalignment builds up over time, starting from the second close encounter of the binary and becoming more pronounced in the merger phase. This misalignment is much larger than the smallness parameters $\eta_{2,3}\sim\mathcal{O}(10^{-5})$, controlling the weak coupling conditions of the initial data. In subsection \ref{subsec:paper2:resultsWCCs} we will provide evidence showing that a suitable local weak coupling condition remains small during the whole evolution of the binary and hence, in our setting, the Horndeski theories should be valid (and predictive) classical EFTs. 
The large misalignment that we observe in figures \ref{fig:g2_0.02_gw} and \ref{fig:g3_gw} is a cumulative effect arising from the locally small differences between GR and Horndeski, and it gets enhanced whenever the system enters the strong field regime, which happens in each close encounter of the eccentric binary and in the merger phase. This is expected since the corrections to GR are sourced by spacetime and scalar curvature and those become more important precisely in the strong field regime. 
Therefore, eccentric binaries seem to be useful to potentially detect deviations from GR sourced by curvature through the built up of small cumulative effects and their enhancement in the close encounters. It is conceivable that linearising the Horndeski theories around GR may allow one to compute some of the misalignment (at least for some small enough couplings) during the merger phase since its duration is relatively short and secular effects may not be an issue. However, it seems unlikely that such an approach would be able to capture the cumulative large deviations that arise from successive close encounters of an eccentric binary, such as in the examples considered here. The relatively long times that we have evolved the binaries require a full non-perturbative treatment of the theory to avoid potential secular effects.

For the $G_2$ theory \eqref{eq:choiceG2}, a positive $g_2$ coupling induces a delay of the waveform when compared to GR, whilst a negative $g_2$ gives rise to an advancement of the signal. On the other hand, for the $G_3$ theory \eqref{eq:choiceG3} the effect is the opposite: a positive $g_3$ coupling leads to an advancement of the signal while a negative $g_3$ leads to a delay when compare to the GR waveform. In general, the observed misalignment between GR and Horndeski seems to be a rather generic effect that does not depend on the specifics of the theory. Of course, the details such as the amount or the sign of the deviations will depend on the details theory under consideration. Therefore, we are tempted to conjecture that gravitational strain computed in general Horndeski theories of gravity that do not admit equilibrium hairy black holes but with dynamical long-lived scalar clouds surrounding black holes will be misaligned with respect of the GR signals. Finally we note that the peak amplitude of the waveforms seems to be very similar across all theories and couplings. We will point out in the section \ref{sec:paper2:conclusions} how this misalignment may be potentially detected in gravitational wave observations.

Note that the final state of GR and Horndeski simulations for $g_2\le0.04$ seems to have the same exact mass and only tiny differences in spin (see Table \ref{tab:runs}). This is counter-intuitive given the differences that the gravitational waveforms exhibit and it could be related to the fact that the trajectories of the black holes in the two theories visibly differ in the intermediate stages of the binary, but coincide again near the merger phase (see figure \ref{fig:AH_xy}). The physical mechanisms behind this observation may be related to the frequency shifts analysed in subsection \ref{subsec:paper2:mismatch}. The fact that the initial and final state coincide and yet the waveforms are different indicates that, at least for equal mass non-spinning binaries, the degeneracy between the class of Horndeski theories that we have considered and GR is broken. This suggests that the degeneracy between GR and Horndeski may also be broken for unequal mass non-spinning configurations. It would be interesting to study the effects of the intrinsic spins in alternative theories of gravity. 
\begin{figure}[t]
\centering
\includegraphics[width=0.98\textwidth]{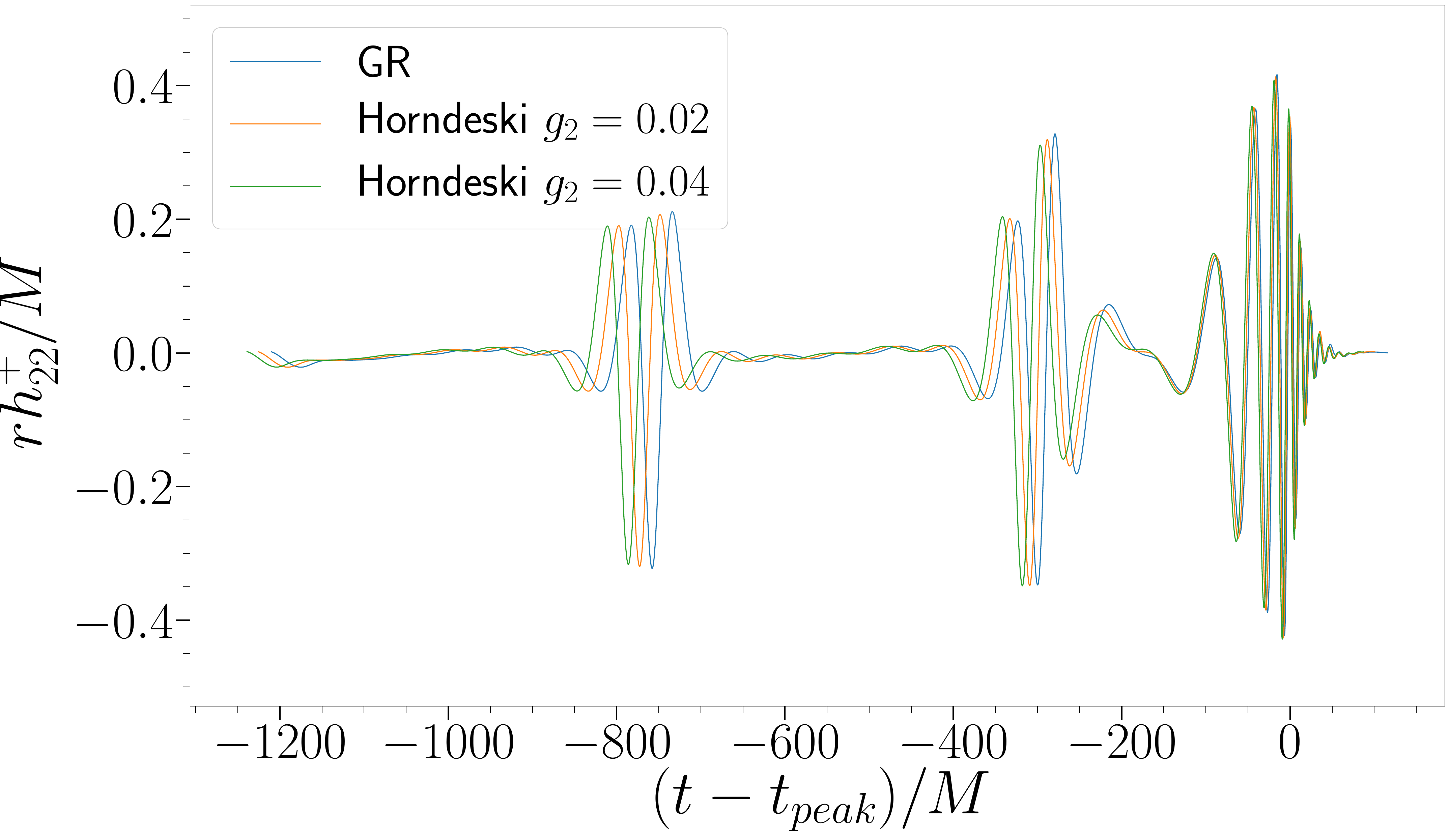}
\caption{\label{fig:g2_0.02_gw_aligned}Comparison of re-aligned $h^+_{22}$ between Horndeski theory with $g_2=0.02$, $g_2=0.04$ and GR. The waves were aligned so the peak of the amplitude of the complex strain coincides.}
\end{figure}

When comparing the waveforms between different theories, one might alternatively want to align the main peaks\footnote{The time of merger may also be estimated from the time when the common apparent horizon forms, but due to the inaccuracy of detecting this precisely, the peak of the amplitude of the strain is a more suitable measure.}. However, clearly the misalignment would not disappear; it would simply be translated along the time axis. This can be seen in figure \ref{fig:g2_0.02_gw_aligned}, where the misalignment is now seen at the early encounters of the inspiral. This shows that the gradual phase shift is a physical effect that cannot be ignored by a constant phase or time shift and does not depend on how one does the comparison. For long lived inspirals beyond the strong field regime simulated with NR, the effect would be enhanced and the misalignment would be present regardless of the time or phase shift considered. An analysis in the frequency domain done in section \ref{subsec:paper2:mismatch} will corroborate this point.

\subsection{\done{Scalar cloud}}\label{subsec:paper2:scalar_cloud}

\begin{figure}[t]
\centering
\includegraphics[width=0.9\textwidth]{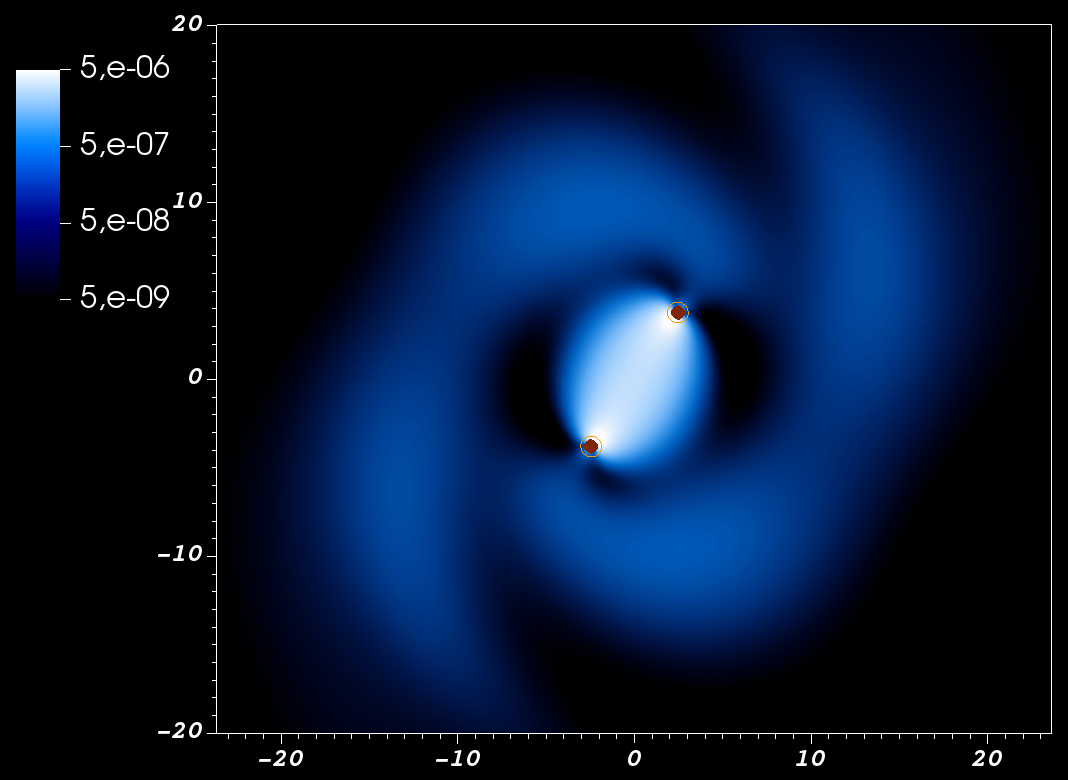}
\caption{Energy density (in blue) of the scalar field surrounding the binary black holes for the Horndeski theory with $g_2=0.005$ at a representative instant of time during the inspiral phase, $t\sim450$. The apparent horizon of the black holes is shown in orange. The region where the weak coupling conditions are larger than one is depicted in brown. Clearly this region is contained well inside the apparent horizon, as required.}
\label{fig:EnergyDensity_and_WFC}
\end{figure}

In figure \ref{fig:results_eta2} we display a snapshot of a binary for the $G_2$ theory with $g_2=0.005$ at a representative instant of time before the merger. This figure shows that the energy density of the scalar field (in blue) is localised in the region near the black holes, being largest near the horizons. It is in these regions where the spacetime and scalar field curvatures are largest, even though the WCCs remain small on and outside the black holes. 

For the Horndeski theories that we considered, the accumulation of non-linear effects is possible due to the presence of long lived scalar cloud surrounding the black holes. This scalar cloud survives all the way up to and well beyond the merger. This is due to the presence of a mass term in the scalar potential, since it is well-known (see e.g., \cite{Press:1972zz,Barranco:2012qs,Barranco:2017aes,Hui:2019aqm,Clough:2018exo,Ikeda:2020xvt}) that the effective potential that the scalar field ``sees'' has a wall that makes it difficult for it to escape to infinity. A scalar mass parameter of $m=0.5$ is comparable to what has been seen to give rise to long lived scalar clouds \cite{Ikeda:2020xvt}, though the effects observed in this chapter did not require any fine tuning. We have also carried out simulations of binaries with massless scalars and the absence of a significant scalar cloud trivially removes any long term effects of the scalar field on the evolution. 
\begin{figure}[t]
\centering
\includegraphics[width=0.98\textwidth]{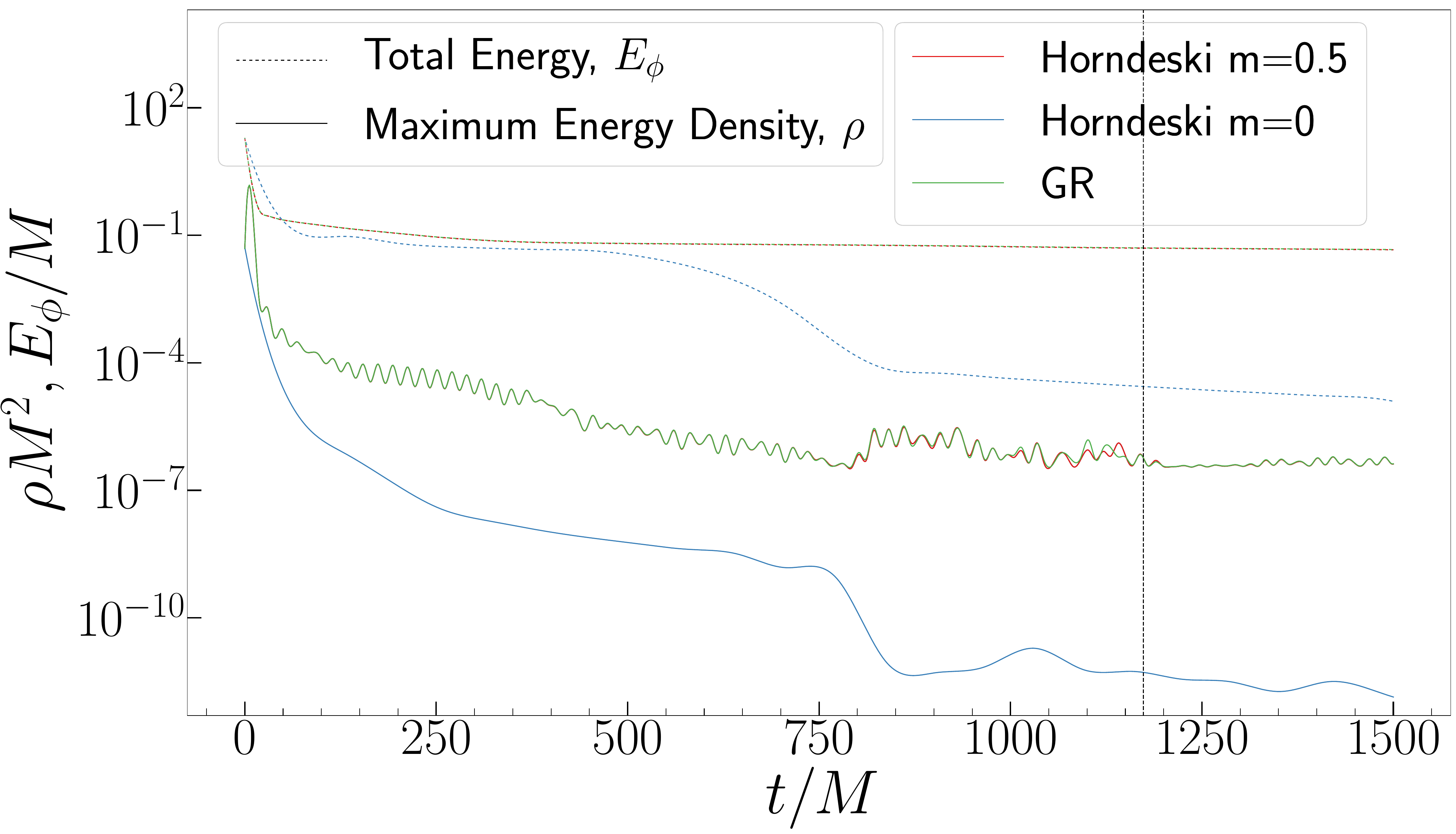}
\caption{Total energy of the scalar field $E_\phi$ and maximum value of energy density $\rho$ on the spacetime (excluding black holes, by removing from the volume of integration the interior of apparent horizons) for Horndeski with $g_2=0.005$ and a massive scalar field (in red), the same theory with a massless field (in blue) and GR with a massive scalar field (in green). A dashed black line is used to indicate the estimated merger time for the Horndeski run with a massive scalar field.}
\label{fig:energy_density_g2}
\end{figure}

In figure \ref{fig:energy_density_g2} we display the evolution of the total energy of the scalar field $E_\phi$ and the evolution of the maximum of the energy density $\rho$ for GR (green), for the $G_2$ theory with $g_2=0.005$ (red) and for the same Horndeski theory but with a massless scalar field (blue). These quantities are defined as $\r=n^\m n^\n T_{\m\n}$, where $T_{\m\n}$ the energy-momentum tensor of the scalar field, and $E_\f=\int_V \r\sqrt{\g} dV$, where $V$ is the spatial volume on a given spacelike hypersurface. After the initial gravitational collapse, most of the scalar field is absorbed by the black holes, but in the massive scalar cases, a long lived scalar cloud forms around the black holes. After the first close encounter of the eccentric binary $(t\sim 400 M)$, the maximum energy density of the scalar cloud is of order $10^{-5}M^{-2}$ for the massive scalar cases (GR and Horndeski), and it decreases very slowly with time. This long lived cloud makes it possible for the scalar field to interact with itself and with the geometry and give rise to the build up of significant differences in the physical observables such as the gravitational strain.

On the other hand, in the massless case, figure \ref{fig:energy_density_g2} shows that a much larger amount of scalar field is absorbed by black holes during the collapse phase. Furthermore, both the total energy of the scalar field and its energy density show a pronounced dip at $t\sim 850M$, namely in the second close encounter of the binary, indicating that any leftover amount of scalar field in the vicinity of the black holes gets absorbed. Beyond this point, the energy density of the scalar field is less than $10^{-10}M^{-2}$ while the total energy is of the order of $10^{-5}M$ (corresponding to scalar waves radiated to infinity), and both continue to steadily decrease with time. By the time the merger takes place the maximum energy density of the scalar field is compatible with numerical error. Therefore, we conclude that in the massless case, after the second close encounter of the binary, there is basically no significant amount of scalar field left in the neighbourhood of the black holes (as expected from no-hair theorems \cite{Hui:2012qt,Sotiriou:2013qea,Sotiriou:2014pfa,Maselli:2015yva}) to give rise to any sizeable effect, at least for the duration of our simulations. As a consequence, no noticeable differences between GR and Horndeski are observed in the massless scalar field case.

Comparing Horndeski with GR in figure \ref{fig:energy_density_g2} shows that local differences (in time) in the energy density between GR and Horndeski for the massive scalar field are not significant for most of the binary, including the two close encounters; only during the merger phase one can see some small differences of order $10^{-5}M^{-2}$. These results are expected if the weak coupling conditions are satisfied. Furthermore, the fact that the energy density of the massive scalar field around the black holes is small during the highly dynamical stages of the binary is necessary but not sufficient to ensure that the WCCs are satisfied.

\subsection{\done{Weak coupling conditions during the evolution}}\label{subsec:paper2:resultsWCCs}

The results reported in subsection \ref{subsec:paper2:waves} can only be trusted as long as the Horndeski theories that we consider are valid (truncated) EFTs. In this subsection we provide evidence that for the initial data and couplings that we considered in this chapter, the local WCCs from equations \eqref{eq:paper1:WFC_2} and \eqref{eq:paper2:lengthWCC} are satisfied at all times, thus ensuring the predictivity of the EFTs. 
\begin{figure}[t]
\centering
\includegraphics[width=0.98\textwidth]{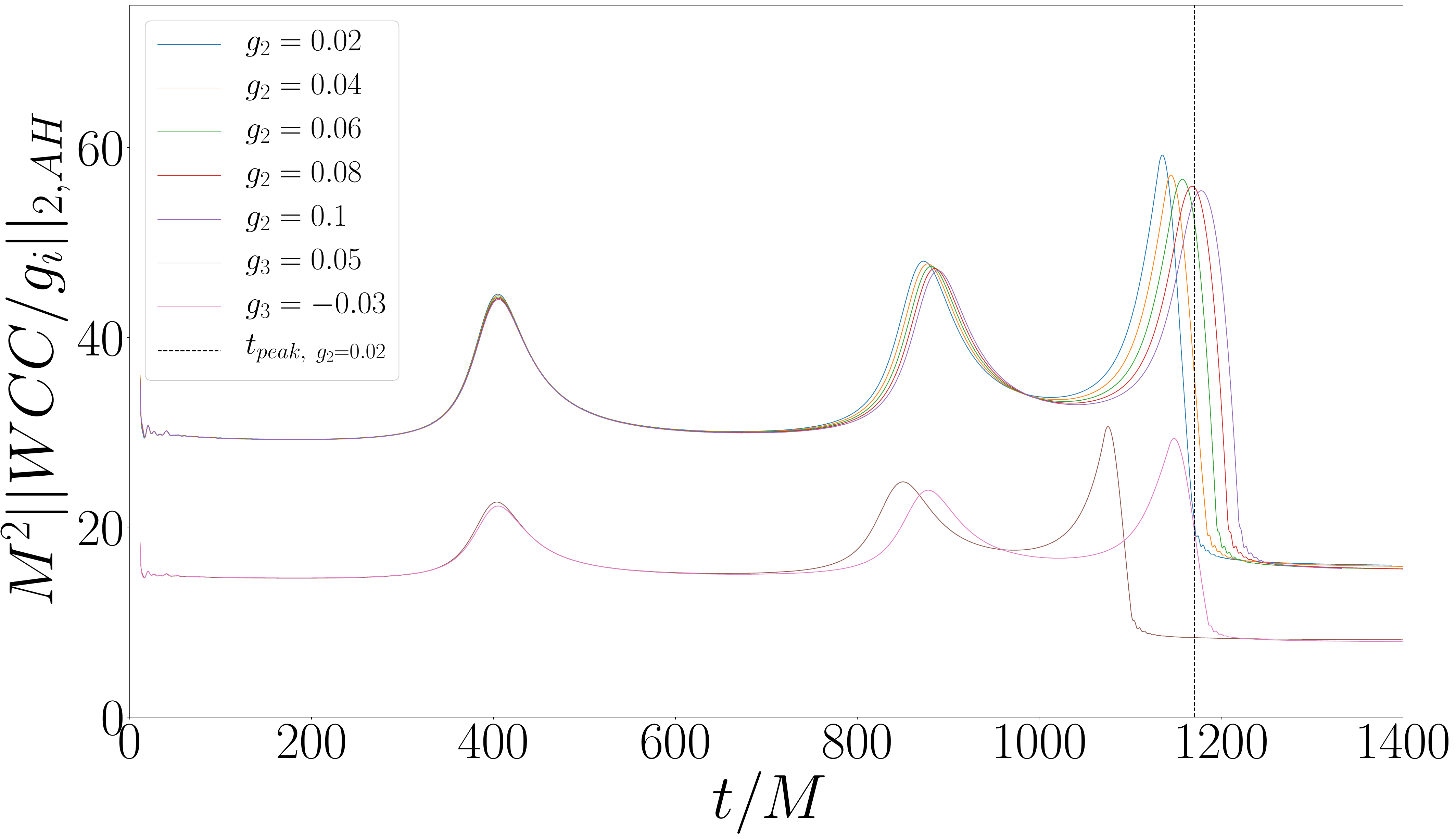}
\caption{$L^2$ norm of the weak coupling conditions \eqref{eq:paper1:WFC_2} integrated over the apparent horizon and normalised by the coupling constant $g_2$ or $g_3$. For the binary that we have evolved, this shows that $M^2 WCC/|g_2| \lesssim 50$ and $M^2 WCC/|g_3| \lesssim 20$, which in turn implies that $|g_2| \lesssim 0.02$ and $|g_3| \lesssim 0.05$ to guarantee that the WCC \eqref{eq:paper1:WFC_2} is roughly less than one. The dashed black line corresponds to the peak of the amplitude of the strain for $g_2=0.02$.}
\label{fig:WFC_spherical_L2}
\end{figure}

In figure \ref{fig:WFC_spherical_L2} we display the $L^2$ norm of the WCCs \eqref{eq:paper1:WFC_2} integrated on the black holes' apparent horizons, as a function of time for an eccentric binary evolved with the $G_2$ theory with different values of the coupling constant $g_2$ and one value $g_3$ coupling for the $G_3$ theory. Excluding the interior of black holes, the apparent horizons are where the WCCs have the largest values in the whole domain. This plot shows that the weak coupling condition \eqref{eq:paper1:WFC_2} remains approximately constant during the evolution, except in the close encounters of the binary and the final merger phase. The latter events correspond to the peaks in figure \ref{fig:WFC_spherical_L2} that can be seen at $t\sim400M$, $t\sim850M$ and $t\sim1100M$, when the system enters the strong field regime. The constancy of \eqref{eq:paper1:WFC_2} during the inspiral phase is related to the fact the energy density of the scalar field in the vicinity of the black hole remains approximately constant during this phase. The fact that the WCCs exhibits local maxima at the close encounters indicates that in an eccentric binary, we probe the strong field regime during various phases of the binary and not only near and during the merger phase as in a circular binary. It is interesting to see that when one normalises the WCCs \eqref{eq:paper1:WFC_2} by the coupling constant $g_{2,3}$, the curves for the $g_2$ couplings collapse onto a single curve, except in regions where the system is in the strong field regime. This indicates that the WCC depends on the coupling constant in a trivial way (linearly) when the system is not in the strong field regime.

Figure \ref{fig:WFC_spherical_L2} shows that for our choice of initial data, $M^2 WCC/|g_2| \lesssim 50$ and also $M^2 WCC/|g_3| \lesssim 20$ at all times. This implies that if we want the WCCs \eqref{eq:paper1:WFC_2} to be roughly less than one at all times, and hence guarantee that that the Horndeski theory is a valid EFT throughout the evolution, then one must choose $|g_2|\lesssim 0.02$ or $|g_3|\lesssim 0.05$. For values larger than these, the WCCs become comfortably larger than one at different (or all) stages of the binary. Not to confuse validity of the EFT with well-posedness of the evolution equations, after the initial collapse stage, even for large values of the coupling $g_2$ well-beyond the regime of validity of EFT (e.g. $g_2=0.1$), the equations of motion of the scalar field remain hyperbolic throughout the inspiral and merger phases as long as the scalar density is small enough near the black holes. 

When evaluating the WCCs \eqref{eq:paper1:WFC_2} on the apparent horizon to produce figure \ref{fig:WFC_spherical_L2}, one has to be careful as we are actually dealing with different trapped surfaces. Due to the slicing condition used, each black hole has a trapped surface that during merger shrinks to the puncture, while a larger common apparent horizon forms, surrounding the previous ones (see appendix \ref{chapter:ahfinder} for details on apparent horizon formation during binaries). This implies that if one is computing the WCCs \eqref{eq:paper1:WFC_2} on the trapped surfaces collapsing to the punctures, it will result in unreasonably large values. To get around this gauge issue, we interpolate the data for the WCC of the original black hole apparent horizons just before merger with the data for the common apparent horizon just after it forms, excluding the unphysically large values right at the merger. The details of how one does the interpolation and which data points are excluded do not affect significantly the bounds $M^2 WCC/g_2 \lesssim 50$ and $M^2 WCC/g_3 \lesssim 20$.

We close this subsection emphasising that our assessment of the regime of validity of EFT is qualitative at best and, up to $\mathcal{O}(1)$ factors, the unity value of the WCCs is a mere order of magnitude; a more detailed study is needed in order to precisely identify this regime for the cases that we have considered. The conditions \eqref{eq:paper1:WFC_2} are only local; over time, the small effects accumulate giving rise to large deviations in some non-local observables such as the waveforms. In the context of complex scalar field with a Mexican hat type of potential, Reall and Warnick \cite{Reall:2021ebq} proved that, for sufficiently long times, the truncated EFT will inevitably deviate from the UV theory. Therefore, one has to be cautious when using a truncated EFT for very long times compared to the UV mass scale, even if the local weak coupling conditions hold (see also \cite{Davis:2021oce}).

\subsection{\done{Mismatch}}\label{subsec:paper2:mismatch}

\begin{figure}[t]
\centering
\includegraphics[width=0.98\textwidth]{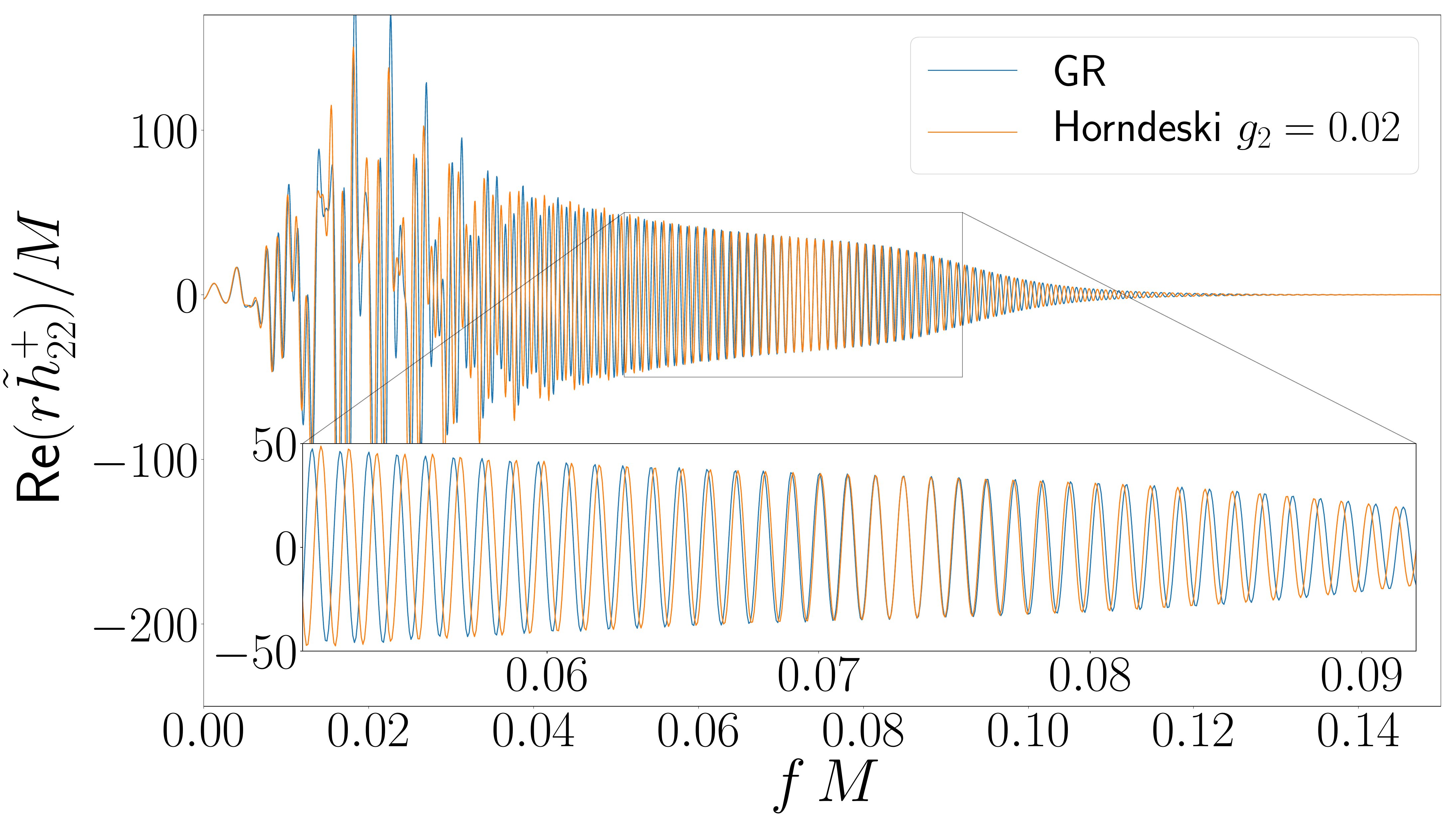}
\caption{Real part of $\ti{h}^+_{22}$, the discrete Fourier transform of $h^+_{22}$ for positive frequencies. Notice that for frequencies around $f\sim0.07M^{-1}$, Horndeski and GR align, but for lower and higher frequencies they separate in phase in opposite directions. This effect cannot be mitigated by a constant time or phase shifts of the waveform.}
\label{fig:FFT_strain_GR_vs_g2}
\end{figure}
\begin{figure}[t]
\centering
\includegraphics[width=0.98\textwidth]{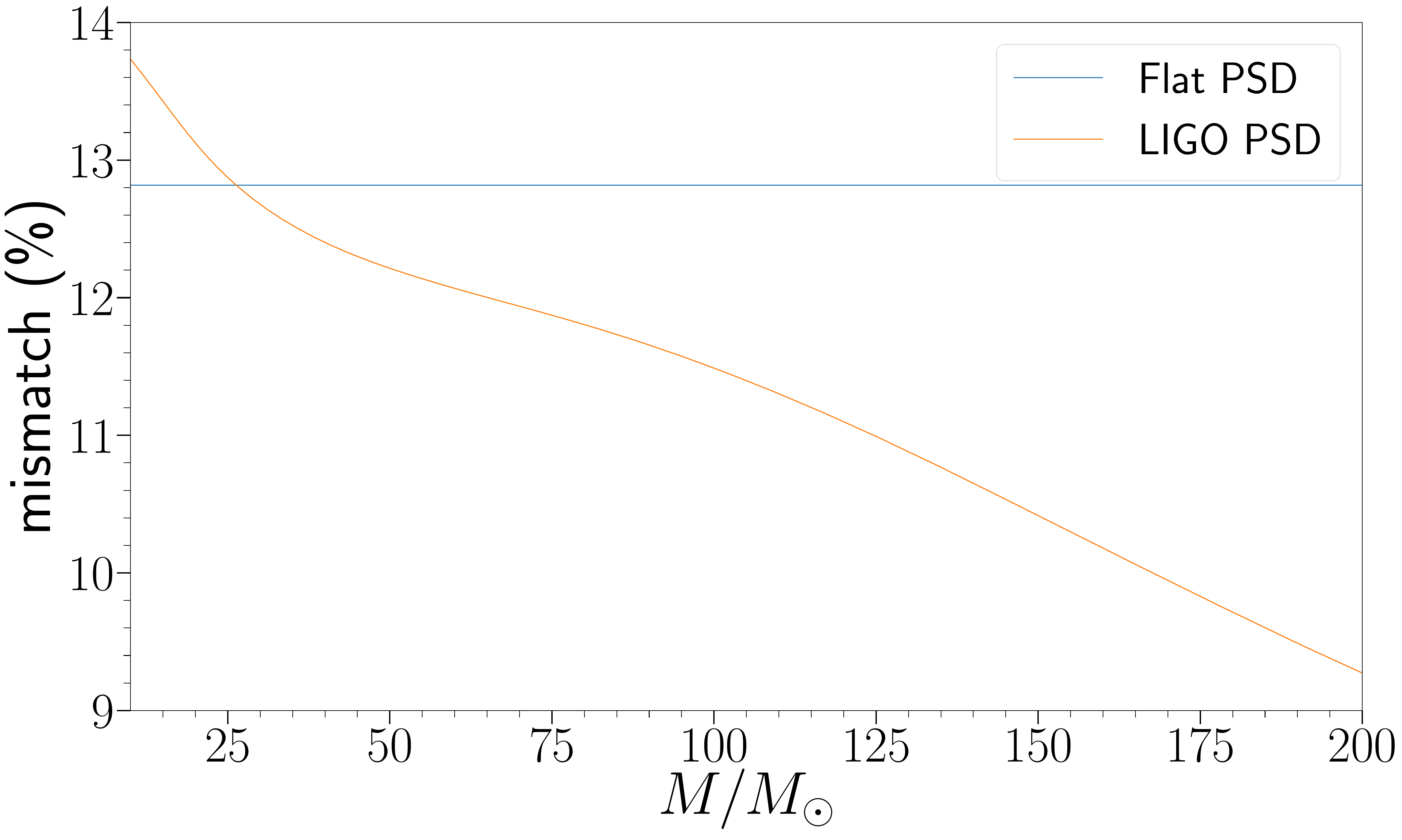}
\caption{Mismatch for $h^+_{22}$ between GR and Horndeski for $g_2=0.02$, as a function of the final black hole mass (in units of solar masses, $M_\odot$). As power spectral densities, we used the updated Advanced LIGO sensitivity design curve (\textit{aLIGODesign.txt} in \cite{aLIGOupdatedantsens}, which imposes $f_{\text{min}}=5$ Hz) and a flat noise mismatches ($S_n=1$). This allows us to estimate a range for expected mismatches of $10-13 \%$.}
\label{fig:mismatches_g2_vs_GR}
\end{figure}

In this subsection we discuss our results for the mismatch between the GR and Horndeski waveforms. We start focusing on comparing the $G_2$ theory with $g_2=0.02$ to GR, since this is an example of the limiting coupling that still satisfies the WCCs. Hence, the results for the mismatch presented should be understood as upper bounds. The mismatch depends on the coupling constants in the expected way, and the results are qualitatively the same for the $G_3$ theory.

In figure \ref{fig:FFT_strain_GR_vs_g2} we compare the frequencies of the real part of $\ti{h}^+_{22}$, the discrete Fourier transform of the $(\ell,m)=(2,2)$ mode of the strain, extrapolated to null infinity, as described in section \ref{subsec:paper2:strain}. Interestingly, this figure shows that in spite of both theories having approximately the same amplitudes for each frequency in the spectrum, the spectrum of the phase of the complex-valued Fourier transform differs. In range of medium frequencies, i.e., $f\sim 0.07-0.08M^{-1}$, GR and Horndeski theory agree very well. However, for both lower and higher frequencies, a significant discrepancy can be clearly seen. This effect cannot be mitigated by a constant time or phase shift of the time-domain waveform and hence we conclude that it is a physical effect. This discrepancy of both the high and low frequencies suggests that Horndeski theory exhibits both an inverse and a direct energy cascades. It would be interesting to confirm if this is indeed the case and quantify these cascades. Note that because the weak cosmic censorship holds in our scenarios, there is a natural UV cutoff for the frequencies that are accessible to external observers. As long as this cutoff is at lower energies, i.e., larger distances, than the UV cutoff of the theory, then the EFT should be valid; the fact that the WCCs hold in our case, indicates that this is indeed the case.
\begin{figure}[t]
\centering
\includegraphics[width=0.98\textwidth]{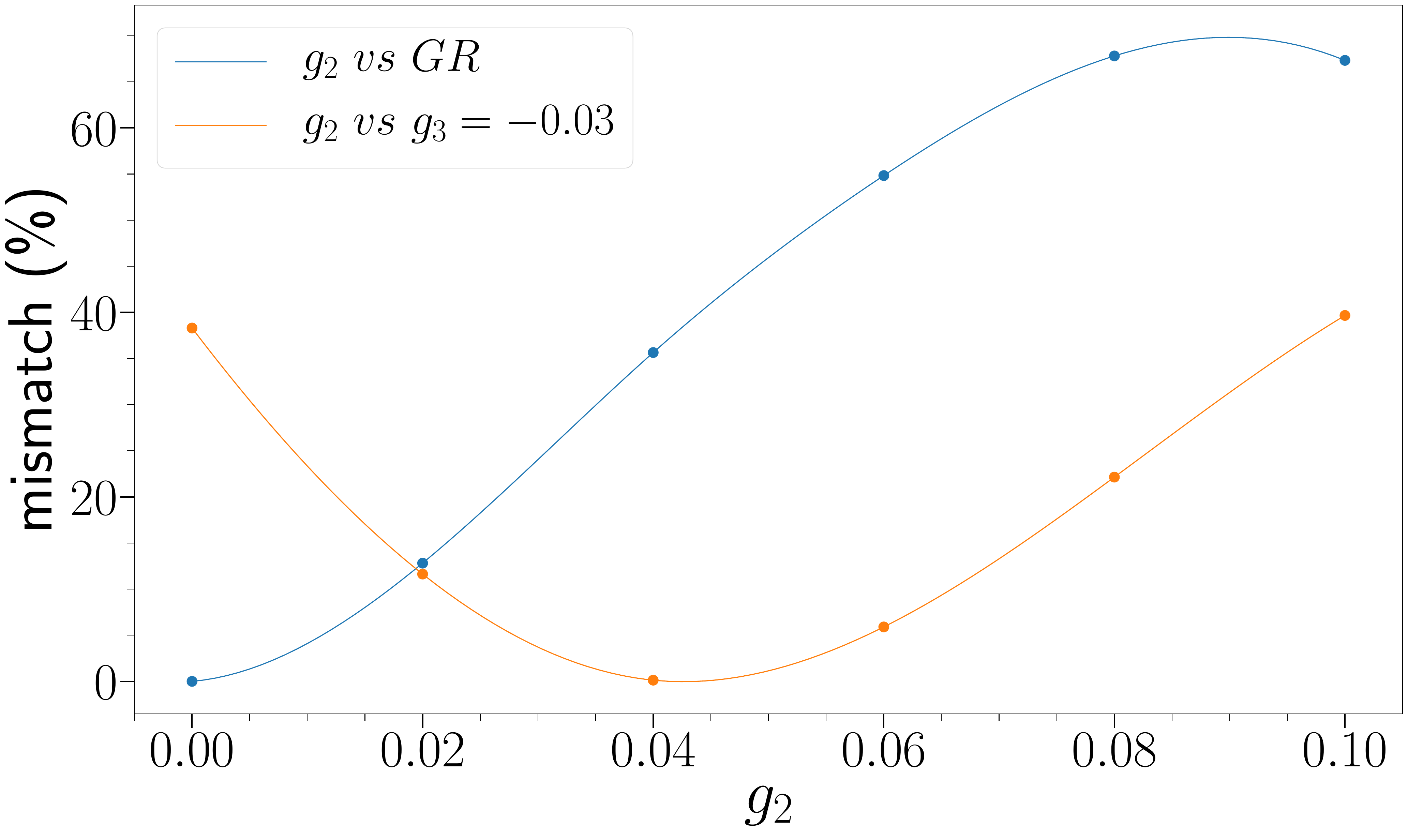}
\caption{Mismatch between the $G_2$ Horndeski theory for several values of $g_2$ and GR (blue curve) and between the same $G_2$ theory and the $G_3$ theory with $g_3=-0.03$ (yellow curve), for a flat PSD. The two Horndeski theories appear to be degenerate for certain values of the couplings.}
\label{fig:mismatch_g2_vs_GR_and_g3}
\end{figure}

In figure \ref{fig:mismatches_g2_vs_GR}, we quantify the mismatch for a detector setup receiving the plus polarisation of the strain, extrapolated to null infinity, between GR and the Horndeski theory. We limit ourselves to  the $(\ell,m)=(2,2)$ mode, $h^+_{22}$, as this is the dominant mode by an order of magnitude when compared to higher modes. We use the updated Advanced LIGO sensitivity design curve (\textit{aLIGODesign.txt} in \cite{aLIGOupdatedantsens}, which imposes $f_{\text{min}}=5$ Hz) and flat noise ($S_n=1$) following the procedure described in section \ref{subsec:paper2:mismatch_theory}. We compute the mismatch for black hole masses in the typical range of stellar mass black holes binaries observed so far, $M\in[10,200]M_\odot$ \cite{LIGOScientific:2021djp}, where $M_\odot$ is one solar mass. As mentioned in section \ref{subsec:paper2:initial_data}, the parameter $g_2$ is dimensionful, with units of $M^2$, and hence this value also varies as we probe different masses in figure \ref{fig:mismatches_g2_vs_GR}. This figure shows that the mismatch varies between $\sim13\%$ at the low mass end and $\sim10\%$ at the high mass end\footnote{This mismatch would be reduced if the minimisation included variation over the binary parameters, such as mass and spin.}. To confirm accuracy of these results, the mismatch between two different resolutions of the same GR evolution ranges between $0.3-0.5\%$ for the same mass ranges. On the other hand, as a reference, for a signal to noise ratio of 25, similar to GW150914 \cite{LIGOScientific:2016aoc}, the minimum expected mismatch for detection is about $0.6\%$ \cite{Chatziioannou:2017tdw,Purrer:2019jcp}. Additionally, Lindblom et al. \cite{Lindblom:2008cm} estimated that a mismatch of $3.5\%$ would result in a $10\%$ lower detection rate. Therefore, the large mismatches obtained for big enough values of the couplings suggest that if the underlying theory of gravity was Horndeski with a massive scalar field, some events happening in the past years would have gone undetected if the black holes had sufficient scalar field surrounding them.

We additionally explore the degeneracy between $G_2$ and $G_3$ Horndeski theories considered in this chapter by evaluating the mismatch between the waveforms obtained for different values of the coupling $g_2$ and a fixed value of $g_3=-0.03$ respectively.\footnote{We thank the anonymous referee for suggesting this calculation.} The results are shown in figure \ref{fig:mismatch_g2_vs_GR_and_g3}. We can see that the waveforms in the two theories appear to be nearly degenerate for $g_2\approx 0.04$ and $g_3=-0.03$. For these values of the couplings, the mismatch is about $0.1\%$, significantly below any detectability threshold or numerical errors. As an additional remark, notice how the mismatch between the $G_2$ theory and GR  starts decreasing for $g_2\gtrsim 0.1$. This effect happens because for these values of $g_2$, the delay of the Horndeski waveform corresponds to more than half one period of the GR wave in the region close in the frequency domain, resulting in more aligned peaks and hence smaller mismatch. Hence, we expect the mismatch to oscillate as we vary  $g_2$ due to this effect, with a local minima bounded above zero since the waveforms are significantly different in spite of aligned peaks.

\subsection{\done{Circular binaries}}\label{subsec:paper2:circular_binaries}
\begin{figure}[h]
\centering
\includegraphics[width=0.45\textwidth]{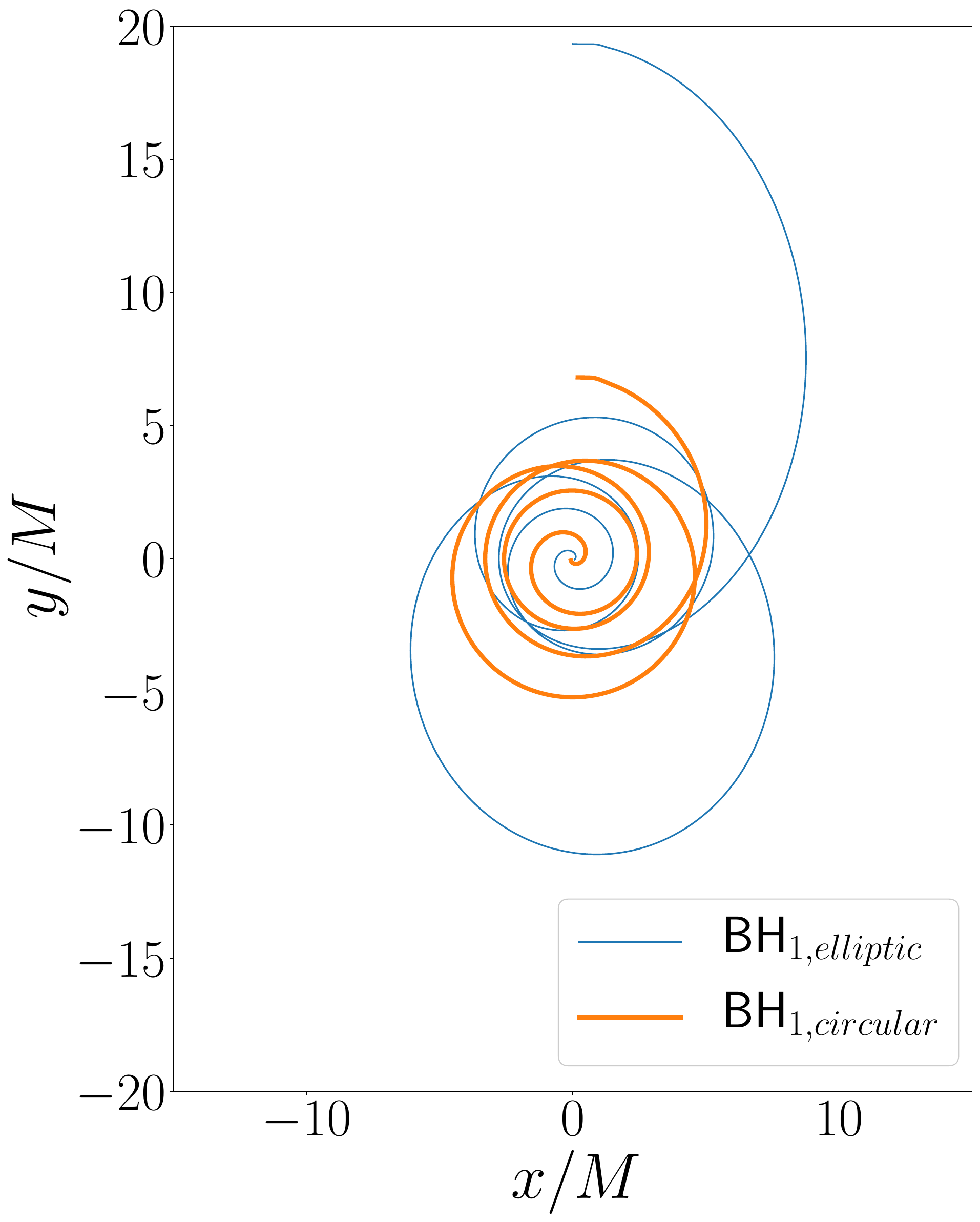}
\caption{Comparison of the orbit of one of the black holes of the GR binary, between the previously analysed elliptic orbits and the circularised binary. Both have approximately 5 orbits.}
\label{fig:extra_circular:xy}
\end{figure}

In this subsection we analyse the impact of the elliptic orbits in the results obtained before. One could wonder if the gravitational wave effects observed increase or not: on the one hand, one could say that elliptic binaries have closer encounters enhancing the strong gravity regime, but on the other hand, circular binaries spend more time overall closer to each other. With the spare computational resources available, we setup not a fully circular binary, but a more circularised binary (without any fine-tuning of the initial speed). To effectively use computational resources, we reduced the initial separation of the scalar bubbles to obtain roughly the same amount of circularised orbits as the elliptic orbits. The separation was changed from $D=40$ to $D=14$. The smaller and more circularised orbits can be seen in figure \ref{fig:extra_circular:xy}.
\begin{figure}[h]
\centering
\includegraphics[width=0.98\textwidth]{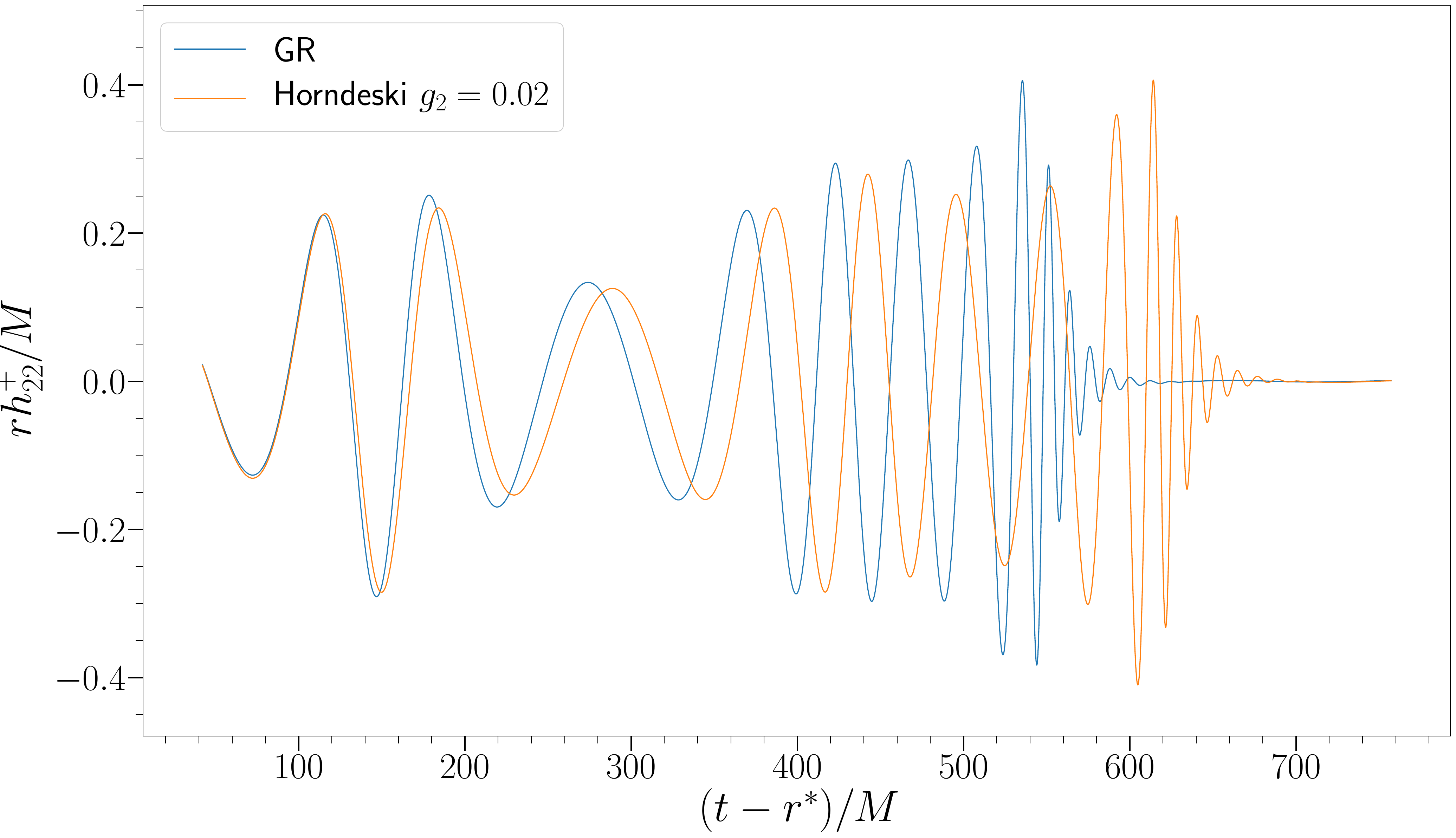}
\caption{Comparison of gravitational wave between Horndeski theory with $g_2=0.02$ and GR in retarded time, $u=t-r^*$, where $r^*$ is the tortoise radius. Displaying the $(\ell,m)=(2,2)$ mode of the plus polarisation of the strain, $h^+_{22}$, extrapolated to null infinity.}
\label{fig:extra_circular:gw}
\end{figure}
\begin{figure}[h]
\centering
\includegraphics[width=0.98\textwidth]{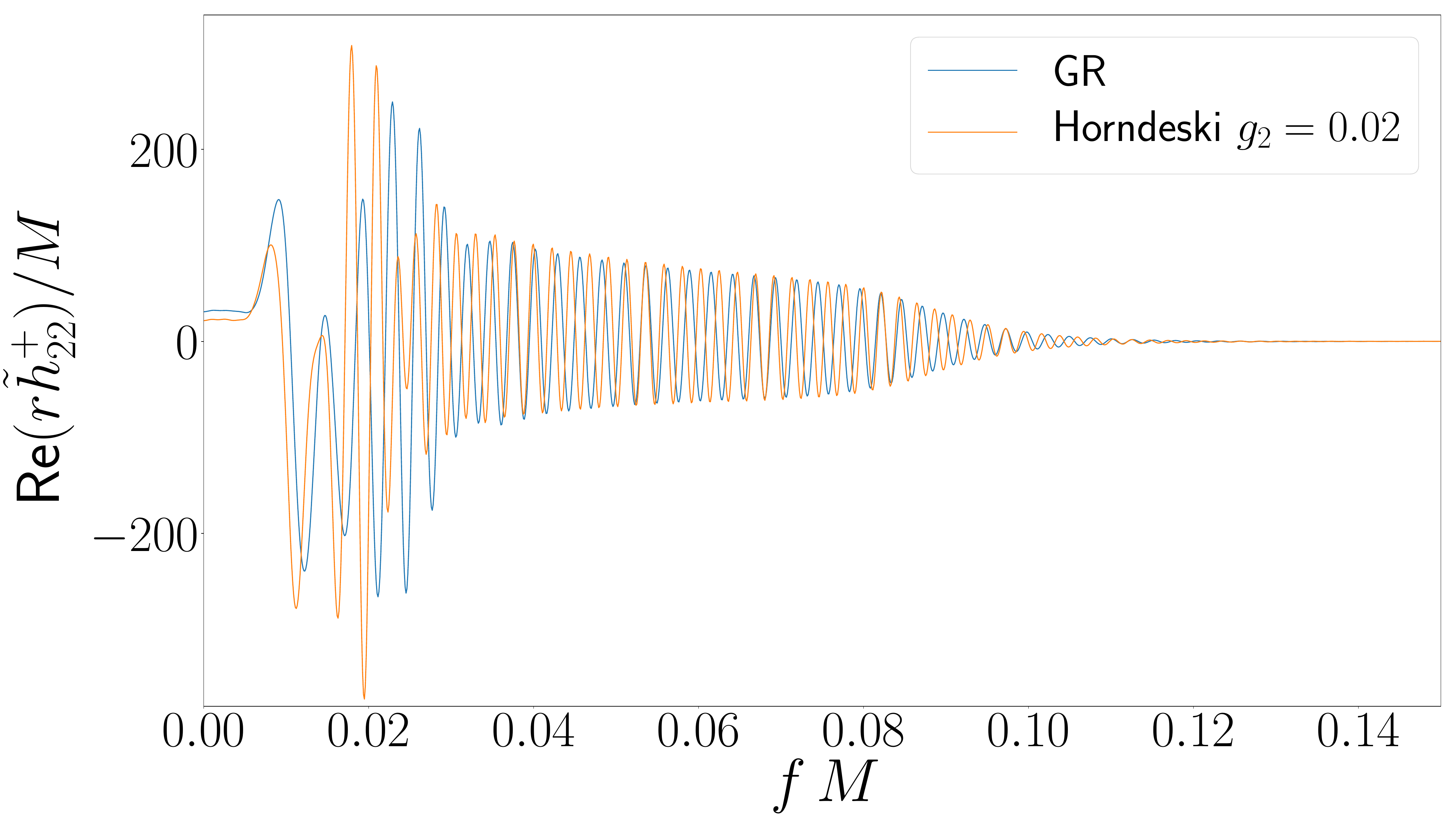}
\caption{Real part of $\ti{h}^+_{22}$, the discrete Fourier transform of $h^+_{22}$ for positive frequencies.}
\label{fig:extra_circular:fft}
\end{figure}
\begin{figure}[h]
\centering
\includegraphics[width=0.98\textwidth]{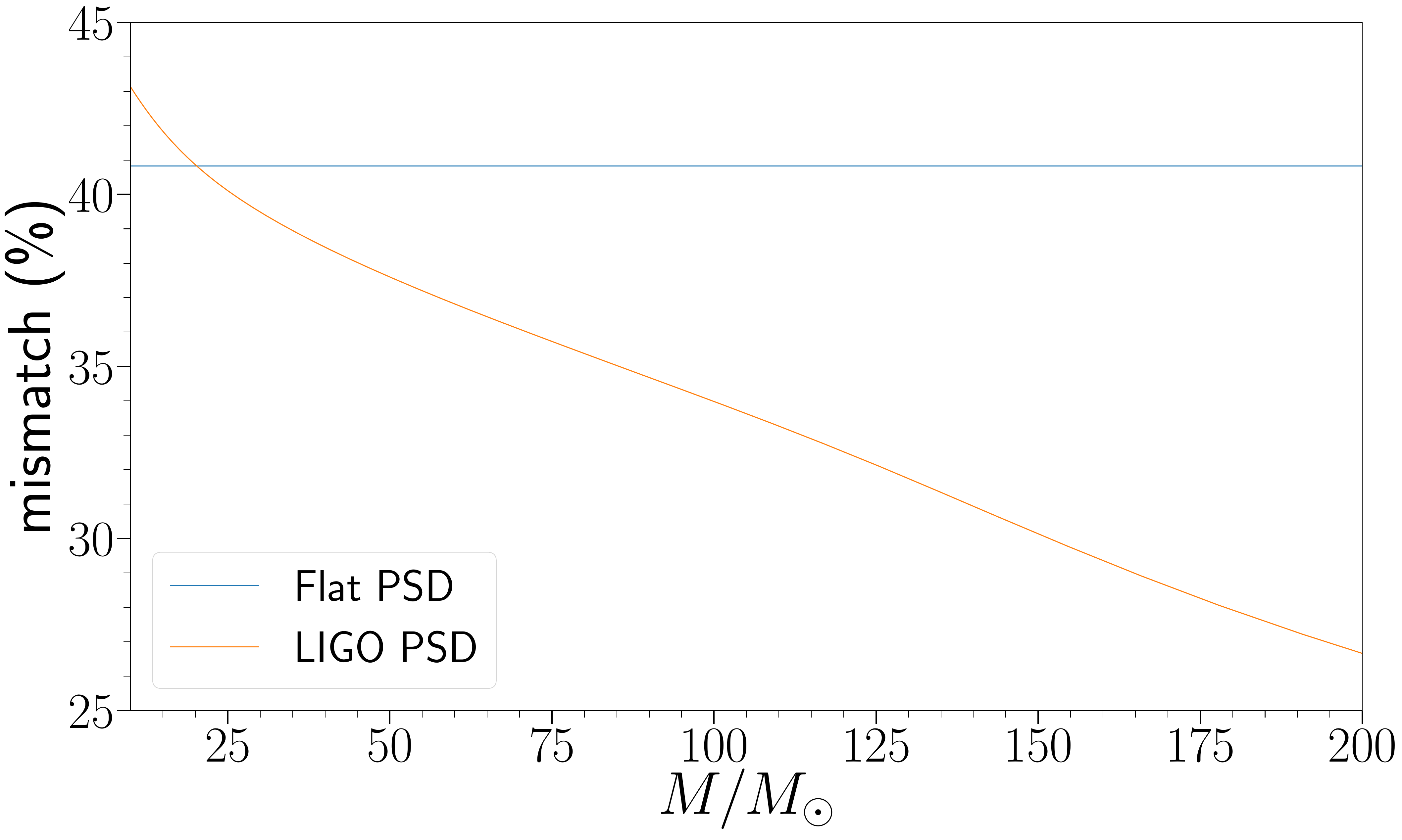}
\caption{Mismatch for $h^+_{22}$ between GR and Horndeski for $g_2=0.02$, as a function of the final black hole mass (in units of solar masses, $M_\odot$). As power spectral densities, we used the updated Advanced LIGO sensitivity design curve (\textit{aLIGODesign.txt} in \cite{aLIGOupdatedantsens}, which imposes $f_{\text{min}}=5$ Hz) and a flat noise mismatches ($S_n=1$). This allows us to estimate a range for expected mismatches of $30-40\%$.}
\label{fig:extra_circular:mismatch}
\end{figure}

At this smaller separation, to ensure the bubbles were minimally affected by the presence of each other, each bubble was made thinner and with bigger amplitude, $(A,r_0,\w)=\br{0.314,2,0.3}$, as to ensure the mass of each bubble would remain the same ($M\approx0.52$). The boost applied to the scalar momentum $\P$ was $|\ve{v}|=0.4$, resulting in a speed of the black hole after collapse of $0.076$, still a value much smaller than $1$, the speed of light. The scalar mass was kept at $m=0.5$. For GR, the resulting ADM mass of the superposed boosted system is $M = 1.0272\pm0.001$. For comparison of the results, we evolved a Horndeski simulation with $g_2=0.02$, matching what was done for figure \ref{fig:g2_0.02_gw}. For this coupling, we have $\eta_2\sim\mathcal{O}(10^{-4})$, opposed to $\mathcal{O}(10^{-5})$ as for the wider bubbles, which indicates more likeliness of hyperbolicity problems and potentially also bigger deviations in the gravitational waveform.

As analysed for the elliptic case, we display in figures \ref{fig:extra_circular:gw}-\ref{fig:extra_circular:mismatch} the $(\ell,m)=(2,2)$ mode of the plus polarisation, $h^+$, extrapolated to null infinity; the real part of its discrete Fourier Transform, and the mismatch between GR and Horndeski for different masses, as detailed in section \ref{subsec:paper2:mismatch}.

Qualitatively, the effects observed are as expected: a positive $g_2$ leads to a positive phase shift in the waveform. The effects are quite significant for this more circular binary, to the extent that mismatches reach $\sim40\%$\footnote{Similarly to the previous section, the mismatch would be reduced if the minimisation included variation over the binary parameters, such as mass and spin.}. At these values one can claim the mismatch computation does not even make sense anymore, as we are simply treating two very different waveforms. It would be interesting to run a full long circular binary (10 or more orbits) and see the full extent that a small coupling (ten times smaller for instance, such that no questions of weak coupling condition violations are raised) can still have in the orbit dynamics.

\section{\done{Conclusions}}\label{sec:paper2:conclusions}

In this chapter we have studied eccentric and circularised black hole binary mergers in certain cubic Horndeski theories \eqref{eq:choiceG2}--\eqref{eq:choiceG3} with a massive scalar field with mass parameter $m=0.5$. We have chosen initial data and small enough coupling constants such that a certain local weak coupling condition \eqref{eq:paper1:WFC_2} is satisfied at all times during the evolution. This condition monitors the size of the Horndeski terms in the equations of motion compared to GR terms, and the fact it holds, ensures that the EFTs are in their regime of validity and hence we can trust their predictions. 

One of the goals of this chapter was to identify potential deviations from GR in some physical quantities that Horndeski theories of gravity may exhibit. We have observed that locally small deviations from GR build up over time and get enhanced whenever the system enters the strong field regime. In the case of the eccentric binaries, this happens during the successive close encounters of the black holes and in the final merger phase, while in circular binaries the effect continuously accumulates. Since the modifications of GR are locally small, large deviations may still arise in non-local observables, such as gravitational waveforms, through a cumulative build up. This cumulative effect gets reflected in the gravitational waveforms as large shifts with respect to the analogous waveforms computed in GR coupled to a massive scalar field with the same mass and angular momentum.

Whilst the details, such as its sign and size, of the observed shift in the waveforms depend on the details of the theory and value of the coupling constants, the effect seems to be generic, at least within the class of Horndeski theories that we have explored here. We conjecture that the same effect should be present in a subset of more general Horndeski theories. We have quantified the misalignment of the $(\ell,m)=(2,2)$ mode of the plus polarisation of the strain, $h^+_{22}$, for one of the Horndeski theories that we have considered. We find that the spectrum differs both for low and high frequencies. Furthermore, for large enough values of the couplings, still in the regime of validity of the EFT, we find that the mismatch is around $10-13\%$ for elliptic binaries and more than $30\%$ for circularised binaries, in the whole mass range of current detections, without including variations over mass and spin. This is quite significant and it suggests that if the underlying theory of gravity differs from GR, some events where the black holes have sufficient scalar field surrounding them may have been and continue to go undetected. For smaller values of the couplings, the mismatch would be smaller.

The misalignment that we have observed is a cumulative effect and hence it only occurs if the black holes are surrounded by a long-lived scalar cloud. In our case this is possible because of the mass term in the scalar potential, which ensures that a non-trivial scalar energy density remains in the vicinity of the black holes for very long periods of time, thus allowing the scalar field to interact with itself and with the geometry. We have also considered massless scalars, but in this case we do not observe any significant difference between Horndeski and GR. This is expected because the theories considered do not admit stationary hairy black holes for massless scalar fields (at least in the slowly rotating limit \cite{Hui:2012qt,Sotiriou:2013qea,Sotiriou:2014pfa,Maselli:2015yva}) and, hence, a massless scalar field gets absorbed by the black holes or disperses to null infinity on a time scale much shorter compared to the binary lifetime. In our particular example, the scalar field is essentially completely absorbed in the second close encounter of the binary and by then there has not been enough time to build up any sizeable deviation from GR. 

In this chapter we considered both $G_2$ and $G_3$ Horndeski theories and, as we have already mentioned, even though the initial and final states are the same, both lead to misaligned waveforms with respect to GR. Therefore, at least for equal mass non-spinning binaries, the degeneracy between the class of Horndeski theories that we have considered and GR is broken. However, we do not see any visible difference between the waveforms obtained in the $G_2$ or in the $G_3$ theories. It would be interesting to investigate if (or how) the degeneracy of the waveforms is broken in Horndeski theories of gravity. It would be interesting to extend our studies to unequal mass and spinning binaries to see if the degeneracy with GR and with the various Horndeski theories is broken when considering different mass ratios and non-zero spins. 

We have considered Horndeski theories simply as toy models for EFTs with higher derivatives; in the Horndeski case, the higher derivatives are in the matter (scalar) sector and the equations of motion are of second order. However, more fundamental theories of gravity, such as string theory, predict higher curvature corrections of the Einstein-Hilbert action. In general, such new terms in the action will result in equations of motion of order higher than two. Refs. \cite{Cayuso:2017iqc,Allwright:2018rut,Cayuso:2020lca} have outlined how the strong field regime of such theories may be probed, but it would be very interesting to do so in the context of a black hole binary. Our work suggests that in the weakly coupled regime, where these theories are valid EFTs, some of the problems that may arise in general, such as loss of hyperbolicity or shock formation, can be controlled in a physical situation that probes the strong field regimes such as a black hole binary merger. 

The main goal of the present chapter was to identify what features in the physical observables extracted from black hole binaries in Horndeski theories can allow one to differentiate these theories from GR. Given that the corrections to GR have to be locally small in order for these theories be valid EFTs, non-local observables such as gravitational waveforms are particularly useful because small effects can accumulate and, for long enough times, give rise to large deviations from GR. These or other deviations from alternative theories of gravity are potentially being undetected by current gravitational wave observatories. Therefore, our results stress the importance of modelling waveforms in alternative theories of gravity treating them fully non-linearly. It would be interesting to identify other observables where large deviations show up. In the case of waveforms, until complete waveform templates are built for alternative theories, a potential way to detect the misalignment that we have identified is the following: future space-based gravitational wave observatories such as LISA \cite{LISA:2017pwj,Arun:2022vqj} are expected to be able to detect gravitational waves produced in stellar mass black hole binaries during earlier stages of the inspiral phase. From these waveforms one should be able to extract the parameters of the binary and, by using GR, predict the time of merger of the binary. Some binaries should enter the LIGO band in the final stages of the inspiral and merger phase, thus allowing to contrast the GR prediction for the merger time with the observation; a certain advancement or delay of the merger could be attributed to the fact that GR corrections modify the theory.

\chapter{\done{Black hole binaries in higher derivative effective field theories}}\label{chapter:eft}

\section{\done{Introduction}}

The detection of gravitational waves with modern detectors is a feat of enormous precision. The LIGO interferometer has arms extending $4$ km and the gravitational waves detected cause the distance between its mirrors to change by roughly 1/1000th the diameter of a proton. In order to find these gravitational waves in the data, experiments require theoretical templates for the gravitational waveforms, i.e. patterns to find a matches for, in the data. For precise tests of GR using gravitational waves, we require precise predictions for how a deviation from GR would affect gravitational waves.

On general grounds, one expects that at sufficiently small distances, Einstein's theory will be modified by some form of quantum corrections. Without a preferred UV-complete theory of quantum gravity, effective field theory (EFT) provides a framework for the construction of possible candidates. Some theories emerge from a bottom-up methodology, adding to GR all possible consistent higher derivative terms which can then be constrained by observational data; or a top-down approach, attempting to derive the low energy behaviour of quantum gravity candidates, such as string theory. Either way, from the point of view of EFT, these corrections can be organised in a series expansion involving increasing powers of the curvature tensor, and consequently higher derivatives of the spacetime metric. The argument is that, since in current experiments we are only probing gravity at low energies, we should only be sensitive to a finite number of terms in the otherwise infinite series of corrections to GR. Moreover, the details of the UV completion of gravity should not be important at such low energies. Any of these alternative theories of gravity should be understood as truncated low energy EFT and, as such, they only make sense if the corrections to GR are small. These corrections may be important in certain situations, as small effects may accumulate over time \cite{Figueras:2021abd}.

Given one such theory, it is often non-trivial to find a suitable formulation of GR which does not admit `runaway' solutions that exhibit cascades of energy to the UV, which are unphysical, inconsistent with the regime of validity of EFT and inconsistent with current observations. A general approach to find well-posed formulations of general alternative theories of gravity has been proposed by \cite{Cayuso:2017iqc,Allwright:2018rut}. This proposal is inspired by the M\"uller-Israel-Stewart (MIS) formulation of relativistic viscous hydrodynamics \cite{Muller:1967aa,Israel:1976213,Israel:1976tn,Israel:1979wp}, and in principle can work even for theories with higher-than-second order equations of motion, as recently shown by Cayuso and Lehner \cite{Cayuso:2020lca} for a certain eight-derivative theory of gravity. This method is described in section \ref{sec:gr:fixing_hd_theories}. Higher derivative corrections are just one example of the myriad of possible modifications to GR that have been considered, and the method has been applied as well for scalar-tensor theories of gravity \cite{Bezares:2021yek,Lara:2021piy,Franchini:2022ukz,Gerhardinger:2022bcw}, with two derivative equations of motion, but in formulations of GR where the theory would originally not be well-posed.

In this chapter, we aim to extend the work in spherical symmetry done by Cayuso and Lehner \cite{Cayuso:2020lca}, to evolve binary black holes in a certain eight-derivative higher derivative theory of gravity, applying the method described in section \ref{sec:gr:fixing_hd_theories}. This is part of work in collaboration with Ramiro Cayuso, Pau Figueras and Luis Lehner \cite{Cayuso:2023aht}.

\subsection{\done{Outline}}

The rest of the chapter is organised as follows. In section \ref{sec:eft:toymodels} we extend the work done by Cayuso et al. \cite{Cayuso:2017iqc} with further analysis and other applications of the method considered. In section \ref{sec:eft:methodology}, we describe the target higher derivative theory, application of the fixing procedure to this theory, and numerical techniques required for simulations and data processing. Section \ref{sec:eft:results} describes the results of the chapter. In \ref{subsec:eft:results_toy_models}, we analyse numerically the stability of the fixing procedure in toy models. In section \ref{subsec:eft:single_bh}, we evolve single boosted black holes to confirm the accuracy of the method, before running binary black holes. Finally, section \ref{subsec:eft:results_bbh} shows the waveforms and waveform mismatch of a binary black hole when comparing the EFT with GR. It further verifies consistency of the method and satisfaction of the weak coupling conditions. Section \ref{sec:eft:conclusions} summarises the results and discusses future directions.

\section{\done{Fixing toy models}}\label{sec:eft:toymodels}

\subsection{\done{Fourth order ODE - wave equation toy model}}

Let us illustrate an explicit example of how this procedure can take place. This follows an adaptation of the construction by Cayuso et al. \cite{Cayuso:2017iqc}. Perturbations around a GR background are known to follow wave equations, called gravitational waves (see section \ref{sec:gr:bh_and_gws} for details). With the intent of exploring higher derivative perturbations to GR, consider the following one-dimensional perturbed wave differential equation for the complex scalar field $\f$ in Cartesian coordinates $(t,x)$:
\begin{equation}
    \square \f = -\e\, \pd_t^4\f\,,
\end{equation}
where $\e$ is some real parameter and $\square = \eta^{\m\n}\grad_\m\grad_\n = -\pd_t^2 + \pd_x^2$. This is a linear equation, but can be seen as well as the linearisation of non-linear equations around some solution, e.g. the linearisation of the equation $\square \varphi = -\varphi \, \pd_t^4\varphi$ around a constant solution $\varphi = \e + \f$. Nonetheless, we take $\e$ to be small, representing small deviations from the standard wave equation.

\begin{figure}[h]
\centering
\begin{multicols}{2}
\includegraphics[width=0.5\textwidth]{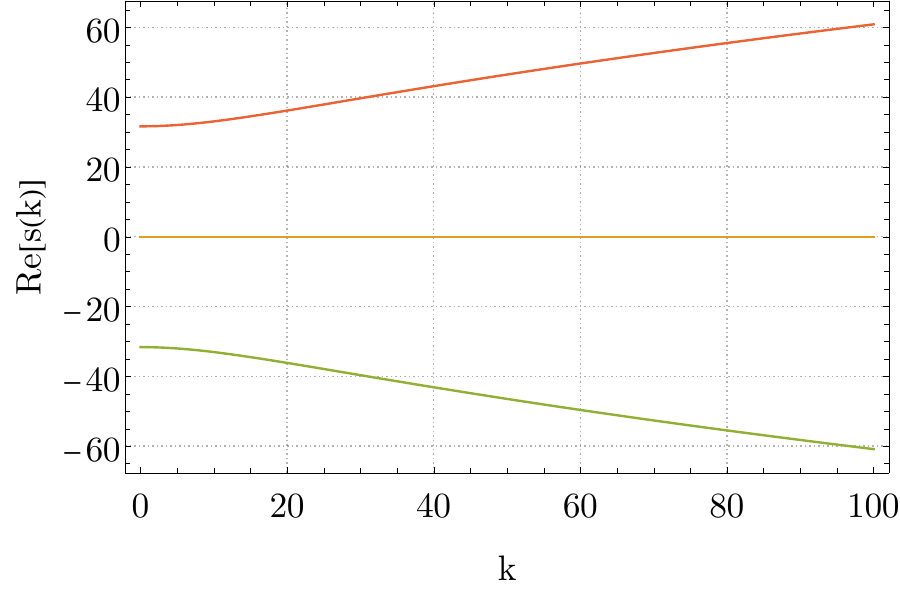}

\includegraphics[width=0.5\textwidth]{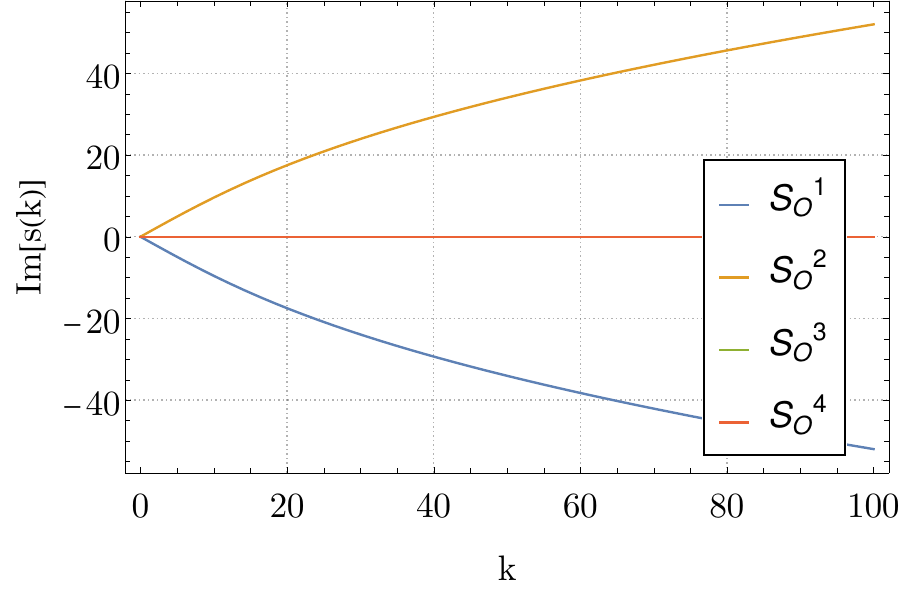}
\end{multicols}
\vspace{-8mm}
\caption{Real and imaginary parts of the dispersion relation of the original system, \eqref{eq:gr:n4_original}, denoted $S_O^{~i}$ for $i=1,2,3,4$, with $\e=10^{-3}$.}
\label{fig:gr:fixing:original_n4}
\end{figure}

To analyse the behaviour of solutions, consider a Fourier mode:
\begin{equation}
    \f(t,x) = A\,e^{st + ikx}\,,
\end{equation}
for some $A,k\in \mathbb{R}$, $s\in \mathbb{C}$. The dispersion relation has 4 solutions:
\begin{equation}\label{eq:gr:n4_original}
    s_{(\pm,\pm)}(k) = \pm \sqrt{\frac{1 \pm \sqrt{1 + 4\e\, k^2}}{2\,\e}}\,.
\end{equation}
This has one solution (both positive branches, $s_{(+,+)}$) with real positive values for any $k$, revealing blowing-up modes. This can be visualised in figure \ref{fig:gr:fixing:original_n4}. This is a problem.

For negative $\e$, the conclusions are similar.

\subsection{\done{Time order reduction}}

\begin{figure}[h]
\centering
\begin{multicols}{2}
\includegraphics[width=0.5\textwidth]{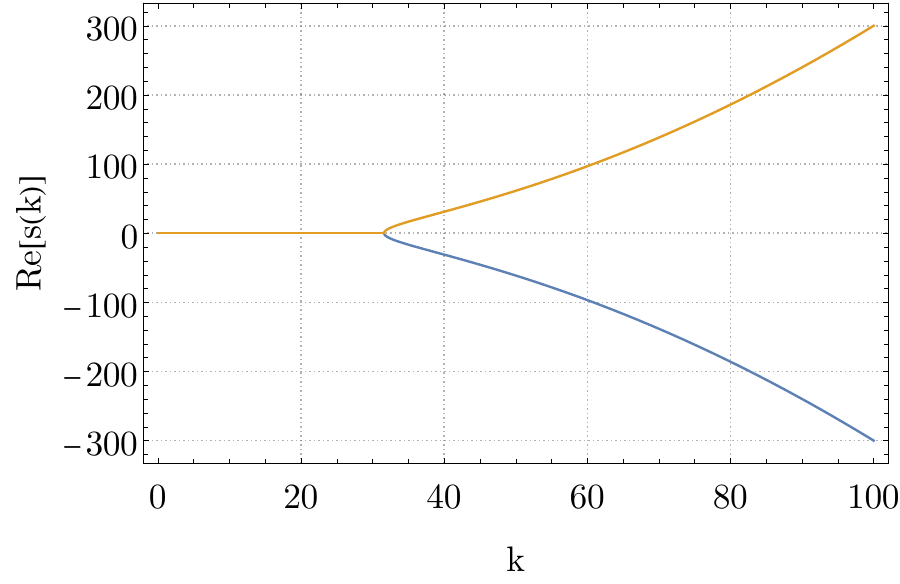}

\includegraphics[width=0.5\textwidth]{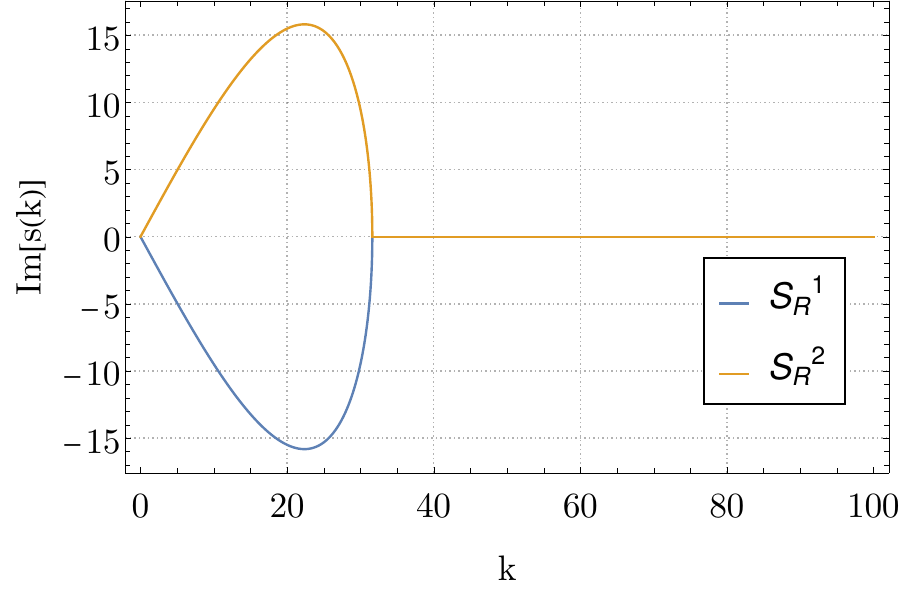}
\end{multicols}
\vspace{-8mm}
\caption{Real and imaginary parts of the dispersion relation of the time reduced system, \eqref{eq:gr:n4_time_reduced}, denoted $S_R^{~i}$ for $i=1,2$, with $\e=10^{-3}$.}
\label{fig:gr:fixing:n4_time_reduced}
\end{figure}

Let us perform time order reduction as described in section \ref{subsubsec:gr:time_reduction}. Notice that to first order in $\e$, $\pd_t^2\f = \pd_x^2\f + \mathcal{O}(\e)$. Hence:
\begin{equation}
    \square \f = -\e\, \pd_x^4\f + \mathcal{O}(\e^2)\,.
\end{equation}
This is the time order reduced equation leaving us with higher order spatial derivatives. If we analyse the dispersion relation for this equation, we find:
\begin{equation}\label{eq:gr:n4_time_reduced}
    s_\pm(k) = \pm\, ik\sqrt{1 - \e\, k^2}\,.
\end{equation}
These solutions are purely imaginary for small $k$, but become real (and positive for one of the solutions) for sufficiently high $k$, high frequencies. This can be easily seen in figure \ref{fig:gr:fixing:n4_time_reduced}. This is better than the original system, because we now can see the meaning of the low frequency behaviour that we want to preserve while getting rid of the spurious high frequency growing modes.

For negative $\e$, the system is stable for all frequencies, but has acausal propagation.

\subsection{\done{``Fixing'' the equations}}

\subsubsection{\done{Naive approach}}

\begin{figure}[h]
\centering
\begin{multicols}{2}
\includegraphics[width=0.5\textwidth]{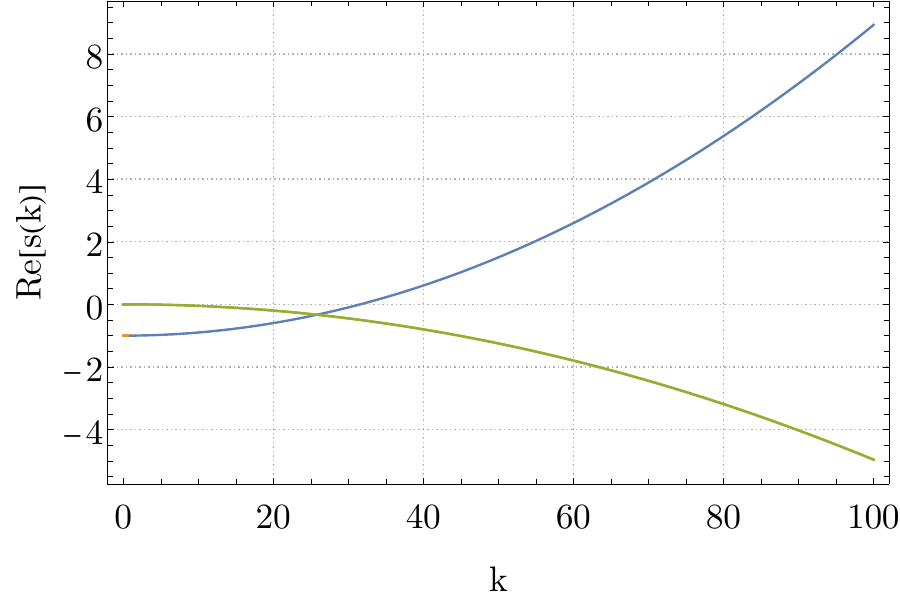}

\includegraphics[width=0.5\textwidth]{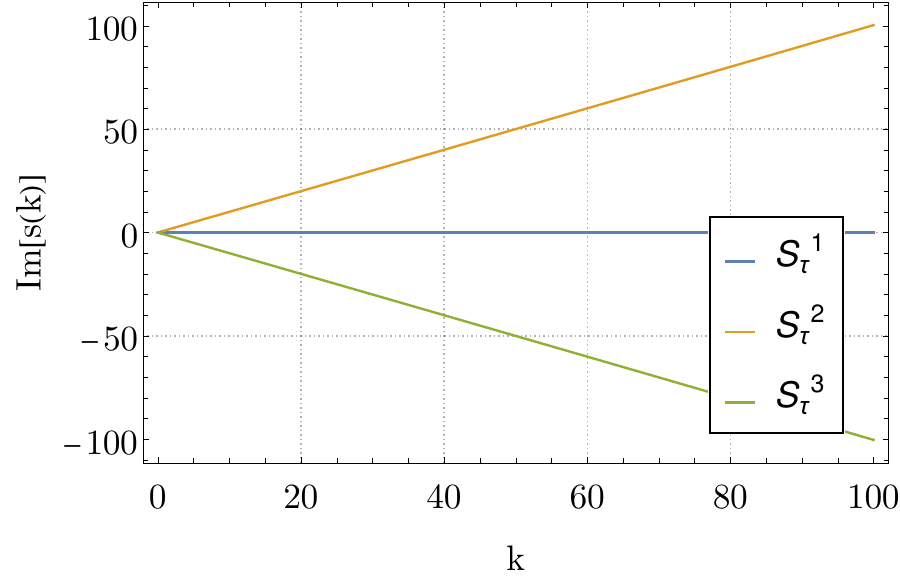}
\end{multicols}
\vspace{-8mm}
\caption{Real and imaginary parts of the dispersion relation of the naive fix of the system \eqref{eq:gr:n4_fix_wrong}, denoted $S_\t^{~i}$ for $i=1,2,3$, with $\e=10^{-3}$ and $\t=1$.}
\label{fig:gr:fixing:n4_fix_wrong}
\end{figure}

We proceed by ``fixing'' the time reduced equation. Introduce a variable $\P$ that replaces second spatial derivatives of $\f$. The simplest equation one could write is:
\begin{equation}\begin{aligned}
    \square \f &= -\e\, \pd_x^2 \P\,,\\\label{eq:fix:damped_Pi}
    \t \pd_t \P &= \pd_x^2\f - \P\,,
\end{aligned}\end{equation}
where $\t$ is a damping timescale. Using the Fourier mode ansatz $\P(t,x)=A\,\b(s,k)e^{st+ikx}$, for some function $\b(s,k)$, this system implies:
\begin{equation}\begin{aligned}
    -s^2 - k^2 &= \e\, \b(s,k) k^2\,,\\
    \t\, \b(s,k)\,s &= -\b(s,k) - k^2\,.
\end{aligned}\end{equation}
We can eliminate $\b$ and obtain the equation:
\begin{equation}\label{eq:gr:n4_fix_wrong}
    (1+\t s)(s^2 + k^2) = \e\, k^4\,.
\end{equation}
This equation has growing modes, as one can easily see that as $k\to\infty$, there is a solution $s\to \br{\frac{\e}{\t}}^{\frac{1}{3}}k^{\frac{4}{3}}$, which is a positive real growing mode. Hence, this modified system does not fix correctly the problem of growing modes. This can be visualised in figure \ref{fig:gr:fixing:n4_fix_wrong}.

Even though the time order reduced equations for negative $\e$ were stable, the naive fix approach actually breaks this stability. Similar to the case analysed above, it has two solutions growing as $s\to \br{\frac{1\pm i\sqrt{3}}{2}}\br{\frac{|\e|}{\t}}^{\frac{1}{3}}k^{\frac{4}{3}}$, where $\br{\frac{1\pm i\sqrt{3}}{2}}$ are two complex cubic roots of negative unity, which have positive real part.

\subsubsection{\done{Educated ansatz}}

\begin{figure}[h]
\centering
\begin{multicols}{2}
\includegraphics[width=0.5\textwidth]{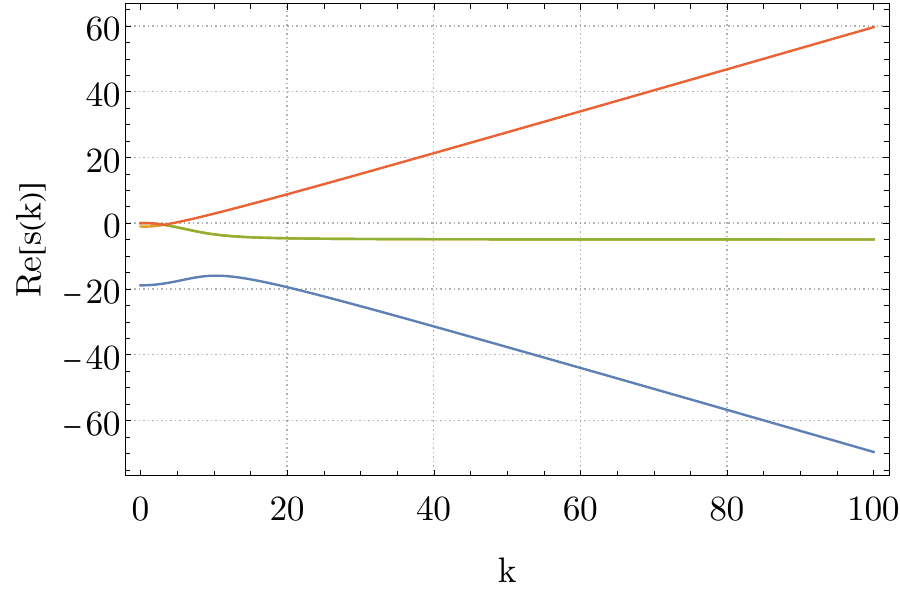}

\includegraphics[width=0.5\textwidth]{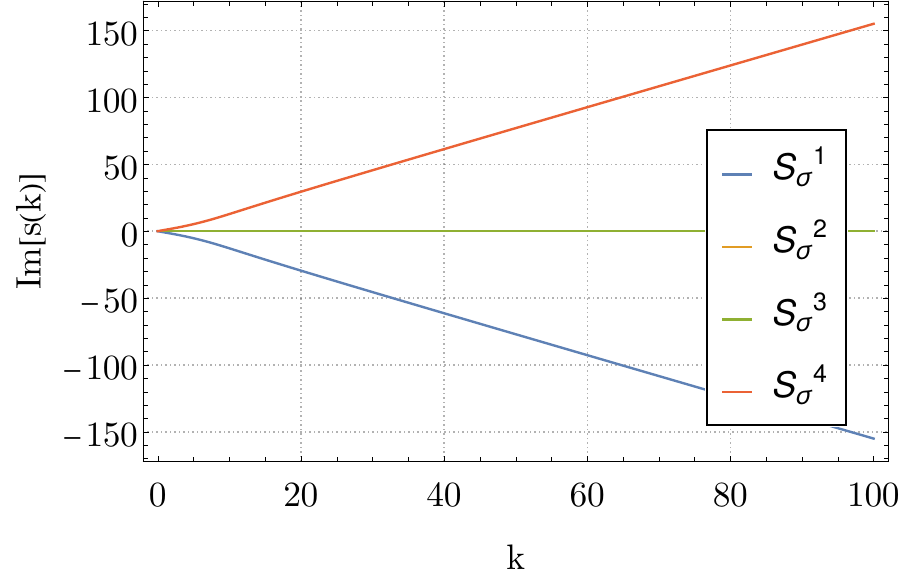}
\end{multicols}
\vspace{-8mm}
\caption{
Real and imaginary parts of the dispersion relation of the fixed system \eqref{eq:fix:box_simple}, denoted $S_\s^{~i}$ for $i=1,2,3,4$, with $\t=1$, $\e=10^{-1}$, $\s=0.05$, \textbf{not} satisfying the stability condition $\s>\e$, and hence having a positive real part.}
\label{fig:gr:fixing:n4_fix_right_sigma_wrong_epsilon}
\end{figure}

We try a better educated fix by using a damped wave equation for $\P$:
\begin{equation}\begin{aligned}
    \square \f &= -\e\, \pd_x^2 \P\,,\\\label{eq:fix:damped_wave_Pi}
    -\s\, \square \P + \t\,\pd_t\P &= \pd_x^2\f - \P\,,
\end{aligned}\end{equation}
where $\s,\t$ are positive real constants. To understand in what sense this is a damped wave equation, if one sets $\b=1$ and looks at the homogeneous equation of \eqref{eq:fix:damped_wave_Pi} (i.e. ignoring the first equation and setting $\f=0$ in the second equation), the dispersion solutions are $s_{\text{hom.}}(k) = -\frac{\t}{2\s} \pm \frac{\sqrt{\t^2 - 4\s(1+k^2\s)}}{2\s}$. For $\s \lessapprox \br{\frac{\t}{2}}^2$ and small $k$, the solutions are real and negative if $k$ is small, while for $\s \gtrapprox \br{\frac{\t}{2}}^2$ or high $k$, they are complex with negative real part. It is important to note that larger $\t$ and smaller $\s$ lead to faster damping. To analyse the new system, we use the same strategy as before to reduce the dispersion equation to a quartic polynomial equation:
\begin{equation}\label{eq:fix:box_simple}
    \br{1+\t s + \s(s^2 + k^2)}(s^2 + k^2) = \e\, k^4\,.
\end{equation}
We can first notice that as $k\to\infty$, this equation implies that $s(k)\to \a\, k$, such that $\s\br{\a^2+1}^2=\e$, with solutions:
\begin{equation}\label{eq:eft:fixXsigmaBox}
    \a_{(\pm,\pm)} = \pm\sqrt{-1\pm\sqrt{\tfrac{\e}{\s}}}\,.
\end{equation}
The next leading order term is a constant, and as $k\to\infty$, $s_{(\pm,\pm)}(k)\to \a_{(\pm,\pm)}\cdot k - \frac{\t}{4\s}$. From this we can see that for $\s < \e$, there are solutions, namely $\a_{(+,+)}$, for which has a real positive part for large $k$. This can be observed in figure \ref{fig:gr:fixing:n4_fix_right_sigma_wrong_epsilon}.

\begin{figure}[h]
\centering
\begin{multicols}{2}
\includegraphics[width=0.5\textwidth]{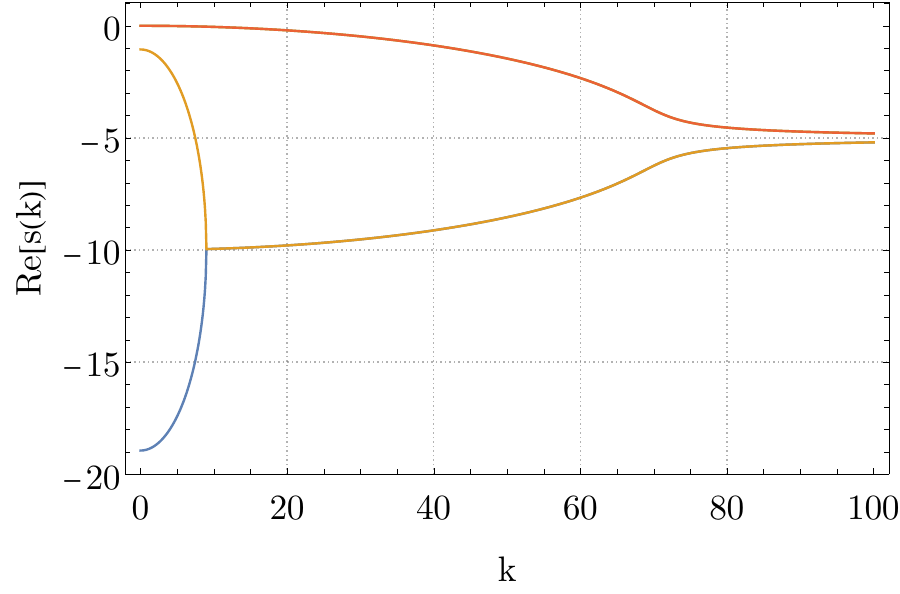}
\includegraphics[width=0.5\textwidth]{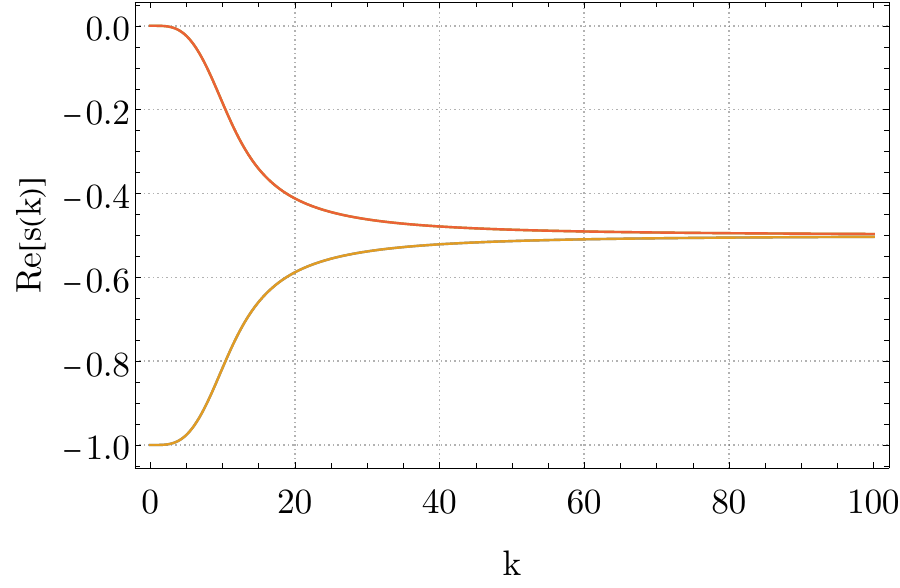}

\includegraphics[width=0.5\textwidth]{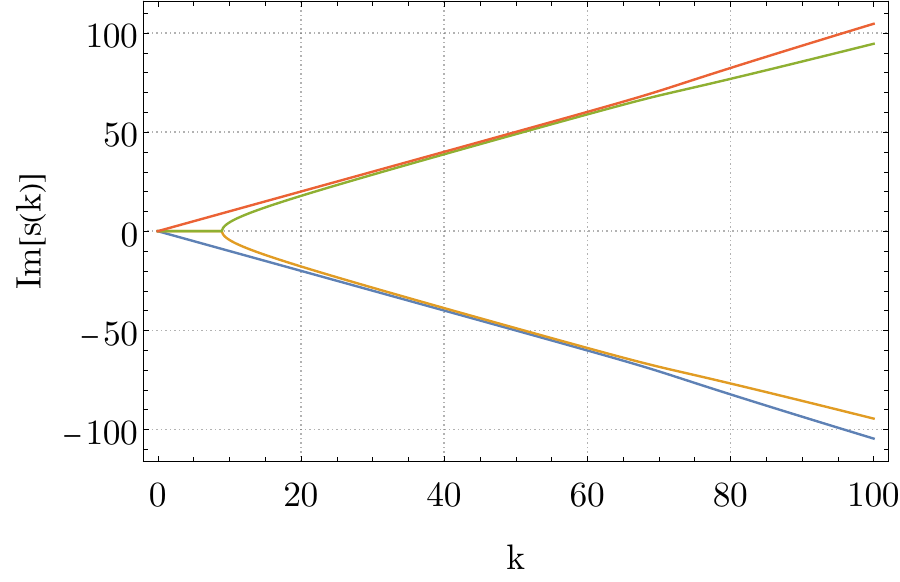}
\includegraphics[width=0.5\textwidth]{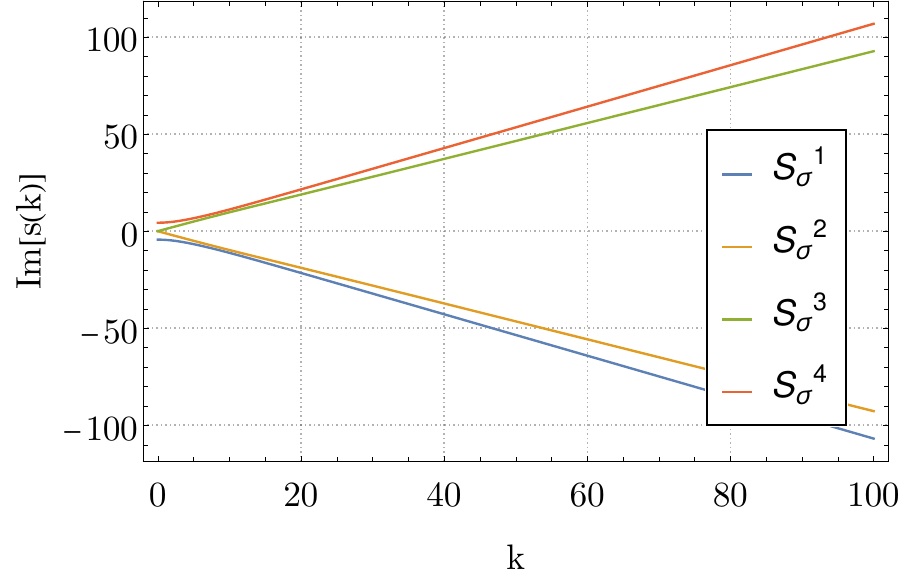}
\end{multicols}
\vspace{-8mm}
\caption{Real and imaginary parts of the dispersion relation of the fixed system \eqref{eq:fix:box_simple}, denoted $S_\s^{~i}$ for $i=1,2,3,4$, with $\e=10^{-3}$ and $\s=0.05$, satisfying the stability condition $\s>\e$. The top figures show the real and imaginary parts for $\t=1$, for which $\s < \br{\frac{\t}{2}}^2$, and the bottom figures the opposite case, with $\t=0.1$.}
\label{fig:gr:fixing:n4_fix_right_sigma_bigger_eps}
\end{figure}

We now analyse the case $\s \ge \e$\footnote{The original paper \cite{Cayuso:2017iqc} presents an alternative proof. Its equation (15) is only applicable to the $s(0)=0$ solutions, which is all that matters since the other 2 solutions $s_\pm(0)$ in equation (14) are already negative. Equation (15) should be corrected to: $s^{(1)}(0)=\pm i,\quad s^{(2)}(0)=0,\quad s^{(3)}(0)=\mp 3i\,\l,\quad s^{(4)}(0)=-12\l\t$, for which $s^{(4)}(0)$ is still correct and the final result is not spoiled.}. For any $k$, take $s(k)=\a+\b\,i$, for $\a,\b\in\mathbb{R}$ dependent on $k$. The imaginary part of equation \eqref{eq:fix:box_simple} is $\b\br{4\a^3\s + \a(4k^2\s-4\b^2\s+2)+3\a^2\t+\br{k^2-\b^2}\t}=0$, which has solutions $\b_1=0$ and $\b_2=\sqrt{k^2+\frac{\a\,\br{2+4\a^2\s+3\a\t}}{4\a\s\t}}$. Replace this back into equation \eqref{eq:fix:box_simple} and we get an equation with only real numbers:
\begin{align}\label{eq:simple_proof_b1}
    \b=\b_1,&\qquad k^2 + \a^4\s + k^4\br{\s-\e} + \a^2\br{1+2k^2\s} + \a^3\t + \a\, k^2\t=0\,,\\\nonumber
    \b=\b_2,&\qquad 64\a^6\s^3+96\a^5\s^2\t+k^4\e\,\t^2+8\a^3\t\br{4\s+8k^2\s^2+\t^2}\\\nonumber
    &\qquad +16\a^4\s\br{2\s+4k^2\s^2+3\t^2} + 4\a^2\br{\s\br{1+4k^4\e\,\s}+\br{2+5k^2\s}\t^2}\\\label{eq:simple_proof_b2}
    &\qquad +2\a\br{\t+4k^4\e\,\s\,\t + k^2\t^3}=0\,.
\end{align}
Equation \eqref{eq:simple_proof_b1} reveals that if $\a>0$ and $\s\ge\e$, then all the terms are positive (or some are zero for $k=0$, for which $\a=\b=s(0)=0$ is one of the solutions), leading to an impossible equality. Similarly, equation \eqref{eq:simple_proof_b2} has only non-negative terms if $\a>0$. Therefore, $\a>0$ is an impossibility, and we thus prove that if $\s>\e$, then $\text{Re}\sbr{s(k)}\le0$ for all $k\in\mathbb{R}$. For completeness, we just note that for $k=0$ there is a double zero $s(0)=0$ and two other solutions $s_\pm(0)=-\frac{\t}{2\s}\pm\frac{\sqrt{\t^2-4\s}}{2\s}$, which is the same as $\s_{\text{hom.}}(0)$ defined above after equation \eqref{eq:fix:damped_wave_Pi}. The later solutions create two branches of behaviour, depending on the condition $\s < \br{\frac{\t}{2}}^2$, for which the real part of $s_\pm(0)$ is the same or not. These cases can be easily seen in figure \ref{fig:gr:fixing:n4_fix_right_sigma_bigger_eps}.

We have created a fixed system, controlled by the parameters $\s$ and $\t$, with a dispersion relation $s(k)$ that has non-positive real part for small and large $k$, preventing the existence of blow-up modes of the original equation if $\s\ge\e$. This allows for stable numerical evolution with the simple cost of adding an auxiliary variable with its own new evolution equation. Nevertheless, by comparing figures \ref{fig:gr:fixing:original_n4}-\ref{fig:gr:fixing:n4_fix_right_sigma_bigger_eps}, one can see that the imaginary low frequency behaviour is reproduced for all dispersion relations (which is the trivial solution $s(k)\to \pm ik$ for $k\to0$), showing that the low energy dynamics are preserved and were not spoiled, as intended.

For negative $\e$, the solutions \eqref{eq:eft:fixXsigmaBox} always have a branch with real positive part, showing the breakdown of the system.

\subsubsection{\done{Alternative approach}}\label{subsubsec:eft:alternative_approach_no_time_order}

We can try to perform fixing of the equations without full time order reduction. One of the ways to explain it is be considering only one iteration of the time order reduction procedure, which results in:
\begin{equation}
    \square \f = -\e\, \pd_t^2\pd_x^2\f + \mathcal{O}(\e^2)\,.
\end{equation}
We can then fix this system in the same way, with a variable $\P$ that approaches $\pd_x^2\f$. This is equivalent to first attempting to apply the fixing procedure to the original equations, using $\P=\pd_t^2\f$, and only after using time order reduction to simplify reduce $\P=\pd_t^2\f = \pd_x^2\f + \mathcal{O}(\e)$.

We then must choose a second order evolution equation for $\Pi$ such as \eqref{eq:fix:damped_wave_Pi}, because a first order equation as \eqref{eq:fix:damped_Pi} does not allow for a straightforward computation of $\pd_t^2\P$. This results in the system:
\begin{equation}\begin{aligned}
    \square \f &= -\e\, \pd_t^2 \P\,,\\\label{eq:fix:damped_wave_Pi_T}
    -\s\, \square \P + \t\,\pd_t\P &= \pd_x^2\f - \P\,,
\end{aligned}\end{equation}

\begin{figure}[h]
\centering
\begin{multicols}{2}
\includegraphics[width=0.5\textwidth]{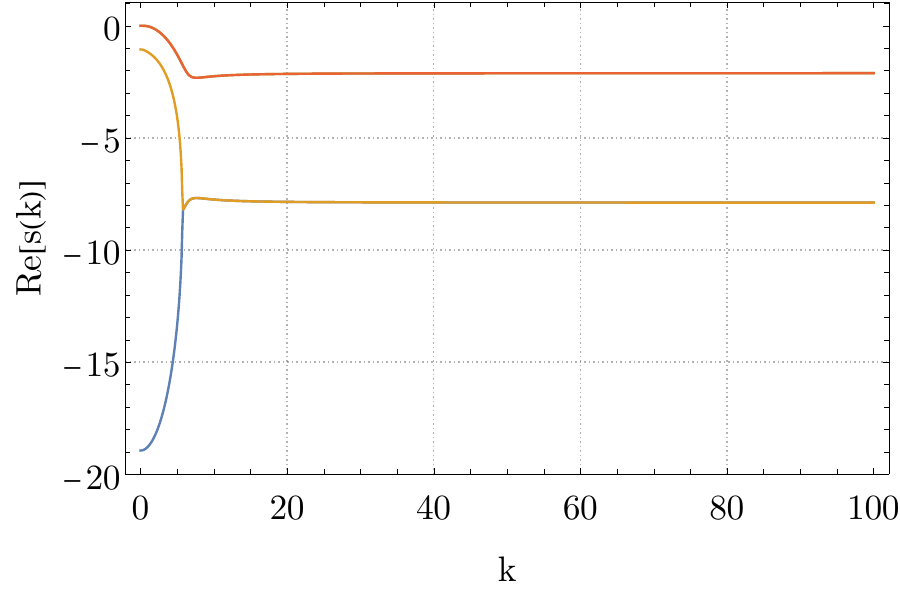}
\includegraphics[width=0.5\textwidth]{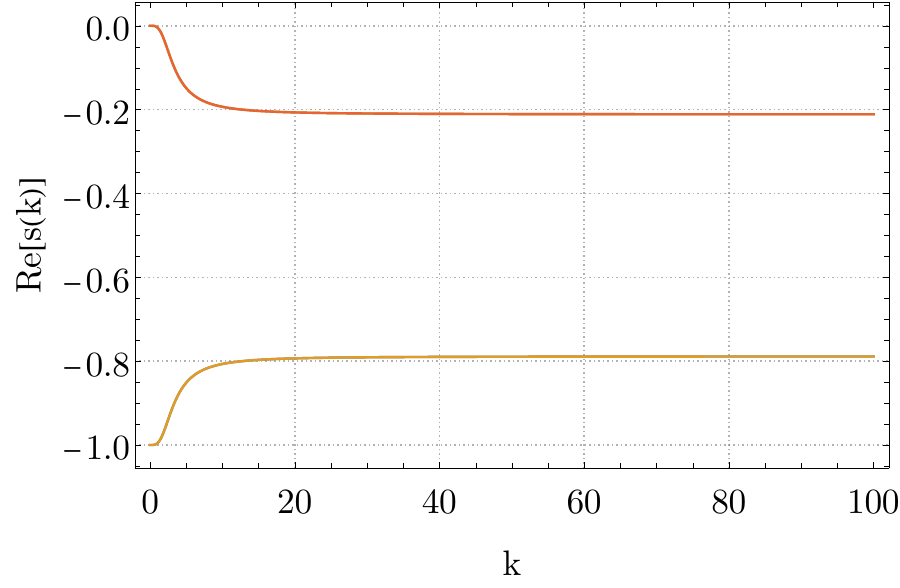}

\includegraphics[width=0.5\textwidth]{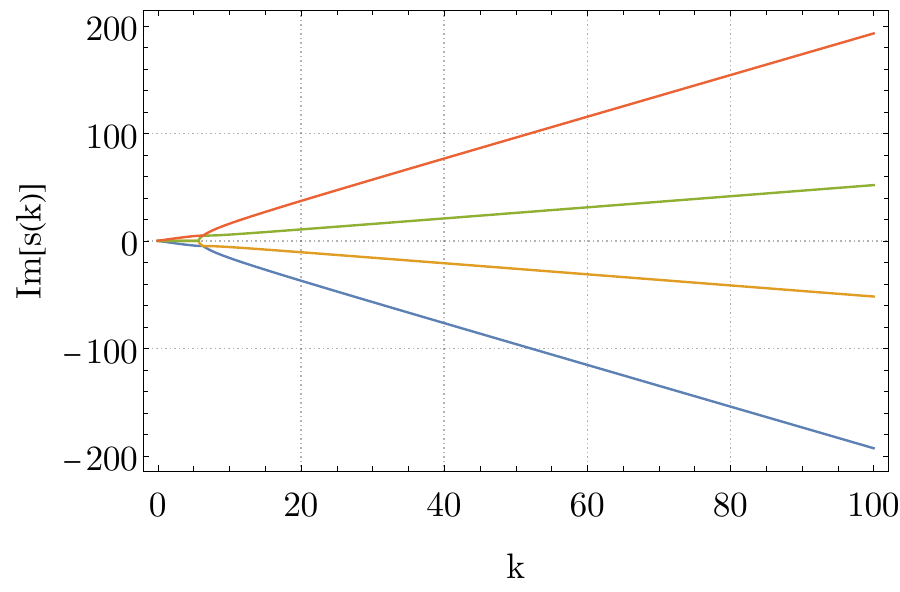}
\includegraphics[width=0.5\textwidth]{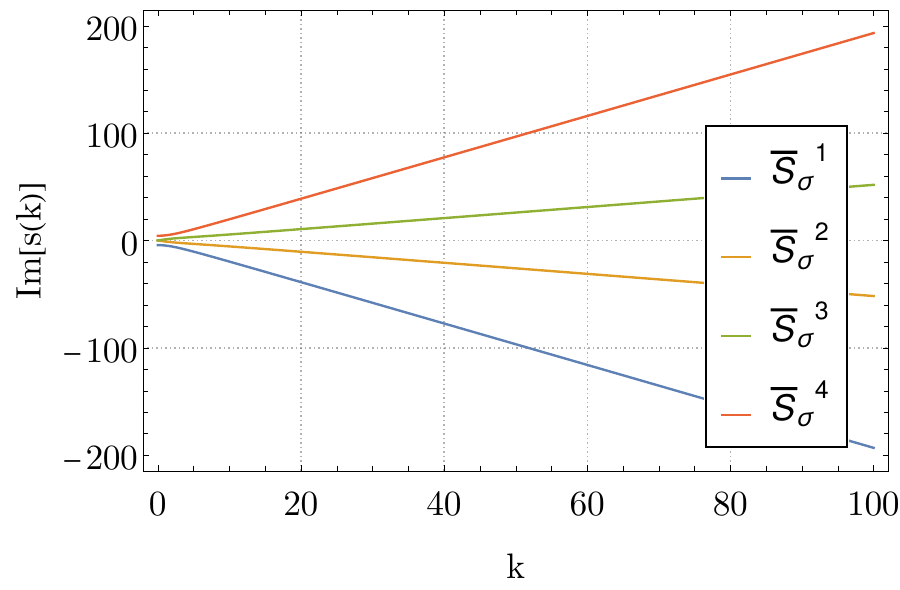}
\end{multicols}
\vspace{-8mm}
\caption{Real and imaginary parts of the dispersion relation of the fixed system \eqref{eq:fix:damped_wave_Pi_T}, denoted $\bar{S}_\s^{~i}$ for $i=1,2,3,4$, with $\e=10^{-1}$ and $\s=0.05$, not satisfying the stability condition $\s>\e$, but yet with stable negative real parts. The top figures show the real and imaginary parts for $\t=1$, for which $\s < \br{\frac{\t}{2}}^2$, and the bottom figured the opposite case, with $\t=0.1$.}
\label{fig:gr:fixing:n4_fix_alternative}
\end{figure}

where $\pd_t^2\P$ in the first equation can always be obtained from a replacement using the second equation. A Fourier analysis of this system results in a stable system for all frequencies for any $\s>0$ and $\e>0$, independent of their relation. This can be proven exactly in the same way as the procedure in equations \eqref{eq:simple_proof_b1}-\eqref{eq:simple_proof_b2}. For completeness, the asymptotic solution is $s(k)\to \a\,k$ with:
\begin{equation}\label{eq:eft:fixTsigmaBox}
    \a_{(\pm,\pm)} = \pm\sqrt{-1-\frac{\e\pm\sqrt{\e^2+4\,\e\,\s}}{2\s}}\,,
\end{equation}
which is always imaginary, because $-\frac{\e+\sqrt{\e^2+4\,\e\,\s}}{2\s}$ is always negative, leading to negative square root, and $-\frac{\e-\sqrt{\e^2+4\,\e\,\s}}{2\s}$ may be positive, but is bounded by $1$ as $\e\to\infty$, also leading to a purely imaginary $\a$. This can be observed in figure \ref{fig:gr:fixing:n4_fix_alternative}.

For negative $\e$ and any $\s>0$, this system is not stable.

\section{\done{Methodology}}\label{sec:eft:methodology}

\subsection{\done{Action and equations of motion}}

Let us follow the setup by Cayuso and Lehner \cite{Cayuso:2020lca}. This consists of using no new degrees of freedom besides the gravitational ones and parameterising new physics as higher powers of the Riemann curvature tensor that preserve unitarity, causality and locality. With no new fields, quadratic terms involving the Riemann tensor can be reduced to the Gauss-Bonnet topological invariant via field re-definitions \cite{Burgess:2003jk}. Cubic terms in the Riemann tensor lead to acausal propagation \cite{Endlich:2017tqa}\footnote{This statement is in fact contentious and a matter of current debate, according to de Rham and Tolley \cite{deRham:2019ctd}.}. The lowest expected terms are then quartic, and with the requirements above they reduce to the 4D action:
\begin{equation}\label{eq:eft:action_full}
    S_\text{EFT} = \int d^4x\, \sqrt{-g}\br{\mathcal{R} - \e\, \mat{C}^2 - \ti{\e}\, \mat{\ti{C}}^2 - \e_-\, \mat{C}\mat{\ti{C}} + \dots}\,,
\end{equation}
where $\mat{C}=\mathcal{R}_{\a\b\g\d}\mathcal{R}^{\a\b\g\d}$, $\mat{\ti{C}}=\mathcal{R}_{\a\b\g\d}\ti{\mathcal{R}}^{\a\b\g\d}$, with $\ti{\mathcal{R}}^{\a\b\g\d}=\e^{\a\b}_{~~\r\s}\mathcal{R}^{\r\s\g\d}$, $\br{\e,\ti{\e},\e_-}=\br{\frac{1}{\L^6}, \frac{1}{\ti{\L}^6}, \frac{1}{\L_-^6}}$, and the $+ \dots$ corresponds to sub-leading corrections. We restrict the analysis\footnote{We also omit the CCZ4 $Z$ terms from the action below to simplify the explanation.} to the case $\ti{\e}=\e_-=0$:
\begin{equation}
    S_\text{EFT} = \int d^4x\, \sqrt{-g}\br{\mathcal{R} - \e\, \mat{C}^2 + \dots}\,.
\end{equation}
Note how the first correction to the Einstein-Hilbert action starts at $\L^{-6}$ corrections, which is only a leading interaction for the case of vacuum \cite{deRham:2019ctd}.

From the action above one can derive the equations of motion,
\begin{equation}
\begin{aligned}\label{eq:EOM_full}
    G_{\m\n} = \e\bigg[ & \mat{C}\br{-8\mathcal{R}_{\m}^{~\a} \mathcal{R}_{\n\a}+8\mathcal{R}^{\a\b}\mathcal{R}_{\m\a\n\b} + 4\mathcal{R}_{\m}^{~\a\b\g}\mathcal{R}_{\n\a\b\g} - \tfrac{1}{2}g_{\m\n}\mat{C} - 4\grad_\m\grad_\n \mathcal{R} + 8\grad_\a\grad^\a \mathcal{R}_{\m\n} } \\
    & - 16\br{\grad_\a \mat{C}}\grad_{(\m}\mathcal{R}_{\n)}^{~\a} + 16\br{\grad_\a \mat{C}}\grad^\a \mathcal{R}_{\m\n} + 8\mathcal{R}_{\m~\n}^{~\a~\b}\grad_\b\grad_\a \mat{C} \bigg]\,.
\end{aligned}
\end{equation}
These equations have up to fourth order derivatives and also non-linear terms in second derivatives, being a good model to test if fixing approach can tame such problems. Given that in vacuum for this theory $\mathcal{R}_{\m\n} = \mathcal{O}(\e)$, to simplify things further, we apply order reduction by replacing any Ricci tensor contributions to the right hand side by $\mathcal{O}(\e^2)$ terms that can be discarded, leaving the equations:
\begin{equation}\label{eq:EOM_4D_simple}
    G_{\m\n} = \e\br{ 4\,\mat{C}\,\mathcal{R}_{\m}^{~\a\b\g}\mathcal{R}_{\n\a\b\g} - \tfrac{1}{2}\mat{C}^2\,g_{\m\n} + 8\,\mathcal{R}_{\m~\n}^{~\a~\b}\grad_\b\grad_\a \mat{C} } + \mathcal{O}(\e^2)\,.
\end{equation}
It is often useful to replace the Riemann with the Weyl tensor, $C_{\alpha\beta\gamma\delta}$, since $\mathcal{R}_{\mu\nu\rho\sigma}=C_{\mu\nu\rho\sigma} + \mathcal{O}(\epsilon)$. Up to $\mathcal{O}(\e)$, it is also true that $\mathcal{C}=C_{\alpha\beta\gamma\delta}\,C^{\alpha\beta\gamma\delta}+ \mathcal{O}(\epsilon)$.
\begin{equation}\label{eq:eft:EOM_4d_weyl}
    G_{\m\n} = \e\br{ 4\,\mat{C}\,C_{\m}^{~\a\b\g}C_{\n\a\b\g} - \tfrac{1}{2}\mat{C}^2\,g_{\m\n} + 8\,C_{\m~\n}^{~\a~\b}\grad_\b\grad_\a \mat{C} } + \mathcal{O}(\e^2)\,.
\end{equation}
%

Cayuso and Lehner \cite{Cayuso:2020lca} provide several ways to fix \eqref{eq:eft:EOM_4d_weyl}, leading to equations with second time derivatives at most. We must take a different approach to solve the problem away from spherical symmetry but in full 4 dimension generality.

\subsection{\done{Constraints on the energy scale parameter $\L$}}\label{subsec:eft:constraints}

To guide the choice of the value $\e=\L^{-6}$, we look at some literature analysing bounds for this particular EFT.

From gravitational wave data, Sennett et al. \cite{Sennett:2019bpc} obtain a bound of $d_\L < 150$ km, with $d_\L = \L^{-1}$, and Silva et al. \cite{Silva:2022srr} improves this bound with $l_{\text{qEFT}} < 50$ km, with $l_{\text{qEFT}} = d_\L$. In geometrised units, the later tighter bound corresponds to $d_\L\lessapprox 33.8M_\odot$, where $M_\odot \approx 1.48$ km is the solar mass. LIGO-Virgo \cite{LIGOScientific:2018mvr, LIGOScientific:2020ibl, LIGOScientific:2021djp} detected final mergers have masses ranging $20-100M_\odot$, leading to a bound of $d_\L \lessapprox 0.34-1.69M$, where $M$ is the total mass of the binary system. This result makes intuitive sense, as the maximum energy scale the EFT should probe should be smaller than the size of the black hole, which is approximately $M$. Finally, this leads to $\frac{\e}{M^6}=\frac{d_\L}{M}^6 \lessapprox 10^{-3}-10^1$. 

The high power relating the energy scale to the action parameter $\e$, makes bounds on $d_\L$ very sensitive. The biggest conclusion to draw is that definitely one should care about $d_\L\ll M$ for consistency of the EFT. 

Using causality arguments, de Rham et al. \cite{deRham:2021bll} argues for the inviability of testing this model. It shows that it is only valid for any $\e<0$ or for $\e>0$ and low azimuthal numbers, with a minimal (potentially much higher) bound of $|\L| > 7 * 10^{-11}eV$, in units where $\hbar=c=1$. Using $\frac{\hbar c}{1 eV} = 1.97*10^7m$, this implies $|d_\L| < 2.8\text{ km} = 1.9M_\odot$. For a black hole mass around $30M_\odot$, this leads to a bound of $d_\L \lessapprox 0.063M$, or $\frac{\e}{M^6} \lessapprox 10^{-7}$. This tight bound shows the claim of inviability to test the model. One should then bear in mind when testing any $\e$ that this is a complete theory and that the objective is, first, to probe effects that might appear in extensions of GR, and second, to test the methods available to potentially probe such theories.

\subsection{\done{Adapting toy models to gravity systems}}

\subsubsection{\done{Alternative systems}}\label{subsubsec:eft:alternative_systems_gravity}

In the previous sections, we understood how to develop a fixed system that can stably evolve the low energy modes of the original equation, seen as an EFT. In gravity settings, we encounter an issue that is not as clear in the wave model above. Gravitational perturbations also act as waves, but there are also stationary solutions such as black holes. Systems like \eqref{eq:fix:damped_wave_Pi_T} lack one property: if there is a stationary solution $\f(x)$, then $\P=\pd_x^2\f$ is not a static solution of the evolution equation for $\P$, meaning that $\P$ is not tracking appropriately the variable it was supposed to, even though any wave-like perturbations of it are being dissipated as intended.

The first thing one might be tempted to do is to remove all spatial derivatives of $\P$ from the fixed equation in \eqref{eq:fix:damped_wave_Pi_T}, evolving instead:
\begin{equation}\begin{aligned}
    &\square \f = -\e\, \pd_t^2 \P\,,\\\label{eq:eft:time_derivatives_only}
    &\s \pd_t^2 \P + \t \pd_t \P = \pd_x^2\f - \P\,.
\end{aligned}\end{equation}
The solution $\P = \pd_x^2\f$ satisfies this equation if $\f$ is a static solution, and perturbations around it also satisfy proper damping. Indeed, a Fourier analysis on the homogeneous equation $\s \pd_t^2\P + \t\pd_t \P = -\Pi$ clearly has solutions $s=-\frac{\t}{2\s}\pm\frac{\sqrt{\t^2-4\s}}{2\s}$, which are real and negative if $\s<\br{\frac{\t}{2}}^2$ and complex with negative real part if $\s>\br{\frac{\t}{2}}^2$.

This system works well for static solutions, but the issue with such system is the lack of advection. Using only the local time coordinate in the fixing equation, instead of covariant quantities, implies the variable $\P$ will not be naturally guided to follow the change in coordinates due to a non-zero shift vector. To ameliorate the situation, we replace the operator $\pd_t$ with an advection operator $\pd_t - \b^i\pd_i$, resulting in the system:
\begin{equation}\begin{aligned}
    &\square \f = -\e\, \pd_t^2 \P\,,\\\label{eq:eft:fix2}
    &\s \br{\pd_t^2 - 2\b^i\pd_{it} + \b^i\b^j\pd_{ij}}\P + \t \br{\pd_t - \b^i\pd_i} \P = \pd_x^2\f - \P\,.
\end{aligned}\end{equation}
Analysing perturbations of the principal part of this equation in one dimension, one gets the solutions 
$s(k)=i k \b^x - \frac{\t}{2\s}\pm\frac{\sqrt{\t^2-4\s}}{2\s}$, which have real negative parts and are hence stable. To reinforce the idea, this is still not done in a covariant way to reduce the amount of spatial terms required that lead the stationary solution astray, as we shall see in the following sections.


\subsubsection{\done{Further improvements}}\label{subsubsection:eft:further_improvements}

There is a lot of potential improvements one can make for added stability in the system. Here we describe some good and some apparently good yet useless things to do.

Comparing the Laplacian version \eqref{eq:eft:c_system_eom_laplacian}, when adapting this to the advection system \eqref{eq:eft:c_system_eom}, one might be tempted to also add the replacement $\t\to\frac{\t}{\a}$ and $\s\to\frac{\s}{\a^2}$, where $\a$ is the lapse function. This proves useless in the sense that very close to the black hole, as the lapse collapses to zero, the ratio $\frac{\t}{\s}\to\frac{\a\,\t}{\s}$ (which governs the control of perturbations, as seen in \ref{sec:eft:toymodels}) goes to zero, effectively not promoting a good damping to the desired solution.

To approximate equation \eqref{eq:eft:fix2} to \eqref{eq:eft:time_derivatives_only}, one might be tempted to replace $\b^i\pd_i \to \eta \b^i\pd_i$, for some $\eta<1$, to reduce the effect of spatial derivative terms while still promoting advection. Using $\eta=1$ indeed leads the the biggest deviation in static solutions, as expected. Yet, when using boosted solutions with $\eta<1$, the additional ``effort'' the equation has to do to keep track of the moving solution leads to poorer tracking due to a time lag. Hence, using $\eta=1$ is still the best approach for the best tracking of the solution.

Additionally, one might infer an instability at low resolutions when using small $\s$ or $\t$. These parameters define timescales of evolution and coarse grids with bigger time evolution steps that might lead to numerical instability. This is very much like the instability found with the $\eta$ time scale parameter in the Gamma driver equation \eqref{eq:ngr:gamma_driver} for the shift vector \cite{Schnetter:2010cz}. As described by Schnetter \cite{Schnetter:2010cz}, a simple solution is to make a spatially varying $\s$ and $\t$, asymptoting to larger values at spatial infinity. It turns out this addition was not required for $\t$, but proved essential for $\s$. This can be done with a radial profile or a profile depending on the conformal factor $\chi$. The latter was implemented, making a smooth transition from whatever $\s$ is used in a simulation to an asymptotic $\s=1$, transitioning after $\chi > 0.92$. This choice requires no fine tuning.

Finally, we know that $\s$ and $\t$ should be as small as possible for good tracking of the auxiliary variable, while at the same time making them too small leads to instabilities in the evolution. The merger is a period of intense dynamics that places a cap on how small these values can be, but arguably, the initial stages of the simulation are equally problematic, if not worse, due to the rapid gauge adjustment of the initial data to moving puncture gauge. To avoid this initial artificial and unphysical dynamics and allow evolution with small $\s$ and $\t$, a simple solution was to ``turn on'' the EFT $\e$ gradually with time. We opted for:
\begin{equation}\label{eq:eft:epsilon_time_dependent}
    \e_t(t) =
    \begin{cases}
      0 & t \le t_0\,,\\
      \e\,\br{\frac{t-t_0}{t_F-t_0}}^2 & t_0 < t \le t_F\,,\\
      \e & t > t_F\,,
    \end{cases}
\end{equation}
for some constants $t_0$ and $t_F$, typically $t_0=10M$ and $t_F=30M$.

\subsubsection{\done{Parameter dependence and stability}}

As advocated in section \ref{subsubsec:gr:fixing}, the fixed equation should be independent of the choice of parameters. For system \eqref{eq:eft:fix2}, we can check for the validity of this principle. It is trivial to understand that bigger $\s$ and $\t$ lead to bigger spatial terms that deviate $\P$ from the physical variable it is tracking, while at the same time being essential for stability. We detail briefly in this section the balance between these two effects.

A fixing equation in dynamical scenarios has to solve in parallel two independent problems: rigorously keeping track of some physical variable and stably damping high frequency modes.

The limitation of effectively damping high frequency perturbations is an obvious constraint on the kind of fixing equation one can ``design'', as seen in the toy model analysis made. But toy models are typically linearised around a constant solution and do not worry about having to accurately keeping track of some dynamic profile in space and time. This tracking of the physical variable involves three different aspects:
\begin{itemize}
    \item coordinate changes, due to non-zero shift vector, $\b^i$;
    \item physical wave propagation in the grid, for example gravitational waves;
    \item shape of the solution changing, for which the auxiliary variable has to adapt its profile to.
\end{itemize}

Covariance of the equations in the Laplacian system or simple advection terms in the advection system described above are useful to tackle such issues. But on top of that, one still faces a balance between two effects: how accurate the tracking is and how fast it adapts to the system dynamics. For instance, small $\t$ leads to slow tracking but to the right solution, while big $\t$ leads to fast tracking but to a less accurate solution. Maximising adaptation to the evolving dynamics leads one to the Laplacian system \eqref{eq:eft:c_system_eom_laplacian}, which fails to match the physical variable precisely, while maximising accuracy of the solution, as it happens in the system with time derivatives only \eqref{eq:eft:time_derivatives_only}, fails to adapt to moving solution dynamics. Ultimately, desiring both, one is led to a system that lands somewhere in between, resulting in the advection system \eqref{eq:eft:fix2}.

\subsection{\done{``Fixing'' the equations - Weyl system}}

Cayuso and Lehner \cite{Cayuso:2020lca} analyse this model in spherical symmetry. With a simplification of the equations, this allows for proceeding with the standard procedure of time order reduction and fixing of the equations. For the full system, as discussed in \ref{subsubsec:eft:alternative_approach_no_time_order}, we may want to proceed to fixing the equations first and apply time order reduction to the fixed system, to avoid expanding all the metric terms of Weyl contractions, as described in \ref{subsubsec:eft:alternative_approach_no_time_order} and as we shall understand below.

One important note is that, besides achieving stability under high frequency perturbations, as done for instance by Franchini et al. \cite{Franchini:2022ukz}, we also want, for pure numerical convenience, to use the fixed system auxiliary variables to remove any higher than second spatial derivative. Inspired by the toy models in \ref{sec:eft:toymodels}, the trivial approach to obtain this is using the Weyl tensor itself as auxiliary variable, resulting in the system:
\begin{equation}
\begin{aligned}\label{eq:eft:weyl_system_eom}
    G_{\m\n} &= \e\br{ 4\,\hat{\mat{C}}\,\Pi_{\m}^{~\a\b\g}\Pi_{\n\a\b\g} - \tfrac{1}{2}\hat{\mat{C}}^2\,g_{\m\n} + 8\,\Pi_{\m~\n}^{~\a~\b}\grad_\b\grad_\a \hat{\mat{C}} }\,.\\
    \sigma\,\nabla^2\Pi_{\mu\nu\rho\sigma} &=\tau\,\mathcal{L}_n\Pi_{\mu\nu\rho\sigma}+C_{\mu\nu\rho\sigma}-\Pi_{\mu\nu\rho\sigma}\,,
\end{aligned}
\end{equation}
where $\Pi_{\m\n\r\s}$ is the auxiliary variable approaching $W_{\m\n\r\s}$, $\mathcal{L}_n$ is the Lie derivative along the timelike normal $n^\m$ orthogonal to spacelike hypersurfaces, and $\hat{\mat{C}}=\Pi_{\m\n\r\s}\Pi^{\m\n\r\s}$ tracks the Kretschmann scalar. Note that we use a second order equation for $\Pi$ as the last term in the metric equation of motion, $\pd_\b\pd_\a \hat{\mat{C}}$ requires second derivatives of $\Pi$. Furthermore, note that indeed this approach involves only computing second derivatives of the metric and $\Pi_{\m\n\r\s}$. Time order reduction comes in when computing $C_{\m\n\r\s}$, following the formulae in \ref{appendix:grchombo:weyl}, where the matter terms are ignored, leading to the Weyl tensor computed only for the vacuum GR theory without leading order corrections in $\e$. The disadvantage of this system is that the Weyl tensor is a rank 4 tensor with only 10 independent components in four dimensions and the obvious way to have an explicit decomposition is using the electric and magnetic parts of the Weyl tensor (see \ref{appendix:grchombo:weyl} for details). Taking this into consideration, along with the considerations in section \ref{subsubsec:eft:alternative_systems_gravity}, we develop the EB system presented in the next section.

\subsection{\done{``Fixing'' the equations - EB system}}\label{subsec:eft:eb_system}

Following \ref{appendix:grchombo:weyl}, one can decompose that Weyl tensor into an electric and magnetic spatial tensors, $E_{ij}$ and $B_{ij}$, using formula \eqref{eq:weyl_decomposition}. Then, one can pick as auxiliary variables $\mathcal{E}_{ij}$ and $\mathcal{B}_{ij}$ that evolve to the physical electric and magnetic parts $E_{ij}$ and $B_{ij}$, such that the auxiliary Weyl tensor and Kretschmann scalar, $\Pi_{\m\n\r\s}$ and $\hat{\mat{C}}$, computed with these auxiliary variables, also track the corresponding physical Weyl tensor and Kretschmann scalar, $C_{\m\n\r\s}$ and $\mat{C}$. The physical electric and magnetic tensors in the CCZ4 formulation can be computed using \eqref{eq:ccz4_Electric} and \eqref{eq:ccz4_magnetic}, the Weyl tensor can be re-constructed using \eqref{eq:weyl_decomposition}, and the Kretschmann scalar can be computed directly, ignoring terms of order $\mathcal{O}(e)$, using \eqref{eq:weyl_squared}.

Following the intuition from \ref{subsubsec:eft:alternative_systems_gravity}, we propose a damped second order system for $\mathcal{E}_{ij}$ and $\mathcal{B}_{ij}$:
\begin{equation}
\begin{aligned}\label{eq:eft:eb_system_eom}
    &G_{\m\n} = \e\br{ 4\,\hat{\mat{C}}\,\Pi_{\m}^{~\a\b\g}\Pi_{\n\a\b\g} - \tfrac{1}{2}\hat{\mat{C}}^2\,g_{\m\n} + 8\,\Pi_{\m~\n}^{~\a~\b}\grad_\b\grad_\a \hat{\mat{C}} }\,.\\
    &\sigma\,\br{\pd_t^2 - 2\b^i\pd_{ij} + \b^i\b^j\pd_{ij}}\E_{ij} + \tau\br{\pd_t - \b^i\pd_i}\E_{ij}=E_{ij}-\E_{ij}\,,\\
    &\sigma\,\br{\pd_t^2 - 2\b^i\pd_{ij} + \b^i\b^j\pd_{ij}}\B_{ij} + \tau\br{\pd_t - \b^i\pd_i}\B_{ij}=B_{ij}-\B_{ij}\,,
\end{aligned}
\end{equation}
which can be further re-written as a first order system. This system is incredibly powerful, as using the electric and magnetic parts of the Weyl tensor can be used to evolve many EFTs, namely the full action \eqref{eq:eft:action_full} or any gravitational part of GR extensions that involves powers of the Riemann tensor.

As a side remark, using equation \eqref{eq:maxwellEB}, other systems making use of these auxiliary variables can be developed, namely if one lifts the convenient restriction of using only up to second spatial order derivatives.

\subsection{\done{``Fixing'' the equations - $\mat{C}$ system}}\label{eft:subsec:c_system}

As a further alternative, one can develop yet another system, which even though not generic and particular to this present theory, is very easy to implement. It is important to remember there is no unique way to fix a theory \textit{à la} Israel-Stewart and all should agree in the infrared regime \cite{Geroch:1995bx, Geroch:2001xs}.

We set the scalar $\mat{C}$ itself as a new independent variable $\hat{\mat{C}}$ damped to the physical one:
\begin{equation}
\begin{aligned}\label{eq:eft:c_system_eom}
    &G_{\m\n} = \e\br{ 4\,\hat{\mat{C}}\,C_{\m}^{~\a\b\g}C_{\n\a\b\g} - \tfrac{1}{2}\hat{\mat{C}}^2\,g_{\m\n} + 8\,C_{\m~\n}^{~\a~\b}\grad_\b\grad_\a \hat{\mat{C}} }\,.\\
    &\sigma\,\br{\pd_t^2 - 2\b^i\pd_{ij} + \b^i\b^j\pd_{ij}}\hat{\mat{C}} + \tau\br{\pd_t - \b^i\pd_i}\hat{\mat{C}}=\mat{C}-\hat{\mat{C}}\,.
\end{aligned}
\end{equation}
This system is less well understood mathematically, as from the PDE point of view, it still has terms with the original Weyl tensor squared, i.e. with second order derivatives multiplied together. Note that this involves only spatial derivatives as one can compute the Weyl tensor using $\mat{C} = 8\br{E_{ij}E^{ij} - B_{ij}B^{ij}}$, ignoring first order terms in $\e$ (see appendix \ref{appendix:grchombo:weyl} for details). In spite of this, it has proven to be stable, as we shall see in the results section. With the addition of a single extra variable and simple equations, this system is an easy add-on to any numerical code. One can reduce it to first order by doing:
\begin{equation}
\begin{aligned}\label{eq:eft:c_system_eom_first_order}
    \pd_t \hat{\mat{C}} &= \P\,,\\
    \pd_t\Pi &= \tfrac{1}{\s}\br{\mat{C}-\hat{\mat{C}} - \t\br{\Pi - \b^i\pd_i\hat{\mat{C}}}} + 2\b^i \pd_{i}\Pi - \b^i \b^j \pd_{ij}\hat{\mat{C}}\,.
\end{aligned}
\end{equation}

We shall refer to this as the advection $\mat{C}$ system, and refer as the Laplacian $\mat{C}$ system to the version using:
\begin{equation}\label{eq:eft:c_system_eom_laplacian}
    \sigma\,\nabla^2 \hat{\mat{C}} = \tau\,\mathcal{L}_n \hat{\mat{C}} +\mat{C}- \hat{\mat{C}}\,.
\end{equation}

\subsection{\done{Excision}}\label{subsec:eft:excision}

We implement excision as explained in section \ref{subsection:paper1:excision} in a slightly modified form explained below. This form of excision proved extremely useful to solve this problem. Given an energy-momentum tensor (we treat the right-hand side of equation \eqref{eq:eft:EOM_4d_weyl} as such), we damped it by a factor $e\cdot T_{\m\n}$, where:
\begin{equation}
\begin{aligned}
    e(\chi,W) &= 1 - \s(\chi; \bar{\chi}, \w_\chi) \s(W; \bar{W}, -\w_W)\,,\\\label{eq:eft:sigmoid}
    \s(x; \bar{x}, \w_x) &= \frac{1}{1 + 10^{\frac{1}{\w_x}\br{\frac{x}{\bar{x}}-1}}}\,,
\end{aligned}
\end{equation}
where $W$ is a weak field measure, $\chi$ is the conformal factor of the metric, $\bar{\chi}$ and $\bar{W}$ are thresholds for the excision cutoff region and $\w_\chi$ and $\w_W$ are smoothness widths. Roughly, $\s(x; \bar{x},\w_x) < 10^{-k}$ for $x>\bar{x}(1+k\,\w_x)$ and $\s(x; \bar{x},\w_x) > 1 - 10^{-k}$ for $x<\bar{x}(1-k\,\w_x)$. The idea is if $\chi < \bar{\chi}$ and $W > \bar{W}$, then $e(\chi,W)$ exponentially approaches 0, and is 1 otherwise. Since we know that the $\chi$ contours track quite well the AH, we can choose an appropriate threshold for $\chi$ to ensure we are only excising (that is, applying $e=0$) inside the AH. During gauge adjustment, the exact contour of $\chi$ that tracks the AH changes. Hence, for accurate tracking, we make use of the analysis of appendix \ref{appendix:grchombo:ah_location} and use a dynamic threshold $\bar{\chi} = p\cdot\chi_{AH,\,t}$ (opposed to a fixed one as in section \ref{subsection:paper1:excision}), where $p$ is a small fraction that forces the contour to be inside of the black hole (most simulations used $p=0.35$) and $\chi_{AH,\,t}$ is based on formula \eqref{eq:chi_over_time}.

To be more specific about the weak field measure $W$, we use $\{\r,S_i,S_{ij}\}$ from the $d+1$ decomposition components of $T_{\m\n}$ to compute $W = \sqrt{\r^2 + S_i S_j \d^{ij} + S_{ij} S_{kl} \d^{ik} \d^{jl}}$. This is not a covariant scalar as $T_{\m\n}T^{\m\n}$, but one just needs some measure to use as threshold\footnote{Using $\r$ may sound like a good idea as well, but this goes to zero at the puncture, making it not ideal by itself.}. We often take $\bar{W} \to 0$ such that $\s(W; \bar{W}\to0, -\w_W)$ is always effectively $1$ and we excise only based on $\chi$. However, we monitor $W$ as a way to keep track of the weak field condition.

\subsection{\done{Weak coupling conditions}}\label{subsec:eft:WCC}

Similarly to the analysis in chapters \ref{chapter:paper1} and \ref{chapter:paper2}, valid EFTs requires systems in the weak field regime. To ensure this is the case, we develop a weak coupling condition that informs about the size of the corrections when compared to GR terms. With that in mind, we use the dimensionless scalar $WCC = \sqrt{\frac{T_{\m\n}T^{\m\n}}{\mat{C}}}$, where $\mat{C}$, the Kretschmann scalar, represents the size of GR terms, and $T_{\m\n}$ corresponds to the $\e$ terms in the right hand side in equation \eqref{eq:eft:EOM_4d_weyl}.

As a mere inspection of this metric, on the horizon scale, $T_{\m\n}T^{\m\n}$ is roughly, up to numerical factors, $\mathcal{O}(\e^2 \mat{C}^4)$, resulting in $WCC\sim \e\, \mat{C}^{3/2}$. For a Schwarzschild black hole, on the horizon scale, $\mat{C} \sim \mathcal{O}(M^{-4})$, meaning that $WCC \sim \e / M^6$, precisely as expected. A precise calculation for the Schwarzschild case yields at the horizon that $WCC \approx 2.7\,\e / M^6$, showing the previous estimation is accurate. Additionally, in isotropic coordinates, the $WCC$ peaks at the horizon, being smaller inside and outside the black hole.

In summary, this analysis leads to the conclusion that as long as $\e / M^6 \ll 1$, we are in the weak field regime of the EFT. Note however that for an equal mass binary evolution, for a final black hole of mass $M$, each individual black hole has mass $M/2$, leading to an effective $\e$ that is $2^6 = 64$ times bigger, or a bound of $\e / m^6 \ll 1$, where $m$ is the smallest mass in the system.

\subsection{\done{From gravitational waveforms to mismatch}}\label{subsec:eft:methodology:gravitational_strain}

Computing the mismatch between two waveforms requires transforming the numerical $\Psi_4$ Weyl scalar into gravitational strain, $h$. The methods used to extract the gravitational wave and compute the strain and the mismatch are described in chapter \ref{chapter:paper2}, sections \ref{subsec:paper2:gravitational_wave_extraction} - \ref{subsec:paper2:mismatch_theory}.

\subsection{\done{Initial data}}\label{subsec:eft:initial_data}

First, we choose the $\hat{\mat{C}}$ system as the easiest approach to evolving this EFT, both for single and binary black holes. For auxiliary variables, specifically in the $\mat{C}$ system \eqref{eq:eft:c_system_eom}, the variable $\mat{C}$ is initialised to the Kretschmann scalar, computed numerically from the metric initial data. The first order auxiliary variable $\Pi = \pd_t \mat{C}$ \eqref{eq:eft:c_system_eom_first_order} is initialised to $0$. This is unphysical, as in puncture gauge there are non-trivial dynamics. This time derivative can be computed using the formulae for the time derivative of the electric and magnetic components (see \ref{appendix:grchombo:weyl}). While experimenting with this, we learnt that changing this could minimally improve initial instabilities, but ultimately was a useless addition, as initialising it to zero converged to the correct value within 2 to 3 timesteps.

For the single black hole case, standard Schwarzschild black hole metrics are not a good representation of the system we want to evolve unless they have non trivial shift vector and are boosted, as they will be for the case of binaries. Hence, we evolve numerical simulations of a boosted Schwarzschild black hole, and wait for it to stabilise in its stationary solution in moving puncture gauge, leading to the the black hole solution that binary black holes will actually evolve. Hence, for initial data in single black holes, we use a boosted black hole solution derived from the conformal transverse-traceless decomposition \cite[p.~73-74]{shapiro} (see also section \ref{subsec:ngr:conformalTT}), which uses an approximate conformal factor solution to the Hamiltonian constraint, valid for small boosts $|\ve{P}|\ll M$, where $\ve{P}$ is the initial momentum of the individual black hole and $M$ is its mass. For binary data, the single black hole initial data is superposed, resulting in non-spinning Bowen-York data.

A few extra remarks on the accuracy of this initial data. First, \texttt{GRChombo} has the \texttt{TwoPunctures} spectral solver \cite{Ansorg:2004ds, Paschalidis:2013oya} integrated. This improves the accuracy of the initial data, but resulted in no significant change in the gravitational waves. Second, one could also realise that for any non-zero $\e$, GR initial data does not satisfy the constraints, due to the ``matter'' terms in them. To test the relevance of this, Ramiro Cayuso developed a spherical symmetry initial condition solver for this EFT, as described by Cayuso and Lehner \cite[IV.A]{Cayuso:2020lca}. In spherical symmetry, the constraints for the conformal factor and $K_{rr}$ can be integrated radially. To find a solution, a shooting method can be implemented from iterating some inner boundary numeric guess, integrated to some outer boundary, where the solution is sufficiently close to a Schwarzschild spacetime such that the expected boundary condition solution is known and enforced. This is valid as we expect the deviations from GR to be of the order of $\e$ and rapidly decay with radius \cite{Cardoso:2018ptl}. Overall, this implies that for $\e / M^6\ll1$, this initial data provides little added accuracy, similar to the comparison between the boosted black hole and \texttt{TwoPunctures} initial data. Moreover, for binary black hole initial data, our approach involves adding a Lorentz boost to the spherically symmetric data and superposing two of such solutions. In spite of the accuracy of the spherically symmetric data, the final boosted superposed binary data resulted in bigger constraints than simply using the GR Bowen-York initial data, even though $\e$ is non-zero. It is important to also recall that the CCZ4 formulation ensures stable control of any of these small constraint deviations.

Single boosted black holes evolved have a mass of $0.5M$ with a boost of $P^x = 0.08M$, representing approximately one of the black holes intended to simulate in binaries. The binary black holes have circularised orbits\footnote{Note that finding the right momenta for an initial circular trajectory is in general a hard problem in NR - simply setting the momentum to the Newtonian approximation will result in elliptic orbits even for well separated initial BHs. One must use the Post-Newtonian approximations \cite{Healy:2017zqj}, and then adjust the momentum manually over several iterations to achieve an initial eccentricity of less than 1\% \cite{Husa:2007rh,Pfeiffer:2007yz,Kyutoku:2020lgg}. We are grateful to Dr Sebastian Khan at Cardiff University for providing the \texttt{GRChombo} team with these initial data values.}, have a mass of $m=0.48847892320123$, a separation of $D=12.21358M$ and initial momentum of $P^i = (0.0841746, 0.000510846, 0)M$ (and opposite for the other black hole).

Finally, we use a time dependent $\e$ as described in \eqref{eq:eft:epsilon_time_dependent} with $t_0=6M$, $t_F=12M$ for single black holes and $t_0=10$, $t_F=30M$ for binary black holes. All simulations use $\e=10^{-5}$ unless made explicit otherwise.

\subsection{\done{Numerical scheme}}

On top of what was already described in section \ref{subsec:ngr:grchombo_params}, for the simulations presented in this chapter, we use a tagging criterion that triggers the regridding based on second derivatives of the conformal factor, forcing certain levels around the location of apparent horizons\footnote{Here, we reinforce that ensuring a good buffer between the finest AMR level and the apparent horizon is very important to avoid energy `leaks' and a monotonous mass drift over time.} and around gravitational wave extraction regions. We use Kreiss-Oliger numerical dissipation with fixed $\s=2$ (this bigger value was useful to stabilise the several systems developed) in all our simulations. As of boundary conditions, we use Sommerfeld boundary conditions and take advantage of the asymmetric symmetry of a boosted black hole to evolve only one quarter of the grid and of the reflective/bitant symmetry of the binary problem to evolve only half of the grid. Sixth order spatial stencils are used in order to improve phase accuracy of the binaries \cite{Husa:2007hp}. For the results presented in this chapter, we have a CFL factor of $1/4$. Single boosted black holes used a coarse level resolution of $\D x = 1$ with 6 additional refinement levels in a computational domain of size $L=384$, while the binaries used $\D x = 4$, with 8 additional refinement levels, and a computational domain of size $L = 1024$.

Unlike most simulations ever evolved in \texttt{GRChombo}, the Gamma driver $\eta$ parameter caused trouble for some of the evolved systems at large radii. This is in fact a simple and known problem related to $\eta$ imposing a time size constraint much like the CFL condition, for which the solution is creating a radial profile reducing $\eta$ for large distances \cite{Schnetter:2010cz}. This is related to the radial profile used for $\s$ as described in section \ref{subsubsection:eft:further_improvements}.

Finally, since the mass of the black hole is affected by $\mathcal{O}(\e)$, the smallness of the values used implied greater care in the perturbations of the ADM mass of the system. AMR boundaries and the evolution scheme contribute to small noise accumulation throughout a binary, which we found to grow above $1\%$ in the ADM mass during a binary, for some large values of $\e$ or particular fixing systems in use. By changing the $\k_i$ parameters of the CCZ4 formulation to $\k_i=(1, -0.8, 1)$, the mass drift improved to $< 3\cdot 10^{-3}M$ per $1000M$ of time evolution for a black hole of mass $0.5M$.

\section{\done{Results}}\label{sec:eft:results}

\subsection{\done{Toy model parameter stability}}\label{subsec:eft:results_toy_models}

As mentioned before, it is essential to trust that the fixing procedure is robust and its details do not affect the physics. Specifically, it is important to verify to what extent the results are invariant to the arbitrary parameters introduced, namely $\s$ and $\t$ of the several toy models discussed before.

With that in mind, we performed simulations of the modified wave equation with the advection fixing method, as shown in equation \eqref{eq:eft:fix2}, further reduced to a first order system. This was a 1D simulation\footnote{We thank Ramiro Cayuso for the development and maintenance of the code.}, with third order spatial derivatives, Runge-Kutta of fourth order for time stepping, Kreiss-Oliger dissipation with $N=2$ and $\s=2$ and CFL condition of $\frac{1}{4}$. The domain had size $L=100$, resolution $\D x= \frac{1}{16}$ and periodic boundary conditions. As initial data, we used a trivial moving wave Gaussian profile $\f=A\,e^{-\frac{1}{2}\br{\frac{x-x_0}{\w}}^2}$, with $A=0.001$, $x_0=50$, $\w=1$. This profile is not a solution of the equation for $\e\neq0$, developing a long oscillatory tail as the dynamics evolve. For these runs, we used $\e=10^{-3}$.
\begin{figure}[h]
\centering
\includegraphics[width=1\textwidth]{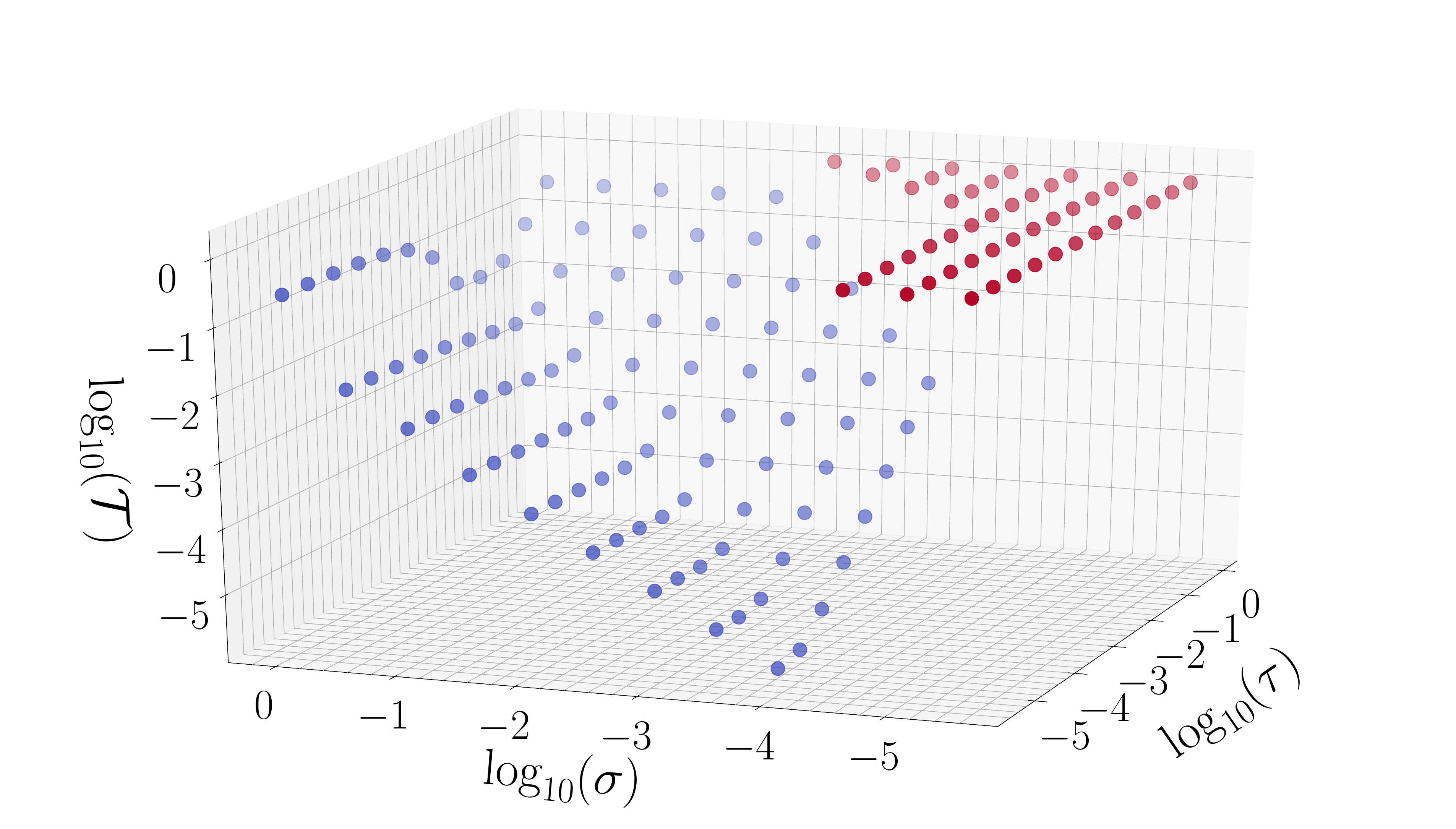}
\vspace{-6mm}
\caption{3D plot of the tracking accuracy (which compares the 1D integral over the domain of auxiliary variable $\Pi$ to the physical variable it tracks $\pd_x^2\f$) for several values of $\s$ ad $\t$, with a fixed $\e=10^{-3}$. In red, we display the simulations that crashed due to instabilities, and in blue the ones that were stable, with a value representing the result of computing the metric $\mathcal{T}(\Pi, \pd_x^2\f)$ in the interval $[0,L]$, using \eqref{eq:eft:tracking_metric}.}
\label{fig:eft:results_fix2_parameter_search}
\end{figure}

To measure accuracy of the tracking of the auxiliary variable $\Pi$ to the physical variable $\pd_x^2\f$, we define a tracking metric $\mathcal{T}$ between two quantities $Q_1$ and $Q_2$ as:
\begin{equation}\label{eq:eft:tracking_metric}
    \mathcal{T}(Q_1, Q_2) = \frac{\int |Q_1(r) - Q_2(r)| dr}{\int Q_1(r)\, dr}\,.
\end{equation}

We evolved the system for $\e=10^{-3}$ for many values of $\s$ and $\t$. We evolved for $7\cdot 10^4$ time steps, after which the long tail becomes comparable to the size of the domain and the evolution becomes unphysical. At the final step, we compute a tracking metric $\mathcal{T}(\Pi, \pd_x^2\f)$ in the interval $[0,L]$.

The result is displayed in figure \ref{fig:eft:results_fix2_parameter_search}. A few remarks:
\begin{itemize}
    \item One can observe an area of red dots, or evolutions that crashed, for very low $\s$. This is approximately at the level of $\s=10^{-4}$, suggesting some stability relationship between $\s$ and $\e$.
    \item As expected, the tracking improves linearly with decreasing $\s$, 
    \item As for $\t$, we can also observe a linear dependence for big values of $\t$, which interestingly flattens to a plateau dependent only on the value of $\s$. 
\end{itemize}

All in all, for all these simple cases, the profile at the final timestep is very similar, up to $0.1\%$ errors, for almost all values of $\s$ and $\t$. Nevertheless, for black hole simulations, the accuracy of tracking does matter and small errors can easily accumulate over time. Hence, we learnt that this system is stable and that, as predicted, we should aim for low $\s$ and $\t$ to get as much accuracy as possible. If they are small enough, we should also in the black hole case get into a regime where the final profile is independent of these parameters.

\subsection{\done{Single black hole}}\label{subsec:eft:single_bh}

\subsubsection{\done{Comparing fixing systems}}

In this section we show how several fixing equations for this system compare in terms of tracking accuracy, leading to the conclusion that the advection $\mat{C}$ system is the best candidate for simulations of single and binary black holes. We compare three versions: the Laplacian system \eqref{eq:eft:c_system_eom_laplacian}, the advection system \eqref{eq:eft:fix2} and the version with only time derivatives \eqref{eq:eft:time_derivatives_only}.

At $t=40M$, the gauge adjustment has stabilised to a fixed profile, and we compute a tracking metric that evaluates the difference between the evolved variable and the physical Kretschmann scalar for this profile. We perform this over a radial ray starting on the apparent horizon and finishing at some `large' radius\footnote{As the tracking gets more accurate with increasing distance from the black hole, the value of this large radius does not affect the result.} $R=20M$, in a direction perpendicular to the direction of motion\footnote{The results are very similar in any direction.}, using $\mathcal{T}(\hat{\mat{C}},\mat{C})$ in the interval $[r_{\text{AH}},R]$.

\begin{figure}[h]
\centering
\includegraphics[width=1\textwidth]{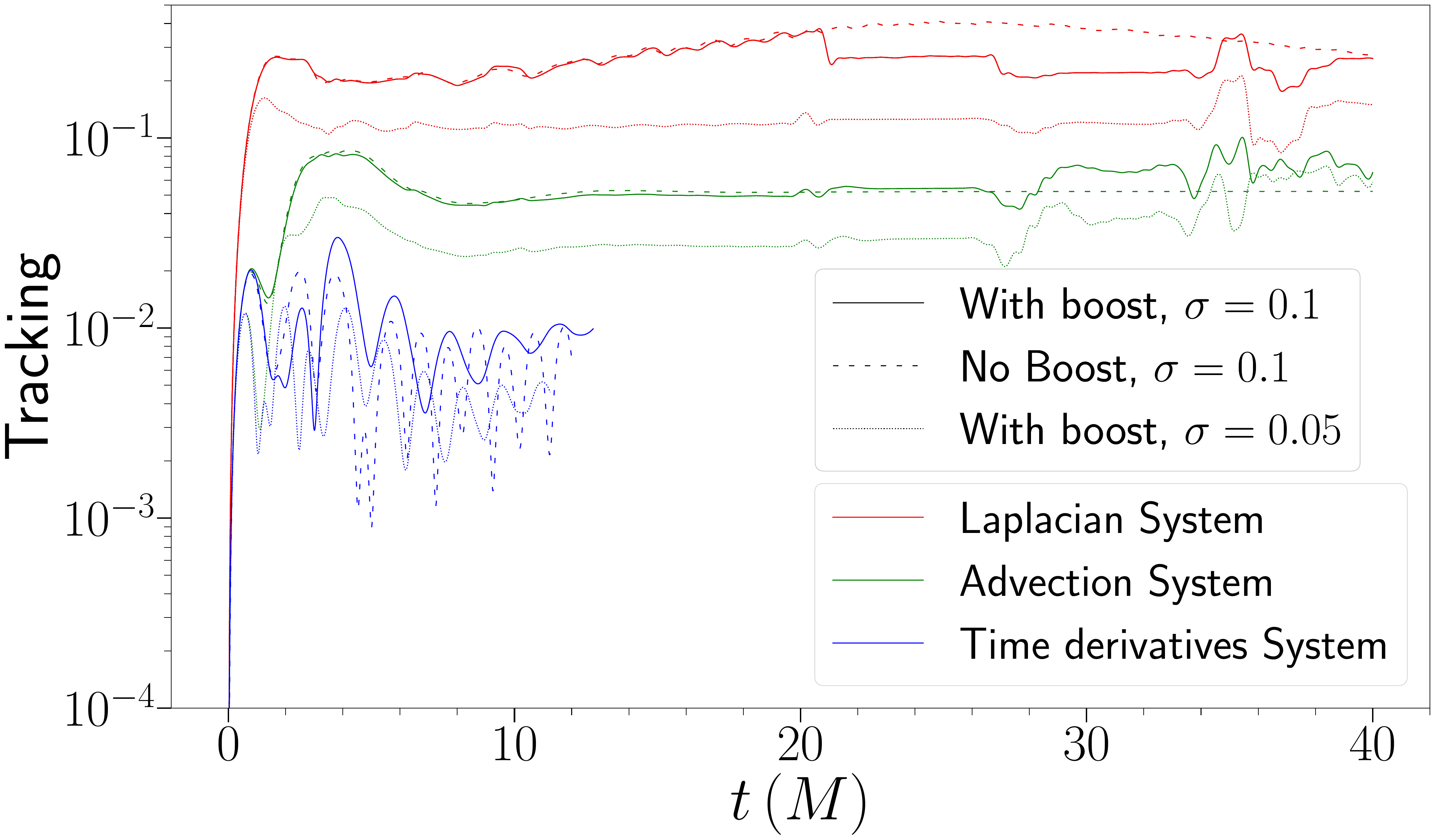}
\vspace{-8mm}
\caption{Evolution of the tracking metric $\mathcal{T}(\hat{\mat{C}},\mat{C})$ in the interval $[r_{\text{AH}},R]$ over time for the Laplacian system \eqref{eq:eft:c_system_eom_laplacian}, the advection system \eqref{eq:eft:fix2} and the version with only time derivatives \eqref{eq:eft:time_derivatives_only}. The solid, dashed or dotted line show several combinations of the presence of boost and values of $\s$. We use $\e=10^{-5}$ and $\t=0.005$.}
\label{fig:eft:results_CvsCphys}
\end{figure}

In figure \ref{fig:eft:results_CvsCphys} we display the tracking metric for the three systems mentioned in 3 types of simulation: with boost $P^x=0.08M$ for $\s=0.1$, with the same boost for $\s=0.05$ and with no boost with the original $\s=0.1$. A few conclusions one can draw from this figure:
\begin{itemize}
    \item the Laplacian system has deviations bigger than $10\%$, while the system with only time derivatives has sub-percent errors. While true, it is important to know that even though the Laplacian system has worse accuracy, it can withstand smaller values of $\s$ while remaining stable. However, this effect is sub-leading and does not alter the conclusion that this system is, in general, worse.
    \item The evolution with or without boost does not affect the result of the Laplacian or advection system.
    \item Halving $\s$ from $\s=0.1$ to $\s=0.05$ approximately halves the value of the tracking accuracy, showing a linear behaviour much like what was seen in section \ref{subsec:eft:results_toy_models}.
    \item The run in blue does not finish at $t=30M$ as at about $t\sim 12-14M$ it crashes. Notice this corresponds to when the system is evolving with non-zero $\e$, as the time dependent $\e$ reaches the value $\e=10^{-5}$ at $t_F=12M$ (see equation \eqref{eq:eft:epsilon_time_dependent}).
\end{itemize}

Finally, we attempted to change $\t$ from $0.005$ to $0.0025$, and this change does not affect the solution (much like the boost and no boost solutions in figure \ref{fig:eft:results_CvsCphys}). This hints that this parameter is already in a region of stability where the solution is independent of its value, an effect also observed in section \ref{subsec:eft:results_toy_models}.

Overall, we conclude the advection system is the best approach out of the systems developed, as the Laplacian system has errors too great that cannot be minimised and the time derivatives system is not stable, though most accurate. Note that the value of $\e$ affects what the lowest values for $\s$ and $\t$ can be while preserving stability.

\subsubsection{\done{Accuracy of advection system}}\label{subsubsec:eft:results:accuracy_of_advection}

\begin{figure}[h]
\centering
\includegraphics[width=.9\textwidth]{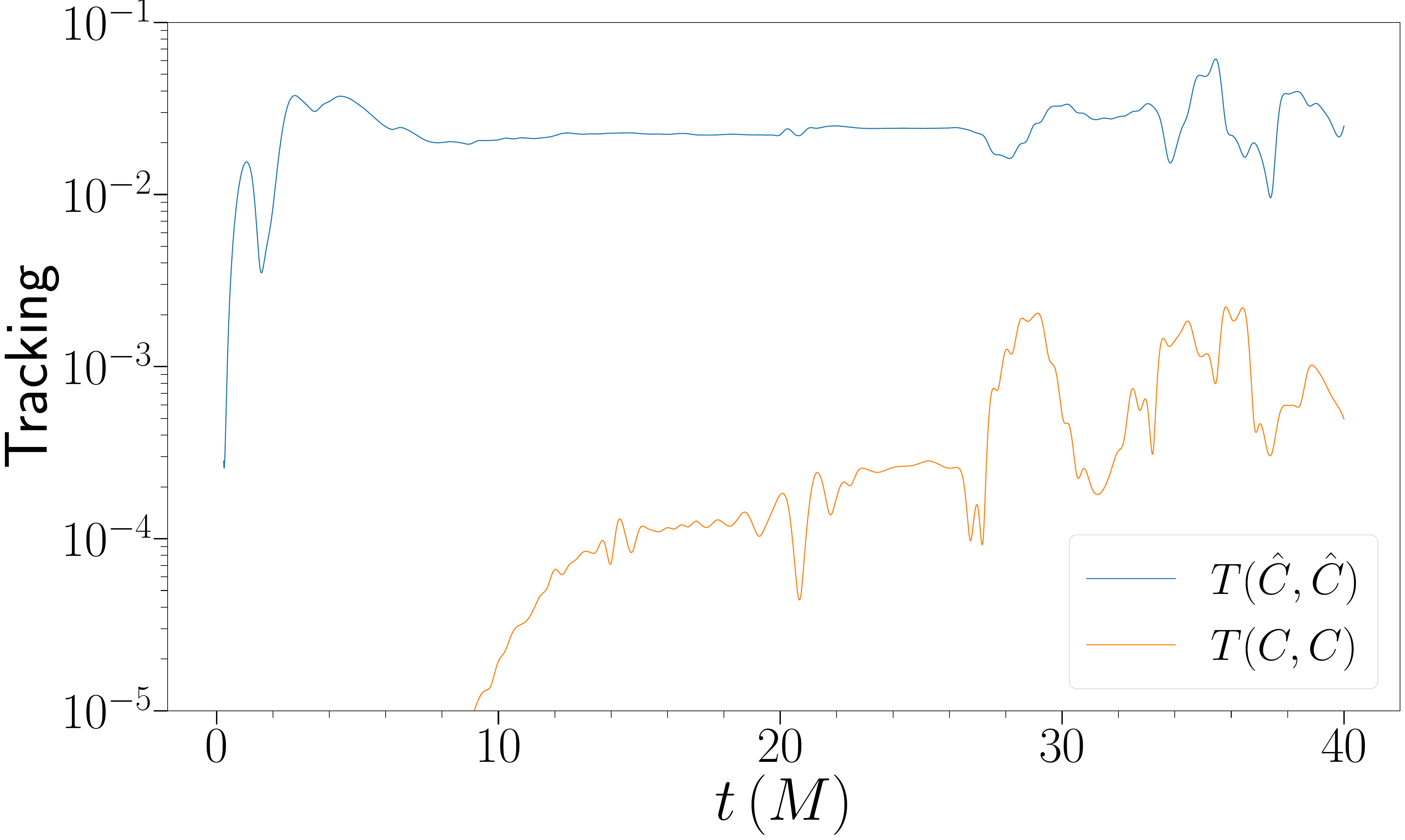}
\vspace{-4mm}
\caption{Evolution over time of the tracking metrics for the advection system \eqref{eq:eft:fix2}, comparing $\mat{C}$ between $\s=0.1$ and $\s=0.05$ (orange), and likewise for $\hat{\mat{C}}$ (blue), using $\mathcal{T}(\mat{C}_{\s=0.1},\mat{C}_{\s=0.05})$ and $\mathcal{T}(\hat{\mat{C}}_{\s=0.1},\hat{\mat{C}}_{\s=0.05})$ in the interval $[r_{\text{AH}},R]$. We use $\e=10^{-5}$ and $\t=0.005$.}
\label{fig:eft:results_CvsCphys_compare_advection_runs}
\end{figure}

One can now ask: given the accuracy of the tracking at the level of $2-5\%$ observed in figure \ref{fig:eft:results_CvsCphys} for the tracking in the advection system, what does this translate to in terms of accuracy of the simulation? To answer this, we compare the physical Kretschmann scalar and the evolved Kretschmann auxiliary variable between the two runs of figure \ref{fig:eft:results_CvsCphys} of a single boosted black hole with $\s=0.1$ and $\s=0.05$. The results are in figure \ref{fig:eft:results_CvsCphys_compare_advection_runs}. One can observe that, even though the auxiliary variable $\hat{\mat{C}}$ differs between runs on the level of $2\%$, the physical variables $\mat{C}$ differ by less than $0.1\%$. In spite of this, seeing clearly that this deviation in the physical variable $\mat{C}$ is growing with time, one can worry that over long ranges of time this effect will accumulate and built up to a large effect. Even though we saw that the solution was independent over changes of $\t$, there is still an undesired sensitivity to $\s$, which one must take into account when comparing results in the EFT to GR. We will analyse this issue in the binary black hole case.

\subsection{\done{Binary black holes}}\label{subsec:eft:results_bbh}

In this section we present the results of our numerical simulations for circular binary black holes in EFT compared to GR, with initial data as described in section \ref{subsec:eft:initial_data}.

\subsubsection{\done{Consistency of the fixing method}}

\begin{figure}[h]
\centering
\includegraphics[width=0.7\textwidth]{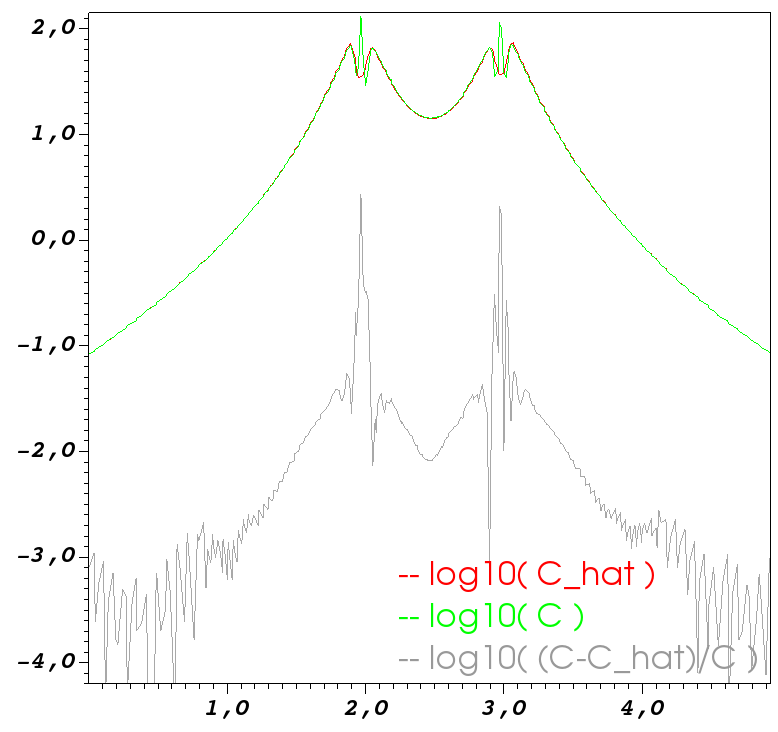}
\caption{Comparison of the physical Kretschmann scalar (in green) with the evolved variable $\hat{\mat{C}}$ tracking it (in red) for a 1D profile of a binary black hole with $\e=10^{-5}$, $\t=0.005$, $\s=0.1$, around the time of merger, $t=2430M$. In grey, we display the relative difference in log scale, $\log_{10}\br{\frac{\mat{C}-\hat{\mat{C}}}{\mat{C}}}$. One can see that the auxiliary variable tracks the physical variable extremely well. In particular, outside each black hole, located at $x=2M$ and $x=3M$, where $x$ is distance measure in the horizontal axis, the accuracy is in the sub-percent level.}
\label{fig:eft:BBH_CvsCphys}
\end{figure}

We start by a quick check that the fixing procedure is working as expected, by inspecting at late times how the auxiliary variable tracks the physical variable. More than just at late times, we perform this comparison in the moment of highest dynamical activity, when the two black holes are merging, about $1M$ apart. The result of a 1D profile passing through each puncture is shown in figure \ref{fig:eft:BBH_CvsCphys}. By visually comparing the green and red line in this plot, one can see great agreement between the auxiliary variable, $\hat{\mat{C}}$, and the physical variable $\mat{C}$. More quantitatively, the difference is below $0.1\%$ for distances bigger than $1M$ from the centre of each black hole, while about $1\%$ difference close to the apparent horizon. Recall these black holes have masses of about $0.5M$ and that in puncture gauge the apparent horizon approximately sits at $r\approx M$ (see section \ref{appendix:grchombo:ah_location} for details).

\subsubsection{\done{Weak coupling conditions during the evolution}}

\begin{figure}[h]
\centering
\includegraphics[width=0.7\textwidth]{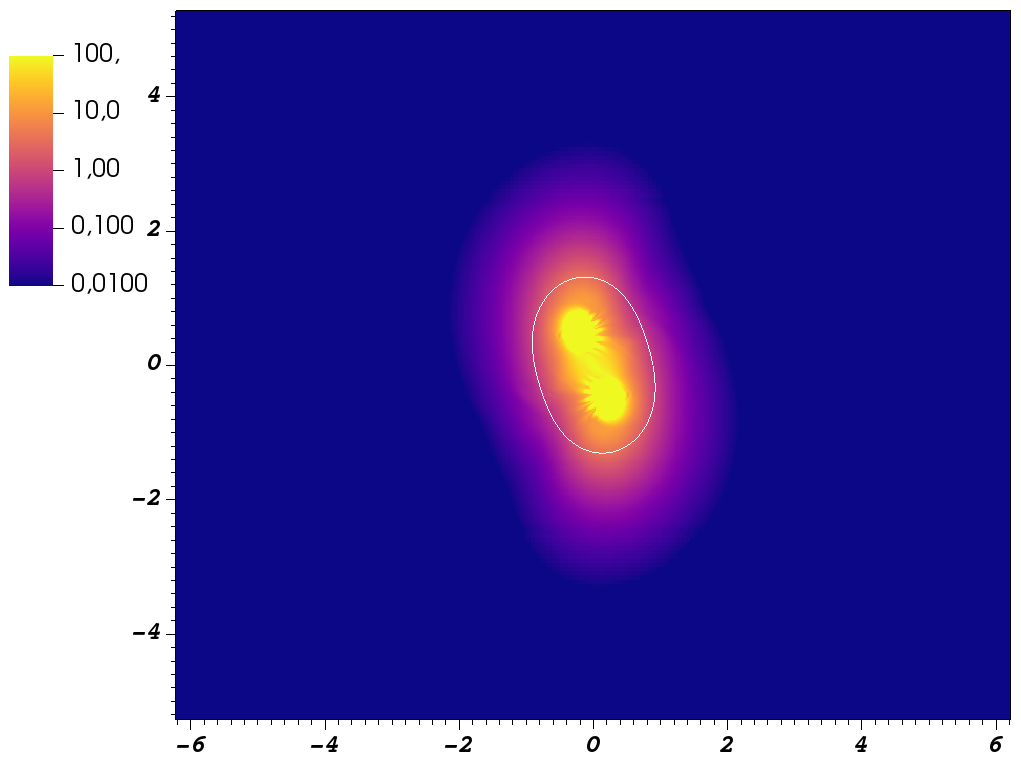}
\caption{Weak coupling condition normalised by $\e$, $\frac{WCC}{\e/M^6} = \frac{1}{\e}\sqrt{\frac{T_{\m\n}T^{\m\n}}{C}}$, for a binary black hole with $\e=10^{-5}$, $\t=0.005$, $\s=0.1$, around the time of merger, $t=2430M$. The white line represents the contour $\chi=0.25$, approximating the location of the common apparent horizon forming. One can see that $WCC/\e$ is in the range $1-10$ in the apparent horizon area delimited in white, and is strictly less than this outside of this region.}
\label{fig:eft:WCC}
\end{figure}

We also know that the results of the theory can only be considered in the EFT truncated regime. To confirm evidence of this, we look at the WCC, as described in section \ref{subsec:eft:WCC}, also at late times at the time of merger, when the black holes at $1M$ apart and the dynamics are the strongest. The result, shown in figure \ref{fig:eft:WCC}, shows that on and outside the apparent horizon, the $WCC$ is smaller or equal to $\mathcal{O}(\e)$, as predicted in section \ref{subsec:eft:WCC}. This shows validity of the EFT for all times of the evolution of the binary. Even though locally its effects are very small, global observables may result in larger deviations from the build up of small effects over time.

\subsubsection{\done{Waveform strain}}

\begin{figure}[h]
\centering
\includegraphics[width=1\textwidth]{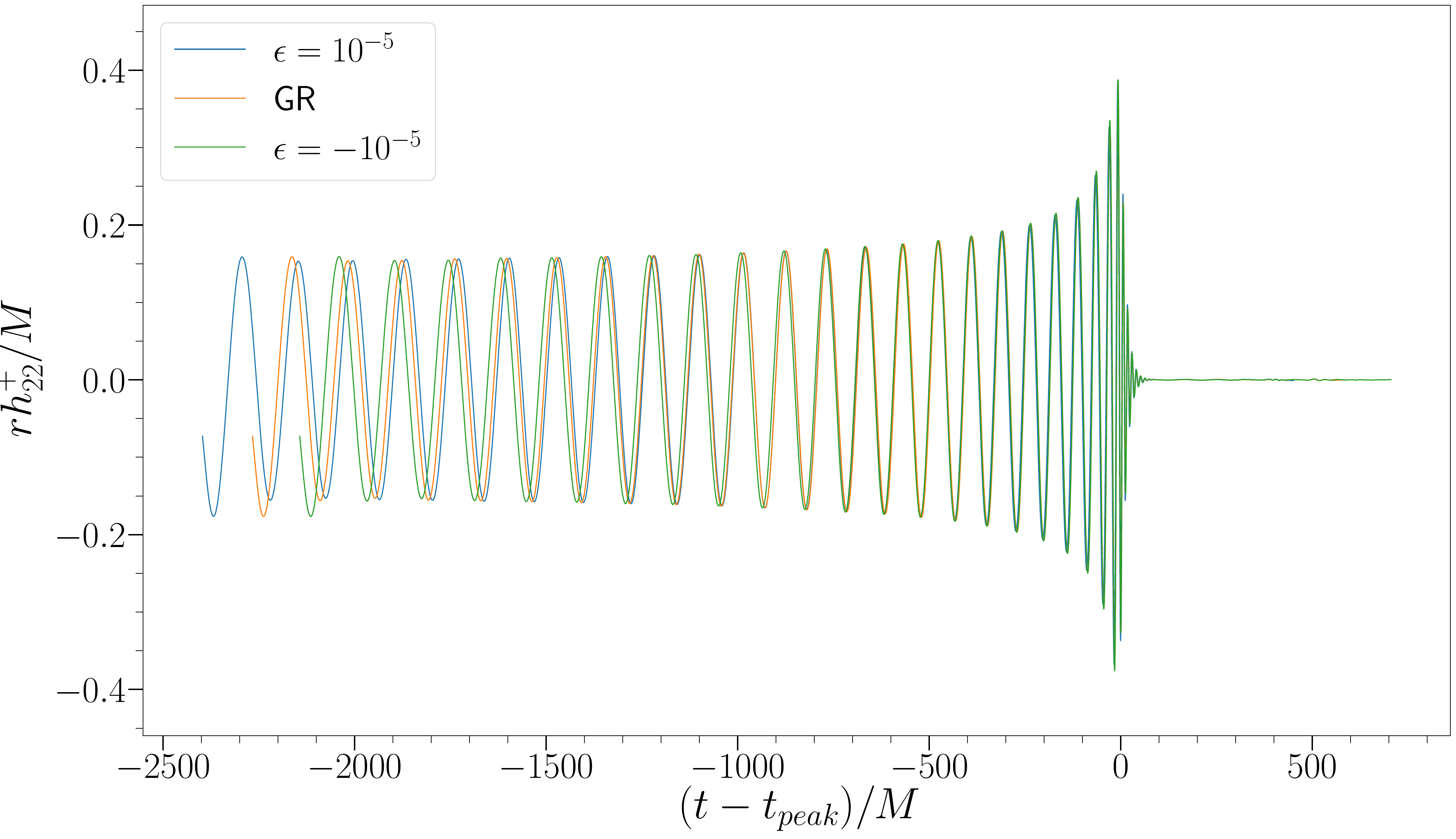}
\vspace{-8mm}
\caption{Comparison of gravitational wave between EFT with $\e=\pm 10^{-5}$ and GR, aligned by the peak of amplitude of the strain. Displaying the $(\ell,m)=(2,2)$ mode of the plus polarisation of the strain, $h^+_{22}$, extrapolated to null infinity. The advection system was used with $\s=0.05$ and $\t=0.005$.}
\label{fig:eft:results_GW}
\end{figure}

\begin{table}[h]
\centering
\begin{tabular}{cccccc}
\hline\hline
\textrm{Coupling}&
\textrm{Final Mass $M_F/M$}&
\textrm{Spin Parameter $a/M_F$}\\
\hline
$\e=-10^{-5}$ & $0.947\pm0.001$ & $0.688\pm0.001$ \\
$\text{GR}$ & $0.943\pm0.001$ & $0.688\pm0.001$ \\
$\e=10^{-5}$  & $0.939\pm0.001$ & $0.688\pm0.001$ \\
\hline\hline
\end{tabular}
\caption{\label{tab:eft:runs}%
Parameters of the final state Kerr black hole for each value of $\e$ and $\s$. The mass and spin are estimated from the final apparent horizon since the ADM quantities are typically significantly noisier. The errors are estimated from the mass drift in the apparent horizon over $500M$ of time after the merger.}
\end{table}

\begin{figure}[h]
\centering
\includegraphics[width=1\textwidth]{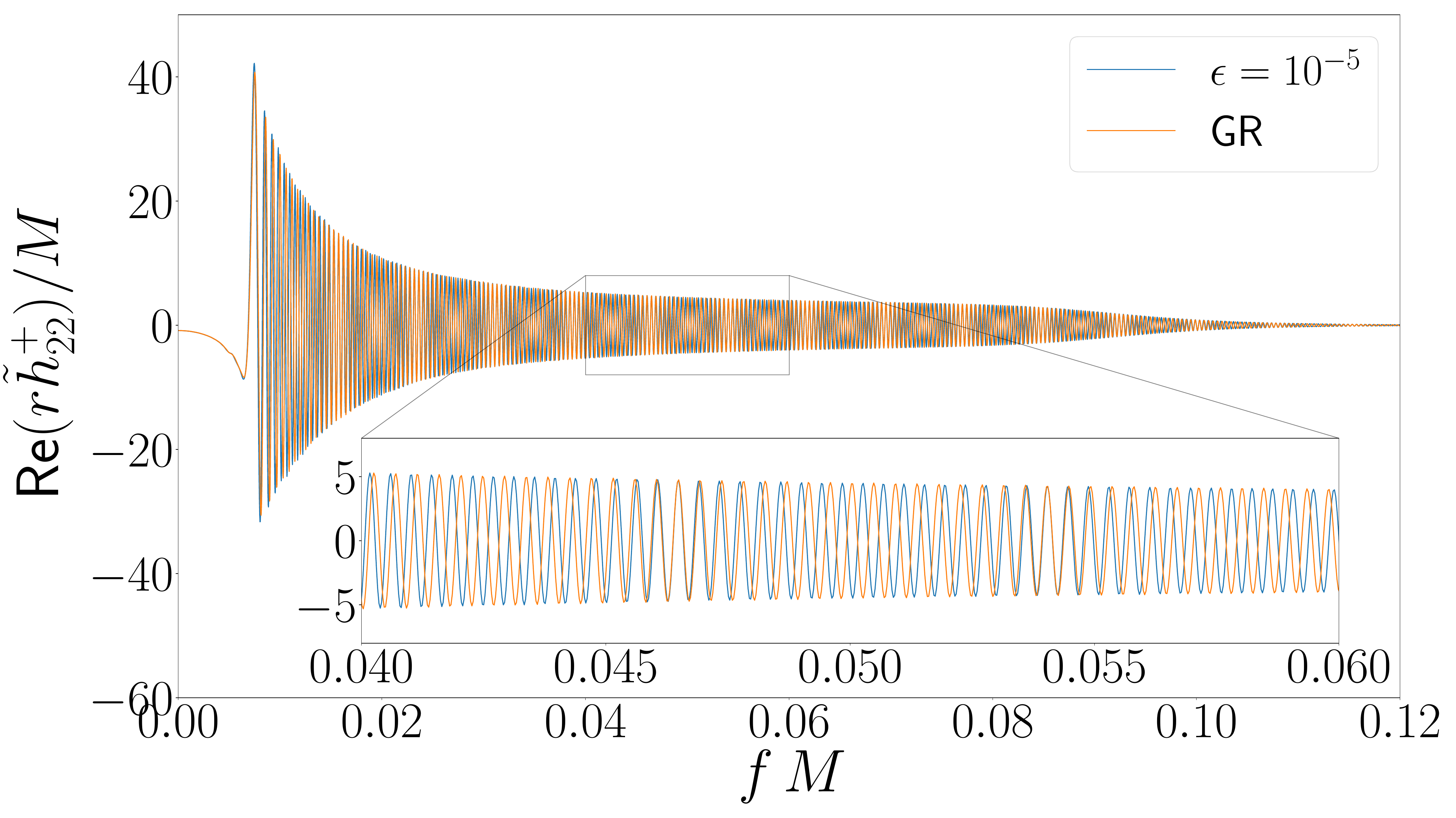}
\vspace{-8mm}
\caption{Real part of $\ti{h}_{22}^+$, the discrete Fourier transform of $h_{22}^+$ for positive frequencies. The case of $\e=10^{-5}$ has faster oscillations in phase, which cannot be mitigated by constant time or phase shifts in the waveform.}
\label{fig:eft:results_FFT}
\end{figure}

We now compare the waveform strain for the circular binaries between GR and the EFT, for different values of $\e$. We extract the mode with highest amplitude, $(\ell,m)=(2,2)$, of the plus polarisation of the strain, $h^+$, extrapolated to null infinity using 6 radii between $50-150M$, as described in \ref{subsec:eft:methodology:gravitational_strain}. Figure \ref{fig:eft:results_GW} shows the gravitational strain for $\e=10^{-5}$, $\e=-10^{-5}$ and GR, all aligned by the time of merger, estimated as the peak of the complex amplitude of the strain. Junk radiation in the first $100M$ of evolution were cropped from the waveform. The advection system was used with $\s=0.05$ and $\t=0.005$. Taking the final state to be approximately a Kerr black hole (valid up to $\mathcal{O}(\e)$), the estimated black hole parameters in table \ref{tab:eft:runs}.

Figure \ref{fig:eft:results_GW} shows the waveforms match to great agreement in the merger, but exhibit, in a similar manner to what was found in the Horndeski case \ref{subsec:paper2:strain}, a frequency change resulting in phase shift noticeable in the earlier stages of the inspiral. This can be confirmed by inspecting the frequency domain of the waveform in figure \ref{fig:eft:results_FFT}, where indeed one can see matching amplitudes but a discrepancy in the complex phase for intermediate frequencies. This cannot be mitigated by constant time or phase shifts in the waveform.

Additionally, it is interesting to notice in figure \ref{fig:eft:results_GW} and table \ref{tab:eft:runs} a linear regime in $\e$, with positive and negative $\e$ displaying symmetric behaviour around GR. This is another positive indication we are in the low energy regime of the EFT.

The analysis of quasi-normal modes and inspection of the ringdown is commented in the next section.

\subsubsection{\done{Mismatch}}

\begin{figure}[H]
\centering
\includegraphics[width=1\textwidth]{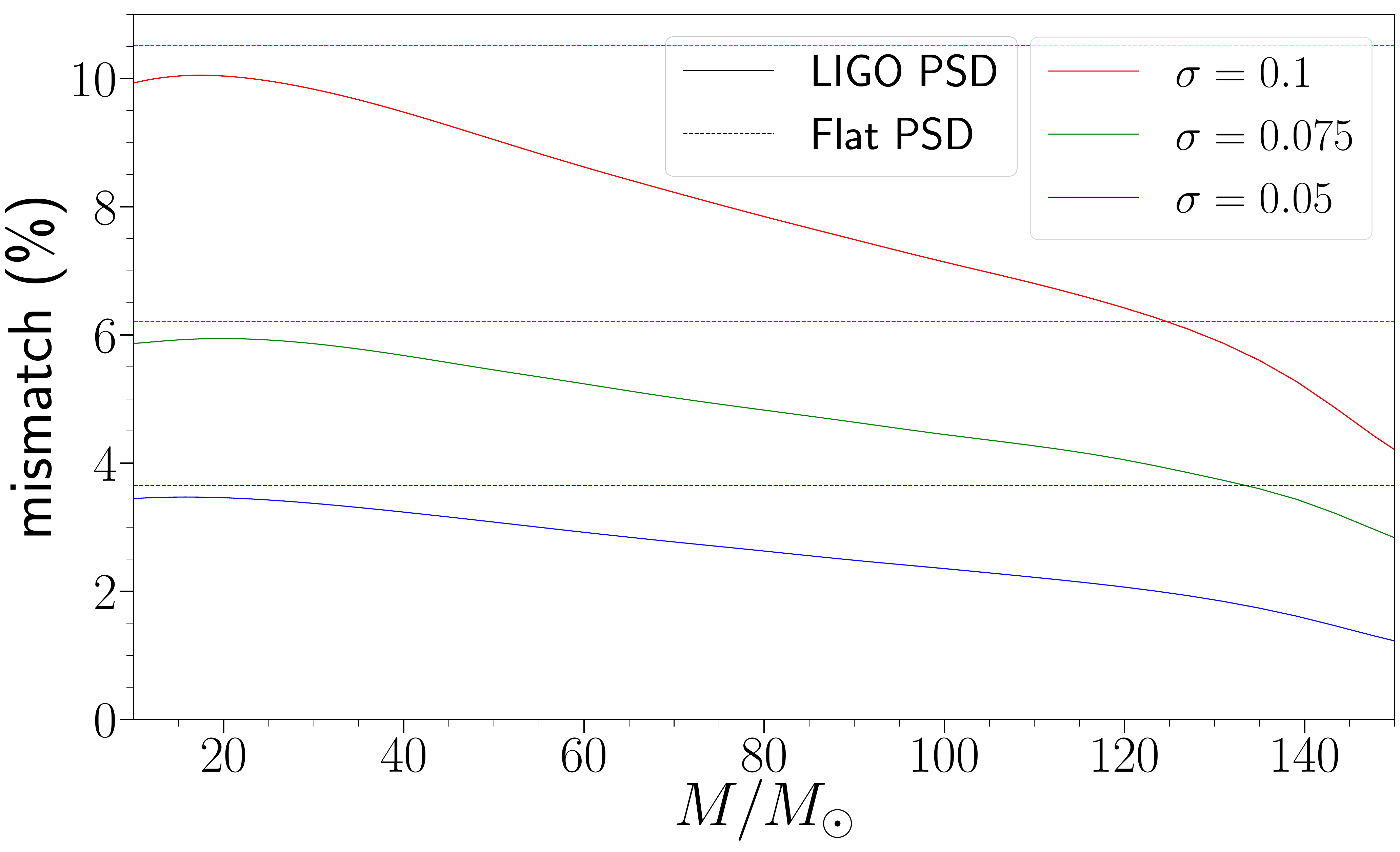}
\vspace{-8mm}
\caption{Mismatch for $h^+_{22}$ between GR and EFT for $\e=10^{-5}$, as a function of the final black hole mass (in units of solar masses, $M_\odot$). As power spectral densities, we used the updated Advanced LIGO sensitivity design curve (\textit{aLIGODesign.txt} in \cite{aLIGOupdatedantsens}, which imposes $f_{\text{min}}=5$ Hz) and a flat noise mismatches ($S_n=1$).}
\label{fig:eft:results_mismatches}
\end{figure}

In figure \ref{fig:eft:results_mismatches}, we quantify the mismatch between the EFT, for $\e=10^{-5}$, and GR for a detector setup receiving the plus polarisation of the strain. Here, we finally quantify the sensitivity to $\s$ discussed in the results section \ref{subsubsec:eft:results:accuracy_of_advection}, by using not only $\s=0.05$, but also two bigger values, $\s=0.075$ and $\s=0.1$. 
Using different values of $\s$ is essential to identifying any physical accumulation of EFT deviations over time and not an artefact of systematic errors related to the poor accuracy of the tracking equation. In terms of gravitational wave strain, related to figure \ref{fig:eft:results_GW}, these bigger values of $\s$ show simply a bigger deviation at early times (not shown to avoid confusion in the plot without much added value). We keep $\t=0.005$ constant, after the conclusion of section \ref{subsec:eft:single_bh} that this choice does not affect the system.

As in chapter \ref{chapter:paper2}, section \ref{subsec:paper2:mismatch}, we limit ourselves to  the $(\ell,m)=(2,2)$ mode, $h^+_{22}$, as this is the dominant mode by an order of magnitude when compared to higher modes. We use the updated Advanced LIGO sensitivity design curve (\textit{aLIGODesign.txt} in \cite{aLIGOupdatedantsens}, which imposes $f_{\text{min}}=5$ Hz) and flat noise ($S_n=1$) following the procedure described in section \ref{subsec:paper2:mismatch_theory}. We compute the mismatch for black hole masses in the typical range of stellar mass black holes binaries observed so far, $M\in[10,150]M_\odot$ \cite{LIGOScientific:2021djp}, where $M_\odot$ is one solar mass.

\begin{figure}[h]
\centering
\includegraphics[width=1\textwidth]{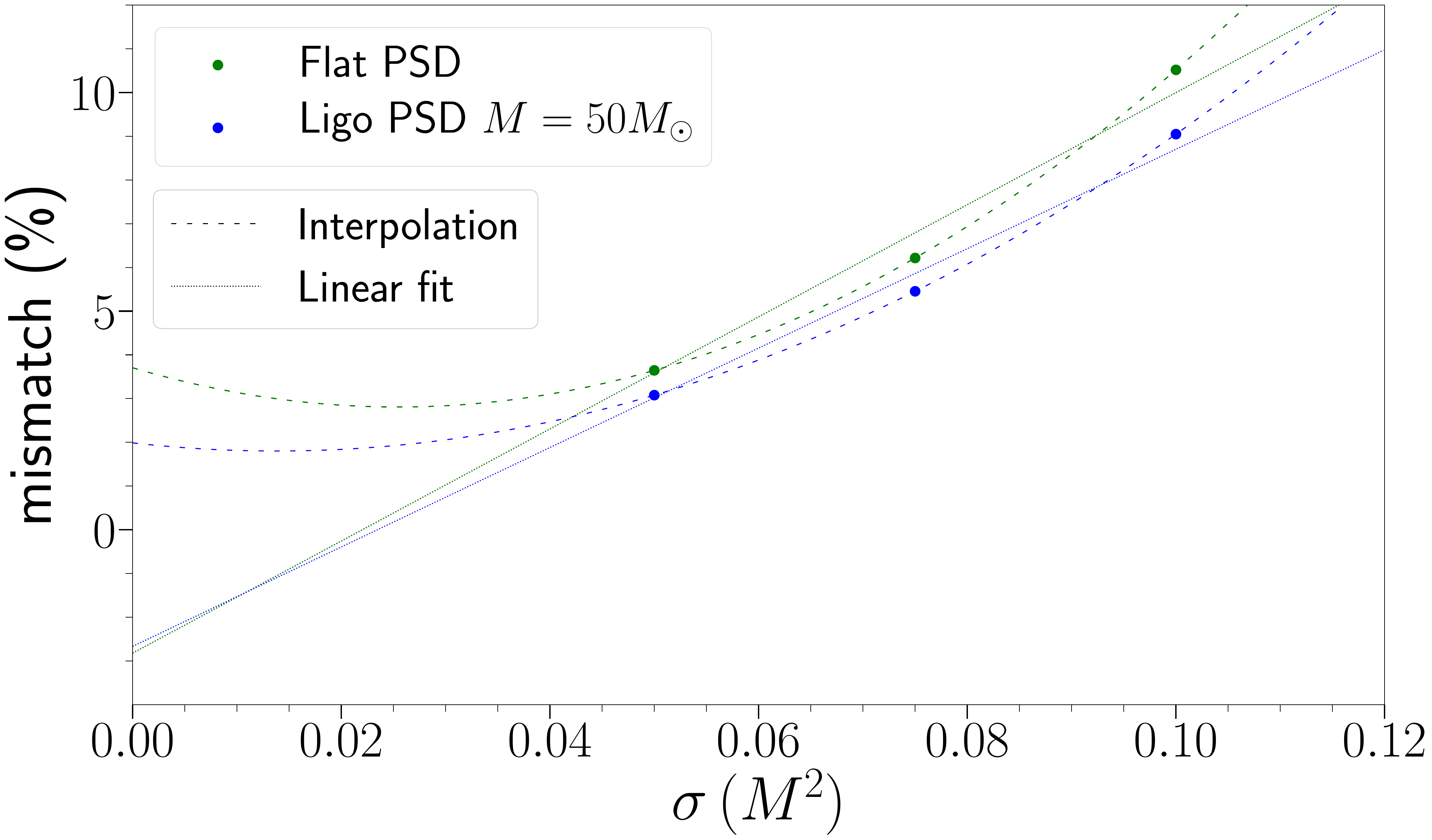}
\vspace{-8mm}
\caption{Mismatch for $h^+_{22}$ between GR and EFT for $\e=10^{-5}$, as a function of the $\s$ parameter of the advection system. We use as power spectral density a flat noise and the LIGO sensitivity curve previously mentioned. As an attempt to extrapolate to $\s=0$, we mix the result of two methods: a quadratic interpolation over 3 points, which we then extrapolate to $\s=0$, and a linear fit, also extrapolated to $\s=0$.}
\label{fig:eft:results_mismatch_extrapolate}
\end{figure}

Figure \ref{fig:eft:results_mismatches} shows that the mismatch varies significantly between different $\s$, revealing that the small effects observed in section \ref{subsubsec:eft:results:accuracy_of_advection} accumulate and dominate over the effects of $\e\neq0$. As an attempt to absorb this systematic error, we extrapolate the mismatch to $\s=0$, for which the advection system would theoretically have perfect tracking. This is shown in figure \ref{fig:eft:results_mismatch_extrapolate}. For extrapolation, given the limited amount of data, we quantify the error by using two methods: a quadratic interpolation and a linear fit, both extrapolated to $\s=0$. This is done for flat noise PSD and LIGO PSD fixed with a final black hole mass of $50M_\odot$. As we can observe, there is a huge uncertainty in the result. This can be aggregated into a final value of the mismatch between EFT for $\e=10^{-5}$ and GR, without errors associated to the fixing procedure, of $\text{mismatch} = -0.39\pm 4.10\%$.

Overall, we conclude that the effects observed are due the sensitivity of the fixing approach to the unphysical parameters introduced, and the effects of $\e\neq 0$ are negligible at this scale, as perhaps expected given the small value of $\e$. We note that it is therefore also unfruitful to attempt to evaluate quasi-normal modes or compare properties of the ringdown, as likely any effects are also artefacts of the fixing procedure. We have learned that in any work involving a fixing procedure, appropriate damping of high frequencies to obtain a stable system is not enough. It is extremely important to check the accuracy of the fixing equations, to check if indeed they reproduce the low energy behaviour of the theory, or are otherwise merely introducing spurious unphysical deviations. To obtain a significant value, the accuracy $\mathcal{T}(\mat{C},\mat{C})=0.1\%$ found in section \ref{subsubsec:eft:results:accuracy_of_advection} has to likely be reduced to $\mathcal{T}\ll\e/M^6$. This can be done by improving the tracking, but also by increasing $\e$ significantly and performing an extrapolation over the unphysical parameters to some asymptotically stable value. Supposing such a procedure works, it appears the price to pay with this new method of successfully evolving theories with higher derivatives is the requirement of evolving multiple binaries in order to guarantee the successful invariance against the choice of fixing procedure.

As a final remark, the results for null convergence condition (NCC) violations identified by Cayuso and Lehner \cite{Cayuso:2020lca}, in which negative values are found in regions outside of the black holes, were verified at all stages of the binaries evolved.

\section{\done{Conclusions}}\label{sec:eft:conclusions}

The goal of this chapter was to simulate binary black holes fully non-linearly in a theory with higher derivative terms in extension to GR, with the hope that small local effects from such deviation to GR can build up over time and leave an imprint in observables as gravitational waves.

After analysing several toy models and their stability properties, with the intuition built from this, we used the fixing method described in section \ref{sec:gr:fixing_hd_theories} to propose multiple methods in which this theory could be evolved in three spatial dimensions with no symmetry assumptions and the CCZ4 formulation. Picking the most straightforward approach, referred to as the $\mat{C}$ system, we analysed its accuracy in reproducing the low energy physics of the original truncated theory.

On of the main goals of the fixing method is to remove UV cascades from the truncated theory without spoiling its low energy physics. Knowing the fixing procedure is not unique, one way in which this manifests, besides simulations not crashing due to high frequency noise, is by demonstrating independence of the result to changes in the unphysical parameters introduced. We analysed several forms of the $\mat{C}$ system and, after picking the most accurate one, referred to as the advection $\mat{C}$ system. We showed invariance over changes of the $\t$ parameter, yet linear sensitivity for its $\s$ parameter. The direct implication of this is that this parameter must be further reduced to enter the regime of insensitivity in which the result will not change regardless of its value. This proved unfeasible without leading to simulations crashing, but we demonstrated accuracy of the solutions on the order of $\mathcal{T} = 0.1\%$ for the smallest possible values of $\s$ this system could use, in the case of single boosted black holes.

When performing binary black hole evolutions, we verified the system remained in the weak regime of EFT even during whole evolution by inspecting the final stages of merger phase. Nonetheless, the small effects originating from inaccuracy (between the auxiliary variable and the physical variable it tracks) dominated over any small effects originating from the low energy behaviour of the theory. In spite of the great success and future possibilities of this method, this brings up the extreme importance of verifying consistency of the methods used. Namely, in our role as theoretical physicists to guide the development of waveform templates of modified theories of gravity it is important to understand if we are evolving the physical gravity theory originally proposed or if the auxiliary parameters introduced make us fool ourselves with unphysical results. After careful analysis, we obtained a mismatch between the EFT, using $\e=10^{-5}$, and GR of $\text{mismatch} = -0.39 \pm 4.10\%$, showing the original apparent deviations from GR observed turned out to be consistent with no deviations. In order to obtain significant results, it is likely the accuracy must be $\mathcal{T} \ll \e/M^6$, but it is likely that the use of this method will require evolving multiple binaries to ensure the results obtained are physical.

The work presented results from on-going work \cite{Cayuso:2023aht}, and there are clear avenues for improvement. The first of which is convergence tests, which, in spite of the trust in \texttt{GRChombo}, are essential to confirm accuracy of the results. These will be presented in \cite{Cayuso:2023aht}. More interestingly, it is relevant to find better ways to improve the $\mat{C}$ system, allowing for more accurate and stable tracking, or allowing for the use of smaller values of $\s$ (and hence more accurate results). Alternatively, using different systems can be a good avenue for improvement as well. Namely, the EB system, proposed in section \ref{subsec:eft:eb_system}, is a very promising direction. It has equivalent simplicity to the $\mat{C}$ system, although requiring the addition of 10 auxiliary variables instead of 1 (the $E_{ij}$ and $B_{ij}$ spatial tensors have 5 independent components each). Moreover, the $\mat{C}$ system is specific to the eight-derivative theory studied here and may not apply to other six or eight derivative corrections. However, the EB system, if successful, may be able to be used in all such theories that need a correction for Riemann-like terms and reduce them to lower order equations (second-order in most cases). On the other hand, if one considers equations that use higher than second spatial derivatives, by incorporating third or fourth spatial stencils in the code, new systems can be devised. In the spirit of EFT and the claim the fixing procedure is not unique, it would be remarkable if one could find multiple accurate formulations, such as some form of the $\mat{C}$ system and some form of the EB system, and show they both lead to the same low energy description.

Ultimately, one is looking for (physical) deviations of GR originated from well motivated extensions of it. This implies analysing quasi-normal modes, tidal effects and the presence of higher modes in the ringdown \cite{Kokkotas:1999bd,Cano:2021myl}. If a more accurate fixing procedure is developed, this may only be possible if one can also use bigger values of $\e$, e.g. $10^{-3}$, while still in the valid regime of EFT according to section \ref{subsec:eft:constraints}. Besides this, analysing more accurately the differences in the present theory for positive and negative $\e$ is also of interest.

Finally, while our studies were restricted to a particular theory and particular fixing method, the generality of the technique can hopefully inspire further improvements and be adopted for better results in other scenarios of this and other theories.

\chapter{\done{\texttt{AHFinder} - apparent horizon finder}}\label{chapter:ahfinder}

For the purpose of the research presented in this thesis, we developed the \texttt{AHFinder}, a public tool now extensively used by the \texttt{GRChombo} collaboration, capable of dynamically tracking apparent horizons in many situations. In this chapter, we briefly describe important mathematical and numerical details of finding apparent horizons. Section \ref{sec:ahfinder:horizons} describes basic concepts about black hole horizons. In section \ref{sec:ahfinder:finding_ahs}, we describe the tools required to find an apparent horizon. Finally, in section \ref{appendix:grchombo:ah_location}, we analyse an approximation of the location of the apparent horizon typically useful for AMR simulations. Details on numerical aspects of \texttt{GRChombo} can be found in section \ref{sec:grchombo_scheme}.

\section{\done{Event horizons and apparent horizons}}\label{sec:ahfinder:horizons}

One of the most important predictions of general relativity is the existence of solutions of the Einstein equations describing spacetimes that contain one or more black holes. Classically, a black hole is a region of spacetime where the gravitational pull is so strong that nothing can escape, not even light. Geometrically, a black hole is a closed region causally disconnected from the exterior spacetime, meaning that outside observers cannot receive any signals generated in its interior (even though anything can enter). Remarkably, black holes can exist even in vacuum, without the presence of any matter. The boundary of the black hole is called an \textit{event horizon}. This is a $D-1$ dimensional surface in the $D$ dimensional spacetime, and the intuitive notion of a black hole, the event horizon at coordinate time $t$, is only its cross section, a spacelike slice of the actual event horizon.

When evolving black hole spacetimes it is of course of extreme utility to know where the black holes are (e.g. to know how their mass evolves, know the formation of a merger in binaries, ensure enough resolution is used, etc.). But, as we mentioned above, finding the event horizon requires knowledge of the entire spacetime, which practically speaking means tracking geodesics through the full evolution of spacetime. In numerical relativity, this is impractical to say the least, and ideally we would like a local measure at a given time slice that could give us the location of the cross section of the event horizon at a particular time. This brings us to the concept of an \textit{apparent horizon}.

An apparent horizon is defined as the outermost marginally trapped surface. Before describing this mathematically, a trapped surface intuitively is a surface where a spherical flash of light rays emitted outwards remains constant in area. Such an experiment would not work in flat spacetime (the flash of light surely increases in area) nor close to a singularity (where light is trapped and reduces in area while falling into the singularity). This is why this notion makes sense as a measure of some form of horizon. More precisely, a trapped null surface is a closed co-dimension 1 hypersurface on a spatial slice of spacetime, such that the outgoing expansion of null geodesics (a quantity we describe below) vanishes on all its points; and the apparent horizon is the outermost such surface. Being a local measure on a spatial slice, in dynamic spacetimes, it may not coincide with the event horizon, as for example an expanding flash of light at a given point in time might be invariably trapped at a later point in time. Given the apparent horizon depends on the slice it is evaluated on, it is a gauge dependent quantity, while the event horizon is a gauge invariant. It can be shown that apparent horizons always lie inside the event horizon in any gauge \cite[p.~221]{alcubierre}, which means their interior is surely causally disconnected from its exterior and that the area one can measure from it is a lower bound of the actual black hole area. Note that the converse is not true: the absence of an apparent horizon does not imply the absence of an event horizon. Both coincide for static spacetimes.

\begin{figure}[h]
\begin{center}
\includegraphics[width=0.6\linewidth]{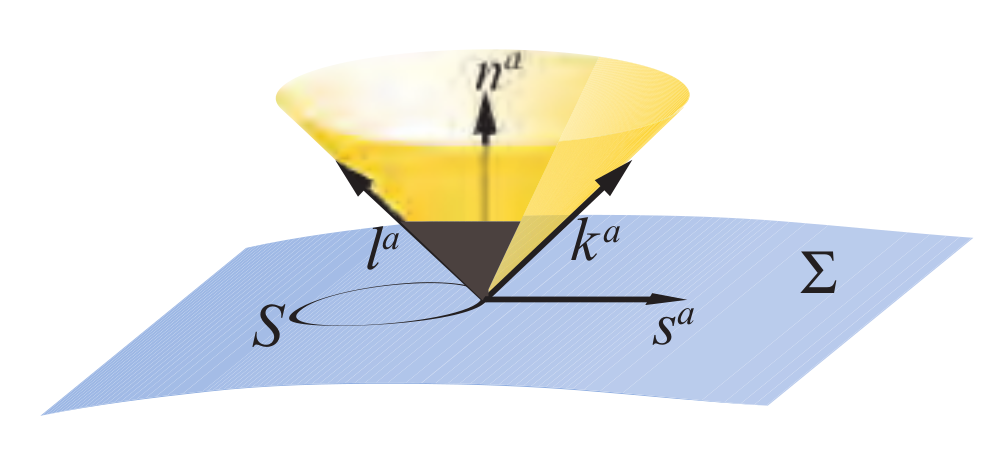}
\caption{A 2 dimensional surface $S$ embedded in $\S$, with a outward pointing normal to $S$, $s^\m$, a normal to $\S$, $n^\m$, and outgoing and ingoing null vectors $k^\m$ and $l^\m$. Figure taken from Baumgarte and Shapiro \cite[p.~236]{shapiro}.}
\label{fig:ah_expansion_diagram}
\end{center}
\end{figure}

To describe this analytically, consider a $d-1$ surface $S$ immersed in the $d$ dimensional spatial hypersurface $\S$. Let $s^\m$ be the spacelike unit outward-pointing normal vector to $S$, and $n^\m$ the timelike unit future pointing normal vector to $\S$, as depicted in figure \ref{fig:ah_expansion_diagram}. The outgoing null vector on $S$ is $k^\m = n^\m + s^\m$. The induced metric on $S$ is $h_{\m\n} = \g_{\m\n} - s_\m s_\n = g_{\m\n} + n_\m n_\n - s_\m s_\n$. 
The expansion of the null lines, $\Q$, is essentially the change in the area elements of $S$ along $k^\m$ \cite[p.~238]{shapiro} \cite[p.~221]{alcubierre}:
\begin{equation}
    \Q = h^{\m\n}\grad_\m k_\n = \br{\g^{ij} - s^i s^j}\br{D_i s_j - K_{ij}} = D_k s^k - K + K_{ij}s^i s^j\,.
\end{equation}
We now define trapped regions as regions where $\Q<0$ everywhere, and the boundary of the outermost such region, where $\Q=0$, as the apparent horizon. It is important to note, as it is a common matter of confusion, that the expansion $\Q$ is a property of a given surface, not a field that can be attributed to each point in spacetime.

\subsection{\done{Apparent horizons in binary black holes}}

\begin{figure}[h]
\begin{center}
\includegraphics[width=0.6\linewidth]{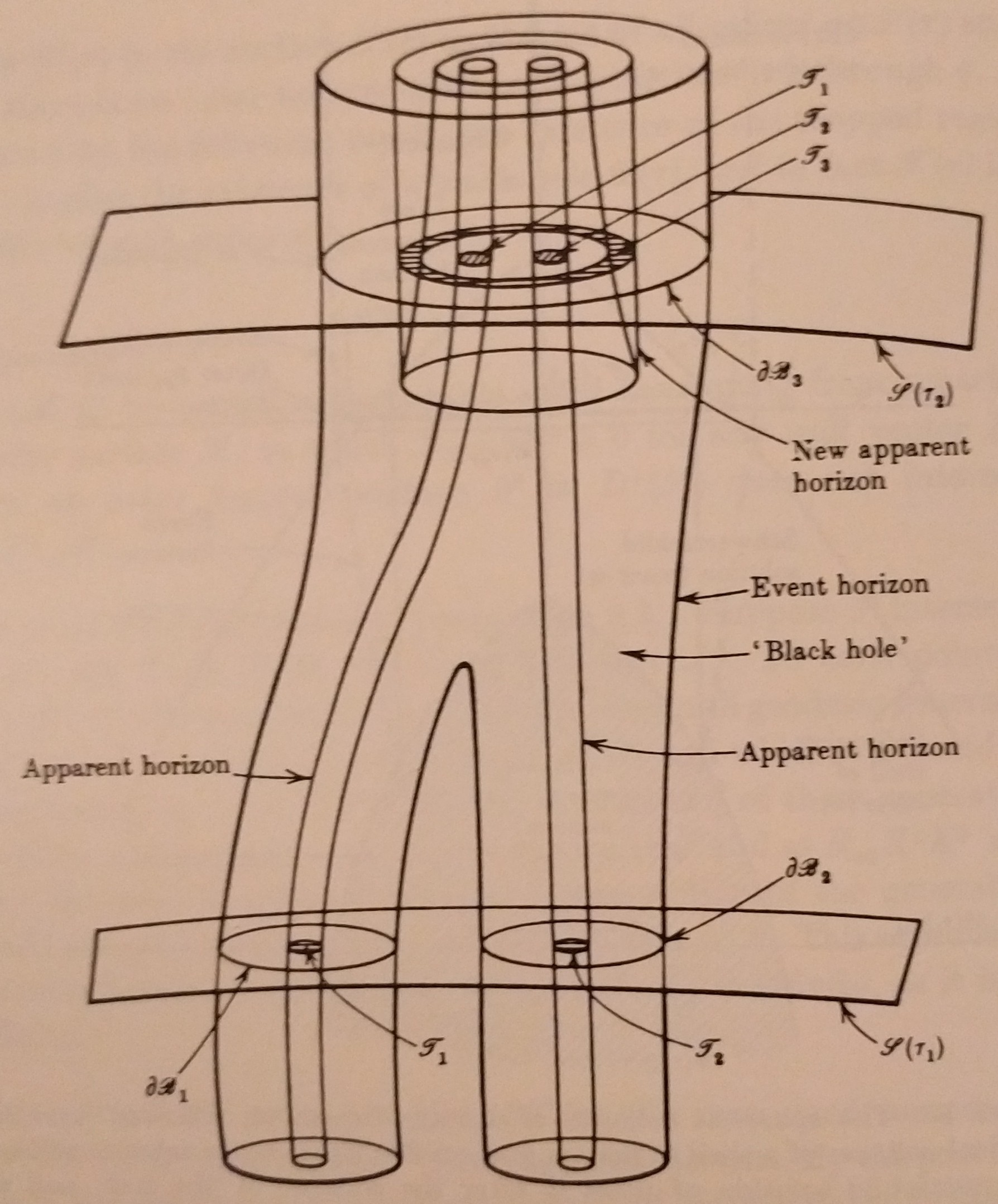}
\caption{Merging of two black holes. At time $\t_1$, there are two apparent horizons inside the event horizon. At time $\t_2$, the event horizon has merged into a single region and a third apparent horizon formed surrounding both previous apparent horizons. Figure taken from Hawking and Ellis \cite[p.~322]{hawking_ellis_1973}.}
\label{fig:eh_vs_ah}
\end{center}
\end{figure}

Apparent horizons do not evolve in a smooth way. One could think that during a black hole merger the apparent horizons of each black hole merge and dynamically form a new joint one. This is not the case. In the typical slicing conditions used, each black hole has a trapped surface that during merger shrinks to the puncture, while a larger common apparent horizon forms, surrounding the previous ones \cite{Pook-Kolb:2019ssg,Thornburg:2006zb,Thornburg:2003sf}. This is not a discontinuity, but an artefact of observing the domain of $\Q$ only via a projection at $\Q=0$. Figure \ref{fig:eh_vs_ah} depicts the emergence of a new apparent horizon (see also Pook-Kolb et al. \cite[fig. 1]{Pook-Kolb:2019ssg}). It can be seen that there are in fact two trapped surfaces forming, an inner one which shrinks to the puncture, and the outermost one, the newly formed apparent horizon of the merger black hole.

\begin{figure}[h]
\begin{center}
\includegraphics[width=0.9\linewidth]{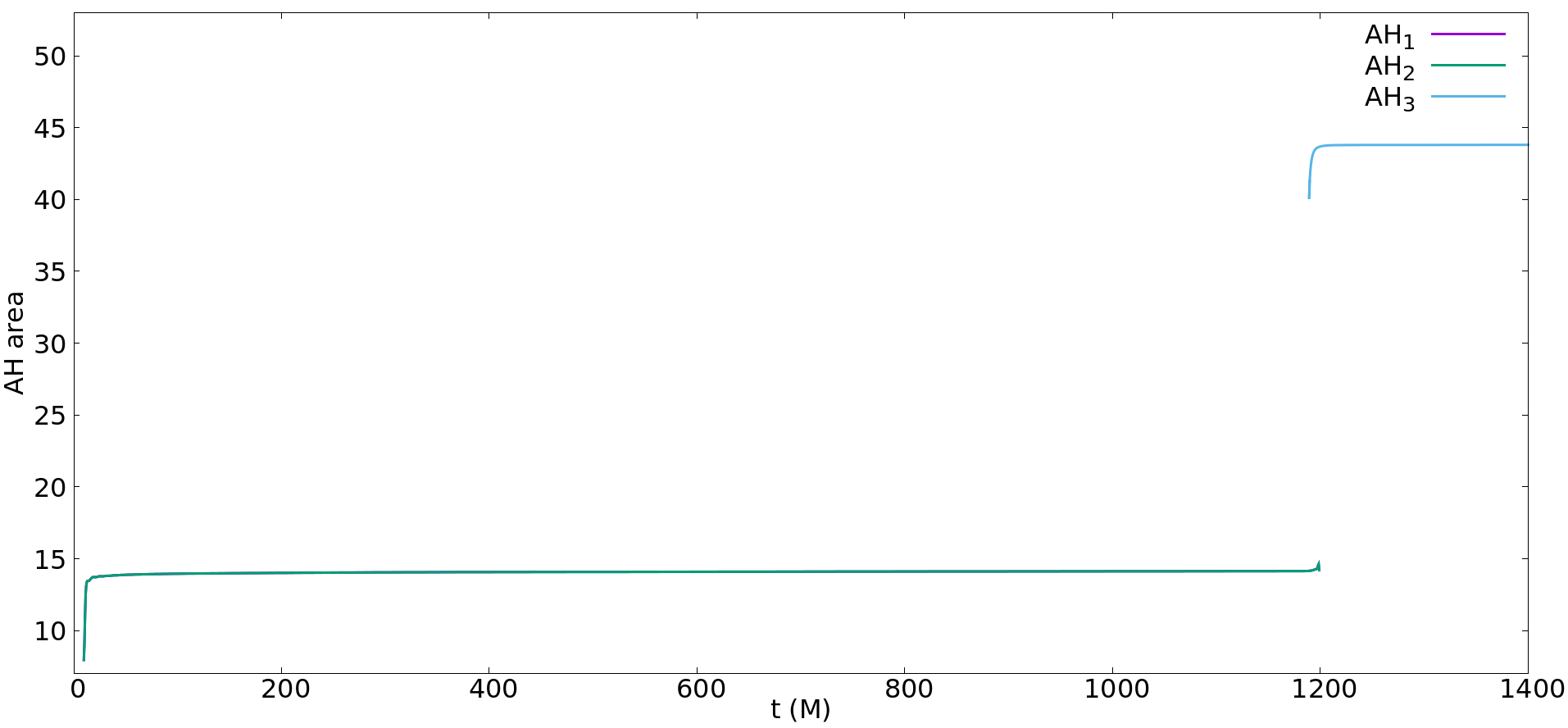}
\includegraphics[width=0.9\linewidth]{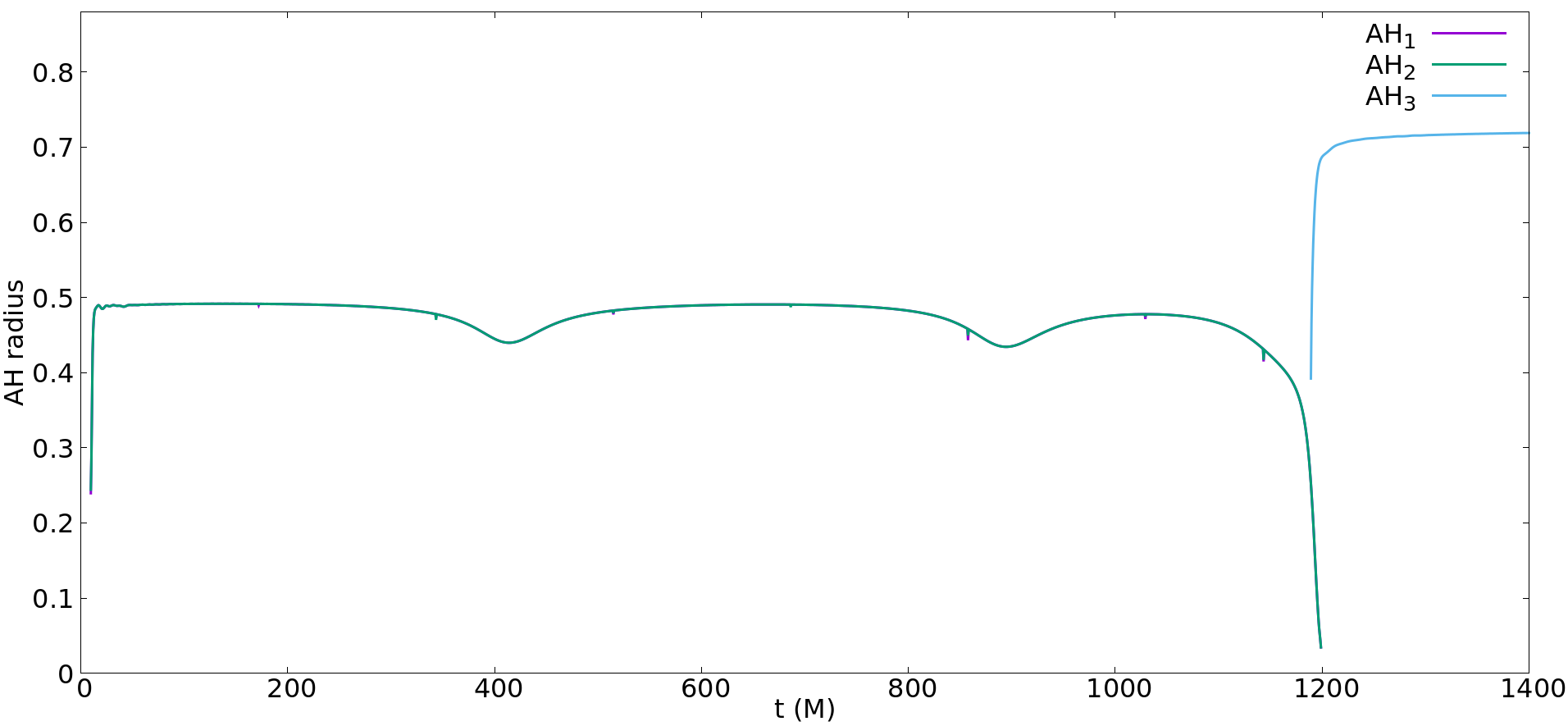}
\caption{Evolution of the apparent horizons for a GR black hole binary as described in chapter \ref{chapter:paper2}. The top figure displays the surface physical area and the bottom figure displays the coordinate radius. One can observe the appearance of a third apparent horizon surrounding the inner two trapped surfaces of each black hole, which in turn shrink to zero. The initial growth around $t\sim0$ is related to the formation of a black hole during gravitational collapse.}
\label{fig:ah_evolution}
\end{center}
\end{figure}

This implies that detecting apparent horizons in dynamical simulations requires tracking more surfaces than the number of black holes. A binary black hole requires looking for the appearance of a third trapped surface which pops up at some point when the original punctures get close enough, with the added care of converging to the outermost one and not some other existing trapped surface\footnote{This typically simply requires setting an initial guess bigger than the expected surface.}. For a triple black hole, one may need five trapped surface trackers. This requires great flexibility from the \texttt{AHFinder}, which ends up extending to the ability to detect appearing apparent horizons during gravitational collapse and even cosmological horizons in expanding universes. In figure \ref{fig:ah_evolution} one can observe evolution of the apparent horizon area for the GR binary evolved in chapter \ref{chapter:paper2}. One can identify: an initial growth in area when the black hole forms after gravitational collapse, and the appearance of a third AH while the two smaller AHs disappear. Note that even though the area remains constant, their coordinate radius is reducing to zero until convergence is no longer possible, accompanied with a distorted metric that keeps the physical area constant. See also figure \ref{fig:ah_discrete_merger}.

\section{\done{Finding apparent horizons}}\label{sec:ahfinder:finding_ahs}

\subsection{\done{Parameterising apparent horizons}}

Finding an apparent horizon entails finding a surface that satisfies $\Q=0$. Numerically, this implies parameterising surfaces and solve $\Q=0$ as a PDE over the surface. A common approach when parameterising apparent horizons relies on level sets, by finding a scalar function which is zero over a given surface, $L(x^i)=0$, such that the unit normal vector $s^i$ can be written as:
\begin{equation}
    s^i = \frac{D^i L}{|DL|}\,,
\end{equation}
where $|DL| = \sqrt{\g^{ij}\br{D_i L} \br{D_j L}}$. This results in:
\begin{equation}
    \Q = \br{\g^{ij} - \frac{\br{D^i L} \br{D^j L}}{|DL|^2}}\br{\frac{D_i D_j L}{|DL|} - K_{ij}} = 0\,.
\end{equation}
This function $L$ can now be parameterised with some height function $H$ via $L(x^i) = h - H(x_S^A)$ for some coordinates $x'^i = (h,x_S^A)$ well-adapted for the problem at hand, where $x_S^A$ parameterises the surface $S$. For $3d$ black holes, one can parameterise ``star-shaped'' shapes (in which we can always find an inner point such that all rays leaving this point intersect the surface only once - see Alcubierre \cite[fig.~6.7]{alcubierre}) by a level set $L(x^i) = r - H(\q,\f)$, using $x_S^A=(\q,\f)$ and where $(r,\q,\f)$ are the usual spherical coordinates, transformed from $x^i$ in the usual way. This corresponds from ``shooting'' rays from some central point with some radius $H$ for each $(\q,\f)$. For this case, $dL = dr - \br{\pd_\q H} d\q - \br{\pd_\f H} d\f$, where $\pd_\q h$ and $\pd_\f H$ (and second derivatives required) are computed numerically, while all other derivatives related to spherical coordinates are known analytically. For a two dimensional black string such as in the work done by Figueras et al. \cite{Figueras:2022zkg}, one can simply pick $L(x^i) = y - H(x)$, where $(x,y)$ is the horizontal and vertical Cartesian coordinates, which corresponds to ``shooting'' rays vertically from the $y=0$ axis. More intricate problems require more generic surfaces, such as reference surfaces \cite{Pook-Kolb:2018igu, Pook-Kolb:2019ssg}, which has not yet been used in \texttt{GRChombo}. For reviews on different algorithms for apparent horizon finding, see \cite{Thornburg:2006zb,Thornburg:2003sf,Gundlach:1997us}.

\subsection{\done{Numerical method}}

\begin{figure}[h]
\begin{center}
\begin{multicols}{2}
\includegraphics[width=\linewidth]{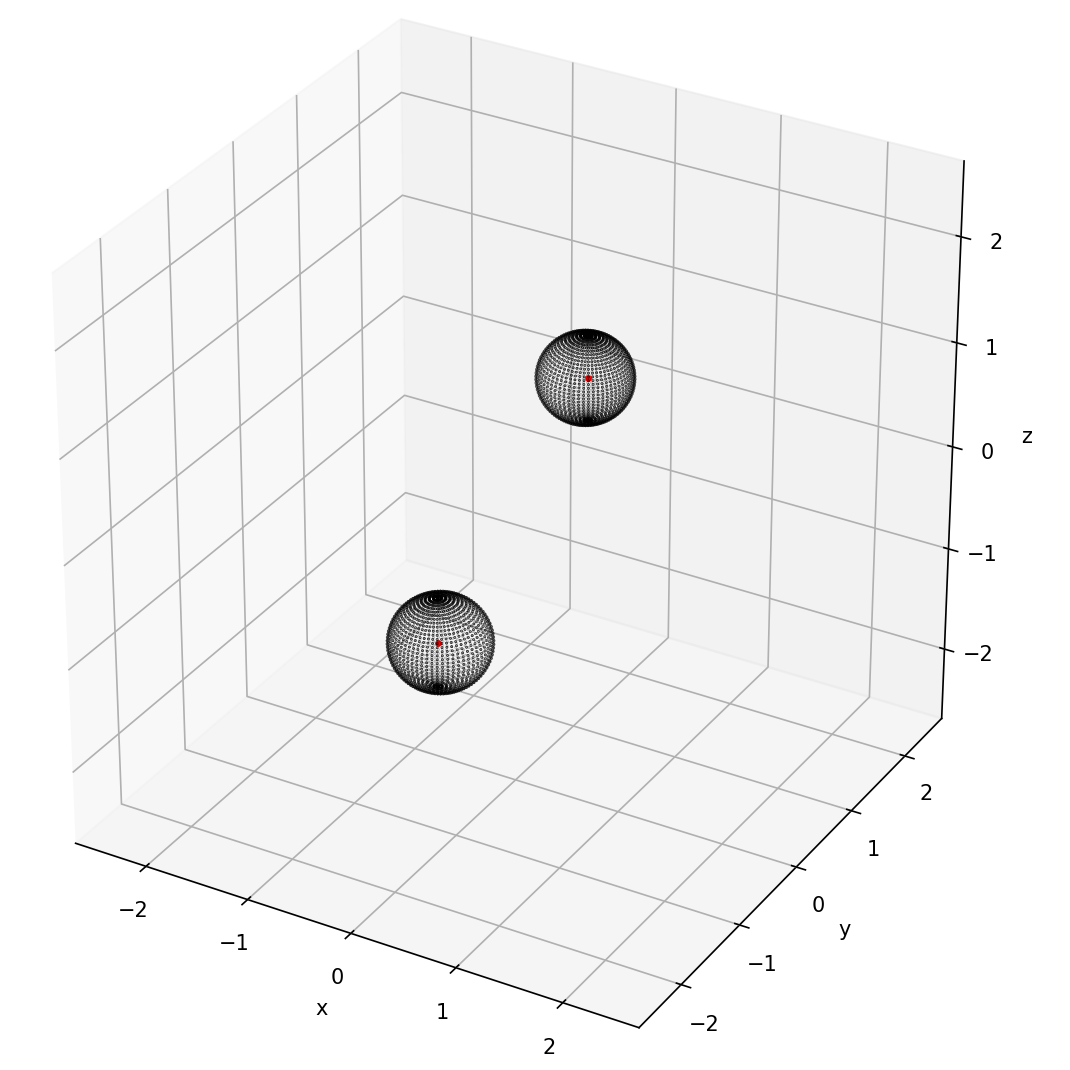}
\includegraphics[width=\linewidth]{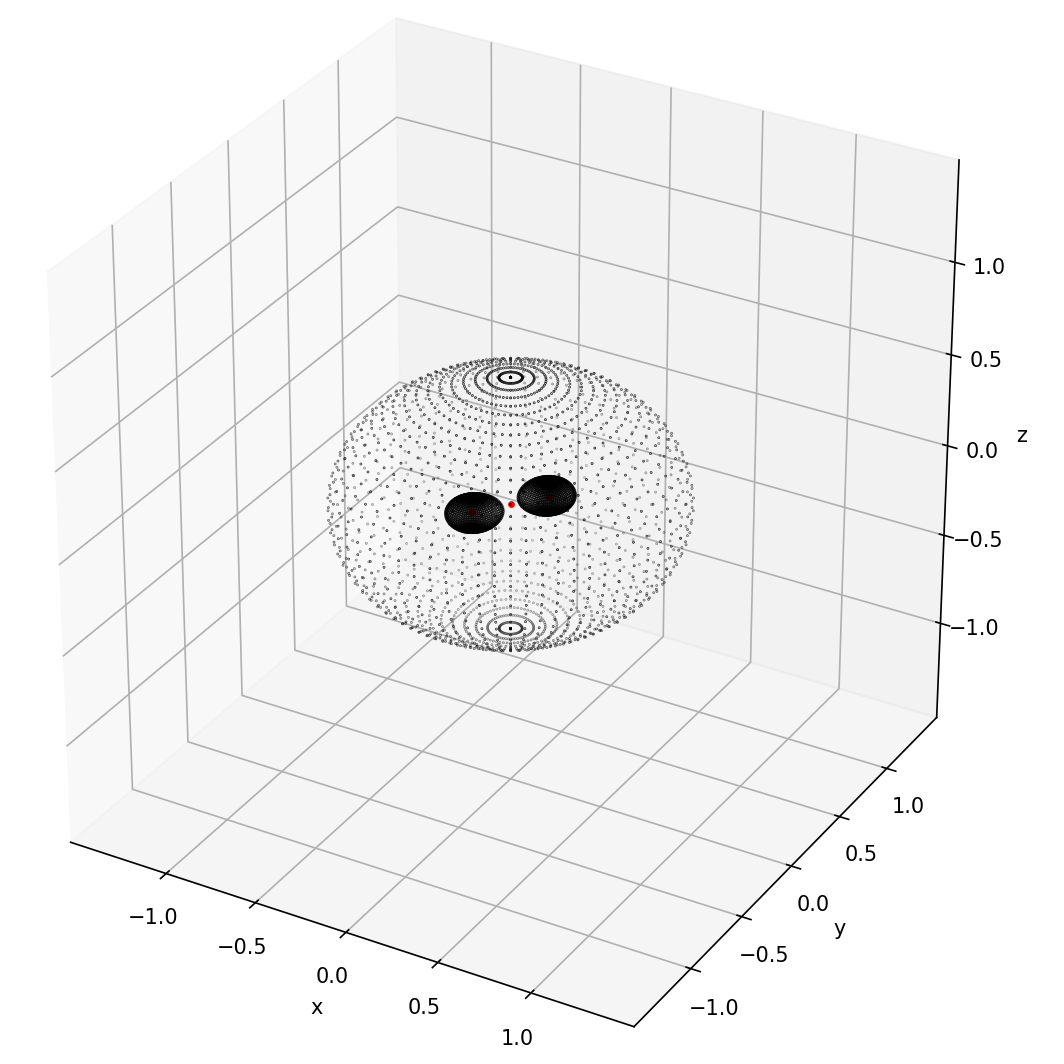}
\includegraphics[width=\linewidth]{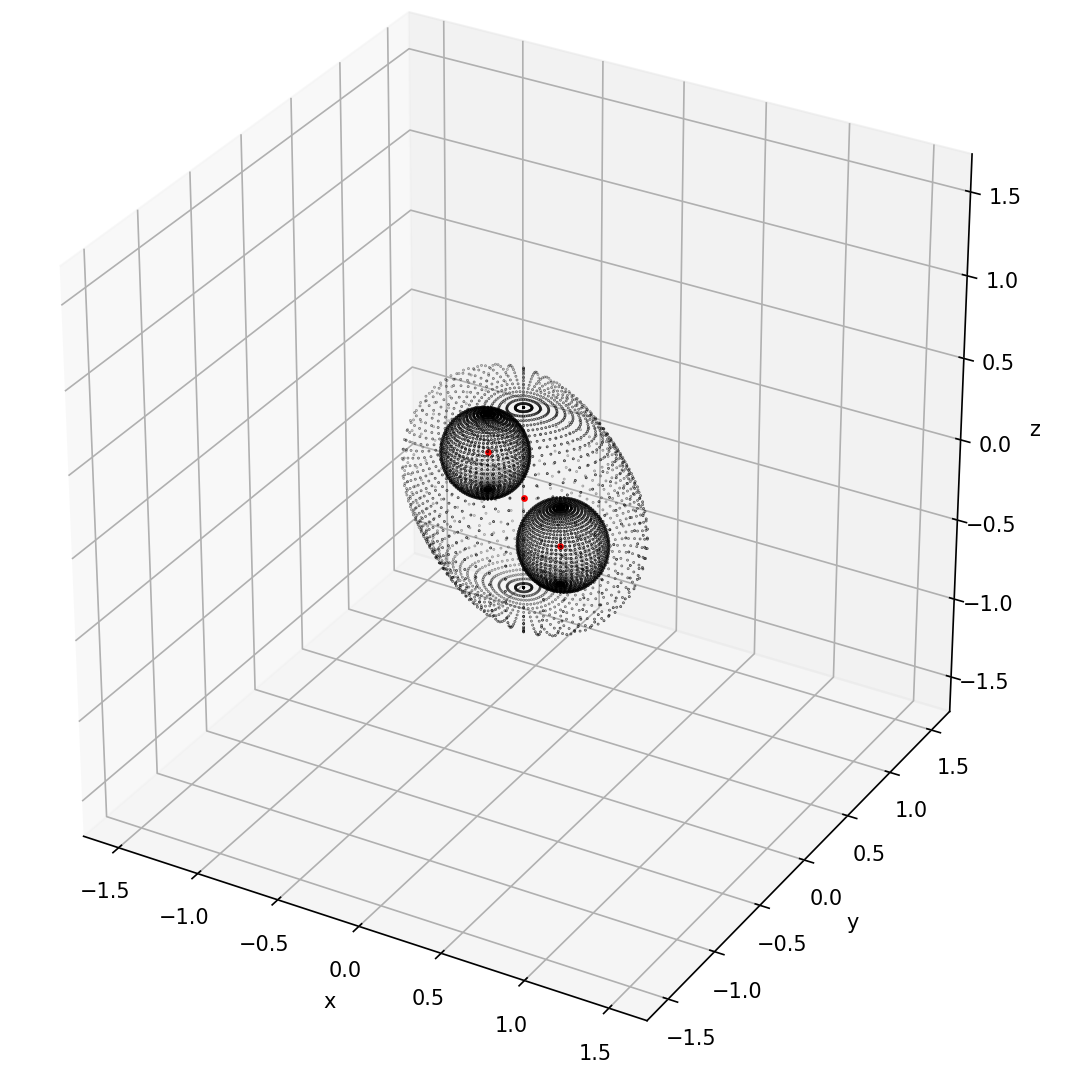}

\includegraphics[width=\linewidth]{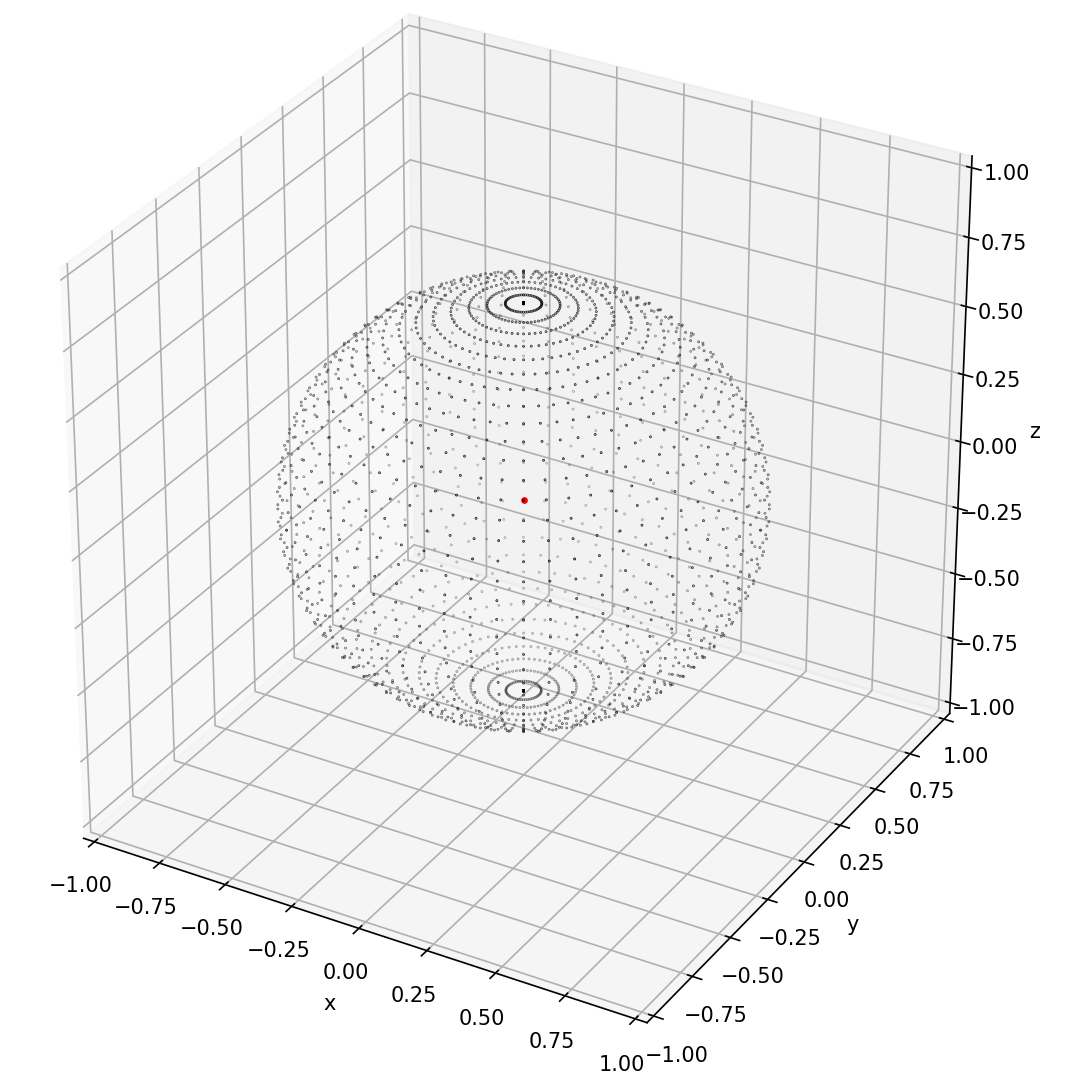}
\end{multicols}
\caption{Evolution of the numerical apparent horizon during an equal mass black hole merger. The points over the surface of each trapped surface is show for $t=36M, 64.5M$ in the top panel and $t=70M, 100M$ in the bottom panel. One can observe the formation of a new trapped surface surrounding the original apparent horizons, which in turn shrink to zero size, leaving a remaining perturbed black hole relaxing to a Kerr black hole.}
\label{fig:ah_discrete_merger}
\end{center}
\end{figure}

Given the parameterisation given in the previous section, the only ingredient missing is a discretisation over the surface $S$, which can be seen in figure \ref{fig:ah_discrete_merger}. Given a set of $N$ points over the surface $S$, one wants to solve the multi-dimensional problem $\ve{\Q}\br{\ve{h}} = 0$, where $\ve{h}\in \mathbb{R}^N$ is  the numerical array of the level set height function at each grid point (simply flatten out all the points into a one dimensional array), and $\ve{\Q}$ is the numerical evaluation of the expansion over the discretisation of the surface $\ve{h}$, using interpolation and numerical derivatives where necessary.

To solve this, one can use any suitable nonlinear solver. The \texttt{AHFinder} uses the nonlinear solver framework (SNES) provided by PETSc \cite{balay1998petsc}, which is based on a quasi-Newton iterative method with an auxiliary linear system solver based on Krylov methods, for determining the Jacobian of the expansion. In short, it uses the typical Newton method iterations:
\begin{equation}
\begin{aligned}
    \ve{h}_{n+1} =&~ \ve{h}_n - \g \D \ve{h}_n\,,\\
    \D \ve{h}_n =&~ \sbr{\grad_{\ve{h}}\ve{\Q}\br{\ve{h}_n}}^{-1} \ve{\Q}\br{\ve{h}_n}\,,
\end{aligned}
\end{equation}
where $\g$ is a constant determining the step of iteration, but where the Jacobian is not inverted. Instead, we compute $\D \ve{h}_n$ via solving the linear system:
\begin{equation}
    \sbr{\grad_{\ve{h}}\ve{\Q}\br{\ve{h}_n}} \D \ve{h}_n = \ve{\Q}\br{\ve{h}_n}\,,
\end{equation}

As an additional note, it is often the case that numerical algorithms may have difficulty finding highly distorted shapes in highly dynamic environments, but, as previously mentioned, often simply re-using the shape of previously found horizons as initial guess for current time slices is enough, a feature integrated in the \texttt{AHFinder}. To estimate the position of the centre $\ve{C}(t_n)$ for a timestep $t_n$, given previous timesteps, one can use first and second numerical derivatives over the previous time steps: $\ve{C}(t_n) = \ve{C}(t_{n-1}) + \sbr{\ve{C}(t_{n-1}) - \ve{C}(t_{n-2})} + \sbr{\ve{C}(t_{n-1}) - 2\,\ve{C}(t_{n-2}) + \ve{C}(t_{n-3})} = \ve{C}(t_{n-1}) - 3\,\ve{C}(t_{n-2}) + 3\,\ve{C}(t_{n-3})$. This is equivalent to an extrapolation of a quadratic fit. One can use only a linear extrapolation if not enough points are available. $\ve{C}(t_n)$ is an approximation of the geometric centre of the surface found, which can be computed in many ways, but for ellipsoid shapes can be very well approximated by computing the average between the maximum and minimum coordinates in each direction for the points that cover the surface $S$.

\begin{figure}[h]
\begin{center}
\includegraphics[width=0.8\linewidth]{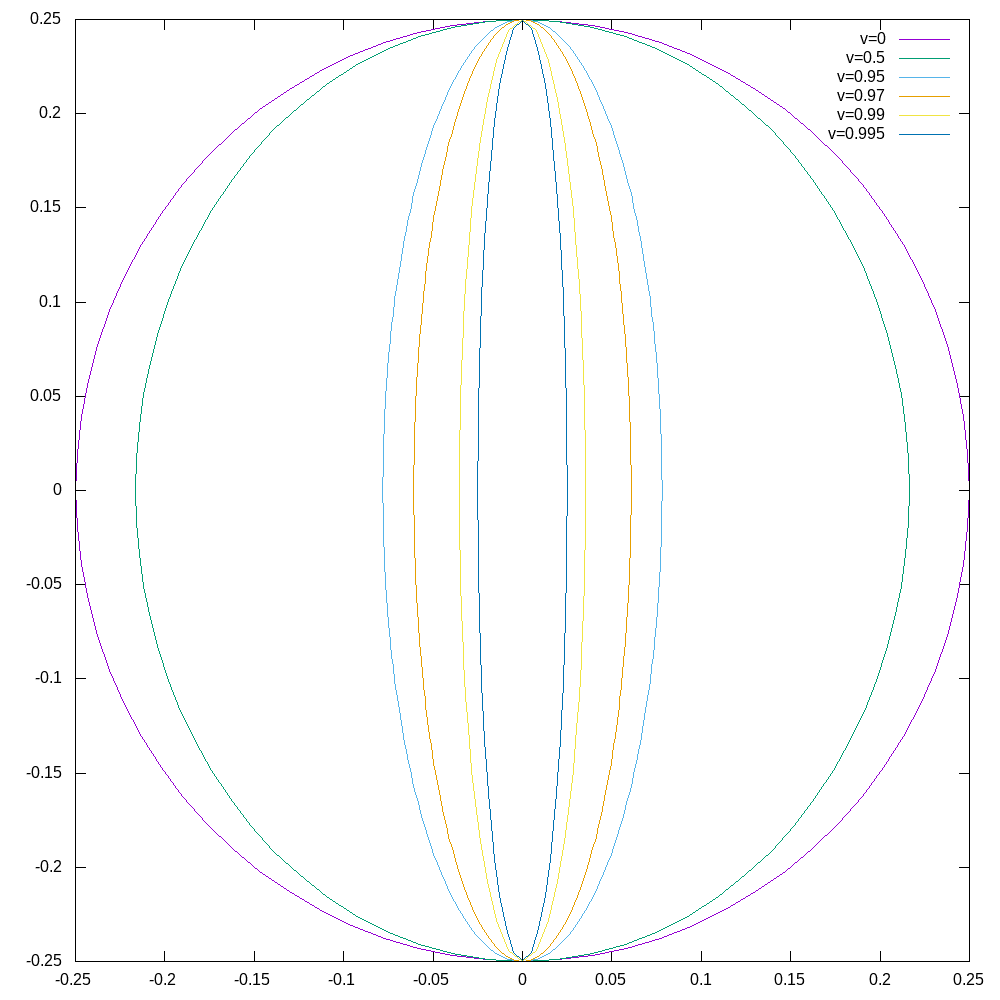}
\caption{Different apparent horizons found for a single black hole with different boosts, with speeds ranging from $v=0$ to $v=99.5\%$ the speed of light.}
\label{fig:ah_boosted}
\end{center}
\end{figure}

\subsection{\done{Optimising convergence}}

With experience of many people using the \texttt{AHFinder}, it was noticeable its sensitivity to the initial guess: initial guesses that were too big were tricky to result in convergence, even though it was obvious a black hole was present. This was the case especially in the case of cosmological horizons \cite{deJong:2023} and mergers (where the initial guess for the common apparent horizon is better put at a large overestimated radius). This can be easily understood by looking at a simple example.

\begin{figure}[h]
\begin{center}
\begin{multicols}{2}
\includegraphics[width=\linewidth]{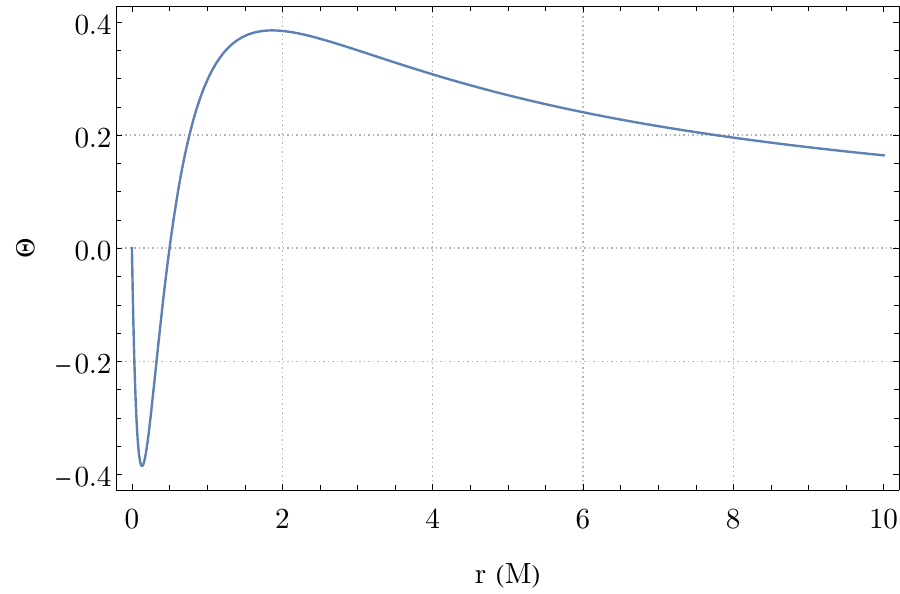}

\includegraphics[width=\linewidth]{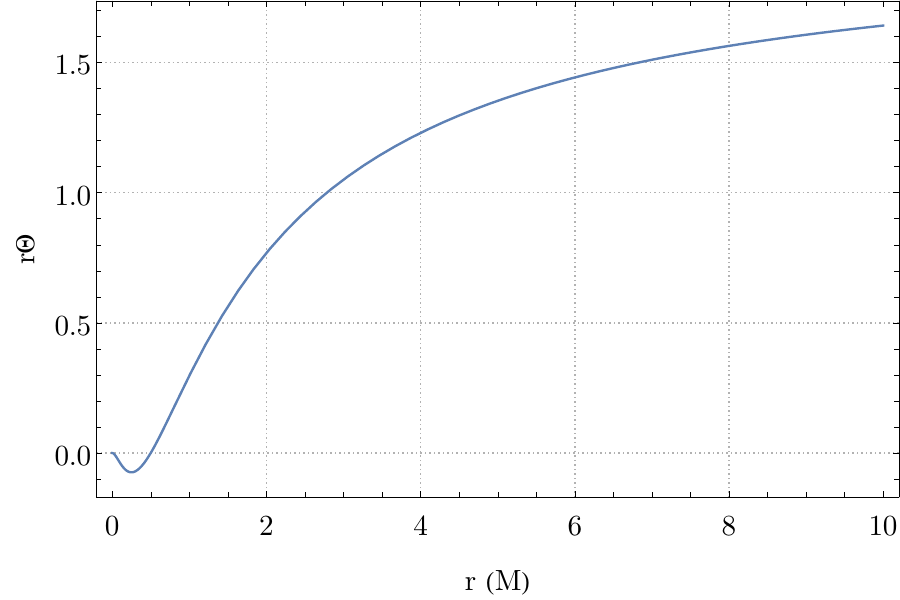}
\end{multicols}
\caption{Plot of the expansion for spherical surfaces in a Schwarzschild spacetime with mass $M$ in isotropic coordinates. On the right, we see instead $r\cdot\Q$ to depict how under Newton's iterative method, this quantity is much better suited to find the $\Q=0$ root.}
\label{fig:ah_expansion}
\end{center}
\end{figure}

Consider the usual $4D$ Schwarzschild spacetime in isotropic coordinates and take the induced spatial interval:
\begin{equation}
    dl^2 = \psi(r)^4\br{dr^2 + r^2d\W^2}\,,
\end{equation}
where $\psi(r) = 1 + \frac{M}{2r}$. Being a spherically symmetric spacetime, assume the black hole to be spherical and take $s^i$ to be the normal vector to spheres: $s^i = \br{\frac{1}{\psi(r)^2},0,0}$, which has unit norm. The resulting expansion is:
\begin{equation}
    \Q = \frac{2\,\br{1 - \frac{M}{2r}}}{r\,\br{1 + \frac{M}{2r}}^3}\,,
\end{equation}
which clearly implies $r=\frac{M}{2}$ to get $\Q=0$, as expected for the location of the horizon in isotropic coordinates. But, as it can be seen from figure \ref{fig:ah_expansion}, at around $r\sim1.87M$ there is a maximum, meaning that any initial guess in Newton's method to the right of this point will lead to divergent results. %
To fix such issue, one can notice in figure \ref{fig:ah_expansion} that the quantity $r\cdot\Q$ has no maximum and converges to the correct root as long as the iteration step $\g$ is not big. This procedure is also mentioned by Thornburg \cite[sec. 8.6.2]{Thornburg:2006zb}.

Hence, the \texttt{AHFinder} solves for $r\Q=0$ for improved convergence. As a matter of fact, the \texttt{AHFinder} provides an easy way to change this and find any quantity, whether that is the shape with $\Q=0$, but also $\chi=\text{constant}$ (often useful) or even $\pd_t \b^x = 0$.

\subsection{\done{Computing observables}}

After finding the apparent horizon, besides knowing its location, it is also important to compute quantities like mass, spin and boost. These can be estimated numerically from the AH shape found numerically. We first transform the Cartesian metric $\g_{ij}$ into a metric $\g'_{ij}$ in the adapted coordinates $x'^i=(h,x_S^A)$:
\begin{equation}
    \g'_{ij} = \frac{\pd x^k}{\pd x'^i}\frac{\pd x^l}{\pd x'^j} \g_{kl}\,.
\end{equation}
The induced metric on the co-dimension 1 surface $S$ over the coordinates $x_S^A$ is:
\begin{equation}
    \g^S_{AB} = \frac{\pd x'^i}{\pd x_S^A}\frac{\pd x'^j}{\pd x_S^B} \g'_{ij}\,,
\end{equation}
where one should use the fact that $\frac{\pd h}{\pd x_S^A} = \frac{\pd H}{\pd x_S^A}$ on the level set. This can now be used to compute the area of the black hole via:
\begin{equation}\label{eq:ahfinder:area}
    A = \oint_S \sqrt{\det\br{\g^S_{AB}}}\, d^2x_S\,.
\end{equation}
For $3d$ spacetimes, one can compute the spin of the black hole using the equator length. For coordinates $x_S^A = (\q,\f)$, fixing $\q=\frac{\p}{2}$, the metric reduces to a single component $\g^S_{\f\f} = \g_{ij}\br{\pd_\f x^i + \pd_\f H\, \pd_r x^i}\br{\pd_\f x^j + \pd_\f H\, \pd_r x^j}$ which can be used to compute the equator length via:
\begin{equation}
    L_{eq} = \int_0^{2\p} \sqrt{\g^S_{\f\f}} d\f\,.
\end{equation}
Assuming $4D$ and Kerr-ness of the black hole (which is only approximately valid for non isolated black holes or for highly dynamic phases, such as the merger, or in the absence of vacuum, if we have matter sources coupling to gravity), one can then compute the dimensionless spin, $s$:
\begin{equation}
    s = \sqrt{1 - \br{\frac{2\p A}{L_{eq}^2}-1}^2}\,,
\end{equation}
for which the angular momentum is $J=sM^2$. The mass can be computed with several formulas:
\begin{equation}
    M = \frac{L_{eq}}{4\p} = \sqrt{\frac{A}{8\p\br{1+\sqrt{1-s^2}}}} =  \sqrt{\frac{A}{16\p} + \frac{4\p J^2}{A}}\,.
\end{equation}
Alternatively, one can compute approximations of the linear and angular momentum vector coordinate independent quantities using a Killing vectors in Minkowski space for boosts and $SO(3)$ rotations. The linear momentum:
\begin{equation}\label{eq:ahfinder:linear_momentum}
    P^l = \tfrac{1}{8\p} \oint_S \sqrt{\det\br{\g^S_{AB}}} \br{\varphi_L^l}^i s^j K_{ij} d^2x_S\,,
\end{equation}
using the generators $\br{\varphi_L^l}^i = \d^{il}$. Similarly the angular momentum $J^l$ can be computed using the rotational Killing vectors $\br{\varphi_A^l}^i = \e^{ilj} x_j$, where $\e^{ilj}$ is the Levi-Civita symbol \cite{Cook:2007wr,Dreyer:2002mx}.

Finally, \texttt{GRChombo} often integrates the Cartoon method, a method to reduce the dimensionality of a spacetime with $SO(n)$ symmetry (see Cook et al. \cite{Cook:2016soy} for details, and applications with \texttt{GRChombo} in Figueras et al. \cite{Figueras:2022zkg, Figueras:2015hkb, Figueras:2017zwa} and Cheung et al. \cite{Cheung:2022rbm}). For those cases, computations like the area \eqref{eq:ahfinder:area} or the linear and angular momentum \eqref{eq:ahfinder:linear_momentum} require including in the integrand the term $\br{x^w\sqrt{\g_{ww}}}^n S_n$, where, following the notation of Cook et al. \cite{Cook:2016soy}, $n$ is the number of Cartoon reduced dimensions, $x^w$ is the Cartoon coordinate, $\g_{ww}$ is the cartoon metric component and $S_n$ is the surface area of a unit n-sphere, $S_n = \frac{2\, \p^{\frac{n+1}{2}}}{\G\br{\frac{n+1}{2}}}$, with $\G$ the Euler Gamma function.

\subsection{\done{Code development}}

A precursor of an apparent horizon finder integrated for \texttt{GRChombo} existed from the work by Tunyasuvunakool \cite{Tunyasuvunakool:2017wdi}, able to find isolated Kerr black holes at specific timesteps (and tweaked for higher dimension problems in specific projects; see also Cook \cite{Cook:2018fxg}). The \texttt{AHFinder} was developed as the integration into the modern and public \texttt{GRChombo} code with user-friendly controlled parameters adaptive to any simulation, but overall working as an easy add-on without the researcher having to know many details. It also integrates standard IO tools to allow seamless restarting from checkpoints, including an interpolation algorithm to allow changing the number of points over the horizon surface, and plotting tools, such as what was used to plot figure \ref{fig:ah_discrete_merger}. The issue with such public integrations is that a lot of use cases must be supported. This lead to many improvements over time that make the \texttt{AHFinder} a robust tool:
\begin{itemize}
    \item the generalisation of the code to finding any quantity contours (not just $\Q=0$) in any type of surface (spheres, cylinders, strings, as long as they can be defined via a level set) in any dimension (currently tested in $2+1$ and $3+1$ dimensions);
    \item the support of Cartoon evolution with reduced dimensions;
    \item the requirement of tracking multiple black holes and estimating future positions of moving black holes for the case of black hole binaries;
    \item the improved convergence by solving $r\Q=0$ and by re-using previously found surfaces as initial guesses;
    \item to adapt to highly parallelisable supercomputing environments, the development of an MPI subcommunicator to process the root finding algorithm in a subset of the computing nodes;
    \item the compatibility with all types of boundary conditions;
    \item the development of tools to integrate custom quantities over the apparent horizon surface, such as the area, angular momentum, or any user-defined custom diagnostic variables.
\end{itemize}

In numerical relativity, we often simulate black holes. When we simulate black holes, it is obviously useful to know where the black holes are. As a consequence, this small project required for the development of chapter \ref{chapter:paper1} led to a multipurpose user-friendly tool used by the \texttt{GRChombo} collaboration in many projects. Future work might revolve around extending current surfaces to reference surfaces, as by Pook-Kolb et al. \cite{Pook-Kolb:2018igu, Pook-Kolb:2019ssg}, making the Cartoon equations compatible with matter, and extending the code for non-zero cosmological constant, such as AdS spacetimes.

\section{\done{Apparent horizon approximation}}\label{appendix:grchombo:ah_location}

In moving puncture gauge, the profile of evolution variables such as the lapse or the conformal factor, and diagnostics such as the location of the horizon are not known very accurately other than from some early studies \cite{Hannam:2006vv,Bruegmann:2006ulg,Baumgarte:2007ht}. The location of the black hole horizon, both in terms of coordinate distance to the puncture and in terms of the conformal factor on its surface, turns out to be quite essential for tasks such as quickly finding the location of the horizon on the grid (often for visualisation purposes), to perform excision (as in sections \ref{subsection:paper1:excision} and \ref{subsec:eft:excision}) or to improve tagging of AMR codes. Regarding the tagging criteria, which refers to the criteria AMR algorithms use for the creation and destruction of AMR levels and boxes within them (see appendix \ref{appendix:grchombo:tagging} for details), it is known that refinement levels close or crossing the apparent horizon can cause drifts in the mass of black holes and even add full orbits in the case of binaries \cite{Radia:2021smk}. Common dynamic tagging criteria may fail to cover the AH, and hence forcing the levels to cover the black hole is clearly something that can help preventing this issue.

\begin{figure}[h]
\centering
\hspace{-30mm}\includegraphics[width=1.2\textwidth]{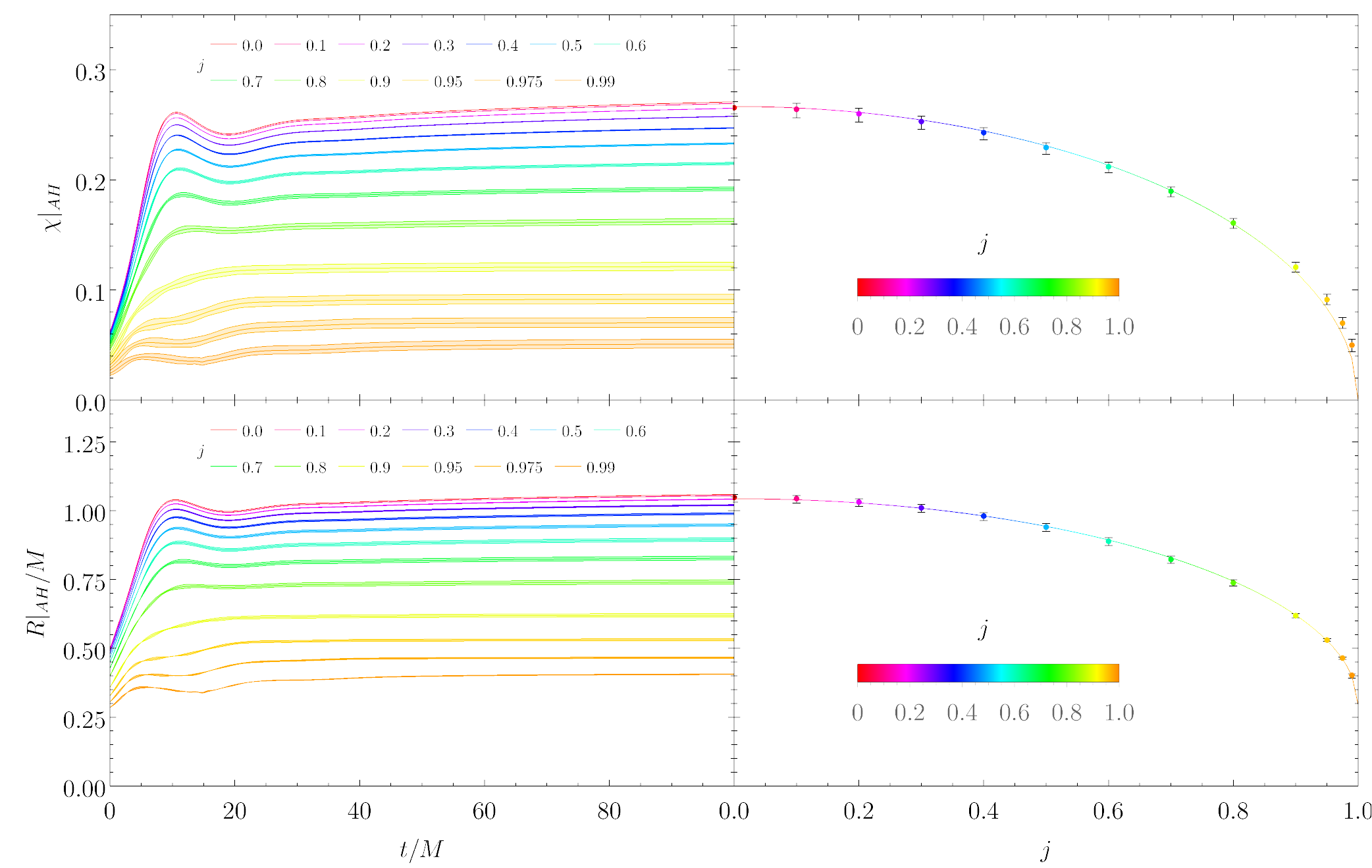}
\caption{Plots illustrating the dependence of the value of the conformal factor $\chi$ and the coordinate radius $R$ on the apparent horizon surface in the moving puncture gauge for different values of the dimensionless spin $j$. For all plots, we use the quasi-isotropic Kerr initial data \cite{Liu:2009al} and the default values of the gauge parameters with $M\eta = 1$. Although we would expect the plots to vary for different gauge parameters (in particular, as $\eta$ is varied, see Bruegmann et al. \cite{Bruegmann:2006ulg}), these plots provide a rough rule-of-thumb. The left panels show the mean value of $\chi$ (top) and $R$ (bottom) as a function of time with the error bands around each curve corresponding to the maximum and minimum on the apparent horizon. The right panels shows the mean value of $\chi$ and $R$ over the interval $t/M \in [40, 100]$ for each $j$ with the error bars corresponding to the minimum and maximum values of $\chi$ over the same interval. Furthermore, we show a fit of the mean value of $\chi$ and $R$ against $j$ which takes the form $\langle \chi \rangle_{\text{AH}} \approx \br{0.266\pm0.003}\sqrt{1 - j^2}$ and $R_{\text{AH}}/M \approx \br{0.294\pm0.006} + \br{0.75\pm0.01}\sqrt{1 - j^2}$.}
\label{fig:grchombo:ah_location:chi_over_time}
\end{figure}

With that in mind, we performed evolutions from quasi-isotropic Kerr black hole initial data \cite{Liu:2009al} with different spins, to find the relation between spin and the conformal factor and coordinate radius of the apparent horizon. This is illustrated in figure \ref{fig:grchombo:ah_location:chi_over_time} and resulted in the fit against the dimensionless spin of the black hole, $j$:
\begin{equation}
\begin{aligned}\label{eq:ah:approximation}
    \langle \chi \rangle_{\text{AH}} &\approx \br{0.266\pm0.003}\sqrt{1 - j^2}\,,\\
    R_{\text{AH}}/M &\approx \br{0.294\pm0.006} + \br{0.75\pm0.01}\sqrt{1 - j^2}\,.
\end{aligned}
\end{equation}

It is important to note that even though $\chi$ varies across the apparent horizon surface, even with spin, the horizon is still topologically spherical, having constant coordinate radius. One other key fact is that this does not depend on the mass of the black hole, making it a very useful metric to identify black holes in a numerical grid. Although we would expect the plots to vary for different gauge parameters (in particular, as $\eta$ is varied, see Bruegmann et al. \cite{Bruegmann:2006ulg}), these plots provide a rough rule-of-thumb. It is valid as a tagging criteria as these do not require precision, and in fact we should gladly take an extra buffer around the black hole as safety margin. This as been used to develop tagging criteria for chapters \ref{chapter:paper1} and \ref{chapter:paper2}.

For techniques like excision, described in \ref{subsection:paper1:excision} and \ref{subsec:eft:excision}, slightly more precision is required (though not exactness). For such purposes, one can use the analytic values for the Kerr BH initial data with mass $M$ and dimensionless spin $j$:
\begin{equation}
\begin{aligned}
    r_{AH,\,t=0} &= \tfrac{M}{4}\br{1 + \sqrt{1 - s^2}}\,,\\
    \chi_{AH,\,t=0} &= \bigg[\tfrac{\br{1-s^2}^{1/6}}{16}\br{\tfrac{1 + \sqrt{1 - s^2}}{2}}^{2/3} \text{for $\theta=0$}, \tfrac{\br{1-s^2}^{1/6}}{16}\br{\tfrac{1 + \sqrt{1 - s^2}}{2}}^{1/3} \text{for $\theta=\frac{\p}{2}$}\bigg]\,,
\end{aligned}
\end{equation}
and make a smooth interpolating function that approximates the apparent horizon location over time:
\begin{equation}\label{eq:chi_over_time}
    \chi_{AH,\,t} = \chi_{AH,\,t\to\infty} + e^{-\l\,t} (\chi_{AH,\,t=0} - \chi_{AH,\,t\to\infty})\,.
\end{equation}
Note that this is a lower bound estimate, meaning that the apparent horizon is guaranteed to be outside of the region delimited by it, given an appropriate choice of $\l$. 

\begin{figure}[h]
\begin{center}
\includegraphics[width=0.8\linewidth]{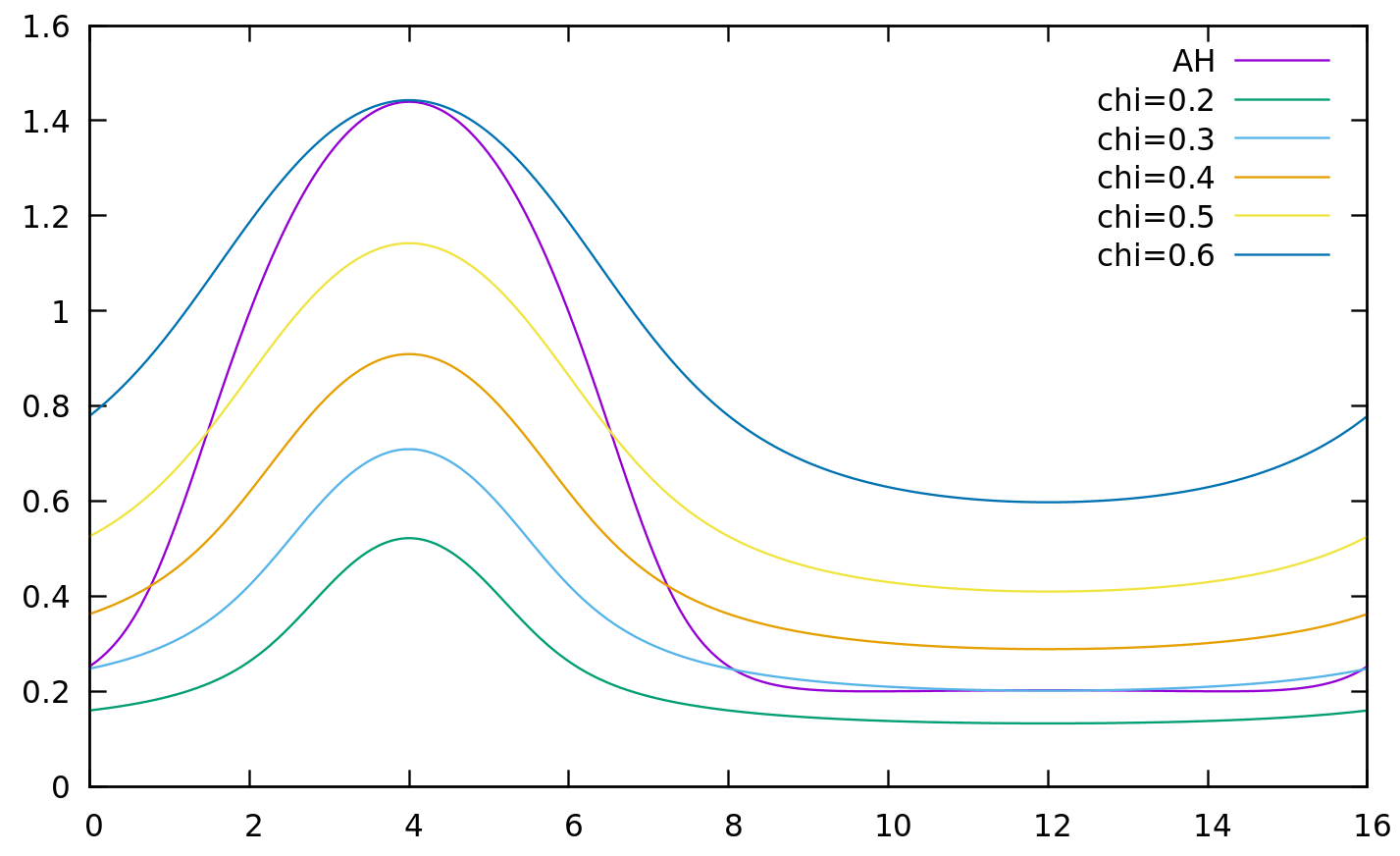}
\caption{Apparent horizon and several $\chi$ contours during the evolution of the Gregory-Laflamme instability of a black string in 5D, done in the context of the work in Figueras et al. \cite{Figueras:2022zkg}. The horizontal and vertical axis are the coordinates of the computational domain for a black string with a four-dimensional mass parameter $M = 1$ and a Kaluza-Klein circle with asymptotic length $L = 16$.}
\label{fig:chi_vs_ah}
\end{center}
\end{figure}

The specific contour of $\chi$ to be used depends strongly on the number of dimensions. The dependence on $\chi$ can be seen for the case of $5D$ black strings in figure \ref{fig:chi_vs_ah}, and more details in Andrade et al. \cite{Andrade:2020dgc}. Being just a rule-of-thumb, a rough estimate can be quickly assessed and it has been used in simulations of several higher dimensional spacetimes such as black rings \cite{Bantilan:2019bvf, Figueras:2017zwa, Andrade:2020dgc, Figueras:2015hkb} and black strings \cite{Figueras:2022zkg}.

Finally, other valid approximation technique involves using a puncture tracker (see section \ref{subsubsection:ngr:observables}) and estimating the size of the black hole from its known or estimated mass. This turns out to be unfeasible in situations involving gravitational collapse, where the initial data has no initial puncture to keep track of, as was the case of chapters \ref{chapter:paper1} and \ref{chapter:paper2}.

\chapter{\done{Conclusions}}\label{chapter:conclusions}

\section{\done{Summary}}

In this thesis, we studied gravitational waves in modified theories of gravity, with the goal of investigating generic effects these theories have on physical observables that may allow for detection of deviations from GR in future observational data. Given the corrections to GR have to be locally small for any modified theory to make sense as valid EFTs, only non-local observables, such as gravitational waves, may have large deviations originated from cumulative small effects over time.

\begin{itemize}
    \item In chapter \ref{chapter:paper1}, we started by exploring the regime of validity of certain cubic Horndeski theories of gravity that have a well-posed initial value problem, with the goal of finding regions in the space of couplings and initial data which have future developments globally compatible with the weak field regime of the theories as valid EFTs and hence suitable to use in binary black holes.

    In this study, we identified that the breakdown of hyperbolicity in the future development can happen for initial data which satisfies strict weakness conditions. Using gravitational collapse of a scalar field bubble, we found Tricomi type transitions for $g_2 > 0$ in the $G_2=g_2 X^2$ theory, and Keldysh type transitions for $g_2<0$ or any value of the sign for the cubic theory $G_3=g_3 X$. We provided analytic justifications and predictions of when these effects occur. Interestingly, even for coupling constants that would break hyperbolicity for low amplitudes of the scalar field, for big enough amplitudes that still satisfies the WCC, the region of validity is unexpectedly extended, as the earlier appearance of a trapped surface shield observers from any pathology of the theory. Numerically this translates to the possibility of excising any problematic region inside apparent horizons from the computational domain.

    This analysis gave an understanding that, while not necessarily present in all theories, many similar models will suffer from the same pathology wherein valid initial data may evolve into regions where the evolution equations turn elliptic or parabolic and the theory and formulation used break down.

    \item Chapter \ref{chapter:paper2} focused on studying binary black holes in cubic Horndeski theories, using couplings and initial data in the regime of validity found via chapter \ref{chapter:paper1}. We identified a mismatch of the gravitational wave strain between Horndeski and GR (coupled to a scalar field) as large as $10-13\%$ for elliptic binaries and more than $30\%$ for circular binaries, in the Advanced LIGO mass range. Being a non-local observable, this is due to a distinct frequency spectrum and, as a consequence an accumulated phase shift over time much smaller than the local deviations Horndeski introduces. These effects are likely generic to many other theories, suggesting that many events might be undetected by current detectors if certain conditions are present, such as the existence of a scalar cloud in the vicinity of the black hole galactic environment, for the case of cubic Horndeski. They can be detected and the couplings can be constrained in future more precise observations if one can detect gravitational waves produced in early stages of a binary, extract its parameters in order to predict the time of merger according to GR, and observe potential deviations between the GR prediction and the observed merger time.

    We confirmed certain weak field conditions are satisfied at all times, including during the merger phase. We concluded that certain cases have the degeneracy between Horndeski and GR broken, even though we observe degeneracy between different Horndeski theories, in the cases $G_2$ and $G_3$ analysed.

    \item In chapter \ref{chapter:eft}, the goal was to simulate fully non-linearly black hole binaries in theories with higher derivative corrections to GR, again, with the hope that small local effects at low energies can build up over time and leave an imprint in non-local observables such as gravitational waves.

    With well-posedness clearly gone out of the window for these theories, we made use of the fixing method described in section \ref{sec:gr:fixing_hd_theories}. After gaining intuition from a toy model adding higher derivatives to a wave equation, we build several systems capable of reproducing the low energy behaviour of the theory while damping high energy frequencies.

    We analysed the accuracy of several fixing systems, their invariance to changes in unphysical parameters and obtain an evolution error of $\mathcal{T}=0.1\%$ for the best model in simulations of single boosted black holes. For binary black holes, this lack of accuracy was a more significant effect than the local changes due to $\e=10^{-5}$, leading to a mismatch of $-0.39 \pm 4.10\%$, compatible statistically with zero.

    Our work demonstrated that phase shifts seem to be generic effects of modified theories, yet emphasises the importance that any research using the fixing procedure must verify if it accurately tracks the physical variables it claims to approach.

    \item Finally, in chapter \ref{chapter:ahfinder}, we presented \texttt{AHFinder}, an open-source and user-friendly tool developed in the context of the \texttt{GRChombo} collaboration to identify apparent horizons in multiple spacetimes with high parallelisation. This includes single or multiple moving black holes and its mergers, cosmological horizons and also higher dimensional black holes, such as black strings, as it supports Cartoon reductions \cite{Cook:2016soy}. The \texttt{AHFinder} can find any contour in any user-defined level set surface in any number of dimensions, and integrate any custom quantities over the converged surfaces.
\end{itemize}

\section{\done{Future work}}

Based on the conclusions drawn above, further work can be carried out in the following topics:
\begin{itemize}
    \item Regarding the analysis done about gravitational collapse in chapter \ref{chapter:paper1}, it would be very relevant to perform the same study for massive scalar fields, as we found this was essential for the findings of \ref{chapter:paper2}. Analysing other $G_2$ and $G_3$ theories is also of interest, as some theories are known not to have hyperbolicity concerns \cite{Bezares:2020wkn}. Finding more general conditions in the initial data considered and other classes of initial data is also of interest, to pursue a more complete map of the Horndeski theory landscape.
    \item Regarding chapter \ref{chapter:paper2}, besides investigating gravitational waves in more Horndeski theories or in systems with spin and unequal mass, it can be very fruitful to analyse binary black holes with initial data other than scalar field bubbles collapsing to black holes. For instance, considering black holes already formed surrounded by scalar field halos may be a physical situation more likely to approximate the real universe. Furthermore, understanding the degeneracy of gravitational waves in modified theory versus GR is of great importance. Additionally, it is important to try to match theoretical predictions as the ones found in this thesis to observational data, in order to (further) constrain the size of the couplings of several Horndeski theories. Finally, besides the work done for Horndeski theories, understanding the detectability of scalar fields (even for minimally coupled scalar fields, as in Barsanti et al. \cite{Barsanti:2022vvl}), opposed to vacuum waveform predictions, is also very pertinent.
    \item We consider that studies made in chapter \ref{chapter:eft} are very promising to study many EFTs that can lead to important discoveries of how GR is extended and how extensions are constrained. It is important to continue the exploration of the fixing method to find new ways of performing the procedure that are both more accurate and that allow the use of bigger coupling constants. It is also relevant, on the one hand, to verify that indeed the form of the system chosen is not unique and not crucial, and on the other hand, to systematise the confirmation of the accuracy of a given new proposed system or of a system previously successful being applied to a new theory.
\end{itemize}

On the computational front, several possibilities would allow to extend this work and open many opportunities to study a wide range of problems in numerical relativity. Namely:
\begin{itemize}
    \item The \texttt{AHFinder} is publicly available, but the task of adding it to the main \texttt{GRChombo} code is in progress, as well as adding documentation for all its use cases and capabilities.
    \item Developing within the \texttt{GRChombo} collaboration a 1D gravity code that is easy to integrate with existing classes can prove miraculously useful and speed up our research tremendously, as much of our research is or can be done in spherical symmetry, for which a 3D code is not required.
    \item Concluding the integration in the public open-source \texttt{GRChombo} code of the 2D Cartoon code started originally by Pau Figueras for studies in higher dimensions \cite{Cook:2016soy,Figueras:2017zwa,Figueras:2015hkb} and used during my research for other projects \cite{boosted_head_ons} can be very useful to speedup 3D simulations with asymmetry, as is the case of head-on mergers, and hence speedup research of various groups as well. Adding matter sources to the Cartoon code is also of interest.
    \item Katy Clough is directing an initiative to aggregate modified gravity codes under the umbrella of \texttt{GRChombo}. Adding the Horndeski and the EFT codes developed for this thesis could be incredibly fruitful to lower the barrier of entry in studying modified gravity theories fully non-linearly.
    \item Finally, though preliminary steps have been taken in joint work with Llibert Aresté Saló, Justin Ripley and Pau Figueras, the addition of the modified generalised harmonic gauge \cite{Kovacs:2020pns, Kovacs:2020ywu} to the \texttt{GRChombo} public code, either in a full $4D$ or in a $3+1$ re-formulation of it in a similar way to the work by Brown \cite{Brown:2011qg,Brown:2009dd}, may be extremely useful in order to create an easier formulation of GR which does not require conformal decompositions. It would also allow the evolution of new theories such as the $4\pd ST$ theory mentioned in section \ref{sec:gr:modified_gravity} and as done in a conformal way in \cite{AresteSalo:2022hua}.
\end{itemize}

In summary, I am confident that the research described in this thesis presents novel and unexpected results, proposes new numerical methods and opens new avenues for the exploration of modified gravity theories and their effect on gravitational waves, with the use of numerical relativity, for us to hopefully find, one day, what Einstein missed.


\part{Extra material}
\blankpage

\begin{appendix}
\counterwithin{figure}{section} 

\chapter{\done{Gravitational collapse in cubic Horndeski theories}}\label{appendix:paper1}


\section{\done{$3+1$ conformal decomposition}}

\subsection{\done{Equations of motion}}\label{appendix:paper1:EOM}

In this appendix we present the conformal 3+1 form of the stress tensor and the scalar equation \eqref{eq:scalar} as we have implemented in our code. 

Consider the usual timelike vector $n^\m$ normal to the spatial hypersurfaces; the projector $\gamma_{\m\n}=g_{\m\n}+ n_\m n_\n$ defines the spatial 3-metric $\gamma_{ij}$ with the corresponding covariant derivative $D_i$. From these, we obtain the following decomposition for the first derivatives of the scalar field:
\begin{align}
    \P &:= \mathcal{L}_n \f = n^\m \grad_\m \f\,, \label{eq:dtphi}\\
    \quad\quad\quad \P_i &:= D_i \f\,,
\end{align}
where $\mathcal{L}_n$ denotes the Lie derivative along $n^\m$. It follows that $\grad_\m\f=\P_\m - n_\m\P$ and $X=\tfrac{1}{2}\br{\P^2 - \P_i\P^i}$. We also decompose the second derivatives of the scalar field, defining the auxiliary variables:
\begin{gather}
\begin{aligned}
    \overline{\mathcal{L}_{\ve{n}}\Pi} &:= n^\mu n^\nu \grad_\mu \grad_\nu \phi = \mathcal{L}_{\ve{n}}\Pi - \Pi_{i} D^{i}\ln{\alpha}\,,\\
    \tau_{i} &:= \gamma_i^\mu n^\nu \grad_\mu \grad_\nu \phi = K_{ij} \Pi^{j} + D_{i}\Pi\,,\\
    \tau_{ij} &:= \gamma_i^\mu \gamma_j^\nu \grad_\mu \grad_\nu \phi = K_{ij} \Pi + D_{(i}\Pi_{j)}\,, 
    \label{eq:auxvars}
\end{aligned}
\end{gather}
and hence $\tau := \tau^{~i}_i = K\Pi + D^i\Pi_i$. Therefore, we get
\begin{equation}
\begin{aligned}
    \nabla_{\mu}\nabla_{\nu}\phi =&~ \overline{\mathcal{L}_{\ve{n}}\Pi}\,n_\mu\, n_\nu - 2\,n_{(\mu}\tau_{\nu)} + \tau_{\mu\nu}\,,\\
    \square\phi =&~ \tau - \overline{\mathcal{L}_{\ve{n}}\Pi}\,,
\end{aligned}
\end{equation}
with $n^\mu \tau_\mu = 0$ and $n^\mu \tau_{\mu\nu}=0$.
In terms of the usual conformal spatial metric $\ti{\gamma}_{ij}:=\chi\gamma_{ij}$ (with $\det(\ti{\gamma}_{ij})=1$) and its associated covariant derivative $\ti{D}_i$, we define the conformal variables for the scalar field as,
\begin{gather}
\begin{aligned}
    \ti{\P}_i &:= \ti{D}_i\f\,, \quad\quad\quad \ti{\tau}_{i} &:= \tau_i\,, \quad\quad\quad \ti{\tau}_{ij} &:= \chi\tau_{ij}\,.
\end{aligned}
\end{gather}
Note that the indices of $\ti{\tau}_{ij}$ are raised with the conformal metric $\ti{\gamma}_{ij}$ so that $\ti{\t} := \ti{\t}^{~i}_i = \t$, and similarly for all other conformal variables. For example, $\ti{\P}^i = \frac{1}{\chi}\P^i$, which implies $X = \tfrac{1}{2}(\Pi^2 - \chi\ti{\Pi}_i\ti{\Pi}^i)$. With these definitions in place, the 3+1 conformal decomposition of the scalar energy-momentum tensor is:
\begin{align}\label{eq:appendix:energy_density}
    \kappa \rho :=&~ \kappa\,n^\m n^\n T_{\m\n} \nonumber \\
    =&~ V - G_2 + \tfrac{1}{2}\big(\Pi^2 + \chi\ti{\Pi}_{i} \ti{\Pi}^{i}\big)\br{1+2\partial_\phi G_3} + \partial_X G_3(\ti{\tau} \Pi^2 -  \chi\ti{\Pi}^{i} \ti{\Pi}^{j} \ti{\tau}_{ij})  + \Pi^2 \partial_X G_2\,, \\\label{eq:appendix:Si}
    \kappa S_{i} :=& -\kappa\,n^\m \g_i^{~n} T_{\m\n} \nonumber \\
    =& -\Pi~\ti{\Pi}_{i} \bigl(1 + \partial_X G_2 + 2 \partial_\phi G_3\bigr) + \partial_X G_3(\chi\ti{\Pi}_{i} \ti{\Pi}^{j} \ti{\tau}_{j} + \Pi ~\ti{\Pi}^{j} \ti{\tau}_{ij} - \ti{\tau} \Pi ~\ti{\Pi}_{i} -  \Pi^2 \ti{\tau}_{i})\,,\\
    \kappa S_{ij} :=&~ \kappa\,\g_i^{~\m} \g_j^{\n}T_{\m\n} \nonumber \\\label{eq:appendix:Sij}
    =&~ \ti{\Pi}_{i} \ti{\Pi}_{j} \bigl(1 + \partial_X G_2 + 2 \partial_\phi G_3\bigr) + \tfrac{1}{\chi}\ti{\gamma}_{ij} \bigl(G_2 - V + X + 2 X \partial_\phi G_3\bigr)\\
    & +\partial_X G_3 \Big[\ti{\tau} \ti{\Pi}_{i} \ti{\Pi}_{j} + 2\Pi~ \ti{\Pi}_{(i} \ti{\tau}_{j)} - 2\ti{\Pi}^{k}\ti{\Pi}_{(i} \ti{\tau}_{j)k} -  \ti{\gamma}_{ij} \ti{\Pi}^{k} (2 \Pi \ti{\tau}_{k} -  \ti{\Pi}^{l} \ti{\tau}_{kl}) \nonumber \\
    &\hspace{1.8cm}+ \overline{\mathcal{L}_{\ve{n}}\Pi}(\tfrac{1}{\chi}\ti{\gamma}_{ij}\Pi^2 - \ti{\Pi}_i\ti{\Pi}_j)\Big]\,, \nonumber
\end{align}
Similarly, the scalar field evolution equation \eqref{eq:scalar} in first order form is given by \eqref{eq:dtphi} and:
\begin{equation}
\begin{aligned}
    &\overline{\mathcal{L}_{\ve{n}}\Pi}\Big[1 + \partial_X G_2 + 2 \partial_\phi G_3 + 2 \ti{\tau} \partial_X G_3 - X^2 \bigl(\partial_X G_3\bigr)^2 - \chi\ti{\Pi}^{i} \ti{\Pi}^{j} \ti{\tau}_{ij} \partial^2_{XX} G_3 - 2X\partial^2_{\phi X} G_3 \\
    &~~~~~~~ + \Pi^2\br{2X\bigl(\partial_X G_3\bigr)^2 + \partial^2_{XX} G_2 + \ti{\tau}\partial^2_{XX} G_3 + 2\partial^2_{\phi X} G_3}\Big] =\\
    =~& \partial_\phi G_2 - \partial_\phi V + \ti{\tau} \sbr{1 + \partial_X G_2 + 2 \partial_\phi G_3 + \ti{\tau} \partial_X G_3 - X^2\bigl(\partial_X G_3\bigr)^2 -2X\partial^2_{\phi X} G_3} \\
    &+\sbr{\partial^2_{XX} G_2 + 2\partial^2_{\phi X} G_3 + 2X\bigl(\partial_X G_3\bigr)^2 + \ti{\tau} \partial^2_{XX} G_3}\chi(2\Pi~ \ti{\Pi}^{i} \ti{\tau}_{i} - \ti{\Pi}^{i} \ti{\Pi}^{j} \ti{\tau}_{ij}) \\
    &- (\partial^2_{\phi X} G_2 + \partial^2_{\phi\phi}G_3) 2 X \\
    &+ \chi\partial^2_{XX} G_3\sbr{(\Pi\ti{\tau}_i - \ti{\Pi}^j\ti{\tau}_{ji})(\Pi\ti{\tau}^i - \ti{\Pi}_k\ti{\tau}^{ki}) - \chi \ti{\Pi}^{i} \ti{\Pi}^{j} \ti{\tau}_{i} \ti{\tau}_{j}} \\\label{eq:Horndeski_final}
    &-\partial_X G_3 \sbr{ G_2 X - 2\chi \ti{\tau}_{i} \ti{\tau}^{i} + \ti{\tau}_{ij} \ti{\tau}^{ij} + X^2 (2 + \partial_X G_2 + 4 \partial_\phi G_3)}\,.
\end{aligned}
\end{equation}
Note that one can obtain the standard 3+1 evolution equations without a conformal transformation by setting $\chi=1$ and dropping any ` $\ti{}$ ' superscripts.

\subsection{\done{Effective metric}}\label{appendix:paper1:effective_metric}

As discussed in section \ref{subsec:paper1:effective_metric}, the quantities $-\a^2 h^{00}$ and $\text{det}\br{h^\m_{~\n}}$ are useful to monitor the hyperbolicity of the scalar equation of motion and determine whether its change of character is of the  Tricomi or Keldysh type. Here we present $-\a^2 h^{00}$ and $h^\m_{~\n}$ in terms of the 3+1 conformal variables, which is how we have calculated them in our code:
\begin{equation}
\begin{aligned}
    h^0_{~i} =&~ \frac{1}{\alpha}\Big\{\Pi~\ti{\Pi}_i\Big[2\,X\,\bigl(\partial_X G_3\bigr)^2 + \partial^2_{XX} G_2 + \ti{\tau}\,\partial^2_{XX} G_3 + 2\,\partial^2_{\phi X} G_3\Big] \\
        &\hspace{0.7cm} + \ti{\tau}_i\br{2\,\partial_X G_3 + \Pi^2\, \partial^2_{XX}G_3}-\partial^2_{XX}G_3\br{\ti{\Pi}^k\,\ti{\tau}_{ki}\,\Pi + \chi\,\ti{\Pi}_i\,\ti{\Pi}^k\,\ti{\tau}_k} \Big\}\,,\\
    h^0_{~0} =&~ \beta^k h^0_{~k} -\a^2 h^{00} \,,\\
    h^i_{~j} =&~ -\beta^i h^0_{~j} - \chi\,\ti{\Pi}^i\,\ti{\Pi}_j\Big[2X\bigl(\partial_X G_3\bigr)^2 + \partial^2_{XX} G_2 + \ti{\tau}\,\partial^2_{XX} G_3 + 2\,\partial^2_{\phi X} G_3\Big]\\
        & + \delta^i_{~j}\Big[ 1 + \partial_X G_2 + 2\, \partial_\phi G_3 + 2\,\ti{\tau}\, \partial_X G_3 - X^2 \bigl(\partial_X G_3\bigr)^2 
         - 2\,X\,\partial^2_{\phi X} G_3\\
        &\hspace{1.2cm} - \chi\,\ti{\Pi}^{k}\, \ti{\Pi}^{l}\, \ti{\tau}_{kl}\, \partial^2_{XX} G_3 + 2\,\chi\,\Pi~\ti{\Pi}^k\,\ti{\tau}_k\,\partial^2_{XX}G_3\Big] \\
        &- 2\,\partial_X G_3\,\ti{\tau}^i_{~j} + \chi\,\partial^2_{XX}G_3\br{\ti{\Pi}^{i}\,\ti{\tau}_{jk}\,\ti{\Pi}^k + \ti{\Pi}_{j}\,\ti{\tau}^{ik}\,\ti{\Pi}_k - \Pi~\ti{\Pi}^{i}\ti{\tau}_{j} - \Pi~\ti{\Pi}_{j}\,\ti{\tau}^{i}}\\
        & -\overline{\mathcal{L}_{\ve{n}}\Pi}\sbr{\delta^i_{~j}\br{2\,\partial_X G_3 + \Pi^2 \partial^2_{XX} G_3} - \chi\,\ti{\Pi}^i\,\ti{\Pi}_j\,\partial^2_{XX}G_3} \,, \\
    h^i_{~0} =&~ \beta^k h^i_{~k } - \alpha^2 \chi\ti{\gamma}^{ik}h^0_{~k} + \a^2 h^{00}\,\beta^i\,,
\end{aligned}
\label{eq:hud}
\end{equation}
and,
\begin{equation}
\begin{aligned}
    -\a^2 h^{00} =&~ 1 + \partial_X G_2 + 2\, \partial_\phi G_3 + 2\,\ti{\tau}\, \partial_X G_3 - X^2 \bigl(\partial_X G_3\bigr)^2 \\
    &- \chi\,\ti{\Pi}^{i}\, \ti{\Pi}^{j}\, \ti{\tau}_{ij}\, \partial^2_{XX} G_3 - 2X\partial^2_{\phi X} G_3 \\
        & + \Pi^2\big[2\,X\bigl(\partial_X G_3\bigr)^2 + \partial^2_{XX} G_2 + \ti{\tau}\,\partial^2_{XX} G_3 + 2\,\partial^2_{\phi X} G_3\big]\,.
\end{aligned}
\end{equation}
From \eqref{eq:hud} one can readily compute $\det(h^\m_{~\n})$. If necessary, the effective metric with both indices up can also be obtained by raising the lower index in \eqref{eq:hud} with the spacetime metric. For completeness, one can write the extra inverse relations with the raised metric, convenient for numerical purposes:
\begin{equation}
\begin{aligned}
    h^{0i} &= -\b^i h^{00} + \chi\, \ti{\g}^{ik} h^0_{~k}\,,\\
    h^{ij} &= \beta^i\beta^j h^{00} - \chi\,\beta^{j}\ti{\gamma}^{ik}h^0_{~k} + \chi\ti{\gamma}^{jk}h^i_{~k}\,.
\end{aligned}
\end{equation}
%

\section{\done{Determinant of the effective metric}}

We can compute $\text{det}\br{h^\m_{~\n}}$ in full generality using Cayley–Hamilton's theorem and Newton's identities. The general case, with both $G_2\neq 0$ and $G_3\neq 0$, is not particularly insightful and in practice it is preferable to directly compute the  determinant of the metric with a lowered index numerically. For use in section \ref{subsection:paper1:case_g3}, in this appendix we provide the explicit expression for the determinant in the case $G_2=0$ and $G_3=g_3\,X$:
\begin{gather}\nonumber
\begin{aligned}
    \text{det}\br{h^\m_{~\n}} =&~ 1 + 6g_3\square\f\,+\\
    & + g_3^2\big[14\br{\Box\f}^2-2\br{\grad_\m\grad_\n\f}\br{\grad^\m\grad^\n\f}\big]\\
    & + g_3^3\bigg[\tfrac{44}{3}\br{\Box\f}^3 - 2\Box\phi\br{2\br{\grad_\m\grad_\n\f}\br{\grad^\m\grad^\n\f} + X^2}\\
    &\hspace{1cm}-4X\br{\grad^\m\f}\br{\grad^\n\f}\br{\grad_\m\grad_\n\f}
      - \tfrac{8}{3}\br{\grad_\m\grad_\n\f}\br{\grad^\m\grad^\r\f}\br{\grad^\n\grad_\r\f}\bigg]
\end{aligned}\\
\begin{aligned}
    & + g_3^4\Big[6\br{\Box\f}^4-4\br{\Box\f}^2 \br{X^2+\br{\grad^\m\grad^\n\f}\br{\grad_\m\grad_\n\f}} - 6X^4 \\
    &\hspace{1cm}-8\Box\f X \br{\grad^\m\f}\br{\grad^\n\f}\br{\grad_\m\grad_\n\f} 
    -8X\br{\grad^\m\f}\br{\grad^\n\f}\br{\grad_\m\grad_\r\f}\br{\grad^\r\grad_\n\f}\\
    &\hspace{1cm} -4\br{\grad^\m\grad_\n\f}\br{\grad^\n\grad_\r\f}\br{\grad^\r\grad_\s\f}\br{\grad^\s\grad_\m\f}\\
    & \hspace{1cm} + \br{\grad^\m\grad^\n\f}\br{\grad_\m\grad_\n\f}\br{-4X^2+2\br{\grad^\r\grad^\s\f}\br{\grad_\r\grad_\s\f}}\Big]\\
    & + g_3^5\Big[2\Box\f X^2\br{2\br{\grad^\m\grad^\n\f}\br{\grad_\m\grad_\n\f}-7X^2}-4\br{\Box\f}^3 X^2 \\
    & \hspace{1cm} + 8X^2\br{\grad^\m\grad_\n\f}\br{\grad^\n\grad_\r\f}\br{\grad^\r\grad_\m\f}\\
    & \hspace{1cm}-16X\br{\grad^\m\f}\br{\grad^\n\f}\br{\grad^\r\grad_\m\f}\br{\grad^\s\grad_\n\f}\br{\grad_\r\grad_\s\f}\\
    & \hspace{1cm} + 8X \br{\grad^\m\f}\br{\grad^\n\f}\br{\grad_\m\grad_\n\f}\br{-\br{\Box\f}^2+ X^2 + \br{\grad^\r\grad^\s\f}\br{\grad_\r\grad_\s\f}}\Big]\\
    & + g_3^6\Big[8\Box\f X^3 \br{\grad^\m\f}\br{\grad^\n\f}\br{\grad_\m\grad_\n\f} \\
    &\hspace{1cm}+ 3X^4 \br{2\br{\grad^\m\grad^\n\f}\br{\grad_\m\grad_\n\f}-5\br{\Box\f}^2} \\
    & \hspace{1cm} + 8X^6 + 8X^3\br{\grad^\m\f}\br{\grad^\n\f}\br{\grad_\m\grad_\r\f}\br{\grad_\n\grad^\r\f} \Big]\\
    & + g_3^7 X^5\br{10\Box\f X - 4 \br{\grad^\m\f}\br{\grad^\n\f}\br{\grad_\m\grad_\n\f}} - 3g_3^8 X^8\,.
\end{aligned}
\label{eq:det_full_g3}
\end{gather}

\section{\done{Solving the Hamiltonian and Momentum constraints}}\label{appendix:paper1:initialData}

In Horndeski, for non-zero $G_2$ and/or $G_3$, even a scalar field with zero momentum will have non-trivial momentum constraints. In vacuum GR, this is not the case, and one can often get away with simple solutions such as $K=\ti{A}_{ij}=0$, since $S_i=0$ for a minimally coupled scalar field. In GR, for the leftover Hamiltonian constraint, several solutions are often introduced: relaxing the conformal factor $\chi$ with the Hamiltonian constraint (this has slow convergence, but works as a simple method); or for instance taking advantage of CCZ4 damping of the constraints assuming that the initial constraints are very small (which is true if the scalar field has low density). The later could work for Horndeski as well, but given the objective of chapter \ref{chapter:paper1} was to explore the evolution of gravitational collapse when varying the initial data parameter space, it was desirable to remove potential sources of error such as constraints in the initial data. For Horndeski, to solve the joint Hamiltonian and Momentum constraints, we used the \textit{conformal transverse-traceless decomposition} (introduced in section \ref{subsec:ngr:conformalTT}). Note how Baumgarte and Shapiro \cite[p.~64]{shapiro} use a different rescaling for $\ti{A}_{ij}$ based on the conformal factor (as mentioned in section \ref{subsec:ngr:conformalTT}).

For the purpose of chapter \ref{chapter:paper1}, we considered spherically symmetric initial data. This has the advantage of reducing the 3 Momentum constraints to 1, in the radial direction. In the case of non-vacuum equations with non trivial $S_i$, the constraints are coupled elliptic equations that we solve numerically in \textit{Mathematica}. Following the naming in \ref{subsec:ngr:conformalTT}, as typical choices for the degrees of freedom, choose the conformal metric to be conformally flat (5 degrees of freedom), choose the trace of the extrinsic curvature to be zero (called maximal slicing, see section \ref{sec:ngr:gauge}), $K=0$ (1 degree of freedom) and the divergence-less traceless part of the extrinsic curvature to be zero, $\ti{A}^{ij}_{TT}=0$ (2 degrees of freedom). This leaves us with 4 equations to solve and 4 degrees of freedom. These are the conformal factor, $\chi$, and the 3 components of the vector potential, $W^i$. Noting that given conformal flatness, $\ti{R}^i_{~j}=0$ and, in Cartesian coordinates, the constraints in 4D reduce to:
\begin{equation}
\begin{aligned}
    \mathcal{H} &= \tfrac{1}{2}\sbr{2\partial^i\partial_i\chi-\tfrac{5}{2\chi}\partial^i\chi\partial_i\chi - \ti{A}_{ij}\ti{A}^{ij}} - \tfrac{\k}{2} \rho =0\,,\\
    \mathcal{M}_i &= \partial^j\partial_jW_i + \tfrac{1}{3}\partial_i\partial^jW_j-\tfrac{3}{2}\ti{A}_{ij}\ti{D}^j\ln{\chi}-\tfrac{\k}{2} S_i = 0\,.
\end{aligned}
\end{equation}
Let us now turn to spherical coordinates. In conformal flatness, the spherical coordinate system $\bar{x}^i = (r,\theta,\phi)$ has its metric in the form:
\begin{equation}\label{eq:metricspherical}
    \ti{\gamma}_{ij}=
        \begin{pmatrix}
            1 & 0 & 0\\
            0 & r^2 & 0\\
            0 & 0 & r^2 \sin{\theta}^2
        \end{pmatrix}\,.
\end{equation}
Let us assume that the vector potential has radial symmetry, $W_\theta=0,~W_\phi=0$, meaning that $W_i = (W_r, 0, 0)$. Any such vector has covariant derivatives as follows\footnote{It applies equivalently to a scalar field second covariant derivatives $\ti{D}_i\ti{D}_j\phi$, using $W_j \equiv \ti{D}_j\phi$.}:
\begin{equation}\label{eq:derWr}
   \ti{D}_i W_j =
    \begin{pmatrix}
        \partial_r W_r & 0 & 0\\
        0 & rW_r & 0\\
        0 & 0 & r\sin{\theta}^2W_r
    \end{pmatrix}\,.
\end{equation}
From this we can conclude the following useful expressions\footnote{See \href{http://mathworld.wolfram.com/SphericalCoordinates.html}{Wolfram MathWorld - Spherical Coordinates} expression 93 for the vector Laplacian. Expressions 53-61 have also proved useful.}:
\begin{equation}
\begin{aligned}
    \ti{D}_iW^i &= \partial_r W_r + \tfrac{2W_r}{r} = \tfrac{1}{r^2}\partial_r\br{r^2 W_r}\,,\\
    \ti{D}^2\phi &= \partial_r^2\phi + \tfrac{2}{r}\partial_r\phi = \tfrac{1}{r^2}\partial_r\br{r^2\partial_r\phi} = \tfrac{1}{r}\partial_r^2\br{r\phi}\,,\\
    \br{\ti{D}^j\ti{D}_jW_i}_r &= \tfrac{1}{r}\partial_r^2\br{rW_r} - \tfrac{2W_r}{r^2} = \partial_r^2W_r + \tfrac{2}{r}\partial_rW_r - \tfrac{2W_r}{r^2} =\\  &=\partial_r\br{\partial_rW_r + \tfrac{2W_r}{r}} = \br{\ti{D}_i\ti{D}^jW_j}_r\,.
\end{aligned}
\end{equation}
From a radial vector potential, one can show that, from equations \eqref{eq:ngr:AijConformalTT}-\eqref{eq:ngr:AijToW}, $\ti{A}_{ij}$ can be written in spherical coordinates as:
\begin{equation}
   \ti{A}_{ij} = \tfrac{4}{3}\br{\partial_r W_r - \tfrac{W_r}{r}}
    \begin{pmatrix}
        1 & 0 & 0\\
        0 & -\frac{r^2}{2} & 0\\
        0 & 0 & -\frac{r^2\sin{\theta}^2}{2}
    \end{pmatrix} = \ti{A}_{rr}
    \begin{pmatrix}
        1 & 0 & 0\\
        0 & -\frac{r^2}{2} & 0\\
        0 & 0 & -\frac{r^2\sin{\theta}^2}{2}
    \end{pmatrix}\,.
\end{equation}
Looking at the form of \ref{eq:metricspherical}, $\ti{A}_{ij}$ can be seen as clearly traceless. One can also see that the contraction of $\ti{A}_{ij}$ with itself yields $\ti{A}_{ij}\ti{A}^{ij}=\tfrac{3}{2}\ti{A}_{rr}^2$.
One can now re-write the constraints to solve in spherical coordinates as:
\begin{align}\label{eq:Hamspherical}
    2\mathcal{H} &= \tfrac{2}{r}\partial_r^2\br{r\chi}-\tfrac{5}{2\chi}\br{\partial_r\chi}^2 - \tfrac{3}{2}\ti{A}_{rr}^2 - \k\rho =0\,,\\\label{eq:Momspherical}
    \mathcal{M}_r &= \tfrac{4}{3}\br{\partial_r^2W_r + \tfrac{2}{r}\partial_rW_r - \tfrac{2W_r}{r^2}}-\tfrac{3}{2\chi}\ti{A}_{rr}\partial_r\chi-\tfrac{\k}{2} S_r = 0\,.
\end{align}
This is a system of two couple equations solving for the conformal factor, $\chi$ and the radial lowered vector potential, $W_r$. Let us now worry about the matter terms. Following equations \eqref{eq:appendix:energy_density}-\eqref{eq:appendix:Si}, and the choice $G_2=g_2X^2$, $G_3=g_3X$ and $V(\f)=\frac{1}{2}m^2\phi^2$ as described for chapters \ref{chapter:paper1}-\ref{chapter:paper2} (see \ref{subsec:paper1:cases_explored} and \cite{Hui:2019aqm}) , one gets:
\begin{equation}
\begin{aligned}
    \k\rho &= \tfrac{1}{2}\br{\Pi^2+\chi\ti{\Pi}_i\ti{\Pi}^i} + g_2\br{-X^2+2X\Pi^2}+\tfrac{1}{2}m^2\phi^2+g_3\br{\ti{\tau}\Pi^2 -\chi \ti{\Pi}^i\ti{\Pi}^j\ti{\tau}_{ij}}\,,\\
    \k S_i &= -\Pi~\ti{\Pi}_i\br{1 + 2g_2X} + g_3\br{\chi\ti{\Pi}_i\ti{\Pi^j}\ti{\tau}_j + \Pi~\ti{\Pi}^j\ti{\tau}_{ij} - \ti{\tau}\Pi~\ti{\Pi}_i - \Pi^2\ti{\tau}_i}\,.
\end{aligned}    
\end{equation}
For the chosen initial data:
\begin{equation}
\begin{aligned}
    X &= \tfrac{1}{2}\sbr{\Pi^2 - \br{\partial_r\phi}^2}\,,\\
    \ti{\tau} &= \chi\ti{D}^2\phi - \tfrac{1}{2}\ti{\Pi}^k\ti{D}_k\chi = \tfrac{\chi}{r}\partial_r^2\br{r\phi}-\tfrac{1}{2}\partial_r\phi\partial_r\chi\,,\\
    \ti{\tau}_i &= \ti{A}_{ij}\ti{\Pi}^j + \ti{D}_i\Pi \Longrightarrow \ti{\tau}_r = \ti{A}_{rr}\partial_r\phi - \partial_r\Pi~~\sbr{\ti{\tau}_\theta=0=\ti{\tau_\phi}}\,,\\
    \ti{\tau}_{ij} &= \ti{A}_{ij}\Pi + \chi\ti{D}_i\ti{\Pi}_j + \ti{\Pi}_{(i}\ti{D}_{j)}\chi - \tfrac{1}{2}\ti{\gamma}_{ij}\ti{\Pi}^k\ti{D}_k\chi ~~\text{[diagonal]}\\
    &\Longrightarrow \ti{\tau}_{rr} = \ti{A}_{rr}\Pi + \chi\partial_r^2\phi + \tfrac{1}{2}\partial_r\phi\partial_r\chi\,,\\
    \ti{\Pi}^i\ti{\Pi}^j\ti{\tau}_{ij} &= \ti{\Pi^r}\ti{\Pi}^r\ti{\tau}_{rr} = \br{\partial_r\phi}^2\ti{\tau}_{rr}\,,\\
    \ti{\Pi}^j\ti{\tau}_j &= \ti{\Pi}^r\ti{\tau}_r = \partial_r\phi \ti{\tau}_r\,,\\
    \br{\ti{\Pi}^j\ti{\tau}_{ij}}_r &= \ti{\Pi}^r\ti{\tau}_{rr} = \partial_r\phi\ti{\tau}_{rr}~~\sbr{\br{\ti{\Pi}^j\ti{\tau}_{ij}}_\theta=0=\br{\ti{\Pi}^j\ti{\tau}_{ij}}_\phi}\,.
\end{aligned}
\end{equation}
There is one left over issue to result: boundary conditions. For $\chi$, one can easily choose $\partial_r\chi\big|_{r=0}=0$ (due to spherical symmetry) and $\chi\big|_{r=\infty}=1$. For $W_r$, one can find its boundary conditions by analysing the boundary conditions we wish for $\ti{A}_{ij}$. These are: $\partial_r\ti{A}_{ij}\big|_{r=0}=0$ and $\ti{A}_{ij}\big|_{r=\infty}=0$. Expanding $\ti{A}_{ij}$ and its derivatives as a function of $W_r$, we get:
\begin{multicols}{2}
    \begin{align}\label{eq:Arr}
        \ti{A}_{rr} &= \tfrac{4}{3}\br{\partial_rW_r - \tfrac{W_r}{r}}\,,\\
        \ti{A}_{\theta\theta} &= -\ti{A}_{rr}\tfrac{r^2}{2}\,,\\
        \ti{A}_{\phi\phi} &= \ti{A}_{\theta\theta}\sin{\theta}^2\,,
    \end{align}
    
    \begin{align}\label{eq:drArr}
        \partial_r\ti{A}_{rr} &= \tfrac{4}{3}\br{\partial_r^2W_r - \tfrac{\partial_rW_r}{r} + \tfrac{W_r}{r^2}}\,,\\\label{eq:drAqq}
        \partial_r\ti{A}_{\theta\theta} &= -\partial_r\ti{A}_{rr}\tfrac{r^2}{2}-\ti{A}_{rr}r\\\nonumber
        &= -\tfrac{2r^2}{3}\br{\partial_r^2W_r + \tfrac{\partial_rW_r}{r} - \tfrac{W_r}{r^2}}\,,\\
        \partial_r\ti{A}_{\phi\phi} &= \partial_r\ti{A}_{\theta\theta}\sin{\theta}^2\,.
    \end{align}
\end{multicols}
We explicitly show 2 ways of deriving the same conclusions:
\begin{enumerate}
    \item First, looking at leading order terms. If we want \ref{eq:Arr} to be zero at infinity, the leading order term $\partial_rW_r$ has to be zero at infinity, and $W_r$ can take any finite value. Then, if we want \ref{eq:drArr} and \ref{eq:drAqq} to be zero at the origin, the leading order term at the origin implies $W_r\big|_{r=0} = 0$.
    \item Second, looking at the full solutions. The solution to $\ti{A}_{rr}=0$ is $W_r = C\,r$ as $r\to\infty$, for some constant $C$. For it to be bounded at infinity, we choose $C = 0$ and hence $W_r\big|_{r=\infty}=0$. The solutions to $\partial_r\ti{A}_{rr}=0$ and $\partial_r\ti{A}_{\theta\theta}=0$ are, respectively, $W_r=A\,r + B\,r\ln{r}$ and $W_r=A'\,r + \frac{B'}{r}$, for constants $A,A',B,B'$. Requiring regularity at $r = 0$, this implies $B=B'=0$ for both cases and hence $W_r = A\,r$ as $r$ tends to 0, i.e. $W_r|_{r=0}=0$ (with $\pd_r W_r$ some arbitrary constant).
\end{enumerate}
We hence conclude that the boundary conditions for this problem are:
\begin{gather}\label{eq:appendix:paper1:bcs}
    \partial_r\chi\big|_{r=0}=0,~~~\chi|_{r=\infty}=1,~~~W_r\big|_{r=0} = 0,~~~W_r|_{r=\infty}=0\,.
\end{gather}
In practical purposes, numerically, we took $r=0$ to be a very small number ($r_0 = 10^{-4}$) and $r = \infty$ a large enough number ($r_\infty = 5000$), solving the equations between $r\in[r_{min} = r_0,~r_{max} = r_\infty]$. 

Having solved the 2 constraint equations \ref{eq:Hamspherical} and \ref{eq:Momspherical}, one has to reconstruct the $\ti{A}_{ij}^C$ in Cartesian coordinates from the tensor in spherical coordinates, $\ti{A}_{ij}^S$, which can be done by a coordinate transformation using the Jacobian matrix from Cartesian coordinates, $x^i=(x,y,z)$:
\begin{equation}
    M^i_{~j}\equiv\frac{\partial \bar{x}^i}{\partial x^j} =
    \begin{pmatrix}
        \frac{x}{r} & \frac{y}{r} & \frac{z}{r}\\
        \frac{xz}{r^2\sqrt{x^2+y^2}} & \frac{yz}{r^2\sqrt{x^2+y^2}} & -\frac{\sqrt{x^2+y^2}}{r^2}\\
        -\frac{y}{x^2+y^2} & \frac{x}{x^2+y^2} & 0
    \end{pmatrix}\,,
\end{equation}
\begin{align}\nonumber
    \ti{A}_{ij}^{C} &= \ti{A}_{kl}^{S}M^k_{~~i}M^l_{~j}=\br{M^T\cdot \ti{A}^{S}\cdot M}_{ij}= 
    \ti{A}_{rr}\begin{pmatrix}
        \frac{3x^2}{2r^2}-\frac{1}{2} & \frac{3xy}{2r^2} & \frac{3xz}{2r^2}\\
        \frac{3xy}{2r^2} & \frac{3y^2}{2r^2}-\frac{1}{2} & \frac{3yz}{2r^2}\\
        \frac{3xz}{2r^2} & \frac{3yz}{2r^2} & \frac{3z^2}{2r^2}-\frac{1}{2}
    \end{pmatrix} =\\\label{eq:appendix:paper1:arr_from_wr}
    &= \ti{A}_{rr}\br{\tfrac{3}{2r^2}x_ix_j-\tfrac{1}{2}\ti{\gamma}_{ij}} = \ti{A}_{rr}\br{\tfrac{3}{2r^2}x_ix_j}^{TF}\,,
\end{align}
where $\ti{A}_{rr}=\frac{4}{3}\br{\partial_r W_r - \frac{W_r}{r}}$, $\ti{\gamma}_{ij}=\delta_{ij}$ in Cartesian coordinates and $x_i$ are lowered with this metric (and hence identical to $x^i$).

\begin{figure}[t!]
\centering
\includegraphics[width=.7\textwidth]{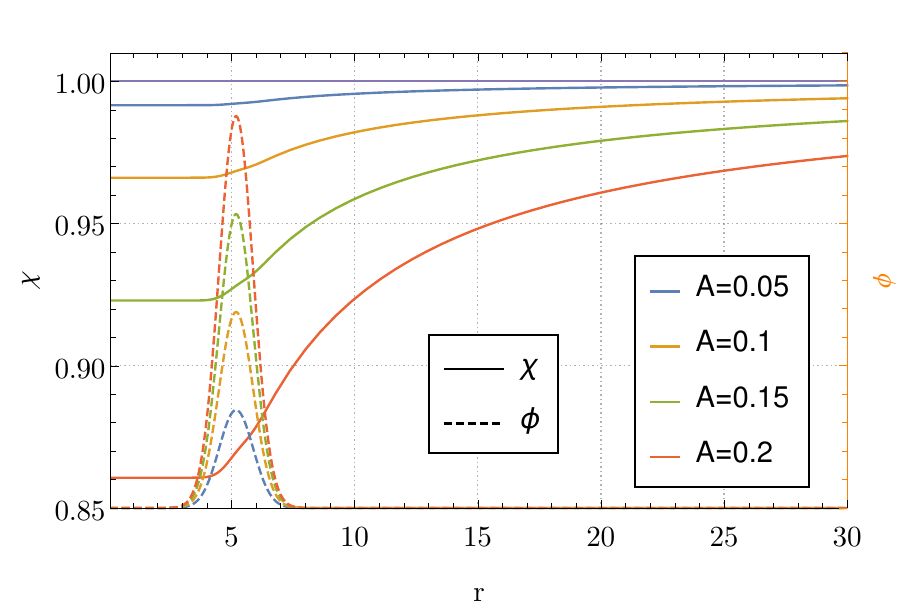}
\vspace{-5mm}
\begin{multicols}{2}
    \includegraphics[width=.51\textwidth]{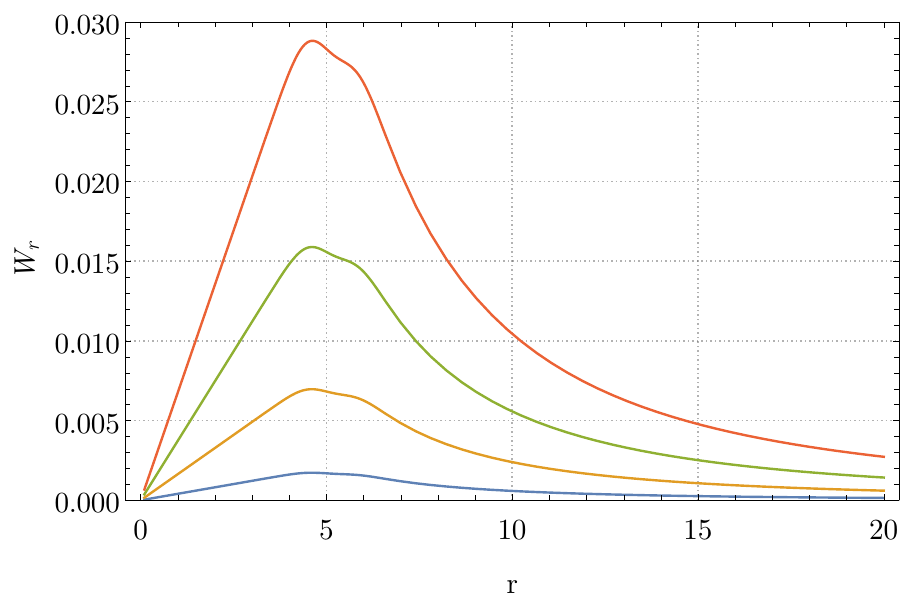}
    \includegraphics[width=.51\textwidth]{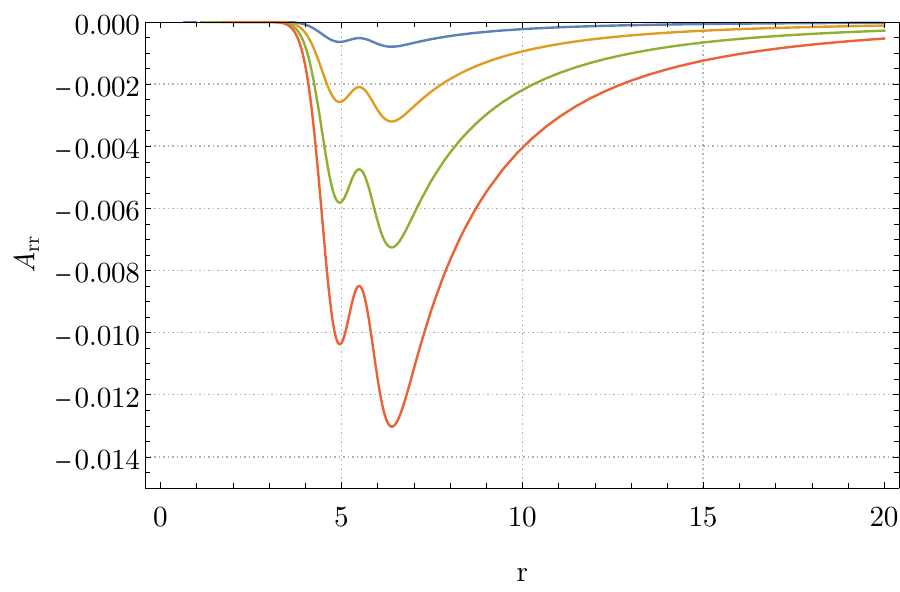}
\end{multicols}
\vspace{-6mm}
\caption{Initial data solutions for $\chi$, $W_r$ and $A_{rr}$ for several values of scalar field profiles (top, in orange, right $y$-axis) with in-going momentum.}
\label{fig:appendix:paper1:initial_data}
\end{figure}

We have now all the tools to obtain our initial data. For some sample parameters, using the initial data of equations \eqref{eq:paper1:phiData}-\eqref{eq:paper1:ini_mom} with $r_0 = 5$, $\w = \tfrac{1}{\sqrt{2}}$ and $g_2=g_3=0$ (we already know the difference for small coupling is small based on figure \ref{fig:paper1:initial_data}), the resulting solutions for varying amplitude $A$ are shown in figure \ref{fig:appendix:paper1:initial_data}.
The behaviour for large radius of $\chi$ follows a $\frac{1}{r}$ asymptote as expected\footnote{This could indeed be used to estimate the mass of the spacetime, but a more accurate method is provided in appendix \ref{appendix:paper1:adm_mass}.}, while $W_r$ seems to decay with $\frac{1}{r^2}$ and $A_{rr}$ with $\frac{1}{r^3}$. The behaviour for small $r$ of $W_r$ follows a linear behaviour $W_r = A\,r$ as also understood when analysing the boundary conditions of the problem.

The way spherically symmetric data is converted into 3D data in our numerical code \texttt{GRChombo} is using interpolation, based on a grid of points from the numerical solution of the coupled constraint equations solved. As a summary, the process for generating initial data is:
\begin{enumerate}
    \item Solve numerically the coupled constraint equations \ref{eq:Hamspherical} and \ref{eq:Momspherical}, with boundary conditions \eqref{eq:appendix:paper1:bcs} in \textit{Mathematica};
    \item Extract a grid of points from the interpolated numerical result;
    \item Solve an interpolation method for its coefficients for the points chosen. We use Cubic Spline interpolation method, outlined in appendix \ref{appendix:grchombo:spline};
    \item Export grid points and interpolation coefficients;
    \item In \texttt{GRChombo}, read grid points and coefficients and perform interpolation in each cell based on equation \eqref{eq:appendix:paper1:arr_from_wr}.
\end{enumerate}

\section{\done{Convergence}}\label{appendix:paper1:convergence}

\begin{figure}[h]
\centering
\includegraphics[width=1\textwidth]{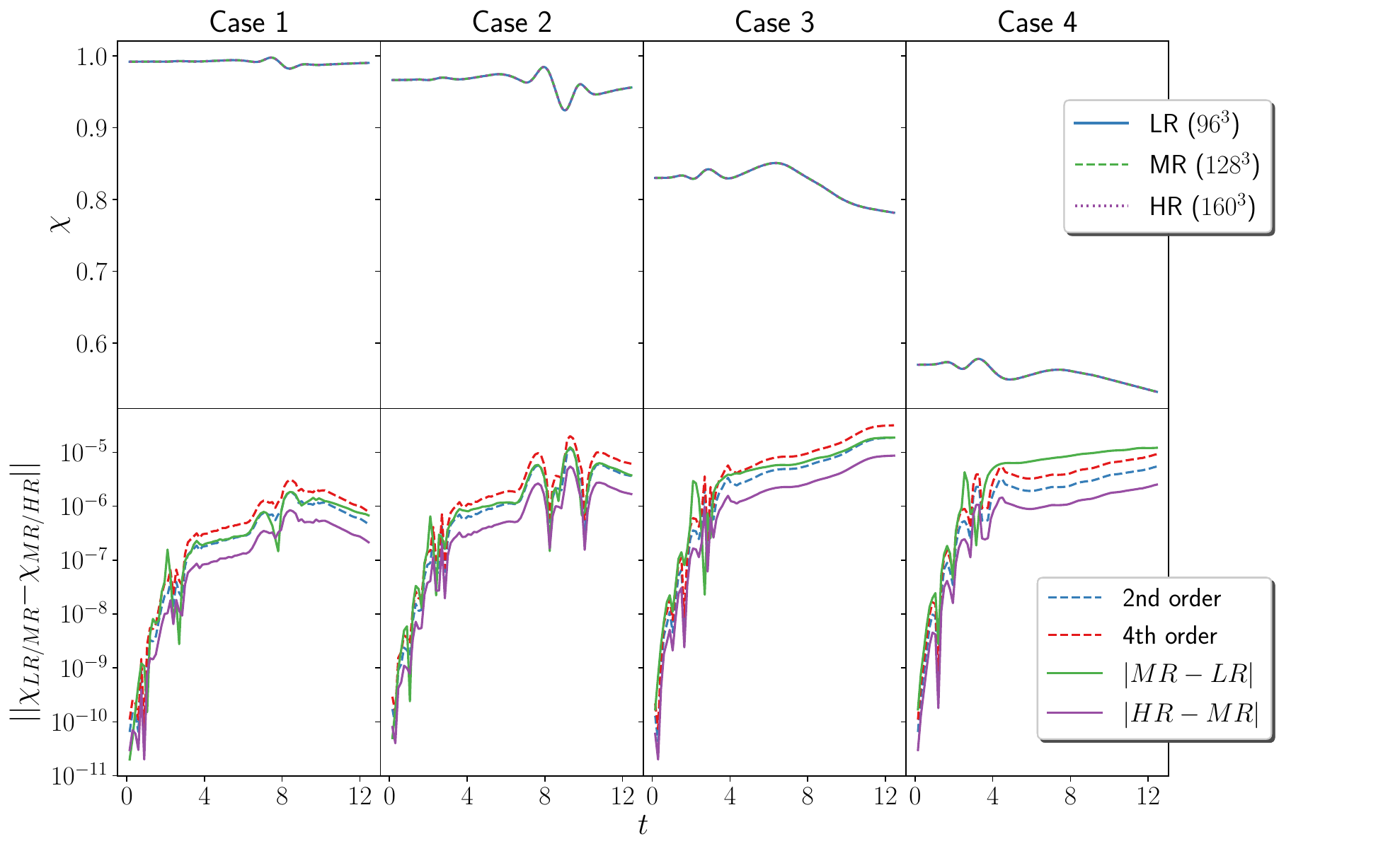}
\vspace{-2mm}
\caption{Convergence test for the $g_2=0.005$ run with different coarse resolutions: low ($LR$: $96^3$), medium ($MR$: $128^3$) and high ($HR$: $160^3$) resolutions, in addition to 7 refinement levels. \textit{Top}: evolution of $\chi$ at a fixed radius of $r=3$. \textit{Bottom}: $|MR-LR|$ and $|HR-MR|$ errors and the expected values for $|MR-LR|$ assuming $2^{\text{nd}}$ and $4^{\text{th}}$ order convergence.}
\label{fig:convergence}
\end{figure}
%
%

In this appendix we provide  details of some of the convergence tests that we have carried out. As an illustrative example, we consider the weak coupling $g_2=0.005$ case presented in \ref{sec:g2_pos_weak_coupling}. To carry out the convergence tests, we used simulations with coarsest level resolutions $\D x = 1$ (low resolution, $LR$), $\D x = 0.75$ (medium resolution, $MR$) and $\D x = 0.6$ (high resolution, $HR$) respectively, all with the same 7 additional levels of refinement. The results of the simulations for the 4 cases analysed are shown in figure \ref{fig:convergence}.  The bottom panel shows the error estimates $|MR-LR|$ (solid green curve) and  $|HR-MR|$ (solid purple curve), and compares them to the expected errors for  $2^{\text{nd}}$ (dashed blue) and $4^{\text{th}}$ (dashed red) order convergence. The latter were obtained from  the $|HR-MR|$ error using the continuum limit of the convergence factor: $\frac{\br{\D x_{LR}}^n - \br{\D x_{MR}}^n}{\br{\D x_{MR}}^n - \br{\D x_{HR}}^n}$. We see that our numerical results are consistent with convergence order between 2 and 4. Notice that it appears that the evolution has not reached a stationary state, but this should not be a concern since the outcome in terms of well-posedness and possible pathologies has already been determined after collapse occurred.

\begin{figure}[h]
\centering
\includegraphics[width=0.99\textwidth]{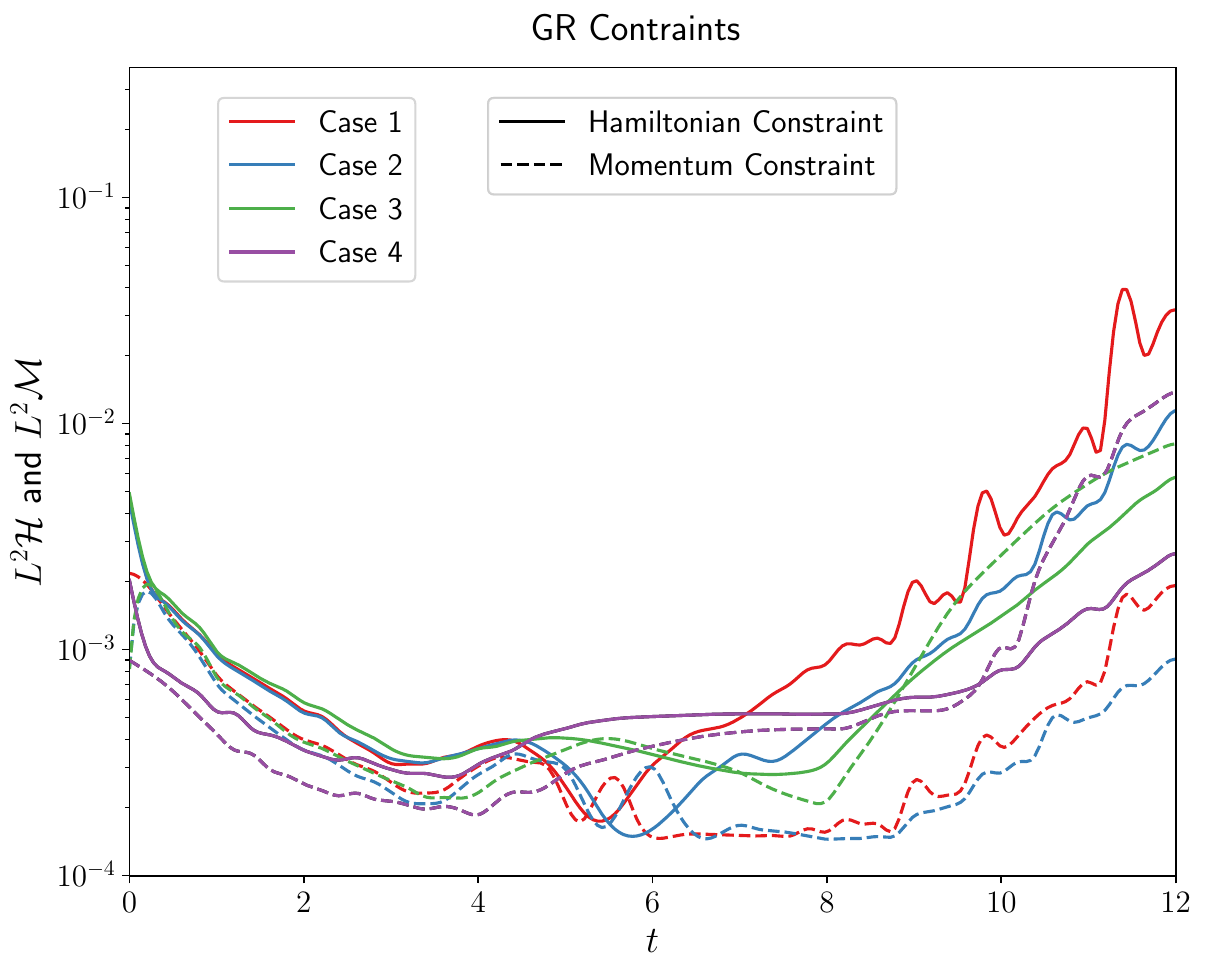}
\vspace{-2mm}
\caption{$L^2$ norm of constraints for the $g_2=0.005$ run  with coarsest level resolution of $\D x=1$ and 7 additional levels of refinement. We normalise
the constraints by the norm of the sum of the absolute value of each term that composes it.}
\label{fig:paper1:constraints}
\end{figure}

We also monitor the behaviour of the Hamiltonian and Momentum constraints for the simulation with $g_2=0.005$ presented in \ref{sec:g2_pos_weak_coupling}. We measure the $L^2$ norm of a quantity $\mathcal{Q}$ by the volume average:
\begin{equation}
	L^2\mathcal{Q} = \sqrt{\tfrac{1}{V}\int_V|\mathcal{Q}^2|dV}\,,
\end{equation}
where $V$ is the volume of the box except the region excised inside black holes (if there are any present). We normalise the constraints by the norm of the sum of the absolute value of each term in the constraints. We show in figure \ref{fig:paper1:constraints} that violations are under the $0.1\%$ level during gravitational collapse. Constraint violations increase at late times due to the following reasons. Firstly,  some of the scalar field (or all of it in Cases 1 and 2) disperses to infinity; as the scalar field propagates towards the boundaries, it moves away from the centre of the grid into coarser refinement levels, and thus resolution is lost. Secondly, as the scalar field disperses or is absorbed by the black hole, matter terms in the constraints become increasingly smaller and, as a consequence, the normalisation factors used also significantly decrease. Therefore, we can conclude that we have a good numerical control over our simulations.

\section{\done{Implementation verification}}\label{appendix:paper1:implementation}

It is not an easy task to ensure the equations of motion evolved in the code are the ones of the original theory intended to study without any errors in derivation or in conversion of equations to code. To ameliorate derivation errors, all the equations of motion, from the 4D equations \ref{subsec:paper1:EOM} to the $3+1$ decomposition \ref{appendix:paper1:EOM}, were derived using the \textit{Mathematica} package \textit{xAct} \cite{xAct}. One has to handle 3 different metrics (the 4D metric, the $3+1$ metric and the $3+1$ conformal metric) in \textit{xAct} for a derivation with no assumptions that achieves a full proof of the equations.

After obtaining the equations, one could use software to convert them into C\textsc{++} code, but this proves unfeasible for two reasons: first, \texttt{GRChombo} is heavily templated and makes use of custom primitives, and second, generated code is often unreadable and hard to edit or inspect manually. Instead, we used \textit{xCoba}, part of the \textit{xAct} package \cite{xAct}, to generate random initial data, evaluate the evolution equations for this data with both the analytical data in \textit{Mathematica} and the numerical code in \texttt{GRChombo}, and ensure the two agree numerically up to machine precision. To perform this, a random polynomial of sufficiently high degree on the Cartesian coordinates $(x,y,z)$ (a combination of terms $\a x^a y^b z^c$ for integers $a,b,c$ and a real constant $\a$) was generated for each variable, such that first and second derivatives can be easily evaluated, as well as computing from these composite variables such as the Riemann tensor. After derivatives are evaluated, one can evaluate the result in a few points and confirm the answers match between the code and \textit{xCoba}. It is important however to preserve physical properties, such as the tracelessness of $\ti{A}_{ij}$ or the unit determinant of $\ti{\g}_{ij}$, which are assumed in the derivation of equations. Hence, polynomials were generated for ${\g_{ij}, K_{ij}, \Q, \hat{\G}, \a, \b, \f, \P, \L}$, which also naturally define valid $\ti{\g}_{ij}$, $\chi$, $\ti{A}_{ij}$ and $K$. Note that this initial data is definitely not constraint satisfying, but this is not a problem, as the CCZ4 equations do not assume the constraints are satisfied and hence we validate the right hand side of all the equations, including the right hand side for $\Q$ and $\Q_i$, without the need of constraint satisfaction.

\section{\done{Other cases of interest}}\label{appendix:paper1:other}

\begin{figure}[H]
\centering
\includegraphics[width=1\textwidth]{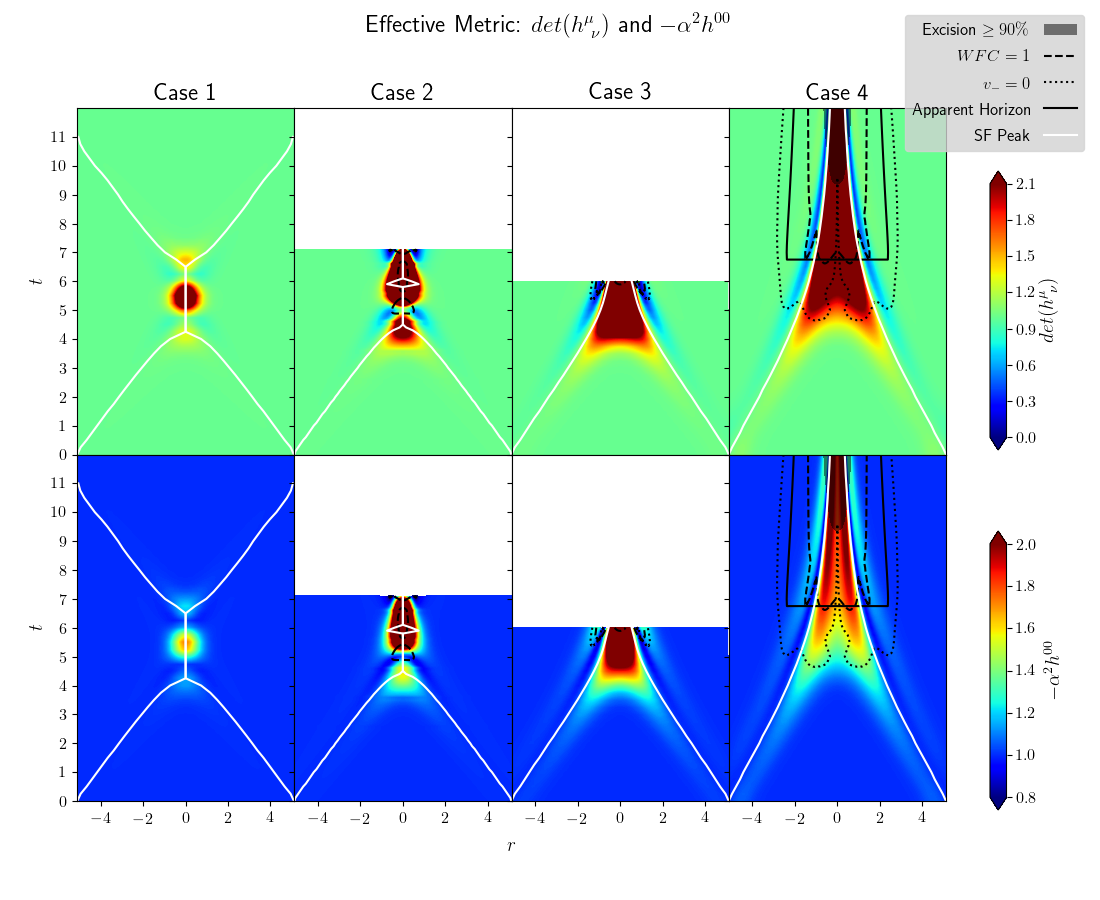}
\caption{$\det (h^\m_{~\n})$ (\textit{top}) and $-\a^2\, h^{00}$ (\textit{bottom}) for an intermediate positive coupling: $g_2=0.2$. The corresponding values of $\eta_2$, from left to right, are: $2\times 10^{-5},8\times 10^{-5},3.9\times 10^{-4},8.7\times 10^{-4}$. For small enough initial data (Case 1) the evolution is perfectly consistent, while it breaks down in a Tricomi-type of transition in Cases 2 and 3. For large enough initial data (Case 4), the pathologies that may develop during the evolution are hidden behind the black hole horizon. In this case,  the weak field condition is small on and outside the black hole horizon.}
\label{fig:appendix:intermediate_g2_pos}
\end{figure}

\begin{figure}[H]
\centering
\includegraphics[width=1\textwidth]{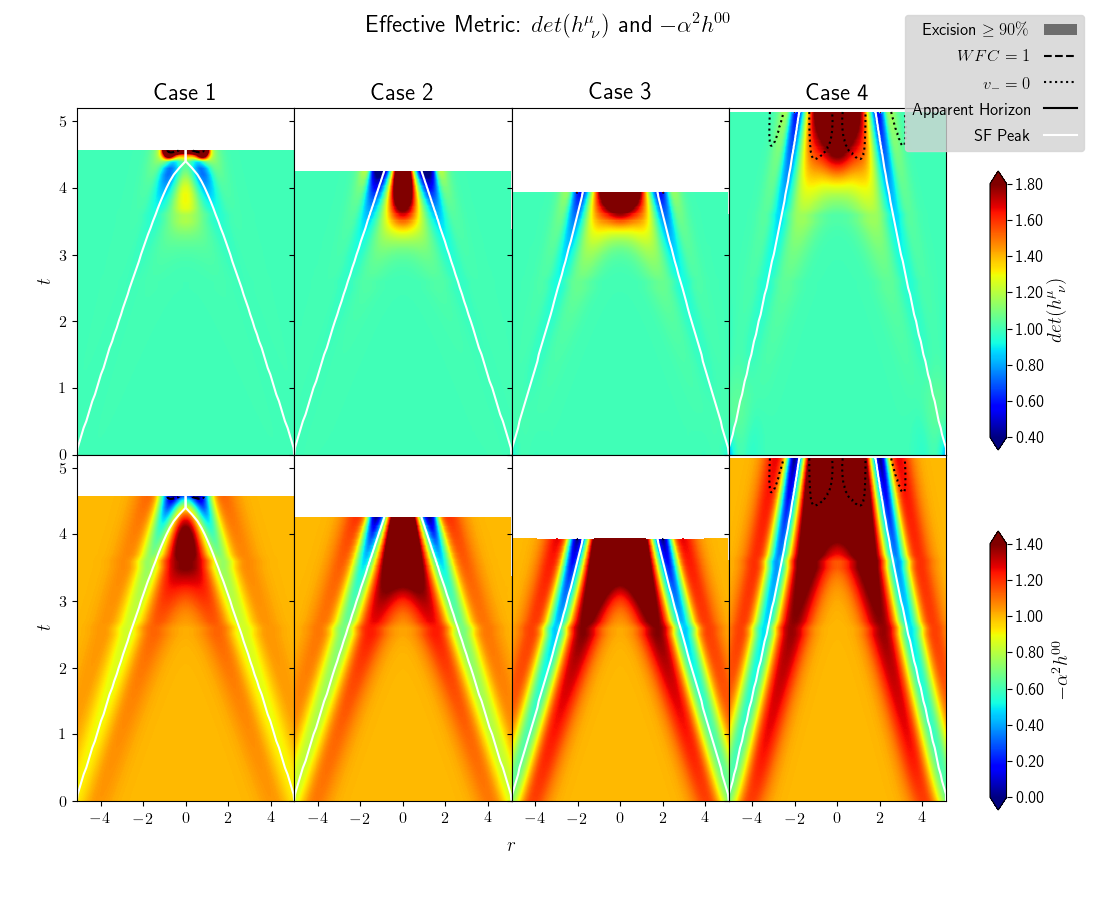}
\caption{$\det (h^\m_{~\n})$ (\textit{top}) and $-\a^2\, h^{00}$ (\textit{bottom}) for $G_3=g_3\,X$ with $g_3=0.4$. The corresponding values of the dimensionless coupling $\eta_3$, from left to right, are: $1.6\times 10^{-5},3.2\times 10^{-5},7\times 10^{-5}, 1\times 10^{-4}$. In all cases the evolution breaks down because $-\a^2h^{00}\to 0$, signalling a Keldysh-type-of transition.}
\label{fig:appendix:strong_g3_pos_det_h00}
\end{figure}

\begin{figure}[H]
\centering
\includegraphics[width=1\textwidth]{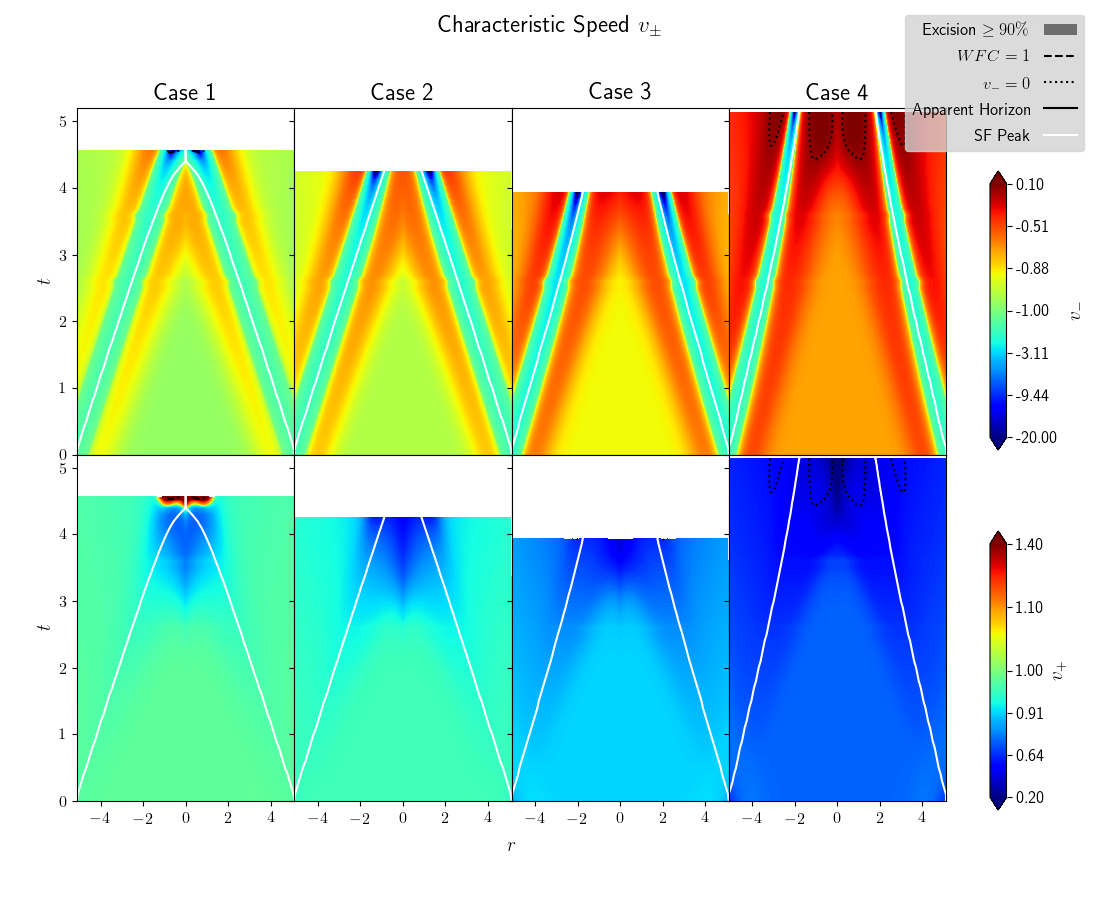}
\caption{Characteristic speeds of the outgoing (\textit{top}) and ingoing (\textit{bottom}) scalar modes for $G_3=g_3\,X$ with $g_3=0.4$. $v_-$ diverges at the transition, but $v_+$ remains finite.}
\label{fig:appendix:strong_g3_pos_speeds}
\end{figure}

In figures \ref{fig:appendix:intermediate_g2_pos}, \ref{fig:appendix:strong_g3_pos_det_h00} and  \ref{fig:appendix:strong_g3_pos_speeds} of this appendix we collect the results of some simulations that are relevant for parts of the discussion in the main text of chapter \ref{chapter:paper1}. 


\section{\done{ADM mass in spherical symmetry}}\label{appendix:paper1:adm_mass}

To calculate the ADM mass, one can use equation (3.128) of Baumgarte and Shapiro \cite[p.~85]{shapiro} (see also Brewin \cite{Brewin:2006qe}):
\begin{gather}
    M_{ADM} = \tfrac{1}{\k}\int_{\pd\S_\infty}\sqrt{\g}\g^{jn}\g^{im}\br{\pd_j\g_{mn}-\pd_m\g_{jn}}dS_i\, ,
\end{gather}
where $\pd\S_\infty$ is the 2-dimensional surface boundary of some spatial slice $\S$, $dS_i=\s_i\sqrt{\g^{\pd\S_\infty}}d^2z$ is the outward-oriented surface element of $\pd\S_\infty$, $z^i$'s are the coordinates on $\pd\S_\infty$, $\g^{\pd\S_\infty}$ the induced metric on it and $\s^i$ the unit normal to it ($\s_i\s^i=1$). To calculate this explicitly in conformal language, realise that in $D=4$: $\sqrt{\g}=\chi^{-\frac{3}{2}}$, $\g^{ij}=\chi \ti{\g}^{ij}$, $\g_{ij}=\frac{1}{\chi}\ti{\g}_{ij}$. For spherical symmetry and conformal flatness: $\s^i\propto\sbr{\frac{x}{r},\frac{y}{r},\frac{z}{r}}$ (which for consistency should be normalised to 1 using the induced metric $\g_{ij}$) and finally $\sqrt{\g^{\pd\S_\infty}}=\frac{r^2 \sin{\q}}{\chi}$. Putting this all together:
\begin{align}\nonumber
    M_{ADM} &= \tfrac{1}{\k}\int_{\pd\S_\infty}\chi^{-3/2}\ti{\g}^{jn}\ti{\g}^{im}\br{\chi\br{\pd_j\ti{\g}_{mn}- \pd_m\ti{\g}_{jn}} - \br{\ti{\g}_{mn}\pd_j\chi - \ti{\g}_{jn}\pd_m\chi}}dS_i =\\
    &=\tfrac{1}{\k}\int_{\pd\S_\infty}\chi^{-5/2}\ti{\g}^{jn}\br{\chi\br{\pd_j\ti{\g}_{in}- \pd_i\ti{\g}_{jn}} - \br{\ti{\g}_{in}\pd_j\chi - \ti{\g}_{jn}\pd_i\chi}}dS^i\, ,
\end{align}
where $dS^i = \frac{\s^i}{\chi} \br{r^2\sin{\q}d\q d\f}$ for some large radius $r$ tending to infinity. Defining $\bar{\s}^i=\sbr{\frac{x}{r},\frac{y}{r},\frac{z}{r}}$ we get the normalised $\s^i=\frac{\bar{\s}^i}{\sqrt{\g_{ij}\bar{\s}^i\bar{\s}^j}}=\frac{\bar{\s}^i\sqrt{\chi}}{\sqrt{\ti{\g}_{ij}\bar{\s}^i\bar{\s}^j}}$ and $\s_i=\g_{ij}\s^j =\frac{\ti{\g}_{ij}\bar{\s}^j}{\sqrt{\chi\ti{\g}_{kl}\bar{\s}^k\bar{\s}^l}}$.
For spherical symmetry and assuming a conformally flat metric ($\ti{\g}_{ij}=\d_{ij}$):
\begin{gather}
\begin{aligned}
    \,M_{ADM} &= \lim_{r\to\infty}\tfrac{r^2}{2\chi^3}\pd_r\chi\, .
\end{aligned}
\end{gather}

\section{\done{Variation of $G_3$ Horndeski term}}\label{appendix:paper1:variation_g3}

Varying actions is something that most physicists should know how to do. But it is the type of exercise that one does not necessarily do that often and there are usually several subtleties. For reference of the bits that are usually hard to make mistakes on, here we show how to calculate the variation with respect to the metric $g^{\mu\nu}$ of the $G_3$ term in the \eqref{eq:cubic_horndeski} metric:
\begin{equation}
    \int d^4x \sqrt{-g} ~G_3(\phi,X)\Box \phi\,.
\end{equation}
Useful identities:
\begin{equation}
\begin{aligned}
    \delta\sqrt{-g} &= -\tfrac{1}{2}\sqrt{-g}g_{\mu\nu}\delta g^{\mu\nu}\,,\\
    \delta g_{\rho\sigma} &= - g_{\rho\mu}g_{\sigma\nu}\delta g^{\mu\nu}\,,\\
    \delta X &= -\tfrac{1}{2}\grad_\mu\phi\grad_\nu\phi\delta g^{\mu\nu}\,,\\
    \delta \Gamma^\rho_{\mu\nu} &= \tfrac{1}{2}g^{\rho\sigma}\sbr{\br{\delta g_{\mu\sigma}}_{;\nu} + \br{\delta g_{\sigma\nu}}_{;\mu} - \br{\delta g_{\mu\nu}}_{;\sigma}}\,,\\
    \delta\grad_\mu\grad_\nu\phi &= -\delta\Gamma^\rho_{\mu\nu}\grad_\rho\phi = -\tfrac{1}{2}\grad^\sigma\phi\sbr{\br{\delta g_{\mu\sigma}}_{;\nu} + \br{\delta g_{\sigma\nu}}_{;\mu} - \br{\delta g_{\mu\nu}}_{;\sigma}}\,.
\end{aligned}
\end{equation}
Using these identities and the fact that $\grad_\mu X = -\grad^\nu\phi\grad_\mu\grad_\nu\phi$:
\begin{align*}
    \delta&\int \sqrt{-g}~G_3(\phi,X)\Box\phi = \text{\{int. by parts\}} = -\delta\int \sqrt{-g}\grad^\mu\phi\br{\partial_\phi G_3 \grad_\mu\phi + \partial_X G_3 \grad_\mu X} =\\
    =&\int\text{\{variation of $\sqrt{-g}$\}}+ \tfrac{1}{2}\sqrt{-g}g_{\mu\nu}\br{\partial_\phi G_3 (\grad\phi)^2 - \partial_X G_3 \grad^\rho\phi\grad^\sigma\phi\grad_\rho\grad_\sigma\phi}\delta g^{\mu\nu}-\\
    &\int\text{\{variation of $g_{\mu\nu}$'s\}}-\sqrt{-g}\br{\partial_\phi G_3 \grad_\mu\phi\grad_\nu\phi - 2\partial_X G_3 \grad_{(\mu}\phi\grad_{\nu)}\grad_\sigma\phi\grad^\sigma\phi}\delta g^{\mu\nu}+\\
    &\int\text{\{variation of $X$'s of $G_3$\}} +\tfrac{1}{2}\sqrt{-g}\grad_\mu\phi\grad_\nu\phi\br{\partial^2_{\phi X} G_3 (\grad\phi)^2 - \partial^2_{XX} G_3 \grad^\rho\phi\grad^\sigma\phi\grad_\rho\grad_\sigma\phi}\delta g^{\mu\nu}-\\
    &\int\text{\{variation of $\grad\grad\phi$\}} -\tfrac{1}{2}\sqrt{-g}\partial_X G_3 \br{\delta g_{\mu\nu}}_{;\rho}\grad^\mu\phi\grad^\nu\phi\grad^\rho\phi\,.
\end{align*}
\vspace{-2mm}
Integrating by parts the last line:
\begin{align*}
    \int-\tfrac{1}{2}\sqrt{-g}\partial_X G_3 \br{\delta g_{\mu\nu}}_{;\rho}\grad^\mu\phi\grad^\nu\phi\grad^\rho\phi = \text{\{int. by parts\}} = -\int \delta g^{\mu\nu}\tfrac{1}{2}\sqrt{-g}\Big[\partial_X G_3\big(\grad_\mu\phi\grad_\nu\phi\Box\phi +&\\
    +2\grad_{(\mu}\phi\grad_{\nu)}\grad_\rho\phi\grad^\rho\phi\big) +\tfrac{1}{2}\sqrt{-g}\grad_\mu\phi\grad_\nu\phi \br{\partial^2_{\phi X}G_3(\grad\phi)^2 - \partial^2_{XX}G_3\grad^\rho\phi\grad^\sigma\phi\grad_\rho\grad_\sigma\phi } \Big]&\,.
\end{align*}
\vspace{-3mm}
All in all:
\begin{align}\nonumber
    \delta\int \sqrt{-g}~G_3(\phi,X)\Box\phi = -\int\delta g^{\mu\nu}\sqrt{-g}&\Big[
    \tfrac{1}{2}\br{\partial_\phi G_3 (\grad\phi)^2 - \partial_X G_3 \grad^\rho\phi\grad^\sigma\phi\grad_\rho\grad_\sigma\phi} -\\
    -\big(\partial_\phi G_3 \grad_\mu\phi\grad_\nu\phi &- \partial_X G_3 \grad_{(\mu}\phi\grad_{\nu)}\grad_\sigma\phi\grad^\sigma\phi\big)-\partial_X G_3\grad_\mu\phi\grad_\nu\phi\Box\phi\Big]\,.
\end{align}

\chapter{\done{Black hole binaries in cubic Horndeski theories}}\label{appendix:paper2}


\section{\done{Convergence}}\label{appendix:paper2:convergence}


In this appendix we provide some details of the convergence tests that we have carried out. As a representative example, we considered the binary in the $G_2$ Horndeski theory with coupling constant $g_2=0.02$, and we performed three simulations with different resolutions to study convergence. Our simulations are evolved with a coarsest level resolution of $\D x = \frac{16}{7}$ (medium resolution), with 8 additional refinement levels and a computational domain of size $1024^3$. To carry out the tests, we used one lower resolution changing $\D x = \frac{8}{3}$ (low resolution) and one higher resolution with $\D x = 2$ (high resolution).
\begin{figure}[h]
\centering
\begin{multicols}{2}
\hspace{-17mm}\includegraphics[width=0.6\textwidth]{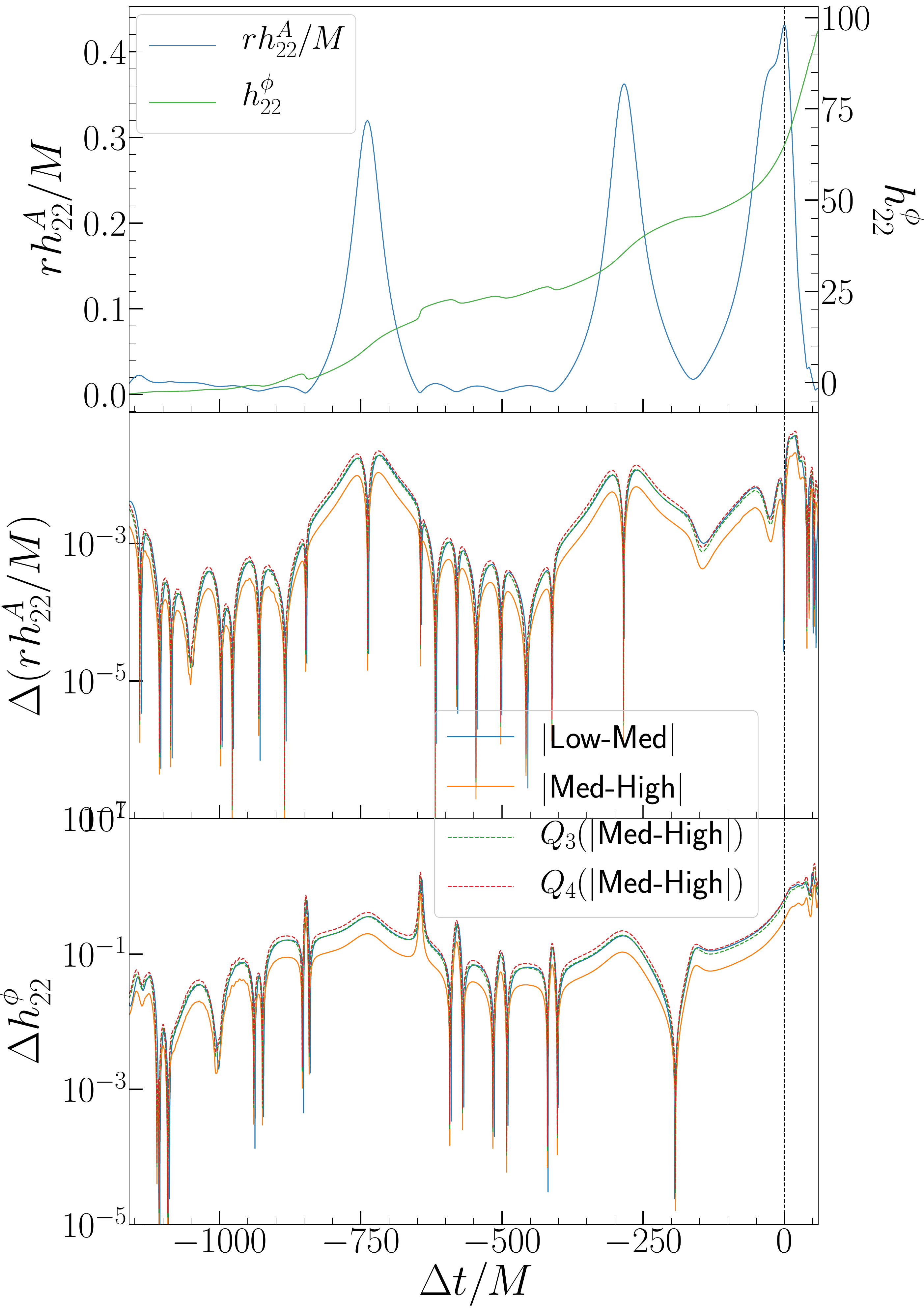}
\caption{Convergence test for $g_2=0.02$ with different coarse resolutions: low ($384^3)$, medium ($448^3$) and high ($512^2$). Convergence performed on the amplitude and phase of the complex strain, $h_{22}$, extrapolated to null infinity. $\D t=0$ is the peak of the amplitude for the highest resolution. This figure indicates consistency with third order convergence.}
\label{fig:GW_strain_convergence}
\end{multicols}
\end{figure}

In figure \ref{fig:GW_strain_convergence} we show the error estimates in the quadrupole mode $h_{22}$ extrapolated to null infinity\footnote{The extrapolation to null infinity has been assumed to work well without any check of the convergence of the wave with extrapolated radii, but indeed this is generically very much the case. Not only do the waves vary little over extraction radii, about $1\%$ in peak amplitude and minimally in phase after tortoise time alignment, but the strain values fit very well to a $\frac{1}{r^*}$ model, as described in section \ref{subsec:paper2:strain}.} between low, medium and high resolutions and the estimates for the expected error assuming third and fourth order convergence. We decompose the complex strain into its amplitude and phase\footnote{We continuously add multiples of $2\p$ to the phase such that we obtain a monotonic function for the phase.}, $h_{\ell m} = h^+_{\ell m} - ih^\times_{\ell m} = h_{\ell m}^A e^{ih_{\ell m}^\f}$. We then interpolate the resulting functions (because different resolutions have points at different times) and compute the expected errors using the continuum limit of the convergence factor of order $n$:
\begin{equation}
    Q_n = \frac{\br{\D x_{Low}}^n - \br{\D x_{Med}}^n}{\br{\D x_{Med}}^n - \br{\D x_{High}}^n}\,.
\end{equation}
This indicates the convergence order of $h_{22}$ is consistent with three. 
\begin{figure}[h]
\centering
\includegraphics[width=0.55\textwidth]{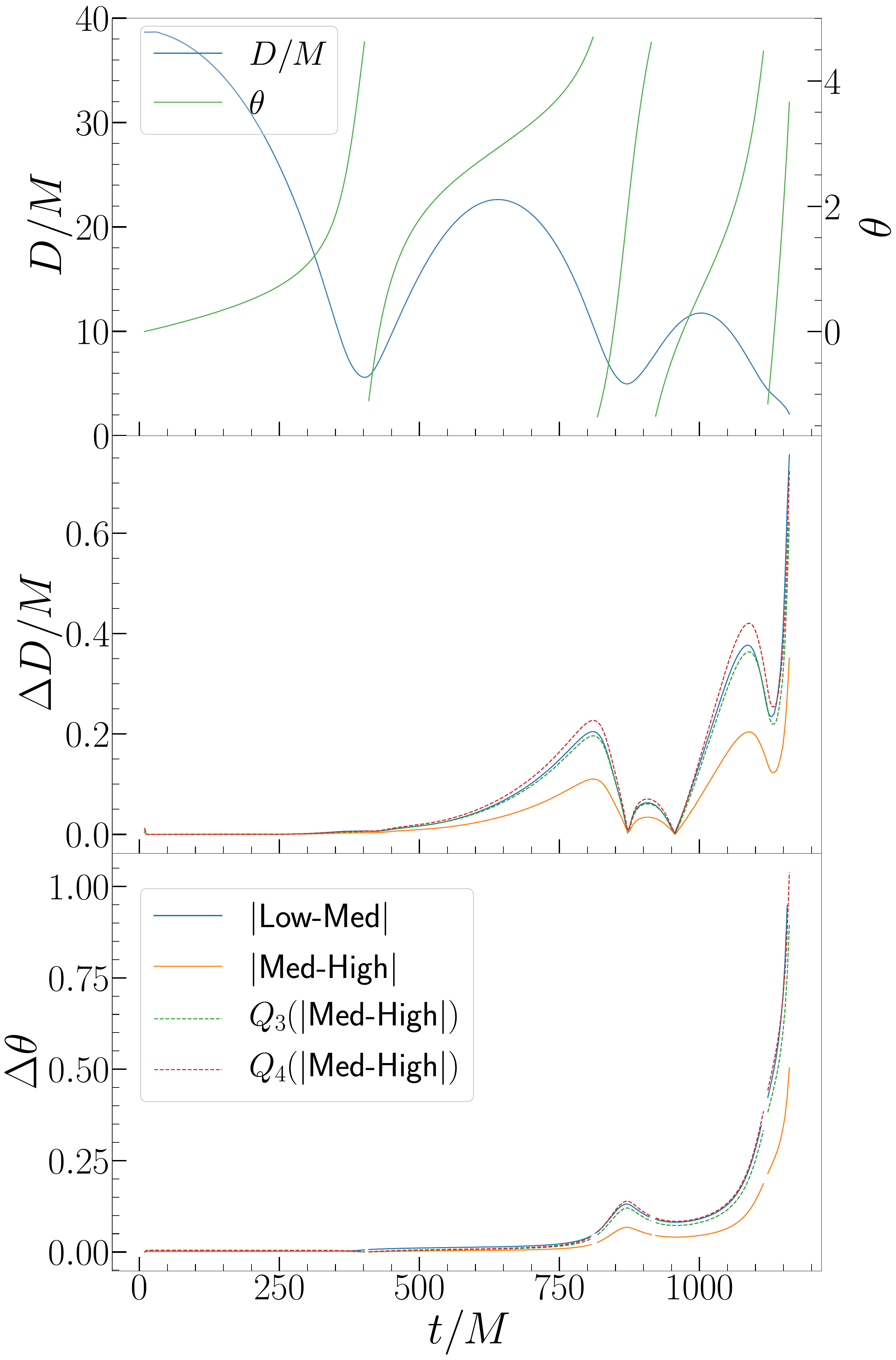}
\caption{\textit{Top panel:} Radial distance $D(t)$ and relative phase $\q(t)$ of the black holes' trajectories as functions of time for the $g_2=0.02$ Horndeski theory. Convergence tests for the radial distance $\D D(t)$ (\textit{middle panel}) and relative phase $\q(t)$ (\textit{bottom panel}). Both of these quantities exhibit between third and fourth order convergence.}
\label{fig:parametrize_2D_convergence}
\end{figure}

We also tested convergence of other variables; for instance, the trajectories of the two black holes, $x^i_{1}(t)$ and $x^i_2(t)$, shown in figure \ref{fig:AH_xy}, can be used to test convergence. We rewrite these trajectories in terms of the radial distance between the black holes,
\begin{equation}
    D(t)=|x^i_1(t) - x^i_2(t)|\,,
\end{equation} 
and the phase relative to the initial positions,
\begin{equation}
    \q(t)=\arccos\sbr{\tfrac{(x^i_1(t) - x^i_2(t))}{D(t)}\cdot\tfrac{(x^i_1(0) - x^i_2(0))}{D(0)}}\,,
\end{equation}
where $\cdot$ here denotes internal product. These quantities for the Horndeski theory are shown in the top panel of figure \ref{fig:parametrize_2D_convergence} for the same binary as in figure  \ref{fig:AH_xy}.
The convergence analysis of these quantities across the three resolutions is shown in the middle and bottom panels of figure \ref{fig:parametrize_2D_convergence}. These figures indicate that both quantities exhibit between third and fourth order convergence.
\begin{figure}[H]
\centering
\includegraphics[width=0.7\textwidth]{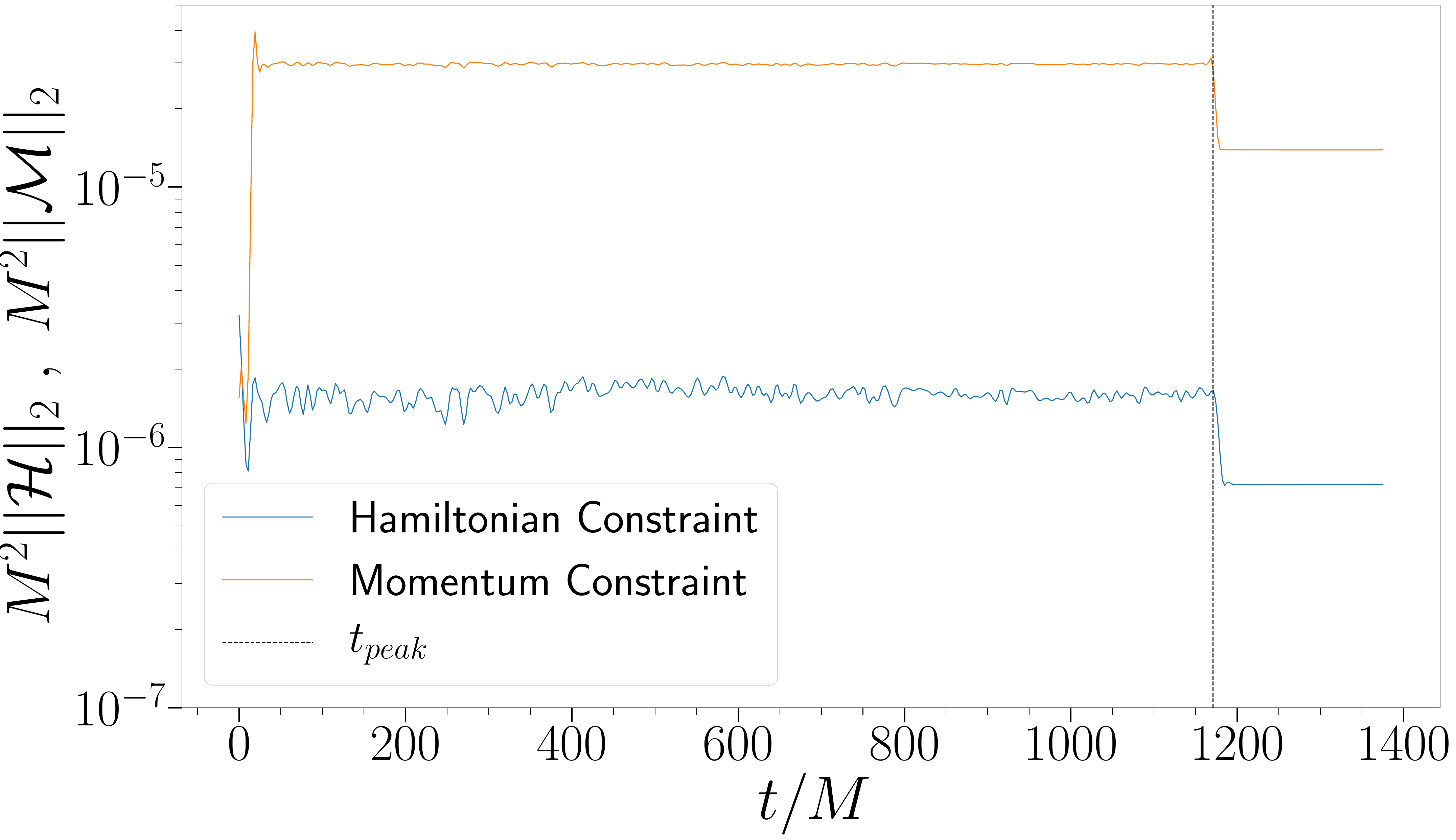}
\caption{$L^2$ norm of the Hamiltonian and the Euclidean norm of the momentum constraints for the medium resolution of the $g_2=0.02$ Horndeski binary.}
\label{fig:paper2:constraints}
\end{figure}

For completeness, in figure \ref{fig:paper2:constraints} we show the $L^2$ norms\footnote{For a given quantity $\mathcal{Q}$, we compute $L^2\mathcal{Q} = \sqrt{\frac{1}{V}\int_V|\mathcal{Q}^2|dV}$.} of the Hamiltonian and the Euclidean norm of the momentum constraints over the full computational domain. This figure shows the constraint violations remain stable at the level of $10^{-6}-10^{-5}M^{-2}$ respectively throughout the whole evolution, with a significant and sudden reduction at the merger. Considering the results of our convergence analysis, we conclude our simulations are stable and in the convergent regime.

\section{\done{Computing accurate gravitational strains}}\label{appendix:paper2:strain_time_vs_frequency}

To compute the gravitational strain from the gravitational wave $\Psi_4$ data spherical harmonics, given formula \eqref{eq:paper2:strain:def}, one may be tempted to compute the strain from a simple double time integration:
\begin{equation}\label{eq:paper2:strain_double_integral}
    h(t) = -\int_{t_0}^t \int_{t_0}^{t'} \Psi_4(t'') dt'' dt'\,,
\end{equation}
where $t_0$ is as close to $-\infty$ as evolved. However, numerical errors in the first and second integral will lead to constants of integration as:
\begin{equation}
    -\int_{t_0}^t \int_{t_0}^{t'} \Psi_4(t'') dt'' dt' \approx h(t) + C_1 t + C_0\,,
\end{equation}
for some constants $C_0,C_1$. $C_0$ has a physical meaning known as the memory effect, while $C_1$ is a noise term from the integral. This can be fixed with a high-pass filter which removes low-frequencies to minimise $C_1$, but a robust way of handling it is by performing the integration in Fourier space, where equation \eqref{eq:paper2:strain_double_integral} becomes:
\begin{equation}
    \ti{h}(f) = -\frac{\ti{\Psi}_4(f)}{\br{2\p f}^2}\,,
\end{equation}
where $\ti{h}(f)$ and $\ti{\Psi}_4(f)$ denote the Fourier transforms of $h(t)$ and $\Psi_4(t)$ for frequency $f$. The details of the Fourier method are described in section \ref{subsec:paper2:strain}. Several parameters are manually chosen but have no effect on the final result:
\begin{itemize}
    \item the size of the Tukey window, as long as not too big to interfere with where the signal is not compatible with noise;
    \item the $t_0$ cutoff to remove junk radiation, as our simulations have very little junk radiation, and changing this does not affect the result;
    \item the final time the wave is extracted at, as long as oscillations are already compatible with noise;
    \item how big the zero-padding at the end of the data set is, which is done to increase the frequency resolution of the discrete Fourier transform;
    \item the high frequency cut-off is also negligible and even though it proved unneeded for our case, it may be added for numerical noise errors if the frequencies removed are bigger than the frequency corresponding to about the size of the black hole;
\end{itemize}

\begin{figure}[h]
\centering
\includegraphics[width=1\textwidth]{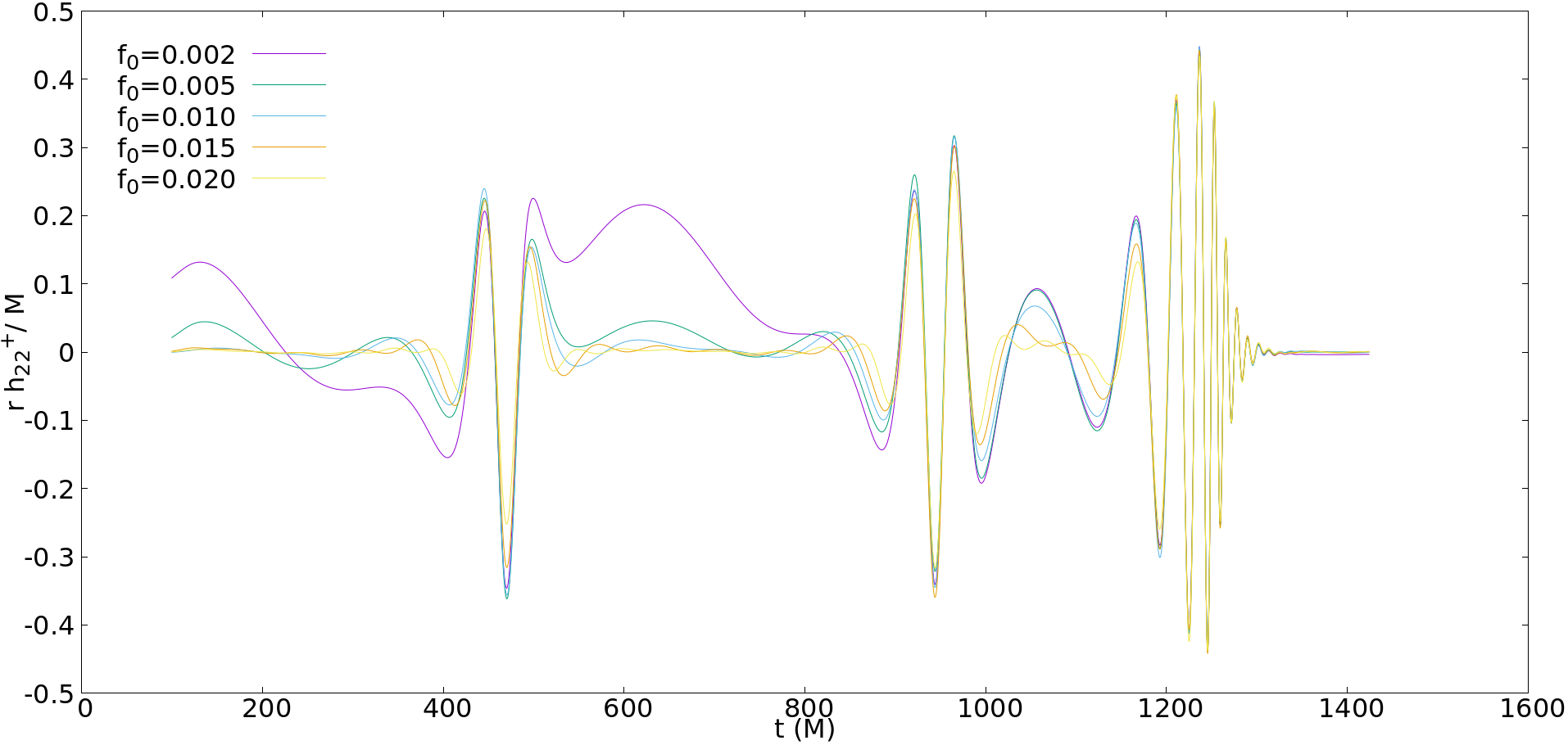}
\caption{Strain of the GR elliptic binary black hole discussed in section \ref{subsec:paper2:waves} using different low frequency cutoffs when performing the double time integration in the frequency domain.}
\label{fig:strain_cutoff}
\end{figure}

The only parameter that has to be carefully chosen is the low frequency cutoff. This parameter can be chosen in many ways, but it proven enough to simply use direct inspection. Take the example of the GR case displayed in section \ref{subsec:paper2:waves}. For cutoff frequencies $f_0 = [0.002, 0.005, 0.01, 0.15, 0.2]$ the resulting strain is displayed in figure \ref{fig:strain_cutoff}. Very low frequencies, $f=0.002$, lead to spurious errors. Very high cutoffs, $f=0.02$, damps errors but also physical effects. For this problem, a cutoff in the range $f\in[0.005,0.015]$ is ideal, balancing between spurious errors and no physical effect damping. If this is kept fixed, all results should remain consistent. 

\chapter{\done{Black hole binaries in higher derivative effective field theories}}\label{appendix:eft}


\section{\done{EFT numerical code}}\label{appendix:eft:implementation}

In this appendix we describe relevant details in the construction of the code to evolve the EFT described in chapter \ref{chapter:eft}. The aim was to build a code that was generic to evolve this and many other EFTs, with and without matter, in many potential future scenarios, hence requiring great flexibility.

\texttt{GRChombo} is already compatible with generic matter, which implies the implementation of the Horndeski equations for chapters \ref{chapter:paper1} and \ref{chapter:paper2} was relatively straightforward. One simply had to make use of the $3+1$ conformal decomposition described in appendix \ref{appendix:paper1:EOM}, equations \eqref{eq:appendix:energy_density}-\eqref{eq:appendix:Sij}.

The right-hand side of the EFT equation of motion, \eqref{eq:eft:EOM_4d_weyl} can be interpreted as `matter' from which one computes matter decomposition terms, $\r, S^i, S_{ij}$. But the $3+1$ decomposition and further conformal decomposition turns out to be a process much more laborious than it is for the Horndeski case, and not generic if one chooses to slightly change the EFT. The advantage of this procedure is speed, in the sense that it reduces the equation to its simplest form using directly the evolution variables directly accessible from the numerical grid.

Faced with this main difficulty and prioritising flexibility, we took a less standard approach: instead of \textit{a priori} decomposing the $4D$ stress-energy tensor into $3+1$ conformal form, the code dynamically takes the $3+1$ conformal evolution variables to compute $3+1$ non-conformal variables, which in turn are used to compute $4D$ quantities, e.g. the spacetime Riemann tensor or the spacetime Christoffel symbols. This can then be used to compute the full stress-energy tensor, which is numerically decomposed in $3+1$ form, $\r, S^i, S_{ij}$, and fed into the BSSN/CCZ4 right-hand side.

This method has one further advantage: the computation of each new quantity ``along the way'' can be individually verified as correct, such that adapting the code to new theories is a matter of putting together pieces of a puzzle already pre-built. The main disadvantage is speed, not only because the right-hand side of the EFT is in itself heavier to compute than GR alone, but because building $4D$ tensors to then decompose them is more computational work than what is typically done. To tackle this, on a given cell at a given timestep, individual quantities computed, such as the spatial extrinsic curvature $K_{ij}$ built using equation \eqref{eq:bssn:kij_definition}, are stored and re-used if any other computation requires it again. For instance, computing the extrinsic curvature $K_{ij}$ is required to compute the CCZ4 right-hand side (ignoring the matter terms), but also required to the Kretschmann scalar in the EFT via the electric and magnetic components of the Weyl. As a more intricate example, the computation of $\pd_j \ti{\G}^k$ is required for the electric part of the Weyl, when computing $R_{ij} + D_{(i}\Q_{j)}$, and also when computing the spacetime Ricci tensor, as the CCZ4 formulation requires the term $\grad_\m Z_\n$ and $Z_i$ depends on $\ti{\G}^k$. This makes computations faster and comparable to the standard \texttt{GRChombo} code. As suggested above as advantage of the method, we implemented verifications of every sub-computation using the same method described in \ref{appendix:paper1:implementation}, making use of \textit{xAct} package of the \textit{Mathematica} software, initialised with random variables.

As a final note, to reduce human error during implementation (the last thing we want is evolving the wrong theory due to a wrong index or minus sign or factor of 2 in the code), we extended the existing \texttt{GRChombo} tensor algebra tools to include contractions, traces, external products, dot products and covariant derivatives of any tensor of any rank and any dimension, such as three dimensional spatial tensors or $4D$ spacetime tensors. This was made possible using C\texttt{++} functions templated over both the tensor rank and dimension. We hope to make these tools public in the future.

As a final remark, an alternative functional method may be the use of the modified generalised harmonic gauge \cite{Kovacs:2020pns, Kovacs:2020ywu} in the \texttt{GRChombo} public code. Preliminary steps have been taken in join work with Llibert Aresté Saló, Justin Ripley and Pau Figueras (see also Aresté Saló et al. \cite{AresteSalo:2022hua}). This may enable a full $4D$ or a $3+1$ non-conformal reformulation of the code, in a similar way to the work by Brown \cite{Brown:2011qg,Brown:2009dd}. If this is possible, it may be extremely useful in order to create an easier formulation of GR which does not require conformal decomposition when starting a new research project on a new EFT.

\section{\done{Diffusion}}

The interior of black holes turns out to be considerably more challenging in non-standard settings. Even with the excision technique of section \ref{subsec:eft:excision}, divergences inside the AH often creep in, typically for variables that involve second derivatives in the equations of motion \cite{Figueras:2015hkb}. To solve this, we can introduce a dissipative diffusion operator of the form:
\begin{equation}
    c_L \D_x^2 \br{\sqrt{\frac{2}{D(D-1)}\sum_{i,j,k}\br{\pd_k\ti{\g}_{ij}}^2}}\s(\chi; \bar{\chi}, \w_\chi)\br{\grad^2 \ti{\g}_{ij}}^{TF}\,,
\end{equation}
where $c_L$ is some constant, $\D_x$ the grid spacing, $\ti{\g}_{ij}$ the conformal spatial metric of the CCZ4 formulation, and $\s(\chi; \bar{\chi}, \w_\chi)$ as defined in equation \eqref{eq:eft:sigmoid} plays the role of activating this operator only for $\chi < \bar{\chi}$. $\grad^2 \ti{\g}_{ij}$ is some notion of the spatial Laplacian of the metric being diffused, for which one can apply a simple flat space operator $\d^{kl}\pd_k\pd_l\ti{\g}_{ij}$. $\br{\cdot}^{TF}$ stands for ``trace-free'' and needs to be applied to the conformal metric diffusion operator to ensure the determinant of this metric remains $1$. Replacing the variable inside the Laplacian term, a similar operator should be applied to other variables that have second derivatives in the equation of motion: the lapse, the shift, the conformal factor, and any other variable a given EFT fix might introduce (e.g. $\hat{\mat{C}}$ for the $\mat{C}$ system of section \ref{eft:subsec:c_system}). The philosophy is that since the theory itself breaks down near the singularity, we can replace it with something that ensures that our evolution variables remain smooth and under control. In this case, the coefficient $\D_x^2$ ensures only small wavelengths that the numerical grid cannot resolve get damped, and that the CFL condition does not get affected.

This technique proved useful during experimentation with single boosted and non-boosted black holes with the several systems described in section \ref{sec:eft:methodology}. However, it was not required in order to successfully evolve any of the final simulations presented, which made use of a small value of $\e=10^{-5}$. Yet, given divergences often appear inside the black hole, this technique can be a useful last resource in an attempt to stabilise the simulation.

\chapter{\done{\texttt{GRChombo} - code development work}}\label{appendix:grchombo}

\section{\done{Tagging criteria}}\label{appendix:grchombo:tagging}

In AMR, creation and destruction of grid boxes and new finer refinement grids is determined by the tagging of cells for refinement, which in turns is controlled by a \textit{tagging criterion}. Although this ability to refine regions can be incredibly powerful, in practice it can be difficult to manage the exact placement of refined grids. For instance, the presence of level boundaries in dynamically sensitive regions of spacetime, such as near apparent horizons, plays an essential role in the accuracy of results\footnote{When it comes to resolution, a common rule-of-thumb is to have as minimum at least 40 grid points across the horizon.}. In this section we explain some of the tagging criteria developed for the purposes of chapters \ref{chapter:paper1}-\ref{chapter:eft}. For details on how the tagging of cells leads to formation of new grids, see Radia et al. \cite{Radia:2021smk}. We have adopted mainly the following types of tagging criteria:
\begin{itemize}
    \item \textit{$\chi$ tagging}: realising that the conformal factor has steeper curves closer to black holes, one can tag based on:
    \begin{equation}
        \D x \sqrt{\d^{ij}\d^{kl}\br{\pd_i\pd_k\chi}\br{\pd_j\pd_l\chi}} > \s_\chi\,,
    \end{equation}
    where $\D x$ is the grid spacing and $\s_\chi$ is a custom refinement threshold. The criterion is multiplied by the grid spacing such that higher levels (smaller $\D x$) require bigger derivatives of $\chi$ to be activated. Second derivatives are used because the first derivative of $\chi$ is actually zero at the puncture. Alternative, one can use first derivatives of $\chi^{-1}$ \cite{Kunesch:2018jeq}.
    \item \textit{$\f$ tagging}: in the presence of scalar fields, often the conformal factor may not be steep enough in spite of large gradients of scalar field. For this, we add a similar criterion using derivatives of the scalar field and its momentum:
    \begin{equation}
        \D x \br{\tfrac{1}{\s_\f}\sqrt{\d^{ij}\d^{kl}\br{\pd_i\pd_k\f}\br{\pd_j\pd_l\f}} + \tfrac{1}{\s_\Pi}\sqrt{\d^{ij}\d^{kl}\br{\pd_i\pd_k\Pi}\br{\pd_j\pd_l\Pi}}} > A\,,
    \end{equation}
    where $\s_\f$ and $\s_\Pi$ are user defined constants, and $A$ is the initial amplitude of the scalar field (typically both $\f$ and $\Pi$ are proportional to $A$, so this makes the criterion robust to changes of initial data).
    \item \textit{Puncture tagging}: even though mathematically  the region inside apparent horizons is causally disconnected from the exterior, numerically, artefacts from discretisation can leak errors that propagate superluminally. Having refinement levels across the horizon turns out to introduce significant inaccuracies, typically manifesting as an unphysical drift in the black hole mass and, in the case of binaries, a loss of phase accuracy and/or a drift in the BH trajectory (see Radia et al. \cite{Radia:2021smk}). If using a puncture tracker and if the mass of the black holes are known (see section \ref{subsubsection:ngr:observables}), this can be fixed by enforcing tagging of cells within the horizon plus a buffer radius.
    \item \textit{AH tagging}: for the cases where a puncture tracker is not as easily feasible (e.g. gravitational collapse, where the initial data has no initial puncture), one can use the apparent horizon (its centre and average radius) to estimate where to force levels to regrid. In practice, when all we want is to roughly cover the black hole, a much easier and equivalent method is to use contours of $\chi$, which turn out to be extremely good proxies for the location of the apparent horizon without the need to know their location or masses \textit{a priori}. For details, see section \ref{appendix:grchombo:ah_location}.
    \item \textit{GW tagging}: when we want to extract gravitational waves far away from the centre of the grid, we force tagging cells that are at a distance smaller than the wave extraction radius if the resolution is lower than the desired extraction resolution.
    \item \textit{Fixed tagging}: for specific dynamic environments such as gravitational collapse have quick dynamics that can lead to too much regridding, spurious unwelcome numerical noise and inconsistency between different runs. In such cases, similar to a many-box-in-many-boxes approach, one can specify fixed regions to be tagged, as a square around the gravitational collapse region. One can also force certain levels to appear at a certain stage, e.g. when the collapse is close enough to forming a small black hole that requires more resolution.
    \item Other tagging criteria, such as asymmetric grid tagging or truncation error tagging, have been developed for \texttt{GRChombo}, but have not been used for the work developed in this thesis. See details in Radia et al. \cite{Radia:2021smk}.
\end{itemize}

One or more tagging criteria can be used per simulation. On top of this, the the minimum and maximum size of boxes, how frequently to actually tag and regrid each level, how many cells need to be tagged in a region to proceed to regridding and buffer regions of forced tagging around a given identified tagged cell\footnote{This is useful to set mesh boundaries further apart from each other and reduce errors from high frequency resonances bouncing off neighbouring boundaries \cite{Radia:2021smk}.} are some of the parameters a user can tune. Overall, AMR is a powerful tool that can deal with extremely dynamic scenarios, but it also has a significant overhead of experimentation to get fundamental pieces such as regridding to work well.

\section{\done{Weyl tensor}}\label{appendix:grchombo:weyl}

Given a Riemann tensor $\mathcal{R}_{\m\n\r\s}$ related to a metric $g_{\m\n}$ in $D>3$ dimensions, the Ricci tensor is its trace, $\mathcal{R}_{\m\n}=\mathcal{R}^\a_{~\m\a\n}$, and the Ricci scalar is $\mathcal{R} = \mathcal{R}_\m^{~\m}$. The symmetries of the Riemann tensor are:
\begin{equation}
    \mathcal{R}_{\mu\nu\alpha\beta} = \mathcal{R}_{[\mu\nu][\alpha\beta]}\,,\quad
    \mathcal{R}_{\mu\nu\alpha\beta}=\mathcal{R}_{\alpha\beta\mu\nu}\,,\quad
    \mathcal{R}_{[\mu\nu\alpha]\beta} = 0\,,
\end{equation}
as well as the Bianchi and contracted Bianchi identities:
\begin{equation}\label{eq:bianchi_riemann}
    \nabla_{[\rho}\mathcal{R}_{\mu\nu]\alpha\beta}=0\,,\quad \nabla^\mu \mathcal{R}_{\mu\nu\alpha\beta} = \nabla_\a \mathcal{R}_{\b\n} - \nabla_\b \mathcal{R}_{\a\n}\,,\quad \nabla^\mu \mathcal{R}_{\m\n} = \tfrac{1}{2} \nabla_\n \mathcal{R} \,.
\end{equation}
The Weyl tensor is complementary to the Ricci tensor, being defined as the trace free part of the Riemann tensor, preserving its symmetries\cite{alcubierre}:
\begin{equation}\label{eq:weyl-definition}
	C_{\m\n\r\s} := \mathcal{R}_{\m\n\r\s} - \tfrac{2}{D-2}\br{g_{\m[\r}\,\mathcal{R}_{\s]\n} - g_{\n[\r}\,\mathcal{R}_{\s]\m}} + \tfrac{2}{(D-1)(D-2)}g_{\m[\r}g_{\s]\n}\,\mathcal{R}\,.
\end{equation}
It is easy to check the extra symmetry: $C^\mu_{\phantom{\mu}\alpha\mu\beta} = 0$. The analogue of the contracted Bianchi identity for the Weyl tensor is:
\begin{equation}\label{eq:bianchi_weyl}
    \nabla^\rho C_{\rho\sigma\mu\nu} = \tfrac{2(D-3)}{(D-2)}\left(\nabla_{[\mu} \mathcal{R}_{\nu]\sigma} + \tfrac{1}{2(D-1)}\,g_{\sigma[\mu} \nabla_{\nu]}\mathcal{R}\right)\,.
\end{equation}

It is well known that the Riemann tensor has $\frac{D^2\br{D^2-1}}{12}$ independent components and the Ricci tensor $\frac{D(D+1)}{2}$. Hence, the Weyl has the leftover $\frac{D(D+1)(D+2)(D-3)}{12}$ independent components. In $D=4$ dimensions, the Riemann tensor has 20 independent components, while both the Ricci and the Weyl tensor have each 10 independent components. 

Additionally, the Weyl tensor is invariant under conformal transformations:
\begin{equation}
    g_{\m\n}\to\ti{g}_{\m\n}=\W~ g_{\m\n} ~\Longrightarrow~ C^\m_{~\n\r\s} = \ti{C}^\m_{~\n\r\s}\,.
\end{equation}

\subsection{\done{Electric and magnetic decomposition}}

In $D=4$, to calculate the Weyl tensor from $d=3$ dimensional quantities, it is usual to use its decomposition into its electric and magnetic parts, defined respective as:
\begin{equation}
\begin{aligned}
    E_{\m\n} &=n^\a n^\b C_{\a\m\b\n}\,,\\
	B_{\m\n} &=n^\a n^\b C^*_{\a\m\b\n}\,,
\end{aligned}
\end{equation}
where $C^*_{\a\m\b\n}$ is the dual Weyl tensor defined by:
\begin{equation}
	C^*_{\m\n\r\s} = \tfrac{1}{2}C_{\m\n\a\b}\e^{\a\b}_{~~~\r\s}\,,
\end{equation}
where $\e_{\a\b\m\n}$ is the Levi-Civita tensor, computed with $\e_{\a\b\m\n}=\sqrt{|g|}~\overline{\e}_{\a\b\m\n}$, where $\overline{\e}_{\a\b\m\n}$ as the Levi-Civita symbol.

Both the electric and the magnetic part are symmetric, traceless and spacelike, in the sense that $n^\m E_{\m\n} = 0 = n^\m B_{\m\n}$. One can recover the Weyl tensor from these using \cite[p.~290]{alcubierre}:
\begin{equation}\label{eq:weyl_decomposition}
	C_{\m\n\r\s}=2\br{l_{\m[\r}E_{\s]\n}-l_{\n[\r}E_{\s]\m} - n_{[\r} B_{\s]\b}\e^{\b}_{~\m\n} -  n_{[\m}B_{\n]\d} \e^{\d}_{~\r\s}}\,,
\end{equation}
where
\begin{equation}
	l_{\m\n} := \g_{\m\n} + n_{\m}n_{\n} = g_{\m\n} + 2n_\m n_\n\,.
\end{equation}

According to \cite[p.~290]{alcubierre}, the standard decomposition in the ADM formalism for $D=4$ is:
\begin{equation}
\begin{aligned}\label{eq:alcubierre_E_and_B}
    E_{ij}^{ADM} &= R_{ij} + KK_{ij} - K_{im}K^{m}_{~~j} - \tfrac{\kappa}{4}\sbr{S_{ij} + \tfrac{1}{3}\gamma_{ij}\br{4\rho-S}} - \tfrac{2}{3}\L\,\g_{ij}\,,\\
    B_{ij}^{ADM} &= \epsilon_{i}^{~mn}\sbr{D_m K_{nj} - \tfrac{\kappa}{4}\gamma_{jm}S_n}\vspace{-15mm}\,,
\end{aligned}
\end{equation}
where, $\e_{\b\m\n} = n^\a \e_{\a\b\m\n}$ is the projected Levi-Civita tensor. The above formulas assume the Hamiltonian and Momentum constraints are satisfied, and do not include the Z4 vector. The full CCZ4 expression is:
\begin{align}\nonumber
    E_{ij} &= E_{ij}^{ADM} - \tfrac{1}{3}\g_{ij}\mathcal{H} + \br{D_{(i} \Q_{j)} - K_{ij}\Q}^{TF} =\\\label{eq:ccz4_Electric}
    &= \br{R_{ij} + D_{(i}\Q_{j)} + (K-\Q)K_{ij} - K_{im}K^{m}_{~~j} - \tfrac{\k}{4}S_{ij}}^{TF}\,,\\\label{eq:ccz4_magnetic}
    B_{ij} &= B_{ij}^{ADM} - \tfrac{1}{2}\e_{ij}^{~~m}\mathcal{M}_m = \e_{i}^{~mn}D_m K_{nj}- \tfrac{1}{2}\e_{ij}^{~~m}\br{D_n K^n_{~m} - D_m K} = \epsilon_{(i}^{~mn}D_m K_{j)n}\,,
\end{align}
where $\mathcal{H}$ and $\mathcal{M}_i$ are the Hamiltonian and Momentum constraints (discussed in section \ref{sec:ngr:ccz4}). Notice as well how these forms, unlike \eqref{eq:alcubierre_E_and_B}, are explicitly symmetric and traceless without recurring to the constraints. Being spatial, symmetric and traceless, each have 5 independent components, together totalling the 10 independent components of the Weyl. This decomposition cannot be done directly for the Riemann tensor and is only valid for $D=4$. Notice as well how in the CCZ4 evolution equations (see section \ref{ccz4_summary}), the typical quantities showing up are $K-2\Q$ and $R_{ij}-2D_{(i}\Q_{j)}$ and not $K-\Q$ and $R_{ij}-D_{(i}\Q_{j)}$ as above. This complicates computation slightly, but it happens as we are doing normal projections of the Weyl tensor and not of the Riemann tensor (for which those factors of 2 appear).

To understand why these tensors are called the electric and magnetic components of the Weyl tensor, we can do a spatial and normal projection of \eqref{eq:bianchi_weyl}, which leads to Maxwell-like evolution equations for the electric and magnetic parts of $C_{\mu\nu\alpha\beta}$ \cite{Friedrich:1996hq}:
\begin{equation}
\begin{aligned}\label{eq:maxwellEB}
    \pounds_n E_{\alpha\beta} &=  \epsilon_{\rho\sigma(\alpha} D^{\sigma}B^{\rho}_{\phantom{\rho}\beta)} -5\,E_{(\alpha}^{\phantom{(\alpha}\lambda} K_{\beta)\lambda} +2\,K\,E_{\alpha\beta}  + E_{\rho\sigma}\,K^{\rho\sigma}\,\gamma_{\alpha\beta} - 2\,a_\rho\,\epsilon^{\rho\lambda}_{\phantom{\sigma\lambda}(\alpha }\,B_{\beta)\lambda} \,,\\
    \pounds_n B_{\mu\nu} & = - \epsilon_{\rho\sigma(\alpha} D^{\sigma}E^{\rho}_{\phantom{\rho}\beta)} -5\,B_{(\alpha}^{\phantom{(\alpha}\lambda} K_{\beta)\lambda} +2\,K\,B_{\alpha\beta}  + B_{\rho\sigma}\,K^{\rho\sigma}\,\gamma_{\alpha\beta} + 2\,a_\rho\,\epsilon^{\rho\lambda}_{\phantom{\sigma\lambda}(\alpha }\,E_{\beta)\lambda}\,,
\end{aligned}
\end{equation}
and the constraints,
\begin{equation}
\begin{aligned}\label{eq:constraintEB}
    D^\beta E_{\beta\alpha} &=- \epsilon_\alpha^{\phantom{\alpha}\rho\sigma} B_\rho^{\phantom{\rho}\lambda}\,K_{\sigma\lambda}\,, \\
    D^\beta B_{\beta\alpha} &= \epsilon_\alpha^{\phantom{\alpha}\rho\sigma} E_\rho^{\phantom{\rho}\lambda}\,K_{\sigma\lambda}\,. 
\end{aligned}
\end{equation}
Here $a^\nu = n^\mu\nabla_\mu n^\nu$ is the acceleration vector and $D_\mu$ is the covariant derivative compatible with the induced metric $\gamma_{\mu\nu} = g_{\mu\nu} + n_\mu\,n_\nu$. It is worth noting that equations \eqref{eq:maxwellEB} and \eqref{eq:constraintEB} should be entirely equivalent to the wave-like equation that one can derive from \eqref{eq:bianchi_riemann} \cite{Friedrich:1996hq}.

\subsection{\done{Kretschmann scalar}}

The Kretschmann scalar, defined as $\mathcal{C} = \mathcal{R}_{\m\n\r\s}\mathcal{R}^{\m\n\r\s}$, is a commonly used scalar as it is non-zero in vacuum, unlike the Ricci scalar. For the Kerr black hole of mass $M$ and spin parameter $a$, it takes the value of $\mathcal{C} = \frac{48M^2}{\br{r^2+a^2\cos^2\q}^6}\br{r^6 - 15a^2 r^4 \cos^2\q + 15a^4 r^2 \cos^4\q - a^6\cos^6\q}$, which reduces to $\mathcal{C} = \frac{48M^2}{r^6}$ for the Schwarzschild black hole, with $a=0$.

One straightforward way to compute it is using the Weyl tensor. Defining the equivalent scalar using the Weyl tensor, $W = C_{\m\n\r\s}C^{\m\n\r\s}$, the Kretschmann scalar can be computed using \cite{Cherubini:2002gen}:
\begin{equation}
    \mathcal{C} = W + \frac{4}{D-2}\mathcal{R}_{\m\n}\mathcal{R}^{\m\n} - \frac{2}{(D-1)(D-2)}\mathcal{R}^2\,.
\end{equation}

Regarding $\mathcal{R}_{\m\n}$ and $\mathcal{R}$, they are either zero in vacuum or they can be computed from the stress energy tensor $T_{\m\n}$ from the equations of motion. Notice that if using formulations such as the CCZ4 formulation, the relation between the Ricci tensor and the stress energy tensor is not the same as GR (see equation \eqref{eq:gr:ccz4_eom}), and the Ricci tensor and scalar may not be zero even in vacuum if the constraints are not identically satisfied.

To compute $W$, in $D=4$ one can make use of the electric and magnetic parts:
\begin{equation}\label{eq:weyl_squared}
    W = 8\br{E_{ij}E^{ij} - B_{ij}B^{ij}}\,.
\end{equation}
Away from $D=4$, one can use the Gauss-Codazzi equation, Codazzi-Mainardi equation and the double normal project of the Riemann \cite{alcubierre} to reconstruct the Riemann tensor fully from $3+1$ quantities and compute the Kretschmann scalar from it.
Additionally, to compute the full Riemann tensor, one can also compute the Weyl tensor using equation \eqref{eq:weyl_decomposition} and then use \eqref{eq:weyl-definition}, where again the Ricci tensor and scalar can be obtained from the equations of motion.

\subsection{\done{Weyl scalar $\Psi_4$}}

In the Newman-Penrose formalism \cite{Newman:1961qr}, one introduces a complex null tetrad $(l^\mu,k^\mu,m^\mu,\bar{m}^\mu)$, where we follow the notation of Alcubierre \cite{alcubierre} in order to avoid confusion with the normal $n^{\mu}$ to the foliation (see \ref{adm_summary}). The Newman-Penrose scalar, or Weyl scalar, $\Psi_4$ is defined by
\begin{equation}
    \Psi_4 = C_{\alpha\beta\gamma\delta} k^\alpha \bar{m}^\beta k^\gamma \bar{m}^\delta\,,
\end{equation}
which can be shown to reduce to \cite{alcubierre}
\begin{equation}
    \Psi_4 = (E_{ij} - i B_{ij})\bar{m}^i\bar{m}^j\,.
\end{equation}
We use the approach described by Baker et al. \cite[sec. V A, step (a)]{Baker:2001sf} to construct a null tetrad with the inner products:
\begin{equation}
    -l_\alpha k^\alpha = m_\alpha \bar{m}^\alpha = 1,
\end{equation}
and all other inner products vanishing. Following Bruegmann et al. \cite{Bruegmann:2006ulg} and Fiske et al. \cite{Fiske:2005fx}, we omit the null rotations in order to bring the tetrad into a quasi-Kinnersley form \cite[sec. V A, step (b)]{Baker:2001sf}.

\section{\done{Scaling tests}}\label{appendix:grchombo:scaling}

\texttt{GRChombo} has been extensively used as a high performance code. During progress made for this thesis, it was necessary to demonstrate it can scale efficiently on large super-computing systems.

There are two common metrics used to measure performance: strong scaling and weak scaling. Strong scaling refers to how much faster a simulation gets using more and more resources. An ideally paralysed program would double in speed if we double resources. In practice, both the computational workload is not infinitely parallelisable or as the parallel components become faster and faster, the serial sections of the code keep running with the same performance. Weak scaling refers to the ability to run a more computationally expensive setup using proportionally more resources. In an ideally scenario, doubling the number of grid points can achieve the same speed of evolution if we double the resources. Weak scaling is referred to as such as it is typically easier to obtain when compared to strong scaling, as the parallel sections of the code are kept constant as we increase both computational work and resources.

In the following sections we show that \texttt{GRChombo} exhibits excellent weak scaling properties for a vast range of jobs sizes.

\subsection{\done{Overview}}

The scaling tests described in this section were performed in one of the largest and newest supercomputers of the world: MareNostrum4. It has 48 racks housing 3456 nodes with a grand total of $165,888$ processor cores and 390 Terabytes of main memory. Each node has 2 sockets Intel Xeon Platinum 8160 CPUs with 24 cores each with 2.10 GHz for a total of 48 cores per node, and 96GB of main memory. For all tests the code was compiled with the Intel 2017 compilers, AVX-512 vectorisation and hybrid MPI/OpenMP parallelisation.

In the tests presented here we fix the box size to $16$ cells in each direction. With load balance, boxes are distributed across MPI processes and, for each  process, work is further distributed between OpenMP threads (see details in section \ref{subsubsection:ngr:parallelization}). Due to memory constraints, we run the simulations with 4 OMP threads per task and we keep this constant in all the jobs to avoid running out of memory.

We run small, medium and large simulations for evolutions ranging from quick single black hole experimentation to production high resolution black hole binary convergence tests. These make full use of AMR, with a hierarchy of 7 levels of refinement (8 in total) set to a 2:1 ratio. Some of these levels are enforced by hand while some others are adjusted dynamically and automatically by our code at every coarse timestep according to some tagging criterion that estimates the local numerical error. In performing the tests, the measurements presented include all necessary ghost cell communication and interpolation, and full regridding and load balancing operations. We excluded the initial data calculation as it takes at most one minute and is only done once in a simulation with a wall-clock time of several days. We have also turned off the HDF5 outputs, since they are infrequent in a real run whilst they could have a misleading impact in the short run of a test. Similarly, some diagnostics such as finding the apparent horizon (AH) of black holes and extracting gravitational waves that we expect to run either infrequently or not at all in the real simulations were not included. All other diagnostics that we expect to need in subcycling steps of a real simulation, such as measurement of hyperbolicity and stability conditions of the equations, are included since they represent an accurate reflection of the type of jobs described in our proposal. Finally, we set the Courant factor to $0.2$.

\subsection{\done{Strong scaling}}\label{sec:strong}

GRChombo has excellent strong scaling as long as there are enough boxes for all the MPI processes. In this section we will show the problem described in chapter \ref{chapter:paper1} can be strong-scaled to a large number of cores. For three types of jobs (small, medium and large), we perform a fit to Amdahl's law of theoretical speedup with the parallel fraction of the algorithm as a free parameter. In all cases we find extremely good agreement with Amdahl's law, which indicates that our small, medium and large jobs are in the regime of strong scalability.

\subsubsection*{\done{Small jobs test}}

The small jobs needed were varied, but typically they were mostly related to parameter searches. In order to have an accurate reflection of a typical ``small'' job that tests the relevant aspects of the simulation, we have chosen to study gravitational collapse in Horndeski theory as in chapter \ref{chapter:paper1}. This problem involves the dynamical generation of new relevant length scales, such as the size of the black hole, and hence it makes full use of AMR. In addition, for the theory that we consider, the r.h.s. of the evolution equations is significantly more involved than in standard general relativity. Therefore, we can also test the efficiency in the computation of a very complex r.h.s., similar to what we encounter in the rest of the chapters.

\begin{figure}[H]
\begin{center}
 \includegraphics[width=0.49\linewidth]{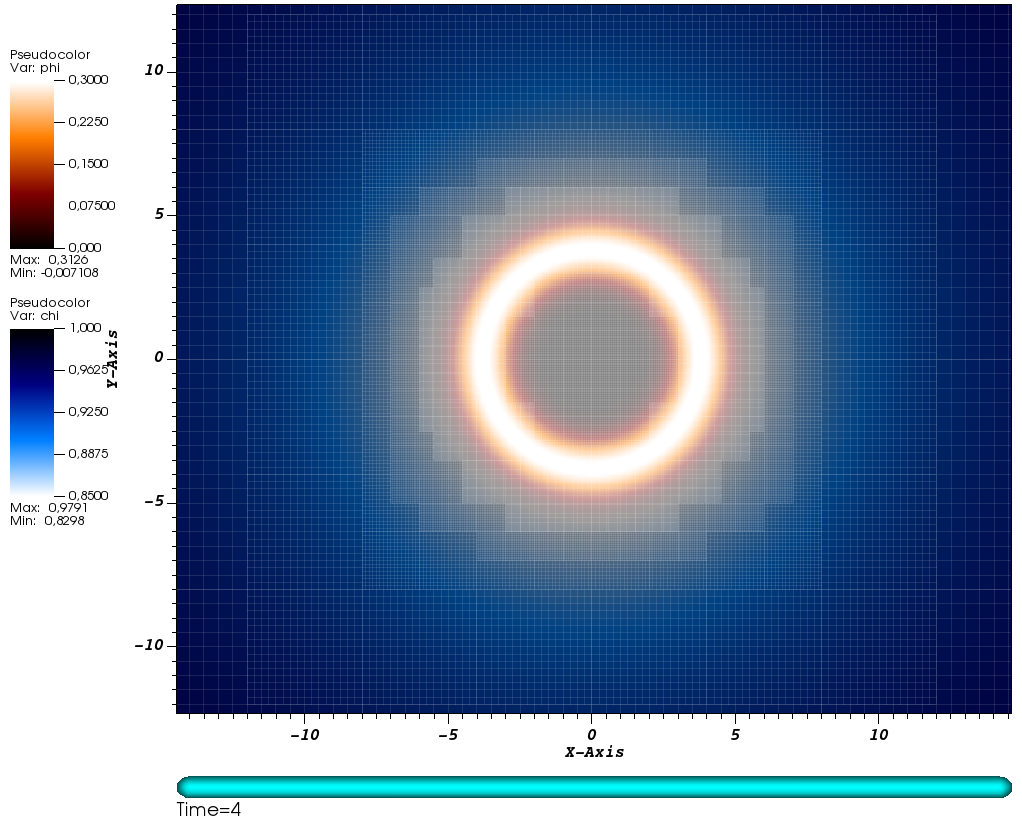}
 \includegraphics[width=0.49\linewidth]{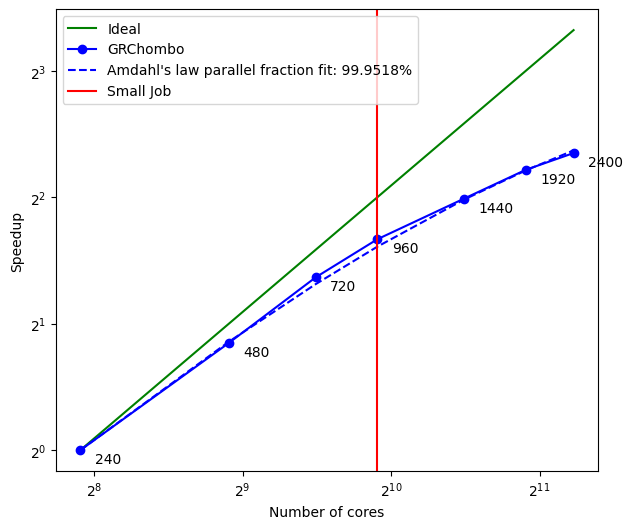}
  \caption{\textit{Left}: Illustration of the job performed: gravitational collapse in Horndeski theory and mesh structure at the last time step of the test. \textit{Right}: strong scaling test for small sized jobs, showing a useful speedup up to 2400 cores. The typical core count for a small job is shown in red.}
  \label{fig:strong_small}
\end{center}
\end{figure}

For this type of job, the coarsest grid size is $64^3$ for a computational domain of $L=64M$ in each direction. We evolve a spacetime of mass $0.5M$ as in Case 3 of section \ref{sec:g2_pos_weak_coupling} for 20 time steps of the coarsest level, including all subcycling steps on the finer levels. The initial data was constraint satisfying and thus constitutes a physically correct proof of principle. The results of the strong scaling tests for the ``small'' jobs are shown in figure \ref{fig:strong_small} and the precise numbers are given in Table \ref{tab:strong_small}.

\begin{table}[H]
\begin{center}
\begin{tabular}{|c|c|c|c|}
\hline
Number of cores & Average Speed ($M$/hr) & Speedup & Ideal speedup \\
\hline\hline
240 & 11.9 & 1.00 & 1.00 \\
480 & 21.5 & 1.80 & 2.00 \\
720 & 30.8 & 2.58 & 3.00 \\
\textbf{960} & \textbf{37.9} & \textbf{3.18} & \textbf{4.00} \\
1440 & 47.4 & 3.97 & 6.00 \\
1920 & 55.5 & 4.65 & 8.00 \\
2400 & 60.8 & 5.10 & 10.00 \\
\hline
\end{tabular}
\caption{Results for the strong scaling tests for small jobs. The typical small job described in our proposal is highlighted in boldface.}
\label{tab:strong_small}
\end{center}
\end{table}

\subsubsection*{\done{Medium jobs test}}

\begin{figure}[H]
\begin{center}
  \includegraphics[width=0.49\linewidth]{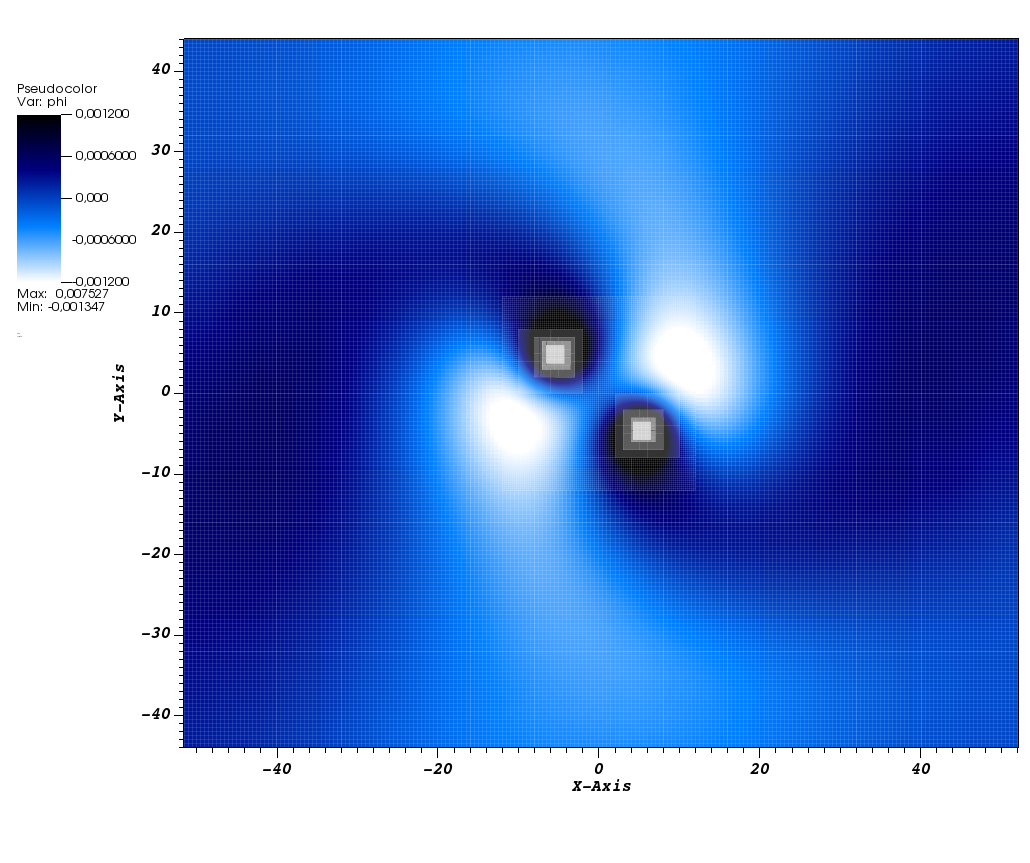}
  \includegraphics[width=0.49\linewidth]{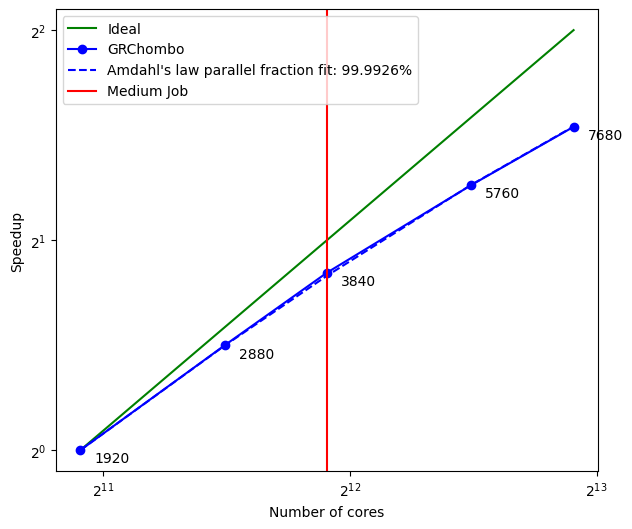}
  \caption{\textit{Left}: Illustration of the job performed: scalar environment of a black hole binary and mesh structure. \textit{Right}: strong scaling test for medium sized jobs, showing a useful speedup up to 7680 cores. The typical core count for a medium job is shown in red.}
  \label{fig-strong_medium}
\end{center}
\end{figure}

The medium sized jobs are also varied, but a typical example is the study of scalar field environments around binary black holes (BHs). An illustrative example of the case we ran is shown in figure \ref{fig-strong_medium} (\textit{left}). Unlike the smaller gravitational collapse example of our previous scaling test, these studies necessitate a larger region of resolution, both around the BHs and asymptotically, which justifies the medium job size proposed.

The coarsest level of refinement has a grid size of $384^3$ for a computational domain of $L=768M$ in each direction. We consider two $0.5M$ BHs surrounded by scalar field with mass  parameter $\mu=0.5 M^{-1}$, with initial separation of $40M$ and tangential velocities of $\sim 0.05$. This constitutes initial data exactly as we intend to use for the results of chapter \ref{chapter:paper2} and hence it is a reliable example of the resources required. We evolve this system for 10 full time steps. The strong scaling is shown in figure \ref{fig-strong_medium}, and the exact results are provided in Table \ref{tab:strong_medium}. To carry out the tests, we halve and double the the number of cores around 3840 cores, which is the typical ``medium'' job size described in our proposal. For this type of job, we could not run the simulation with 960 cores and 4 OMP threads without running out of memory.

\begin{table}[H]
\begin{center}
\begin{tabular}{|c|c|c|c|}
\hline
Number of cores & Average Speed ($M$/hr) & Speedup & Ideal speedup \\ \hline\hline
1920 & 13.5 & 1.00 & 1.00 \\ 
2880 & 19.1 & 1.41 & 1.50 \\ 
\textbf{3840} & \textbf{24.2} & \textbf{1.80} & \textbf{2.00} \\ 
5760 & 32.4 & 2.40 & 3.00 \\ 
7680 & 39.2 & 2.91 & 4.00 \\ \hline
\end{tabular}
\caption{Results for the strong scaling tests for the medium sized jobs. The typical medium job described in the proposal is highlighted in boldface.}
\label{tab:strong_medium}
\end{center}
\end{table}

\subsubsection*{\done{Large jobs test}}

The large sized jobs proposed in our application will be used for high resolution runs and convergence tests for the final production simulations. These jobs make use of the excellent weak scaling properties of GRChombo by increasing both the resources and the resolution in order to get adequate and reasonable runtime.

\begin{figure}[H]
\begin{center}
  \includegraphics[width=0.6\linewidth]{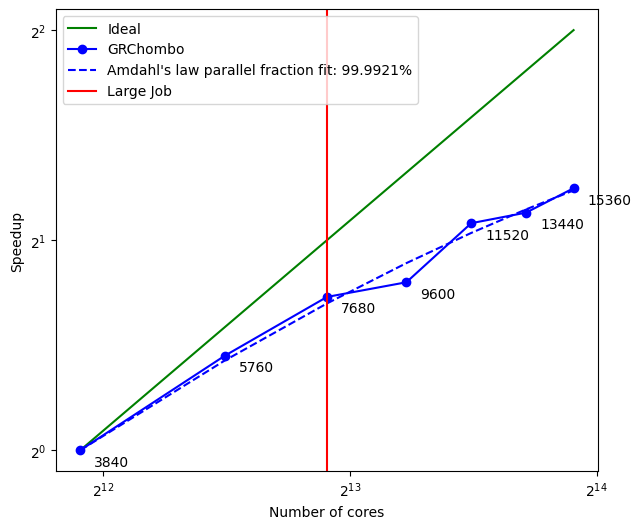}
  \caption{Strong scaling test for large sized jobs, showing a useful speedup up to 15360 cores. The typical core count for a large job is shown in red.}
  \label{fig-strong_large}
\end{center}
\end{figure}

To illustrate this, we ran the same configuration as in the medium jobs test, changing only the coarsest level resolution to $432^3$. This would correspond to a higher resolution run that we expected to use for a convergence test. The strong scaling is shown in figure \ref{fig-strong_large}, and the exact results are provided in Table \ref{tab:strong_large}. We halve and double the number of cores around 7680 cores, which is the typical size for a ``large'' job in our proposal. For this type of job, we could not run simulation with 1920 cores and 4 OMP threads without running out of memory. Although beyond 7680 cores we still have good strong scaling, we observe a more erratic behaviour. The reason for this is that even though the job is such that there are boxes for all ranks, at higher core counts each rank has fewer boxes, which results in an uneven load balance. Increasing the resolution of the coarsest level should give rise to more boxes per level, a more even load balance, and hence better strong scaling. Unfortunately we were not able to check this since we ran out of resources.

\begin{table}[H]
\begin{center}
\begin{tabular}{|c|c|c|c|}
\hline
Number of cores & Average Speed ($M$/hr) & Speedup & Ideal speedup \\ \hline\hline
3840 & 16.5 & 1.00 & 1.00 \\
5760 & 22.5 & 1.37 & 1.50 \\
\textbf{7680} & \textbf{27.4} & \textbf{1.66} & \textbf{2.00} \\
9600 & 28.7 & 1.74 & 2.50 \\
11520 & 34.9 & 2.12 & 3.00 \\
13440 & 36.1 & 2.19 & 3.50 \\
15360 & 39.2 & 2.37 & 4.00 \\ \hline
\end{tabular}
\caption{Table of results for the strong scaling tests for large sized jobs. The typical large job described in our proposal is highlighted in boldface.}
\label{tab:strong_large}
\end{center}
\end{table}

\subsection{\done{Weak scaling}}
\label{sec:weak}

\begin{figure}[H]
\begin{center}
  \includegraphics[width=0.7\linewidth]{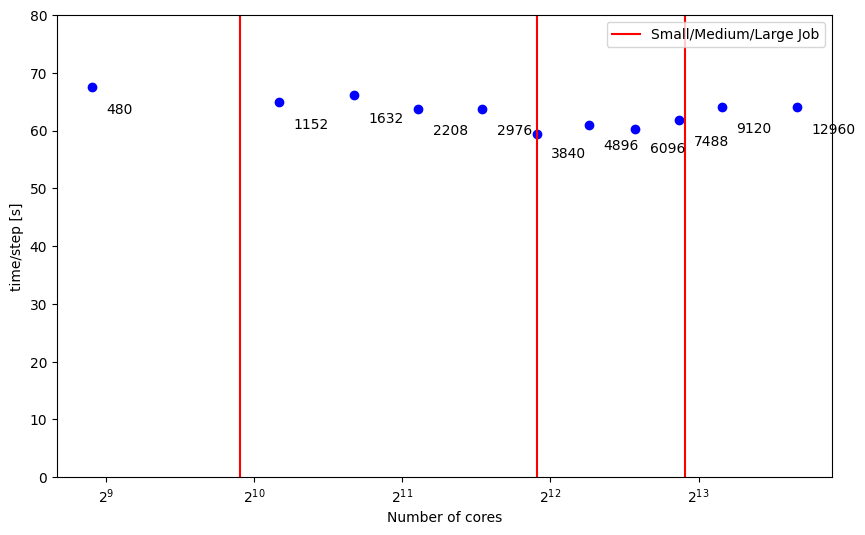}
  \caption{Results for the weak scaling test, showing excellent weak scaling for the range of jobs (small, medium and large) described in our proposal.}
  \label{fig-weak}
\end{center}
\end{figure}

To test the weak scaling of GRChombo, we use the same configuration as for the medium jobs strong scaling test on 3840 cores as a starting
point. We then decrease and increase the resolution while adjusting the number of nodes and the grid size to keep the number of cells per node fixed. We show the results of the weak scaling test in figure \ref{fig-weak}. In Table \ref{tab:weak} we show the speedup per full coarse time step as well as the ratio between grid size and core count to demonstrate that the work per process is kept approximately constant as the number of cores changes. 

We see extremely good weak scaling around small, medium and large jobs, with the performance even improving when increasing both the resources and the resolution. We attribute this effect firstly to the dynamical addition of finer levels. At low resolutions, coarser levels need larger boxes around the regions of interest, ending up covering more physical space. On the other hand, higher resolutions are able to have more precise contours of the deeper refinement levels around the regions of interest and hence refine smaller regions, which results in a slight efficiency boost. Another reason for this effect is that with more resolution there are more boxes per level and hence it is simply easier to obtain an evenly distributed load balance of boxes over ranks, which boosts the performance.

\begin{table}[H]
\begin{center}
\begin{tabular}{|c|c|c|c|}
\hline
Number of cores & Average Speed (s/step) & $\sqrt[3]{\textrm{Grid size}}$ & $\sqrt[3]{\textrm{Grid size / core}}$ \\ \hline\hline
480 & 67.6 & 192 & 24.52 \\ 
1152 & 64.9 & 256 & 24.42 \\
1632 & 66.1 & 288 & 24.46 \\
2208 & 63.8 & 320 & 24.57 \\
2976 & 63.8 & 352 & 24.47 \\
3840 & 59.5 & 384 & 24.52 \\
4896 & 60.9 & 416 & 24.50 \\ 
6096 & 60.4 & 448 & 24.52 \\ 
7488 & 61.8 & 480 & 24.53 \\ 
9120 & 64.1 & 512 & 24.51 \\ 
12960 & 64.0 & 576 & 24.52 \\ \hline
\end{tabular}
\caption{Results for the weak scaling tests.}
\label{tab:weak}
\end{center}
\end{table}

\section{\done{Cubic spline interpolation}}\label{appendix:grchombo:spline}

When generating initial data for chapters \ref{chapter:paper1}, \ref{chapter:paper2} and \ref{chapter:eft}, it was common to them all that spherically symmetric data was generated in one dimension in some external software, and had to be ported to \texttt{GRChombo} 3D code. Excluding the aspects of transforming spherically symmetric data to Cartesian volume coordinates (details in section \ref{appendix:paper1:initialData}), simply exporting 1D data (discretised with some articular numerical grid) into 3D (with some other numerical grid requiring points at various radii) required the usage of a method like Cubic Spline for interpolation.

There are several methods of interpolation. \texttt{GRChombo} uses fourth order Lagrange interpolation to compute stencils. But when the intention is to interpolate a discrete set of points approximating some function, interpolating all the data across domains with different behaviour with a single high order polynomial leads to poor results and wild oscillations. A common technique is polynomial interpolation, in which the set of points is divided in chunks (e.g. 2 points), a polynomial of some degree is interpolated between them (with some boundary conditions to match with the surrounding chunks) and the overall function becomes defined by piecewise polynomials. A common method of this kind, known for generating smooth continuous curves, is cubic spline. The main advantages of cubic spline are: the smoothness of the generated curves; the spacing between the interpolated points can be arbitrary and uneven; the interpolated function passes by design in points being interpolated (opposed to, for example, fitting methods); uses cubic polynomials and hence is reasonably cheap and stable to compute, and finally it requires knowledge of the function alone without its derivatives.

The outline of the method for one dimensional data is as follows\footnote{Cubic Spline is easily applied to multiple dimensions, in fact it is quite a common method practised in computer graphics for 2D and 3D data (see \href{https://en.wikipedia.org/wiki/Cubic_Hermite_spline}{https://en.wikipedia.org/wiki/Cubic\_Hermite\_spline}).}: divide all the set of points into pairs of adjacent points. Using linear interpolation would require a lot of points along each curve to obtain enough precision. Furthermore, it is not smooth. If we require smoothness of the first and second derivatives across segments with only a reasonable amount of points, we need to use a cubic polynomial for each segment, resulting in Cubic Spline. More technically, there are many ways to describe the method. Below follows a particular one that we found to be particularly clean and intuitive. 

Suppose there is a function $f(x)$ that we want to recover via interpolation, and we discretise it with $n+1$ points, $\br{x_i,y_i=f(x_i)}$ for $(i=1,\dots,n+1)$ (these are ordered yet arbitrary points in the domain of $f$). On every interval $[x_i,x_{i+1}]$, we'll interpolate with an auxiliary set of functions $f_{i,i+1}=f_{x\in [x_i,x_{i+1}]}$ for each subdomain $[x_i,x_{i+1}]$ with $(i=1,\dots,n)$ defined as the cubic polynomial \eqref{eq:f_i} and subject to the conditions \eqref{eq:cond1}-\eqref{eq:cond4}:
\begin{align}\label{eq:f_i}
    f_{i,i+1}(x)&=a_i + b_i(x - x_i) + \tfrac{C_i}{2}(x-x_i)^2 + \tfrac{d_i}{3!}(x-x_i)^3\,,
\end{align}
\vspace{-15mm}
\begin{multicols}{2}
    \begin{align}\label{eq:cond1}
        f_{i,i+1}(x_i)&=y_i\,,\\\label{eq:cond2}
        f_{i,i+1}(x_{i+1})&=y_{i+1}\,,
    \end{align}
    
    \begin{align}\label{eq:cond3}
        f_{i,i+1}'(x_i)&=f_{i-1,i}'(x_i)\,,\\\label{eq:cond4}
        f_{i,i+1}''(x_i)&=f_{i-1,i}''(x_i)\,,
    \end{align}
\end{multicols}
where $a_i,b_i,C_i,d_i$ are constants for $(i=1,\dots,n)$. Since with $n+1$ points there are $n$ segments, this results in $4n$ parameters to fix. Then there are 2 zeroth-order constraints per segment, \eqref{eq:cond1} and \eqref{eq:cond2}, plus 2 for each point connecting segments (all except the extremes), \eqref{eq:cond3} and \eqref{eq:cond4}. This total $2n+2(n-1)=4n-2$ constraints. This results in two missing conditions, arising from our lack of knowledge of the derivatives at the extremes. To complete the set of constraints, we choose second derivatives at the extremes to be zero: $f_{1,2}''(x_1)=C_1=0$ and $f_{n,n+1}''(x_{n+1})=C_{n}+d_n(x_{n+1}-x_n):=C_{n+1} = 0$. This is often referred to as a ``no bending'' condition at the endpoints, as it extends the first derivative along the extremes with the same slope it had before.

The $n-1$ conditions \eqref{eq:cond4} result in $d_{i} = \frac{C_{i+1} - C_i}{x_{i+1} - x_{i}}\,,~~(i=1,\dots,n)$, where $C_{n+1}=0$ from the no bending condition at $x_{n+1}$. This results in making the second derivative for any segment a simple linear interpolation:
\begin{gather}
    f_{i,i+1}''(x) = C_i + (C_{i+1}-C_i)\frac{x-x_i}{x_{i+1}-x_i} = \frac{C_{i+1}(x-x_i) + C_i(x_{i+1}-x)}{x_{i+1}-x_i}\,,
\end{gather}
thus eliminated the 3rd order coefficients. Integrating twice and redefining the constants $a_i$ and $b_i$, we can re-write \ref{eq:f_i} as:
\begin{gather}
    f_{i,i+1}(x) = \frac{1}{x_{i+1}-x_i}\sbr{\tfrac{C_i}{6}(x_{i+1}-x)^3 + \tfrac{C_{i+1}}{6}(x-x_i)^3} + A_i(x_{i+1}-x) + B_i(x-x_{i})\,.
\end{gather}
Using constraints \ref{eq:cond1} and \ref{eq:cond2} we get:
\begin{equation}
\begin{aligned}
    A_i &= \frac{y_i}{x_{i+1}-x_i}-\frac{C_i}{6}(x_{i+1}-x_i)\,,\\
    B_i &= \frac{y_{i+1}}{x_{i+1}-x_i}-\frac{C_{i+1}}{6}(x_{i+1}-x_i)\,.
\end{aligned}
\end{equation}
So, overall we have:
\begin{align}\label{eq:appendix:spline:f0}
    f_{i,i+1}(x) &= \frac{C_{i}}{6}\sbr{\frac{(x_{i+1}-x)^3}{x_{i+1}-x_i} - (x_{i+1}-x)(x_{i+1}-x_i)} +\\\nonumber & ~~~~ + \frac{C_{i+1}}{6}\sbr{\frac{(x-x_i)^3}{x_{i+1}-x_i} - (x-x_i)(x_{i+1}-x_i)} + \frac{y_{i+1}(x-x_i)+y_i(x_{i+1}-x)}{x_{i+1}-x_i}\,,\\\label{eq:appendix:spline:f1}
    f_{i,i+1}'(x) &= -\frac{C_{i}}{2}\sbr{\frac{(x_{i+1}-x)^2}{x_{i+1}-x_i} - \frac{x_{i+1}-x_i}{3}} + \frac{C_{i+1}}{2}\sbr{\frac{(x-x_i)^2}{x_{i+1}-x_i} - \frac{x_{i+1}-x_i}{3}} + \frac{y_{i+1}-y_{i}}{x_{i+1}-x_i}\,,\\\label{eq:appendix:spline:f2}
    f_{i,i+1}''(x) &= \frac{C_{i+1}(x-x_i) + C_i(x_{i+1}-x)}{x_{i+1}-x_i}\,.
\end{align}
We have left the job of finding the $C_i$ second order coefficients. This is done by using \ref{eq:cond3}, the first derivative condition, together with the no bending conditions $C_1 = 0 = C_{n+1}$. The former results in:
\begin{gather}
    C_{i-1}(x_i-x_{i-1})+2\,C_i(x_{i+1}-x_{i-1})+C_{i+1}(x_{i+1}-x_i) = 6\br{\frac{y_{i+1}-y_{i}}{x_{i+1}-x_i}-\frac{y_{i}-y_{i-1}}{x_{i}-x_{i-1}}} \equiv \Delta_i\,,
\end{gather}
for $i\in(2,\dots,n)$. We now have to solve the linear algebra problem of these $n-1$ coupled equations, together with the condition $C_1=0:=\D_1$. Using the vectors $\ve{C}=(C_1,\dots,C_n)^T$ and $\ve{\Delta}=(\Delta_1,\dots,\Delta_n)^T$, we have:
\begin{gather}
    A\,\ve{C} = \ve{\Delta}\,,\\
    A = \scriptscriptstyle{
    \begin{pmatrix}
        1 & 0 & 0 &  \dots & 0 & 0 & 0\\
        (x_2-x_1) & 2\,(x_3-x_1) & (x_3-x_2) & \dots & 0 & 0\\
        0 & (x_3-x_2) & 2\,(x_4-x_2) & \dots & 0 & 0\\
        \vdots & \vdots & \vdots & \ddots & \vdots & \vdots\\
        0 & 0 & 0 & \dots & 2\,(x_{n}-x_{n-2}) & (x_{n}-x_{n-1})\\
        0 & 0 & 0 & \dots & (x_n-x_{n-1}) & 2\,(x_{n+1}-x_{n-1})\\
    \end{pmatrix}}\,.
\end{gather}
The matrix $A$ is a tridiagonal matrix which can be efficiently solved with tridiagonal matrix inversion algorithms (e.g. \href{https://en.wikipedia.org/wiki/Tridiagonal_matrix_algorithm}{https://en.wikipedia.org/wiki/Tridiagonal\_matrix\_algorithm}). Recovering the final vector $\ve{C}$, then the function $f(x)$ and its first and second derivatives can be approximated by the piece-wise functions \eqref{eq:appendix:spline:f0}-\eqref{eq:appendix:spline:f2}.

\section{\done{Higher order Richardson extrapolation}}\label{appendix:sec:richardson_extrapolation}

Standard Richardson extrapolation is a common technique when looking for the converging or asymptote value of some sequence, some value $f^* = \lim_{h\to0}f(h)$, when we are only able to evaluate $f$ for several values of $h$ and having to extrapolate to $h=0$. It is often useful when computing derivatives is unfeasible or imprecise, and when a fitting method is not well suited or has not enough points. It allows one to get rid of first order errors, leaving us with second order perturbations, as we shall see. For the research of this thesis, it was not suited for several extrapolations needed as extrapolating the Weyl-4 scalar for gravitational waves, as these carry a significant amount of noise that strongly interferes with the accuracy of Richardson extrapolation, but it proved very useful for measures like the ADM mass at infinity, by measuring it at finite radii. It also proved useful to get a higher accuracy extrapolation, since getting rid of first order errors may not be good enough. Finally, Richardson extrapolation is also at the core of convergence testing, where we test numerical evolutions by checking the expected behaviour as we approach infinite resolution (or zero grid spacing). Below we describe the standard version of Richardson extrapolation and a higher order version which was used for extrapolations like the ADM mass of the spacetime.

Take any quantity that follows a known power law or constant asymptote behaviour as it approaches infinity. This could be the ADM mass, which is computed in spherical shells of increasing radius and takes the limiting value $M_{ADM}$ at spatial infinity; or the Weyl-4 scalar, which decays as $\Psi_4(r) \propto \frac{1}{r}$. Re-scale it to obtain a function, $A(r)$, that asymptotes to a constant at infinity (the ADM mass already does; for the Weyl-4 scalar, use $A(r)=r\cdot \Psi_4(r)$). We will now treat the smallness parameter $h=\frac{1}{r}$, such that $h\to0$ as $r\to\infty$. Expanding $A(r)$ around infinity:
\begin{gather}
    A(r) = A_{\infty} + \frac{B}{r} + \frac{C}{r^2} + \frac{D}{r^3} + O\br{r^{-4}}\,.
\end{gather}
The goal is to find out the value of $A_\infty$ (i.e. the ADM mass, the asymptotic decay of the gravitational wave, etc.). Using 2 points $r_1$ and $r_2$ (assume $r_1 > r_2$), we can easily see that:
\begin{gather}\label{eq:2point}
    \frac{A(r_1)-\frac{r_2}{r_1}A(r_2)}{1-\frac{r_2}{r_1}} = A_\infty -\frac{C}{r_1 r_2} + O\br{r^{-3}} = A_\infty + O\br{r^{-2}}\,.
\end{gather}
What did we achieve? Based on computing 2 points, we achieved a ratio which allows us to compute $A_\infty$ up to precision of $\mathcal{O}(r^{-2})$ instead of $\mathcal{O}(r^{-1})$, significantly improving the result provided we started with an error which is already small.

If we want to improve this result by achieving $\mathcal{O}(r^{-3})$ precision, we can do so by using 3 points. Using 3 points such that $r_1 > r_2 > r_3$, one can eliminate one further term, result in:
\begin{gather}\label{eq:3point}
    \frac{A(r_1)-\frac{r_2}{r_1}\br{\frac{1-\frac{r_3}{r_1}}{1-\frac{r_3}{r_2}}}A(r_2)+\frac{r_3}{r_1}\br{\frac{1-\frac{r_2}{r_1}}{\frac{r_2}{r_3}-1}}A(r_3)}{1-\frac{r_2+r_3}{r_1} + \frac{r_2 r_3}{r_1^2}} = A_\infty + \frac{D}{r_1 r_2 r_3} + O\br{r^{-4}} = A_\infty + O\br{r^{-3}}\,.
\end{gather}
As expected, \ref{eq:3point} reduces to \ref{eq:2point} if we set $r_3=0$.
\subsection{\done{General order decay}}\label{appendix:grchombo:general_order_decay}
It is often known that the quantity $A(r)$ that asymptotes to some $A_\infty$ decays not as $\tfrac{1}{r}$ but some higher power $\tfrac{1}{r^o}$. For this more general order decay, results \eqref{eq:2point} and \eqref{eq:3point} become:
\begin{gather}
    A_o(r) = A_{\infty} + \frac{B}{r^o} + \frac{C}{r^{o+1}} + \frac{D}{r^{o+2}} + O\br{r^{-(o+3)}}\,,
\end{gather}
\begin{gather}
    \frac{A_o(r_1)-\br{\frac{r_2}{r_1}}^o A_o(r_2)}{1-\br{\frac{r_2}{r_1}}^o} = A_\infty -\frac{C}{r_1 r_2}\br{\frac{r1-r2}{r1^o-r2^o}} + O\br{r^{-(o+2)}} = A_\infty + O\br{r^{-(o+1)}}\,,
\end{gather}
\begin{gather}
    \frac{A_o(r_1)-\br{\frac{r_2}{r_1}}^o\br{\frac{1-\frac{r_3}{r_1}}{1-\frac{r_3}{r_2}}}A_o(r_2)+\br{\frac{r_3}{r_1}}^o\br{\frac{1-\frac{r_2}{r_1}}{\frac{r_2}{r_3}-1}}A_o(r_3)}{1-\br{\frac{r_2}{r_1}}^o\br{\frac{1-\frac{r_3}{r_1}}{1-\frac{r_3}{r_2}}} + \br{\frac{r_3}{r_1}}^o\br{\frac{1-\frac{r_2}{r_1}}{\frac{r_2}{r_3}-1}}} = A_\infty + O\br{r^{-(o+2)}}\,.
\end{gather}
As expected, we can recover the previous formulas by setting $o=1$.
\subsection{\done{Convergence testing}}\label{appendix:grchombo:convergence}
In convergence testing, this technique can help figuring out the limiting value of numerically evolves variables if you have 2 or 3 different resolutions available. But more often, it is desirable to double check the convergence rate of the code, the order of decay $o$ of section \ref{appendix:grchombo:general_order_decay}. For that, start by re-writing $A(r)$ explicitly with a smallness parameter $h$ representing the grid spacing $\D x$:
\begin{gather}
    A_o(h) = A_{\infty} + B\,h^o + O\br{h^{o+1}}\,.
\end{gather}
Notice that, given $h_1 < h_2 < h_3$:
\begin{gather}
    c_o:=\frac{A_o(h_3) - A_o(h_2)}{A_o(h_2)-A_o(h_1)} = \frac{h_3^o - h_2^o}{h_2^o - h_1^o}\,\br{1 + O\br{h_3}}\,.
\end{gather}
For a constant rate of resolution improvement where $h_2/h_1 = r$ and $h_3/h_2 = r$, then $c_o\to r^o$. In general, give each resolution and the values of $A_o$ for each, one can solve the previous equation and find $o$, the order of convergence of the evolution scheme. Numerical noise often brings a lot of error to this measurement, and it is often enough to use the factor $c_o=\frac{h_3^o - h_2^o}{h_2^o - h_1^o}$ and the product $c_o\cdot\br{A_o(h_2) - A_o(h_1)}$ as a predictor of $A_o(h_3) - A_o(h_2)$. By computing this predictor using different values of $o$ when computing $c_o$ (2,3,4 if one expects second, third, fourth order convergence), one can qualitatively see what convergence rate a given solution follows. For examples of this, see appendixes \ref{appendix:paper1:convergence} and \ref{appendix:paper2:convergence} related to convergence of simulations of chapters \ref{chapter:paper1} and \ref{chapter:paper2}.

\chapter{\done{Numerical relativity formulations}}\label{appendix:ngr_formulations}

This appendix is a quick review of the main formulations in numerical relativity, mainly the BSSN and CCZ4 formulations used in \texttt{GRChombo}, supporting the summaries provided in \ref{adm_summary}-\ref{ccz4_summary}.


\section{\done{ADM formulation}}\label{sec:ngr:adm}

\subsection{\done{$d+1$ decomposition}}

The ADM decomposition\footnote{Originally derived by Arnowitt-Deser-Misner \cite{Arnowitt:1962hi} and re-written to its final form by York \cite{York:1978gql}.} starts with a foliation: splitting spacetime in successive non-intersecting $d$ dimensional spacelike surfaces, each identified by a scalar interpreted as a global time function, $t$, which may not coincide with the proper time of any particular observer. A spacetime that allows such foliation is called globally hyperbolic and each time slice is called a Cauchy hypersurface, $\S_t$.

Introducing spatial coordinates $x^i(t)$ labelling each hypersurface, define:
\begin{itemize}
    \item $dl^2 = \g_{ij}dx^idx^j$, defines the proper distance $dl$ in each hypersurface measured by an induced spatial metric $\g_{ij}$,
    \item $d\t = \a(t,x^i)dt$, for proper time $\t$ defines the lapse function, $\a$,
    \item $x^i(t+dt) = x^i_t - \b^i(t,x^i)dt$, defines the shift vector, $\b^i$.
\end{itemize}

The lapse function encodes our freedom to pick the time evolution of each slice and describes the proper time distance between hypersurfaces for observers at rest relative to the slices (moving in the normal direction to each). A value less than one means that proper time runs slower than our time coordinate $t$, and it does not change the physical interactions that occur. The shift vector allows re-parameterisation of the spatial coordinates, which is often essential to keep numerical stability as observers labelled by $x^i$ move relative to fixed points.
\begin{figure}[h]
\centering
\includegraphics[width=.6\textwidth]{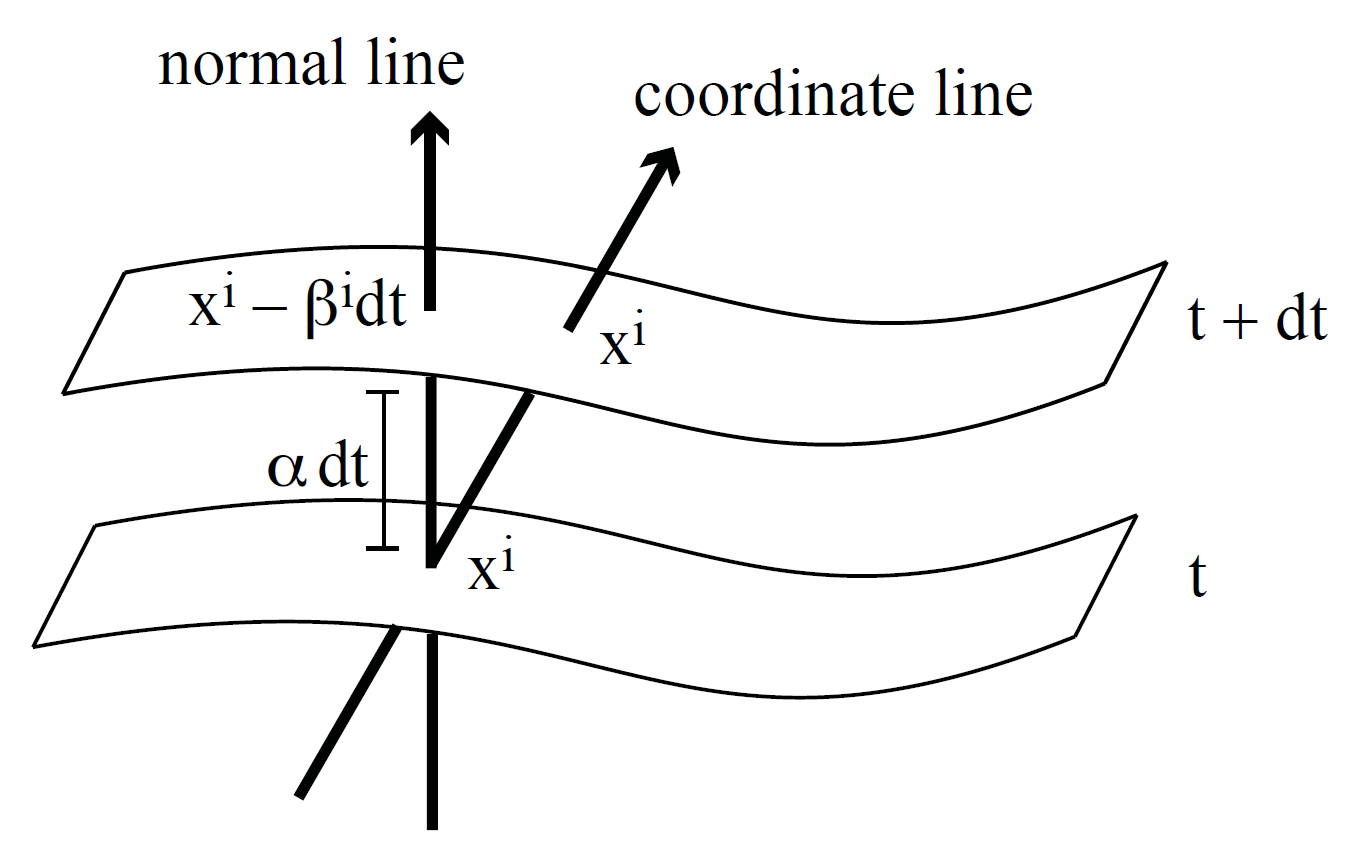}
\caption{Two adjacent hypersurfaces showing the meaning of the lapse function $\a$ and the shift vector $\b^i$. Figure taken from Alcubierre \cite[p.~66]{alcubierre}.}
\label{fig:ngr:lapse_shift}
\end{figure}
Using vector addition one can with this re-write the spacetime interval $ds$ and the spacetime metric as:
\begin{align}\nonumber
    ds^2 &= -\a^2dt^2 + \g_{ij}\br{dx^i+\b^idt}\br{dx^j + \b^jdt} = \\\label{eq:ngr:spacetime_interval}
    &= \br{-\a^2 + \b_i \b^i}dt^2 + 2\b_i dx^idt + \g_{ij}dx^idx^j\,.
\end{align}
\begin{gather}
    g_{\mu\nu}=
        \begin{pmatrix}
            -\alpha^2+\beta_i\beta^i & \beta_i\\
            \beta_j                 & \gamma_{ij}
        \end{pmatrix}\,,
        \quad\quad\quad
        g^{\mu\nu}=
        \begin{pmatrix}
            -\frac{1}{\alpha^2}         & \frac{\beta^i}{\alpha^2}\\
            \frac{\beta^i}{\alpha^2}  & \gamma^{ij}-\frac{\beta^i\beta^j}{\alpha^2}
        \end{pmatrix}\,.
\end{gather}
Note the metric determinant results in $\sqrt{-g}=\alpha\sqrt{\gamma}$. The unit and timelike normal vector to each slice of constant time, $n^\m = \frac{-\grad^\m t}{|\grad_\m t|}$ is\footnote{The minus sign ensures $n^\m$ is future directed, aligned with increasing $t$.}:
\begin{gather}
    n_\m = \br{-\a,0}\,, \quad\quad\quad n^\m = \br{\tfrac{1}{\a},-\tfrac{\b^i}{\a}}\,,
\end{gather}
where $n_\m n^\m = -1$. Finally, the normal vector allows to define a projector operator from spacetime tensors to the spatial hypersurface:
\begin{gather}\label{eq:ngr:projector}
   P_\m^{~\n} = \d_\m^{~\n} + n_\m n^\n = \g_\m^{~\n}\,.
\end{gather}
Note we associated this with the induced metric $\g_{ij}$, which is consistent with a constant time spacetime interval \eqref{eq:ngr:spacetime_interval}. This can by checked by verifying that:
\begin{gather}
    P_\m^{~\a}P_\n^{~\b}g_{\a\b} = g_{\m\n} + n_\m n_\n = \g_{\m\n}\,.
\end{gather}

\subsection{\done{Extrinsic curvature}}

\begin{figure}[h]
\centering
\includegraphics[width=.6\textwidth]{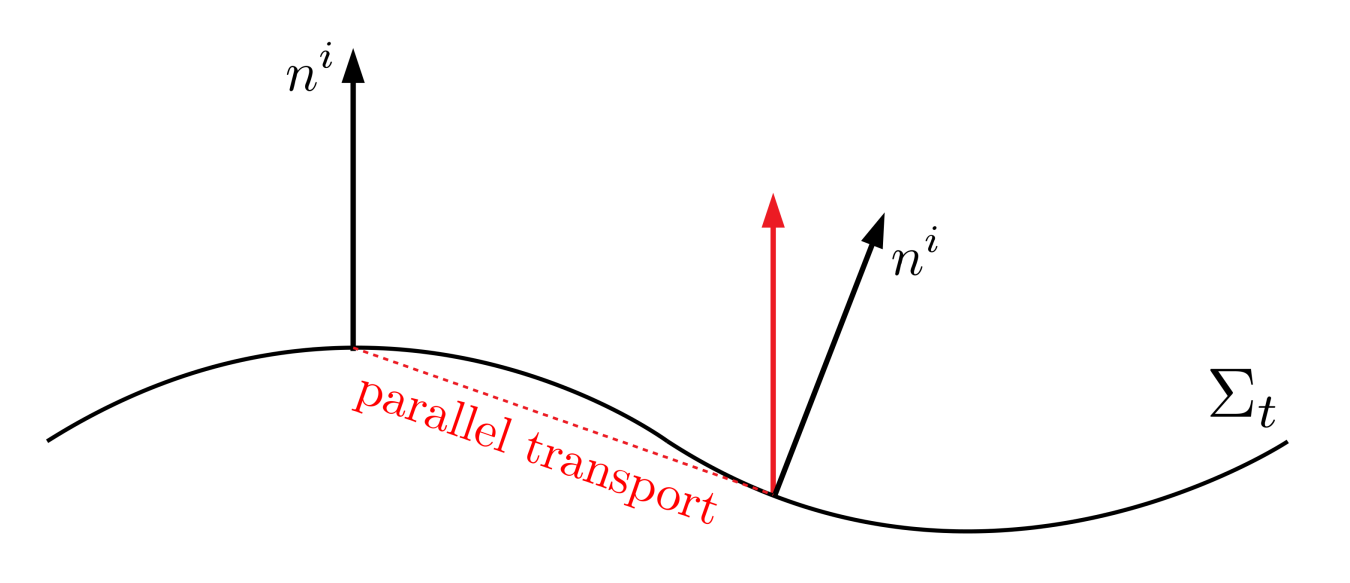}
\caption{Representation of extrinsic curvature. The normal vector is deformed as we parallel transport it along a given hypersurface. Figure taken from Aurrekoetxea \cite[p.~30]{JosuThesis}}
\label{fig:ngr:extrinsic_curvature}
\end{figure}

As a set of second order PDEs, Einstein equations require information about time derivative of the spatial metric $\g_{ij}$. This can intuitively be seen as curvature, but it is important to distinguish between two types of curvature. Intrinsic curvature is curvature within each hypersurfaces, associated with the $d$ dimensional Riemann tensor. Extrinsic curvature: related to how each slide is embedded in the full spacetime, describes how much the normal vector $n^\m$ is deformed from point to point along each slice. As shown in figure \ref{fig:ngr:extrinsic_curvature}, the parallel transported normal is deformed as we changed positions. This change in a given direction is parameterised by $\grad_\m n_\n$. When projected back to the spatial surface with the projector \eqref{eq:ngr:projector}, this defines the extrinsic curvature on a given slice\footnote{The extrinsic curvature is often also defined as the Lie derivatives of the spatial metric along the normal direction $n^\m$.}:
\begin{equation}\label{eq:ngr:K4Ddef}
    K_{\m\n} = -\g_\m^{~\a}\g_\n^{~\b}\grad_\a n_\b = -\br{\grad_\m n_\n + n_\m n^\a \grad_\a n_\n}\,.
\end{equation}

The time components of $K_{\m\n}$ can be safely ignored, as by definition it is a purely spatial tensor where $n^\m K_{\m\n} = 0$. Computing this explicitly using the necessary Christoffel symbols (\cite[p.~409]{alcubierre}), one gets:
\begin{equation}\label{eq:ngr:kij_def}
    K_{ij} = \tfrac{1}{2\a}\br{-\pd_t\g_{ij} + D_{(i}\b_{j)}}\,,
\end{equation}
where $D_i$ represents the $d$ dimensional covariant derivative with respect to the induced metric $\g_{ij}$. The trace of the extrinsic curvature is written as $K = K_i^{~i}$. Equation \eqref{eq:ngr:kij_def} provides our first evolution equation:
\begin{equation}\label{eq:ngr:gij_eq}
    \br{\pd_t-\b^k\pd_k} \g_{ij} = -2\a K_{ij} + \g_{ij}\pd_j \b^k + \g_{kj}\pd_i\b^k\,.
\end{equation}
The metric and extrinsic curvature $\br{\g_{ij},K_{ij}}$ are the $d(d+1)$ fundamental variables of our initial value problem measuring the state of the gravitational field over time, equivalent to the positions and velocities of classical mechanics.

\subsection{\done{Evolution and constraint equations}}

In order to complete the system, we have to obtain the evolution equation for $K_{ij}$. The Einstein field equations (see \ref{gr_summary}) must be decomposed into spatial and normal directions, using projections of $n^\m$ and $\g_\m^{~\n}$. This is a long calculation which will only be summarised here, more details can be found in Alcubierre \cite[sec. 2.4-2.5]{alcubierre}.

\begin{itemize}
    \item Normal Projection:
    \begin{equation}\label{eq:ngr:ham_eq}
        n^\m n^\n\br{G_{\m\n}+\L g_{\m\n} - \tfrac{\k}{2}T_{\m\n}} = \tfrac{1}{2}\br{R + K^2 - K_{ij}K^{ij}} - \tfrac{\k}{2} \r - \L = 0\,,
    \end{equation}
    \item Mixed Projections:
    \begin{equation}\label{eq:ngr:mom_eq}
        \g^{i\m}n^\n\br{G_{\m\n}+\L g_{\m\n} - \tfrac{\k}{2}T_{\m\n}} = D_j\br{K^{ij} - \g^{ij}K} - \tfrac{\k}{2}S^i = 0\,,
    \end{equation}
    \item Spatial Projections:
    \begin{align}\nonumber
        &\g_i^{~\m}\g_j^{~\n}\br{G_{\m\n}+\L g_{\m\n} - \tfrac{\k}{2}T_{\m\n}} = 0 \Leftrightarrow\\\nonumber
        \br{\pd_t-\b^k\pd_k} K_{ij} =&~ 2K_{k(i}\pd_{j)}\b^k -D_iD_j\a + \a\br{R_{ij} + K\,K_{ij} - 2K_{ik}K^k_{~j}} \\\label{eq:ngr:kij_eq}
        & + \tfrac{\a\,\k}{2}\br{\g_{ij}\tfrac{S-\r}{d-1}-S_{ij}} - \tfrac{2\,\a}{d-1}\g_{ij}\L\,,
    \end{align}
\end{itemize}
where $R_{ij}$ and $R$ are the $d$ dimensional Ricci tensor and scalar on the hypersurface, respectively, and the stress-energy tensor $T_{\m\n}$ is decomposed into:
\begin{align}\nonumber
    \r &= n^\m n^\n T_{\m\n}\,,\\
    S^i &= -\g^{i\m}n^\n T_{\m\n}\,,\\\nonumber
    S^{ij} &= \g^{i\m}\g^{j\n}T_{\m\n}\,,
\end{align}
where $\r$ is typically denominated energy density.

Equations \eqref{eq:ngr:ham_eq} and \eqref{eq:ngr:mom_eq} are called the Hamiltonian and Momentum constraints, respectively. They are not evolution equations as they do not involve time derivatives. Instead, they are constraints the dynamic variables $\br{\g_{ij},K_{ij}}$ must satisfy at all times, showing us we are not free to specify them arbitrarily. These constraints are also independent of the gauge parameters $\a$ and $\b^i$, except if the matter content $T_{\m\n}$ is not gauge independent but specific to some coordinate observers. 

Equation \eqref{eq:ngr:kij_eq} closes the system with the evolution equation for $K_{ij}$. Together with \eqref{eq:ngr:ham_eq} and \eqref{eq:ngr:mom_eq}, these make up the ADM equations.

A few remarks to close this section. Firstly, note that Einstein equations involve second derivatives of the metric and equation \eqref{eq:ngr:kij_eq} replaces these by $K_{ij}$ making the system of equations first order. Second, to obtain \eqref{eq:ngr:kij_eq}, the constraint equations had to be used. Additionally, it can be shown that if the constraint equations are satisfied on the initial slice, they remain preserved along time. Numerically, this is typically not enforced, but simply monitored as an accuracy indicator (called free evolution). Finally, a check on degrees of freedom: Einstein equations are $\frac{D(D+1)}{2}=\frac{(d+1)(d+2)}{2}$ equations, decomposed into $d+1$ constraint equations and $\frac{d(d+1)}{2}$ evolution equations. The definition of $K_{ij}$ brings \eqref{eq:ngr:gij_eq} to evolve the metric $\g_{ij}$, an additional $\frac{d(d+1)}{2}$ evolution equations, totalling $d(d+1)$ evolution equations. This matches our evolved variables, $\br{\g_{ij},K_{ij}}$, which are $d(d+1)$. In $D=4$, the six degrees of freedom of the metric, taking into account the $D$ constraint equations, leave two physical degrees of freedom corresponding to the polarisations of gravitational waves.

\section{\done{BSSNOK formulation}}\label{sec:ngr:bssnok}

In spite of all the work to develop the ADM equations, these are known to be weakly hyperbolic. This means they are not suited for long numerical simulations, as errors (namely, constraint violations) propagate exponentially, leading to a crash of the code even for simple vacuum spacetimes. The most popular alternative satisfying strong hyperbolicity \cite{Sarbach:2002bt} (when supplemented with appropriate gauge conditions, see details in section \ref{sec:ngr:gauge}) is the \textit{BSSNOK} formulation (Baumgarte, Shapiro, Shibata, Nakamura, Oohara and Kojima), commonly known simply as \textit{BSSN formulation}. The original work was presented in 1987 by Nakamura, Oohara and Kojima \cite{Nakamura:1987zz} and then evolved to the current version, based on the work of Shibata and Nakamura \cite{Shibata:1995we}, and Baumgarte and Shapiro \cite{Baumgarte:1998te}.

The basic features new to BSSNOK explained in this section have to do with introducing 4 changes: a rescaling factor in the metric, splitting traceless components of the extrinsic curvature, new independent variables and multiples of the constraints added to the evolution equations.

\begin{itemize}
    \item \textbf{Conformal rescaling:} to isolate big numerical factors in the spatial metric, $\g_{ij}$, we re-written with a conformal factor, $\chi$, and a conformal metric, $\ti{g}_{ij}$, such that the conformal metric has unit determinant\footnote{The conformal factor can be written in other ways as well, like $\ti{g}_{ij} = \Psi^4 g_ij$. We adopt $\chi$ for consistent with the CCZ4 formulation \ref{sec:ngr:ccz4} and what our code \grchombo uses.}:
    \begin{gather}
        \ti{\g}_{ij} = \chi\,\g_{ij}\,,\\
        \det{\ti{\g}} = 1 \Longrightarrow \chi = \det{\g}^{-\frac{1}{d}}\,.
    \end{gather}
    \item \textbf{Split of the Extrinsic curvature:} with the intention of splitting the evolution equations of transverse and longitudinal modes, the extrinsic curvature, $K_{ij}$, is separated into its trace, $K=K_i^{~i}$, and its trace-free part, $A_{ij}$, which is also conformally rescaled\footnote{This rescaling is not unique and often different sources may use different powers of the conformal power.} into $\ti{A}_{ij}$:
    \begin{align}\label{eq:ngr:aij_conformal}
        \ti{A}_{ij} = \chi A_{ij} = \chi\br{K_{ij} - \tfrac{1}{d}\g_{ij}K}\,.
    \end{align}
    \item \textbf{Conformal connection functions:} three auxiliary variables are introduced to adjust the form of the characteristic matrix that affect hyperbolicity. These are the conformal contracted Christopher symbols, $\ti{\G}^i$ (equation \ref{eq:ngr:ccf}), using the Christoffel symbols associated with the conformal metric, $\ti{\Gamma}^i_{~jk}$, related to the original Christoffel symbols by equation \ref{eq:ngr:ChristoffelsConformal}:
    \begin{align}\label{eq:ngr:ChristoffelsConformal}
        \Gamma^i_{~jk} &= \ti{\Gamma}^i_{~jk} - \tfrac{1}{2\chi}\br{\delta_j^i\partial_k\chi + \delta_k^i\partial_j\chi-\gamma_{jk}\gamma^{il}\partial_l\chi}\,,\\\label{eq:ngr:ccf}
        \ti{\Gamma}^i &= \ti{\gamma}^{jk}\ti{\Gamma}^i_{~jk}=-\partial_j \ti{\gamma}^{ij}\,,
    \end{align}
    where the last simplification uses the fact that $\ti{\gamma}=1$ and holds only for Cartesian coordinates, using the relation $\gamma^{mk}\partial_i\gamma_{mk}=\partial_i\ln{\det{\gamma}}$.
    \item \textbf{Add multiples of the constraints to the evolution equations:} to improve stability, the Ricci is removed from equation \eqref{eq:ngr:kij_eq} using the Hamiltonian constraint, and a divergence term $\pd_j\ti{A}^{ij}$ when computing $\pd_t\ti{\G}^i$ is removed using the Momentum constraints, resulting in equation \eqref{eq:ngr:bssn_gamma_eom}. This affects hyperbolicity of the system without affecting the physics.
\end{itemize}

A few notes follow. First, $\ti{A}_{ij}$ and other conformal objects should be raised and lowered with the conformal metric. For example, $\ti{A}^{ij} = \frac{1}{\chi}A^{ij}$ and $\ti{\g}^{ij} = \frac{1}{\chi}\g^{ij}$. Second, new algebraic constraints such as $\det{\ti{\g}}=1$, $\ti{A}_i^{~i} = 0$ and $\ti{\Gamma}^i =-\partial_j \ti{\gamma}^{ij}$ can, like the constraints, be either enforced or simply tracked. For \grchombo, trace-free property of $\ti{A}_{ij}$ is enforced at every timestep \cite{Andrade:2021rbd}, while the remaining two are well preserved numerically.

Putting all this together, the $d(d+1)$ evolution variables $\br{\g_{ij},K_{ij}}$ become $\br{\chi,\ti{\g}_{ij},K,\ti{A}_{ij},\ti{\G}^i}$, with evolution equations:
\begin{align}
    \br{\partial_t-\b^k\pd_k}&\chi = \tfrac{2}{d}\chi\br{\alpha K-\partial_i\beta^i}\,,\\
    \br{\partial_t-\b^k\pd_k} & \ti{\gamma}_{ij} =  -2\alpha \ti{A}_{ij}+2\ti{\gamma_{k(i}}\partial_{j)}\beta^k-\tfrac{2}{d}\ti{\gamma_{ij}}\partial_k\beta^k\,,\\
    \br{\partial_t-\b^k\pd_k} & K = -\gamma^{ij}D_jD_i\alpha +\alpha\br{\ti{A}_{ij}\ti{A}^{ij}+\tfrac{1}{d}K^2}\\\nonumber
    &~~~~~~ +\tfrac{\a\,\k}{2(d-1)}\br{S+(d-2)\rho}-\tfrac{2}{d-1}\a\L\,,\\
    \br{\partial_t-\b^k\pd_k} & \ti{A}_{ij} = \chi\br{-D_iD_j\alpha +\alpha R_{ij}-\tfrac{\a\,\k}{2} S_{ij}}^{TF}+\alpha \br{K\ti{A}_{ij}-2\ti{A}_{ik}\ti{A}^k_{~j}}\\\nonumber
    &~~~~~~~ +2\ti{A}_{k(i}\partial_{j)} \beta^k -\tfrac{2}{d}\ti{A}_{ij}\partial_k\beta^k\,,\\\nonumber
    \br{\partial_t-\b^k\pd_k} & \ti{\Gamma}^i = -2\ti{A}^{ij}\partial_j\alpha +2\alpha\br{\ti{\Gamma}^i_{~jk}\ti{A}^{kj}-\tfrac{d}{2}\ti{A}^{ij}\tfrac{\partial_j\chi}{\chi}-\tfrac{d-1}{d}\ti{\gamma}^{ij}\partial_jK-\tfrac{\k}{2}\ti{\gamma}^{ik}S_k}\\\label{eq:ngr:bssn_gamma_eom}
    &~~~~~~ -\ti{\Gamma}^j\partial_j\beta ^i +\tfrac{2}{d}\ti{\Gamma }^i\partial _j\beta ^j+\tfrac{d-2}{d}\ti{\gamma }^{ki}\partial _k\partial_j\beta^j + \ti{\gamma}^{kj}\partial _j\partial_k\beta^i\,,
\end{align}
where $(\dots)^{TF}$ denotes 'trace-free'\footnote{Enforcing the trace-free property on a spatial tensor $T_{ij}$ corresponds to $\br{T_{ij}}^{TF}=T_{ij}-\frac{1}{d}\gamma_{ij}T_k^{~k}$.} and where the Ricci scalar, $R$, can be computed from conformal variables using the decomposition into the conformal Ricci tensor, $\ti{R}_{ij}$ and a component dependent only on the conformal factor, $R_{ij}^\chi$:
\begin{align}\label{eq:ngr:ricci}
    R_{ij} &= \ti{R}_{ij} + R^\chi_{ij}\,,\\\label{eq:ngr:ricci:conformal}
    \ti{R}_{ij} & = -\tfrac{1}{2}\ti{\gamma}^{lm}\partial_l\partial_m\ti{\gamma}_{ij}+\ti{\gamma}_{k(i}\partial_{j)}\ti{\Gamma}^k+\ti{\Gamma}^k\ti{\Gamma}_{(ij)k}+\ti{\gamma}^{lm}\sbr{2\ti{\Gamma}^k_{~l(i}\ti{\Gamma}_{j)km}+\ti{\Gamma}^k_{~im}\ti{\Gamma}_{klj}}\,,\\\label{eq:ngr:ricci:chi}
    R^\chi_{ij}&=\tfrac{d-2}{2\chi}\br{\ti{D}_i\ti{D}_j\chi - \tfrac{1}{2\chi}\ti{D}_i\chi\ti{D}_j\chi} + \tfrac{1}{2\chi}\ti{\gamma}_{ij}\br{\ti{D}^k\ti{D}_k\chi   -\tfrac{d}{2\chi}\ti{D}^k\chi\ti{D}_k\chi}\,,
\end{align}
where $\ti{D}_i$ is the covariant derivative associated with the conformal metric, also raised with this metric. The purpose of the new variables $\ti{\G}^i$ can now be better understood, by noticing the term $\pd_j \ti{\G}^k$ in equation \eqref{eq:ngr:ricci:conformal} now reduces the second order derivatives in Ricci tensor to a simple (hyperbolic) Laplacian operator, $\ti{\g}^{lm}\pd_l\pd_m\ti{\g}_{ij}$.

\section{\done{CCZ4 formulation}}\label{sec:ngr:ccz4}

As explained in section \ref{sec:ngr:adm}, constraints conditions satisfied in the initial data can be proven to propagate and be satisfied at all times. With numerical noise from discretisation errors, constraint violations can spoil accuracy by growing and accumulating significantly without being controlled dynamically. For most of the numerical results presented, we have used a refinement of the BSSN system which tackles this issue, called the Conformal Covariant Z4 (CCZ4) formulation \cite{Alic:2011gg, Alic:2013xsa, Bona:2003fj, Bona:2003qn}. The fundamental idea is to convert the constraint equations into dynamically fields that can have its own evolution equation to which we can introduce damping terms. The CCZ4 damped formalism replaces the original field equations \ref{eq:einstein} by introducing a new variable, $Z_\m$:
\begin{equation}\label{eq:ngr:einstein_Z}
    \mathcal{R}_{\mu\nu} + 2\grad_{(\mu}Z_{\nu)} - \kappa_1 \sbr{2n_{(\mu}Z_{\nu)}-\tfrac{2}{D-2}(1+\kappa_2)g_{\mu\nu}n_\sigma Z^\sigma} - \tfrac{2}{D-2}\L g_{\m\n} = \tfrac{\k}{2}\br{T_{\mu\nu}-\tfrac{1}{D-2}g_{\mu\nu}T}\,,
\end{equation}
where $\k_1$ and $\k_2$ are constants. With this modification, the Hamiltonian and Momentum constraint become evolution equations for $Z^0$ and $Z_i$, respectively. It can be shown that effective damping is achieved only with $\k_1>0$ and $\k_2>-1$ \cite{Gundlach:2005eh}. For these values, the constraints evolve according to a damped wave equation, which means that besides damping, they also propagate and leave the numerical grid (if the boundary conditions allow so).

When writing this system as a $d+1$ formalism along the lines of the BSSNOK formulation, the following changes are introduced:
\begin{itemize}
    \item The D-vector $Z_\m$ is decomposed into its spatial components, $Z_i$, and the normal component, $\Q$, defined as:
    \begin{equation}
        \Q = -n^\m Z_\m = \a Z^0\,.
    \end{equation}
    \item The spatial components, $Z_i$, replace the BSSNOK variables $\ti{\G}^i$ with the independent variables $\hat{\G}^i$:
    \begin{equation}\label{eq:ngr:hat_gamma_def}
        \hat{\G}^i = \ti{\G}^i + 2\ti{\g}^{ij}Z_j\,.
    \end{equation}
\end{itemize}

The condition $Z_\m=0$ that recovers the original Einstein equations reduces to the two additional algebraic constraints $\Q=0$ and $\hat{\G}^i = \ti{\G}^i$.
We now have the variables $\br{\chi,\ti{\g}_{ij},K,\ti{A}_{ij},\Q,\hat{\G}^i}$, and the resulting equations of motion are:
\begin{align}
    \br{\pd_t - \b^k \pd_k} & \chi = \tfrac{2}{d}\chi\br{\alpha K -\partial_i\beta^i}\,,\\
    \br{\pd_t - \b^k \pd_k} & \ti{\gamma}_{ij} =  -2\alpha \ti{A}_{ij}+2\ti{\gamma_{k(i}}\partial_{j)}\beta^k-\tfrac{2}{d}\ti{\gamma_{ij}}\partial_k\beta^k\,,\\
    \br{\pd_t - \b^k \pd_k} & K = -\gamma^{ij}D_jD_i\alpha +\alpha\br{R +2D_i Z^i +K(K-2\Theta)}\\\nonumber
    &~~~~~~ -\tfrac{2d}{d-1}\alpha \kappa_1\br{1+\kappa _2}\Theta+\tfrac{\a\,\k}{2(d-1)}\br{S-d\rho}-\tfrac{2d}{d-1}\a\L\,,\\
    \br{\pd_t - \b^k \pd_k} & \ti{A}_{ij} = \chi \sbr{-D_iD_j\alpha +\alpha\br{R_{ij}+2D_{(i}Z_{j)}-\tfrac{\k}{2} S_{ij}}}^{TF}\\\nonumber
    &~~~~~~~ +\alpha \sbr{\br{K-2\Theta}\ti{A}_{ij}-2\ti{A}_{ik}\ti{A}^k_{~j}} +2\ti{A}_{k(i}\partial_{j)} \beta^k-\tfrac{2}{d}\ti{A}_{ij}\partial_k\beta^k\,,\\\label{eq:ngr:theta_CCZ4_evol}
    \br{\pd_t - \b^k \pd_k} & \Theta = \tfrac{\a}{2}\br{R +2D_i Z^i -2K\Theta +\tfrac{d-1}{d} K^2 -\ti{A}_{ij}\ti{A}^{ij}-\k\rho-2\L}\\\nonumber
    &~~~~~~ -Z^i\partial_i\alpha-\alpha \kappa_1\br{2+\kappa_2}\Theta\,,\\\label{eq:ngr:gamma_CCZ4_evol}
    \br{\pd_t - \b^k \pd_k} & \hat{\Gamma}^i = -2\ti{A}^{ij}\partial_j\alpha +2\alpha\br{\ti{\Gamma}^i_{~jk}\ti{A}^{kj}-\tfrac{d}{2}\ti{A}^{ij}\tfrac{\partial_j\chi}{\chi}-\tfrac{d-1}{d}\ti{\gamma}^{ij}\partial_jK-\tfrac{\k}{2}\ti{\gamma}^{ik}S_k}\\\nonumber
    &~~~~~~ -\ti{\Gamma}^j\partial_j\beta ^i +\tfrac{2}{d}\ti{\Gamma }^i\partial_j\beta ^j+\tfrac{d-2}{d}\ti{\gamma }^{ki}\partial _k\partial_j\beta^j + \ti{\gamma}^{kj}\partial _j\partial_k\beta^i-2\alpha \kappa _1 \ti{\gamma}^{ij}Z_j\\\nonumber
    &~~~~~~ +2\ti{\gamma }^{ij}\br{\alpha \partial _j\Theta -\Theta \partial _j\alpha -\tfrac{2}{d}\alpha K Z_j}+2\kappa _3 \br{\tfrac{2}{d}\ti{\gamma }^{ij}Z_j \partial_k \beta^k-\ti{\gamma }^{jk}Z_j \partial _k \beta ^i}\,,
\end{align}
where a new damping parameter $\k_3$ was introduced in equation \eqref{eq:ngr:gamma_CCZ4_evol} as extra damping useful when simulating black hole spacetimes \cite{Alic:2011gg,Palenzuela:2020tga}.

A few remarks follow. First, since in black hole spacetimes with common gauges (see section \ref{sec:ngr:gauge}) the lapse freezes to zero, it is useful to replace $\br{\a\,\k_1}\to\k_1$ \cite{Alic:2013xsa}. Second, during evolution, $Z_i$ is calculated using \eqref{eq:ngr:hat_gamma_def}, where $\ti{\G}^i$ is computed directly from the metric and $\hat{\G}^i$ is an evolved variable. Note as well how equation \eqref{eq:ngr:theta_CCZ4_evol} reduces to the Hamiltonian constraint if we set $\Q=0=Z_i$. Finally, notice how the Ricci tensor always appears in the combination $R_{ij} + 2D_{(i}Z_{j)}$, which in equation \eqref{eq:ngr:ricci} amounts to adding $R_{ij}^{\Theta}$ and replacing $\ti{\G}^i$ by $\hat{\G}^i$, thus not affecting hyperbolicity properties of the system:
\begin{align}
    R_{ij} &+ 2D_{(i}Z_{j)} = \hat{R}_{ij} + R^\chi_{ij} + R^\Theta_{ij}\,,\\
    \hat{R}_{ij} & = -\tfrac{1}{2}\ti{\gamma}^{lm}\partial_l\partial_m\ti{\gamma}_{ij}+\hat{\gamma}_{k(i}\partial_{j)}\hat{\Gamma}^k+\hat{\Gamma}^k\ti{\Gamma}_{(ij)k}+\ti{\gamma}^{lm}\sbr{2\ti{\Gamma}^k_{~l(i}\ti{\Gamma}_{j)km}+\ti{\Gamma}^k_{~im}\ti{\Gamma}_{klj}}\,,\\
    R^\Theta_{ij}&=\tfrac{1}{2\chi}\br{\hat{\Gamma}^k - \ti{\Gamma}^k}\br{\ti{\gamma}_{ki}\pd_j\chi + \ti{\gamma}_{kj}\pd_i\chi - \ti{\gamma}_{ij}\pd_k\chi}\,,
\end{align}
where $\hat{R}_{ij}$ is $\ti{R}_{ij}$ with $\ti{\Gamma}^k$ replaced by $\hat{\Gamma}^k$ and where $R^\chi_{ij}$ is as defined in equation \eqref{eq:ngr:ricci:chi}.

Given the constraint damping properties of CCZ4 formulation, it is common to use initial data which is only approximately constraint solving (without the need for a full initial data solver that generates initial data perfectly satisfying the Hamiltonian and Momentum constraints, a hard task for non trivial spacetimes). CCZ4 terms will quickly damp initial violations before any physical effect takes place.

\section{\done{Other formulations}}

The BSSN and CCZ4 conformal $3+1$ formulations are extensively used in the field. Based on Alcubierre \cite{alcubierre}, we briefly describe other formulations used in the literature.

The \textbf{Hamiltonian formulation} is often useful for analytical considerations, especially in the realm of quantum gravity.

The \textbf{Characteristic formalism} performs foliation of spacetime into null hypersurfaces, instead of spacelike hypersurfaces, typically compactified to infinity. This has advantages for gravitational wave extraction, for setting boundary conditions, evolving fewer variables, and finally, it lacks constraint equations as the ADM formulation. Caustics can easily form in this formalism, making it hard to handle numerically.

The \textbf{Conformal formalism} has a similar spacelike formulation as the $3+1$ formalism, but brings asymptotic null infinity to a finite region in coordinate space. This is advantageous also for gravitational wave extraction without the issue of caustics. Boundary conditions and hyperboloidal initial conditions are two of the challenges of this formulation.

Finally, and perhaps more important, the \textbf{Generalised Harmonic Formulation (GHC)} \cite{Lindblom:2005qh,Pretorius:2004jg,Bantilan:2012vu} is a full $4D$ formulation that imposes a generalised harmonic gauge, which can be written as $\square x^\m = -\G^\m = -g^{\r\s}\G^\m_{~\r\s} = H^\m$ for some generalised gauge functions $H^\m$. It has very well understood well-posedness properties as in this gauge the principal part of the equations behaves as wave equations, $g^{\r\s}\pd_\r\pd_\s g_{\m\n} = F_{\m\n}(g_{\a\b}, \pd_\r g_{\a\b})$ for some functions $F_{\m\n}$ (see the section \ref{subsec:gr:well_posedness} for details on hyperbolicity). It also shares the constraint damping properties of the CCZ4 formulation (for its constraint $C^\m = H^\m + \G^\m = 0$). Moreover, its elegance when compared to the lengthy BSSN/CCZ4 calculations make it a simple and flexible approach. As disadvantages, there is first a less well understood control over the gauge evolution, and second, as this approach cannot deal with physical singularities, the requirement of the use of excision, by excluding the interior of horizons from the computational domain. Excision requires an apparent horizon finder, stable one-sided stencils and boundary conditions in the excision boundary, and a moving grid as black holes move in the computational domain. This was the approach has been used extensively, such as for the first successful numerical binary black hole evolution \cite{Pretorius:2005gq} or higher dimension studies (e.g. \cite{Lehner:2010pn}). Is it important to mention the only known strongly hyperbolic formulation for modified theories of gravity such as general Horndeski theory or Lovelock theories: the \textbf{modified generalised harmonic (MGH)} formulation \cite{Kovacs:2020pns,Kovacs:2020ywu}. This extends the GHC formulation by introducing two auxiliary metrics that break the degeneracy in the characteristics speeds of the physics, the coordinate and the constraint violating degrees of freedom.
\end{appendix}



\setlength{\parskip}{10pt}              


\blankpage
{
\bibliographystyle{bibliography/JHEP}
\bibliography{thesis}
}


\end{document}